%% file: book.tex
\documentclass[graybox,envcountchap,sectrefs]{svmono}

\usepackage{mathptmx}
\usepackage{helvet}
\usepackage{courier}
\usepackage{type1cm}         

\usepackage{makeidx}         
\usepackage{graphicx}        
\usepackage{multicol}        
\usepackage[bottom]{footmisc}

\usepackage{amsmath}   
\usepackage{amssymb}   

\input{definiciok}

\makeindex             

\begin{document}

\author{J.~K.~Asb{\'o}th, L.~Oroszl{\'a}ny, A.~P{\'a}lyi}
\title{A Short Course on \\ Topological Insulators}
\subtitle{Band-structure topology and edge states in one and two dimensions}
\maketitle

\frontmatter

\include{preface}

\include{introduction}

\tableofcontents

\mainmatter


\include{ssh}

\include{berry_chern}

\input{polarization}

\input{thouless_pump}

\input{adiabatic_pumping_current.tex}

\input{qiwuzhang}

\input{2D_Dirac.tex}

\input{bhz_Smatrix}

\input{bhz_bulk.tex}

\input{Landauer-v2}

\bibliographystyle{plain}
\bibliography{top_ins}

\backmatter




\end{document}

%% file: definiciok.tex
\newcommand{\ket}[1]{\left\vert #1\right\rangle}
\newcommand{\bra}[1]{\left\langle #1\right\vert}
\newcommand{\ketbra}[1]{\ket{#1}\bra{#1}}
\newcommand{\braket}[2]{\left\langle #1\left|\vphantom{{#1}{#2}}\right.#2\right\rangle}
\newcommand{\sbraket}[2]{\langle #1 \vert #2\rangle}   
\newcommand{\expect}[1]{\langle{#1}\rangle}

\newcommand{\abs}[1]{\left\vert #1 \right\vert}
\newcommand{\avg}[1]{\langle #1 \rangle}          

\newcommand{\im}{\mathrm{Im}\,}
\newcommand{\Tr}{\mathrm{Tr}\,}

\newcommand{\nvec}[1]{{\mathbf {#1}}}

\newcommand\nvect[1]{\mathbf{#1}}
\newcommand\nab{\nvect{\nabla}}
\newcommand{\nabr}{\nvect{\nabla}_\RR}
\newcommand{\RR}{\nvect{R}}
\newcommand{\hh}{\nvect{d}}
\newcommand{\hhabs}{d}
\newcommand{\kk}{\nvect{k}}

\newcommand{\AAA}{\nvect{A}}
\newcommand{\BB}{\nvect{B}}

\newcommand{\SSS}{\mathcal{S}}        
\newcommand{\CC}{{\mathcal C}}        
\newcommand{\TTT}{\hat{\mathcal{T}}}  

\newcommand{\ssigma}{\boldsymbol{\sigma}}  

\newcommand{\HH}{\hat{H}}               
\newcommand{\Hbulk}{\hat{H}_\text{bulk}} 
\newcommand{\HHbhz}{\hat{H}_\text{BHZ}}  
\newcommand{\HHtri}{\hat{H}_\text{TRI}}  
\newcommand{\Htri}{H_\text{TRI}}         
\newcommand{\hsigma}{\hat{\sigma}}      
\newcommand{\htau}{\hat{\tau}}          
\newcommand{\hs}{\hat{s}}               
\newcommand{\Inv}{\hat{\Pi}}            
\newcommand{\Invi}{\hat{\pi}}           
\newcommand{\PP}{\hat{P}}           

\newcommand{\II}{\mathbb{I}}         
\newcommand{\jjj}{j}                 
\newcommand{\uu}{u}                  
\newcommand{\TRIM}{\text{TRIM}} 
\newcommand{\NF}{{N_F}}                 
\newcommand{\Nbulk}{{N_\text{bulk}}}                 

\newcommand{\wilsonv}{\underline{v}}    
\newcommand{\MM}{{M}}   
\newcommand{\WW}{{W}}   

\newcommand{\bea}{\begin{eqnarray*}}
\newcommand{\eea}{\end{eqnarray*}}
\newcommand{\bne}{\begin{equation*}}
\newcommand{\ede}{\end{equation*}}
\newcommand{\bnen}{\begin{equation}}
\newcommand{\eden}{\end{equation}}
\newcommand{\bean}{\begin{eqnarray}}
\newcommand{\eean}{\end{eqnarray}}
\newcommand{\bnsn}{\begin{subequations}}
\newcommand{\edsn}{\end{subequations}}

\newcommand{\bna}{\begin{array}}
\newcommand{\eda}{\end{array}}
\newcommand{\bnm}{\begin{enumerate}}
\newcommand{\edm}{\end{enumerate}}

\newcommand{\spinor}[2]{\left(\bna{c} #1 \\[1.5ex] #2 \eda\right)}

\newcommand{\ttmatrix}[4]{\left(\bna{cc} #1 & #2 \\ #3 & #4 \eda\right)}

\newcommand{\uci}{m}             
\newcommand{\vuci}{\nvect{\uci}} 
\newcommand{\nuc}{N}             
\newcommand{\idi}{\alpha}        
\newcommand{\eii}{n}             
\newcommand{\chl}{l}             

\newcommand{\rso}[1]{\hat{#1}} 
\newcommand{\ido}[1]{\hat{#1}} 

\newcommand{\ground}{1}      
\newcommand{\excited}{2}     

%% file: preface.tex
%
%

\preface

These lecture notes provide an introduction to some of the main
concepts of topological insulators, a branch of solid state physics
that is developing at a fast pace. They are based on a one-semester
course for MSc and PhD students at the E\"otv\"os University,
Budapest, which the authors have been giving since 2012.

Our aim is to provide an understanding of the core topics of
topological insulators -- edge states, bulk topological invariants,
bulk--boundary correspondence -- with as simple mathematical tools as
possible. We restricted our attention to one- and two-dimensional band
insulators.  We use noninteracting lattice models of topological
insulators, and build these up gradually to arrive from the simplest
one-dimensional case (the Su-Schrieffer-Heeger model for
polyacetylene) to two-dimensional time-reversal invariant topological
insulators (the Bernevig-Hughes-Zhang model for HgTe). In each case we
introduce the model first, discuss its properties, and then
generalize.  The prerequisite for the reader is quantum
mechanics and not much else: solid state physics background is
provided as we go along.

Since this is an introduction, rather than a broad overview, we try to
be self-contained and give citations to the current literature only
where it is absolutely necessary. For a broad overview, including
pointers to the original papers and current topics, we refer the
reader to review articles and books in the Introduction.

Supporting material for these lecture notes in the form of ipython
notebooks will be made available online.

Despite our efforts, the book inevitably contains typos, errors, and
less comprehensible explanations. 
We would appreciate if you could inform us
of any of those; please send your comments
to \verb;janos.asboth@wigner.mta.hu;.

\textit{Acknowledgments}.
We are grateful for enlightening discussions on topoological
insulators with Anton Akhmerov, Andrea Alberti, Carlo Beenakker, and
Alberto Cortijo. We thank the feedback we got from participants at the
courses at E\"otv\"os University, especially Vilmos Kocsis. A version
of this course was given by one of us (J.K.A.) in the PhD programme of
the University of Geneva, on invitation by Markus B\"uttiker, which
helped shape the course.

The preparation of the first version of these lecture notes were
 supported by the grant T\'AMOP4.1.2.A/1-11/0064.  We acknowledge
 financial support from the Hungarian Scientific Research Fund (OTKA),
 Grant Nos. PD100373, K108676, NN109651, and  from the Marie Curie
 programme of the European Union, Grant No. CIG-293834. J.K.A. and
 A.P. were supported by the J\'anos Bolyai Scholarship of the
 Hungarian Academy of Sciences

\vspace{\baselineskip}
\begin{flushright}\noindent
Budapest,\hfill {\it J\'anos K.~Asb\'oth, L\'aszl\'o Oroszl\'any, Andr\'as P\'alyi}\\
September 2015\hfill {\it ~}\\
\end{flushright}

%% file: introduction.tex
\chapter*{Introduction}

The band theory of electric conduction was one of the early victories
of quantum mechanics in the 1920s. It gave a simple explanation of how
some crystalline materials are electric insulators, even though
electrons in them can hop from one atom to the next.  In the bulk of a
band insulator, the electrons occupy eigenstates that form energy
bands.  In a band insulator, there are no partially filled bands:
completely filled bands are separated by an energy gap from completely
empty bands, the gap represents the energy cost
of mobilizing electrons.
In contrast, materials with partially filled bands are conductors,
where there are plane wave states available to transmit electrons
across the bulk at arbitrarily low energy. Although we now know of
situations where band theory is inadequate (e.g., for Mott
insulators), it remains one of the cornerstones of solid states
physics.

The discovery of the Quantum Hall Effect (1980) has
shown that the simple division into band insulators and metals is not
the end of the story, not even in band theory. In the quantum Hall
effect, a strong magnetic field confines the motion of electrons in
the bulk, but the same field forces them into delocalized edge states
on the surface. A two-dimensional metal in strong magnetic
field is thus an insulator in the bulk, but conducts along the
surface, via a discrete number of completely open edge state channels
(in the language of the Landauer--B\"uttiker formalism). The number of
edge state channels was linked to the Chern number, a topological
invariant of the occupied bands (1982).

Over the last twenty years, theoretical progress over artificial
systems has shown that the external magnetic field is not necessary
for an insulator to have robust conducting edge states: instead, the
nontrivial topology of the occupied bands is the crucial
ingredient. The name \emph{topological insulator} was coined for such
systems, and their study became a blossoming branch of solid state
physics. Following the theoretical prediction (Bernevig, Hughes and
Zhang, 2006 \cite{Bernevig-qshe}), 
electronic transport measurements confirmed 
that a thin layer of HgTe is a topological insulator 
(K\"onig et al, 2007 \cite{Konig-qshe}). 
Since that time, a host of materials
have been shown to be three-dimensional topological insulators, and
thin films and quantum wires shown to be two- and one-dimensional
topological insulators \cite{topological_materials2013}.

The intense theoretical interest in topological insulators has led to
signature results, such as the ``the periodic table of topological
insulators'' \cite{tenfold_way}, which shows that similarly to phase
transitions in statistical mechanics, it is the dimensionality and the
basic symmetries of an insulator that decide whether it can be a
topological insulator or not. Although it was derived by different
ways of connecting topological insulators of various
dimensions and symmetries (so-called dimensional
reduction schemes), the mathematically rigorous proof of the periodic table is
still missing. 

The field of topological insulators is very active, with many
experimental challenges and open theoretical problems, regarding the
effects of electron-electron interaction, extra crystalline
symmetries, coupling to the environment, etc.


\subsubsection*{Literature }

To get a quick and broad overview of topological insulators, with
citations for relevant research papers, we recommend the review papers
\cite{Hasan-topinsreview,Zhang-topinsreview,Budich-topinsreview}.  For
a more in-depth look, there are already a few textbooks on the subject
(by Bernevig and Hughes \cite{Bernevig-book}, by Shen
\cite{Shen-book}, and one edited by Franz and Molenkamp
\cite{Franz-book}). To see the link between momentum-space topology
and physics in a broader context, we direct the reader to a book by
Volovik\cite{volovik2009universe}.

There are also introductory courses on topological insulators with a
broad scope. We recommend the lectures by Charles Kane (the video
recording of the version given at Veldhoven is freely available
online), and the online EdX course on topology in condensed matter by
a group of lecturers, with the corresponding material collected at
topocondmat.org.

\subsubsection*{These lecture notes}

Our aim with this set of lecture notes is to complement the literature
cited above: we wish to provide a close look at some of the core
concepts of topological insulators with as simple mathematical tools
as possible. Using one-and two-dimensional noninteracting lattice
models, we explain what edge states and what bulk topological
invariants are, how the two are linked (this is known as the \emph{
  bulk--boundary correspondence}), the meaning and impact of
some of the fundamental symmetries. 

To keep things as simple as possible, throughout the course we use
noninteracting models for solid state systems. These are described
using single-particle lattice Hamiltonians, with the zero of the
energy corresponding to the Fermi energy. We use natural units, with
$\hbar=1$ and length measured by the lattice constant.

%% file: ssh.tex
%
%
%

\chapter{The Su-Schrieffer-Heeger (SSH) model} 
\label{chap:ssh}

\abstract*{ We take a hands-on approach and get to know the basic
  concepts of topological insulators via a concrete system: the
  Su-Schrieffer-Heeger (SSH) model 
of
  polyacetylene. This model describes spinless fermions hopping on a
  one-dimensional lattice with staggered hopping amplitudes. Using the
  SSH model, we introduce the concepts of single-particle Hamiltonian,
  the difference between bulk and boundary, chiral symmetry, adiabatic
  equivalence, topological invariants, and bulk--boundary
  correspondence.  }

We take a hands-on approach and get to know the basic concepts of
topological insulators via a concrete system: the Su-Schrieffer-Heeger
(SSH) model 
describes spinless fermions hopping on a one-dimensional lattice with
staggered hopping amplitudes. Using the SSH model, we introduce the
concepts of single-particle Hamiltonian, the difference between bulk
and boundary, chiral symmetry, adiabatic equivalence, topological
invariants, and bulk--boundary correspondence.

\begin{figure}
\includegraphics[width=\columnwidth]{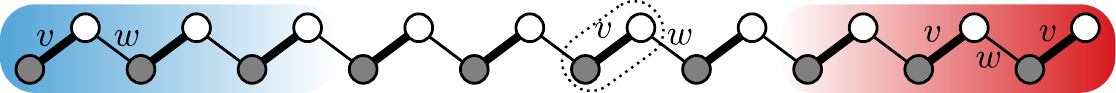}
\caption{Geometry of the SSH model. Filled (empty) circles are sites
  on sublattice $A$ ($B$), each hosting a single state. They are
  grouped into unit cells: the $n=6$th cell is circled by a dotted
  line. Hopping amplitudes are staggered: intracell hopping $v$ (thin
  lines) is different from intercell hopping $w$ (thick lines).  The
  left and right edge regions are indicated by blue and red shaded
  background.
\label{fig:Geometry-of-the-SSH}}
\end{figure}

\section{The SSH Hamiltonian}
\label{sec:ssh-hamiltonian}




The Su-Schrieffer-Heeger (SSH) model describes
electrons hopping on a chain (one-dimensional lattice), with staggered hopping
amplitudes, as shown in Fig.~\ref{fig:Geometry-of-the-SSH}. The chain
consist of $N$ unit cells, each unit cell hosting two sites, one on
sublattice $A$, and one on sublattice $B$. Interactions between the
electrons are neglected, and so the dynamics of each electron is
described by a single-particle Hamiltonian, of the form
\begin{align}
\HH &= 
v \sum_{\uci=1}^{N} \big( \ket{\uci,B}\bra{\uci,A} + h.c. \big) +
w \sum_{\uci=1}^{N-1} \big( \ket{\uci+1,A}\bra{\uci,B}  + h.c. \big).
\label{eq:ssh_hamiltonian_def}
\end{align}
Here $\ket{\uci,A}$ and $\ket{\uci,B}$, with $\uci \in \{ 1,2,\ldots,N\}$,
denote the state of the chain where the electron is on unit cell $\uci$,
in the site on sublattice $A$, respectively, $B$, and $h.c.$ stands
for Hermitian Conjugate (e.g., $h.c.$ of $\ket{\uci,B}\bra{\uci,A}$ is
$\ket{\uci,A}\bra{\uci,B}$). 

The spin degree of freedom is completely absent from the SSH model,
since no term in the Hamiltonian acts on spin. Thus, the SSH model
describes spin-polarized electrons, and when applying the model to
a real physical system, e.g., polyacetylene, we have to always take
two copies of it. In this chapter we will just consider a single copy,
and call the particles fermions, or electrons, or just particles.


We are interested in the dynamics of fermions in and around the ground
state of the SSH model at zero temperature and zero chemical
potential, where all negative energy eigenstates of the Hamiltonian
are singly occupied (because of the Pauli principle). As we will show
later, due to the absence of onsite potential terms, there are $N$
such occupied states. This situation -- called half filling -- is
characteristic of the simplest insulators such as polyacetylene, where
each carbon atom brings one conduction electron, and so we find 1
particle (of each spin type) per unit cell. 

For simplicity, we take the hopping amplitudes to be real and
nonnegative, $v,w \ge 0 $.  If this was not the case, if they carried
phases, $v = \abs{v} e^{i \phi_v}$, and $w = \abs{w} e^{i \phi_w}$,
with $\phi_v,\phi_w \in [0,\pi)$, these phases could always be gauged
  away.  This is done by a redefinition of the basis states:
  $\ket{\uci,A} \to e^{-i (\uci-1)(\phi_v+\phi_w)}$, and $\ket{\uci,B}
  \to e^{-i\phi_v} e^{-i (\uci-1)(\phi_v+\phi_w)}$.

%
%

The matrix for the Hamiltonian of the SSH model,
Eq.~\eqref{eq:ssh_hamiltonian_def},  
on a real-space basis, 
for a chain of $N=4$ unit cells, reads
\begin{align}
H&=\begin{pmatrix}
0 & v & 0 & 0 & 0 & 0 & 0 & 0 \\
v & 0 & w & 0 & 0 & 0 & 0 & 0 \\
0 & w & 0 & v & 0 & 0 & 0 & 0 \\
0 & 0 & v & 0 & w & 0 & 0 & 0 \\
0 & 0 & 0 & w & 0 & v & 0 & 0 \\
0 & 0 & 0 & 0 & v & 0 & w & 0 \\
0 & 0 & 0 & 0 & 0 & w & 0 & v \\
0 & 0 & 0 & 0 & 0 & 0 & v & 0
\end{pmatrix}.
\end{align}

\subsubsection*{External and internal degrees of freedom}

There is a practical representation of this Hamiltonian, which
emphasizes the separation of the external degrees of freedom (unit
cell index $\uci$) from the internal degrees of freedom (sublattice
index). We can use a tensor product basis, 
\begin{align}
\ket{\uci,\alpha} \to \ket{\uci}\otimes\ket{\alpha} \in 
\mathcal{H}_\text{external} \otimes \mathcal{H}_\text{internal}, 
\end{align}
with $\uci =1,\ldots,N$, and $\alpha \in \{A,B\}$. On this basis, with
the Pauli matrices,
\begin{align}
\sigma_0 &= \begin{pmatrix} 1&0\\0&1 \end{pmatrix};&
\sigma_x &= \begin{pmatrix} 0&1\\1&0 \end{pmatrix};&
\sigma_y &= \begin{pmatrix} 0&-i\\i&0 \end{pmatrix};&
\sigma_z &= \begin{pmatrix} 1&0\\0&-1 \end{pmatrix},
\label{eq:ssh-pauli_matrix_def}
\end{align}
the Hamiltonian can be written 
\begin{align}
\HH &= 
v \sum_{\uci=1}^{N} \ket{\uci}\bra{\uci} \otimes {\hsigma_x}  +
w \sum_{\uci=1}^{N-1} \left( \ket{\uci+1}\bra{\uci} \otimes 
\frac{\hsigma_x+ i\hsigma_y}{2} + h.c. \right).
\label{eq:ssh_tv_hamiltonian_def}
\end{align}
The intracell hopping shows up here as an intracell  operator,
while the intercell hopping as a hopping operator. 

\section{Bulk Hamiltonian}
\label{sec:ssh-bulk}

As every solid-state system, the long chain of the SSH model has a
\emph{bulk} and a \emph{boundary}. The bulk is the long central part
of the chain, the boundaries are the two ends, or ``edges'' of the
chain, indicated by shading in Fig.~\ref{fig:Geometry-of-the-SSH}.  We
first concentrate on the bulk, since, in the thermodynamic limit of
$N\to \infty$, it is much larger than the boundaries, and it will
determine the most important physical properties of the
model. Although the treatment of the bulk using the Fourier
transformation might seem like a routine step, we detail it here
because different conventions are used in the literature. More on this
in Appendix~\ref{chap:two_fourier}.


The physics in the bulk, the long central part of the system, should
not depend on how the edges are defined, and so for simplicity we set
periodic (Born-von Karman) boundary conditions. This corresponds to
closing the bulk part of the chain into a ring, with the bulk
Hamiltonian defined as
\begin{align}
\HH_\text{bulk} 
&= \sum_{\uci=1}^{\nuc} \big( v \ket{\uci,B}\bra{\uci,A} + w
\ket{(\uci \,\text{mod}\,N)+1,A}\bra{\uci,B} \big) + h.c..
\label{eq:ssh_bulk_hamiltonian_def}
\end{align}
We are looking for eigenstates of this Hamiltonian, 
\begin{align}
\label{eq:ssh_bulk_schrodinger1}
\HH_\text{bulk} \ket{\Psi_\eii(k)} &= E_\eii(k)\ket{\Psi_\eii(k)},
\end{align}
with $\eii \in \{1,\ldots,2\nuc \}$. 

\subsubsection*{Bulk momentum-space Hamiltonian}

Due to the translation invariance of the bulk, Bloch's theorem
applies, and we look for the eigenstates in a plane wave form.  We
introduce the plane wave basis states only for the external degree of
freedom,
\begin{align}
\label{eq:ssh_planewave_def}
\ket{k} &= \frac{1}{\sqrt{N}} \sum_{\uci=1}^N 
e^{i\uci k} \ket{\uci},& \quad \quad 
\text{for }
k&\in \{ \delta_k, 2\delta_k,\ldots, N \delta_k\}\quad \text{with }
\delta_k = \frac{2\pi}{N},
\end{align}
where the wavenumber $k$ was chosen to take on values from 
 the first Brillouin zone. 
The Bloch eigenstates read  
\begin{align}
\label{eq:ssh_like_Bloch_theorem}
\ket{\Psi_\eii(k)} &= \ket{k} \otimes \ket{u_\eii(k)};&
\ket{u_\eii(k)} &= a_\eii(k) \ket{A} + b_\eii(k) \ket{B}. 
\end{align}
The vectors $\ket{u_\eii(k)} \in \mathcal{H}_\text{internal}$ are
eigenstates of the \emph{bulk momentum-space Hamiltonian} $\HH(k)$
defined as
\begin{align}
\label{eq:ssh_bulk_H_matrixelements_def}
\HH(k) &= \bra{k} \HH_\text{bulk} \ket{k} =  
\sum_{\alpha,\beta\in \{A,B\}} 
\bra{k,\alpha} H_\text{bulk} \ket{k,\beta} \cdot
\ket{\alpha}\bra{\beta};\\
\HH(k) \ket{u_\eii(k)} &= E_\eii(k) \ket{u_\eii(k)}.& 
\end{align}

\subsubsection*{Periodicity in wavenumber}

Although Eq.~\eqref{eq:ssh_like_Bloch_theorem} 
has a lot to do with the continous-variable Bloch theorem, 
$\Psi_{\eii,k}(x) = e^{ikx} u_{\eii,k}(x)$, this correspondence is not
direct. 
In a discretization of the
continuous-variable Bloch theorem, the internal degree of freedom
would play the role of the coordinate within the unit cell, which is
also transformed by the Fourier transform. Thus, the function
$u_{\eii,k}(x)$ cell-periodic, $u_{\eii,k}(x+1)= u_{\eii,k}(x)$, but
not periodic in the Brillouin zone, $u_{\eii,k+2\pi}(x+1) \neq
u_{\eii,k}(x)$. Our Fourier transform acts only on the external degree of
freedom, and as a result, we have periodicity in the Brillouin zone,  
\begin{align}
\label{eq:ssh_internal_periodic}
\HH(k+2\pi) &= \HH(k);& 
\ket{u_\eii(k+2\pi)} &= \ket{u_\eii(k)}.
\end{align}
This convention simplifies the
formulas for the topological invariants immensely. Note, however, that
the other convention, the discretization of the Bloch theorem, is also
widely used in the literature. We compare the two approaches in 
Appendix~\ref{chap:two_fourier}. 


As an example, for the SSH model on a chain of $N=4$ unit cells, 
the Schr\"odinger equation,
Eq.~\eqref{eq:ssh_bulk_schrodinger1}, using 
Eq.~\eqref{eq:ssh_like_Bloch_theorem}, translates to 
a matrix eigenvalue equation,
\begin{align}
\begin{pmatrix}
0 & v & 0 & 0 & 0 & 0 & 0 & w \\
v & 0 & w & 0 & 0 & 0 & 0 & 0 \\
0 & w & 0 & v & 0 & 0 & 0 & 0 \\
0 & 0 & v & 0 & w & 0 & 0 & 0 \\
0 & 0 & 0 & w & 0 & v & 0 & 0 \\
0 & 0 & 0 & 0 & v & 0 & w & 0 \\ 
0 & 0 & 0 & 0 & 0 & w & 0 & v \\
w & 0 & 0 & 0 & 0 & 0 & v & 0  
\end{pmatrix}
\begin{pmatrix}
a(k) e^{ik} \\ b(k) e^{ik} \\
a(k) e^{2ik} \\ b(k) e^{2ik} \\
a(k) e^{3ik} \\ b(k) e^{3ik} \\
a(k) e^{N ik} \\ b(k) e^{N ik}
\end{pmatrix} &=  E(k)
\begin{pmatrix}
a(k) e^{ik} \\ b(k) e^{ik} \\
a(k) e^{2ik} \\ b(k) e^{2ik} \\
a(k) e^{3ik} \\ b(k) e^{3ik} \\
a(k) e^{N ik} \\ b(k) e^{N ik}
\end{pmatrix}.
\end{align}
The Schr\"odinger equation defining the matrix $H(k)$ of the
bulk momentum-space Hamiltonian reads
\begin{align}
H(k) &=
 \begin{pmatrix} 
0 & v + w e^{-ik} \\
v + w e^{ik} & 0
\end{pmatrix};& 
H(k)
 \begin{pmatrix} a(k)\\b(k)\end{pmatrix} 
&= 
E(k)
 \begin{pmatrix} a(k)\\b(k)\end{pmatrix} .
\label{eq:ssh_bulk_H_def}
\end{align}




\subsection{The hopping is staggered to open a gap} 

\begin{figure}
\includegraphics[width=\columnwidth]{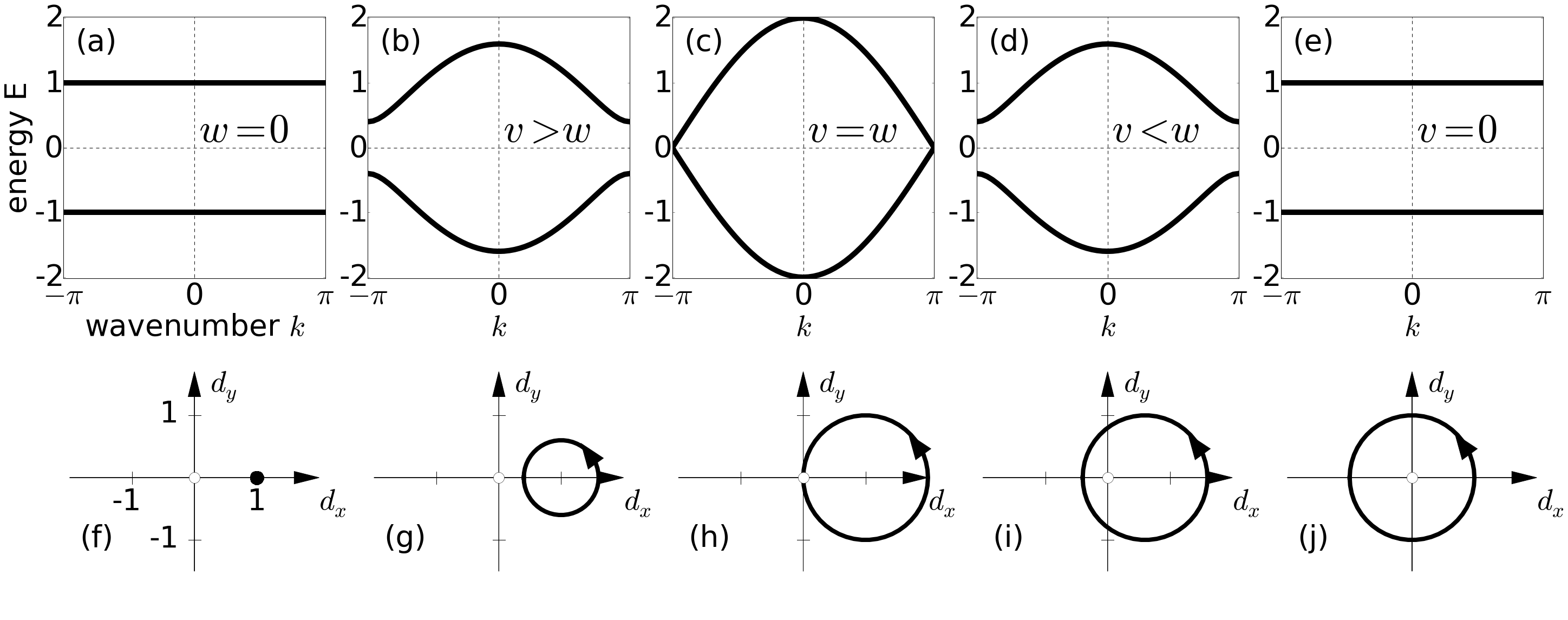}
\caption{Dispersion relations of the SSH model,
  Eq.~\eqref{eq:ssh_dispersion}, for five settings of the hopping
  amplitudes: (a): $v=1,w=0$; (b): $v=1, w= 0.6$; (c): $v=w=1$; (d):
  $v=0.6, w = 1$; (e): $v=0, w = 1$. In each case, the path of the
  endpoints of the vector $\hh(k)$ representing the bulk momentum-space
  Hamiltonian, Eqs.~\eqref{eq:ssh_bulk_hamilt} and
  \eqref{eq:ssh_bulk_h}, are also shown on the $d_x,d_y$ plane, as the
  wavenumber is sweeped across the Brillouin zone, $k=0 \to 2\pi$.
\label{fig:ssh_5_dispersions}.}
\end{figure}

The dispersion relation of the bulk can be read off from
Eq.~\eqref{eq:ssh_bulk_H_def}, using the fact that $\HH(k)^2 = E(k)^2
\hat{\II}_2$. This gives us
\begin{align}
E(k) &= \abs{v+e^{-ik} w} = 
\sqrt{v^2 + w^2 + 2 v w  \cos k}. 
\label{eq:ssh_dispersion}
\end{align}
We show this dispersion relation for five choices of the parameters in
Fig.~\ref{fig:ssh_5_dispersions}.

As long as the hopping amplitudes are staggered, $v\neq w$,
(Figs.~\ref{fig:ssh_5_dispersions} (a),(b),(d),(e), there is an energy gap of
$2\Delta$ separating the lower, filled band, from the upper, empty
band, with
\begin{align}
\Delta = \text{min}_k E(k) = \abs{v-w}.
\end{align}
Without the staggering, i.e., if $v=w$,
(Fig.~\ref{fig:ssh_5_dispersions} (c), the SSH model describes a
conductor. In that case there are plane wave eigenstates of the bulk
available with arbitrarily small energy, which can transport electrons
from one end of the chain to the other. 

The staggering of the hopping amplitudes occurs naturally in many
solid state systems, e.g., polyacetylene, by what is known as the
Peierls instability.  A detailed analysis of this process neccesitates
a model where the positions of the atoms are also
dynamical\cite{solyom2008fundamentals3}. Nevertheless, we can
understand this process intuitively just from the effects of a slight
staggering on the dispersion relation.  As the gap due to the
staggering of the hopping amplitudes opens, the energy of occupied
states is lowered, while unoccupied states move to higher
energies. Thus, the staggering is energetically favourable.

\subsection{Information beyond the dispersion relation}

Although the dispersion relation is useful to read off a number of
physical properties of the bulk of the system (e.g., group
velocities), there is also important information about the bulk that
it does not reveal. Stationary states do not only have an energy and
wavenumber eigenvalue, but also an internal structure, represented by
the components of the corresponding vector $\ket{u_\eii(k)} \in
\mathcal{H}_\text{internal}$. We now define a compact representation
of this information for the SSH model.

The bulk momentum-space Hamiltonian $\HH(k)$ of any two-band model
(i.e., a model with 2 internal states per unit cell), reads
\begin{align}
\label{eq:ssh_bulk_hamilt}
H(k) &= d_x(k) \hat{\sigma}_x + d_y(k) \hat{\sigma}_y + d_z(k) \hat{\sigma}_z =
d_0(k) \hat{\sigma}_0 + \nvect{d}(k) \nvect{\hat{\sigma}}.
\end{align}
For the SSH model, $d_0(k)=0$, and the real numbers
$d_{x,y,z}\in\mathbb{R}$, the components of the $k$-dependent
3-dimensional vector $\hh(k)$, read
\begin{align}
\label{eq:ssh_bulk_h}
d_x(k) &= v + w \cos k;&
d_y(k) &= w \sin k;&
d_z(k) &= 0. 
\end{align}
The internal structure of the eigenstates with momentum $k$ is given
by the direction in which the vector $\hh(k)$ of
Eq.~\eqref{eq:ssh_bulk_h} points (the energy is given by the magnitude
of $\hh(k)$; for details see Sect.~\ref{sec:berry-example-spinhalf}). 


As the wavenumber runs through the Brillouin zone, $k=0\to2\pi$, the
path that the endpoint of the vector $\hh(k)$ traces out is a closed
circle of radius $w$ on the $d_x,d_y$ plane, centered at $(v,0)$. 
For more general 2-band insulators, this path need not be a circle,
but it needs to be a closed loop due to the periodicity of the bulk
momentum-space Hamiltonian, Eq.~\eqref{eq:ssh_internal_periodic}, and
it needs to avoid the origin, to describe an insulator. The topology
of this loop can be characterized by an integer, the \emph{bulk
  winding number} $\nu$. This counts the number of times the loop
winds aroung the origin of the $d_x,d_y$ plane. For example, in
Fig.~\ref{fig:ssh_5_dispersions}(f),(g), we have $\nu=0$, in 
Fig.~\ref{fig:ssh_5_dispersions}(i),(j), we have $\nu=1$, while in 
Fig.~\ref{fig:ssh_5_dispersions}(h), the winding number $\nu$ is undefined.






\section{Edge states}
\label{sec:ssh-edges}

Like any material, the SSH Hamiltonian does not only have a bulk part,
but also boundaries (which we refer to as \emph{ends} or
\emph{edges}). 
The distinction between bulk and edge is not sharply defined, it
describes the behaviour of energy eigenstates in the thermodynamic
limit.  In the case we consider in these lecture notes, the bulk is
translation invariant, and then the we can distinguish edge states and
bulk states by their localized/delocalized behaviour in the
thermodynamic limit. We will begin with the fully dimerized limits,
where the edge regions can be unambiguously defined.  We then move
away from these limits, and use a practical definition of edge states.

\subsection{Fully dimerized limits}

\begin{figure}
\includegraphics[width=\columnwidth]{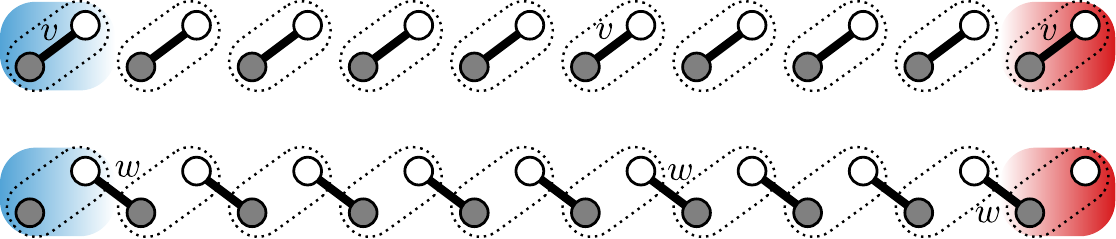}
\caption{Fully dimerized limits of the SSH model, where the chain has
  fallen apart to disconnected dimers.  
In the trivial
  case (top, only intracell hopping, $v=1,w=0$), every energy
  eigenstate is an even or an odd superposition of two sites at the
  same unit cell. In the topological case, (bottom, only intercell
  hopping, $v=0,w=1$), dimers are between neighboring unit cells, and
  there is 1 isolated site per edge, that must contain one zero-energy
  eigenstate each, as there are no onsite potentials.
\label{fig:fully_dimerized}.}
\end{figure}

The SSH model becomes particularly simple in the two fully dimerized
cases: if the intercell hopping amplitude vanishes and the intracell
hopping is set to 1, $v=1, w=0$, or vice versa, $v=0, w=1$.  In both
cases the SSH chain falls apart to a sequence of disconnected dimers,
as shown in Fig.~\ref{fig:fully_dimerized}.  

\subsubsection*{The bulk in the fully dimerized limits has flat bands}

In the fully dimerized limit, one can choose a set of energy
eigenstates which are restricted to one dimer each. These consist of
the even (energy $E=+1$) and odd (energy $E=-1$) superpositions of the
two sites forming a dimer.

In the $v=1, w=0$ case, which we call
\emph{trivial}, we have
\begin{align}
v=1, w&=0: & \HH (\ket{\uci,A}\pm \ket{\uci,B}) 
&= \pm (\ket{\uci,A} \pm \ket{\uci,B}).
\label{eq:ssh_trivial_bulk}
\end{align}
The bulk momentum-space Hamiltonian is $\HH(k)=\hat{\sigma}_x$,
independent of the wavenumber $k$. 

In the $v=0, w=1$ case, which we call \emph{topological}, each dimer is
shared between two neighboring unit cells,
\begin{align}
v=0, w&=1: & \HH (\ket{\uci,B}\pm \ket{\uci+1,A}) &= \pm (\ket{\uci,B} \pm
\ket{\uci+1,A}),
\label{eq:ssh_topological_bulk}
\end{align}
for $\uci=1,\ldots,N-1$.
The bulk momentum-space Hamiltonian now is $\HH(k) =
\hat{\sigma}_x \cos k + {\hsigma}_y \sin k $.  

In both fully dimerized limits, the energy eigenvalues are independent
of the wavenumber, $E(k)=1$. In this so-called flat-band limit,
the group velocity is zero, which again shows that as the chain falls
apart to dimers, a particle input into the bulk will not spread along
the chain.

\subsubsection*{The edges in the fully dimerized limit can host 
zero energy states}

In the trivial case, $v=1, w=0$, all energy eigenstates of the SSH
chain are given by the formulas of the bulk,
Eq.~\eqref{eq:ssh_trivial_bulk}. A topological, fully dimerized SSH
chain, with $v=0,w=1$, however, has more energy eigenstates than those
listed Eq.~\eqref{eq:ssh_topological_bulk}.  Each end of the chain
hosts a single eigenstate at zero energy,
\begin{align}
v=0, w&=1: & \HH \ket{1,A} = \HH\ket{N,B}&=0. 
\end{align}
These eigenstates have support on one site only. Their energy is zero
because onsite potentials are not allowed in the SSH model. These are
the simplest examples of \emph{edge states}.

\subsection{Moving away from the fully dimerized limit} 
\label{subsec:ssh_moving_away}

\begin{figure}[b!]
\includegraphics[width=\linewidth]{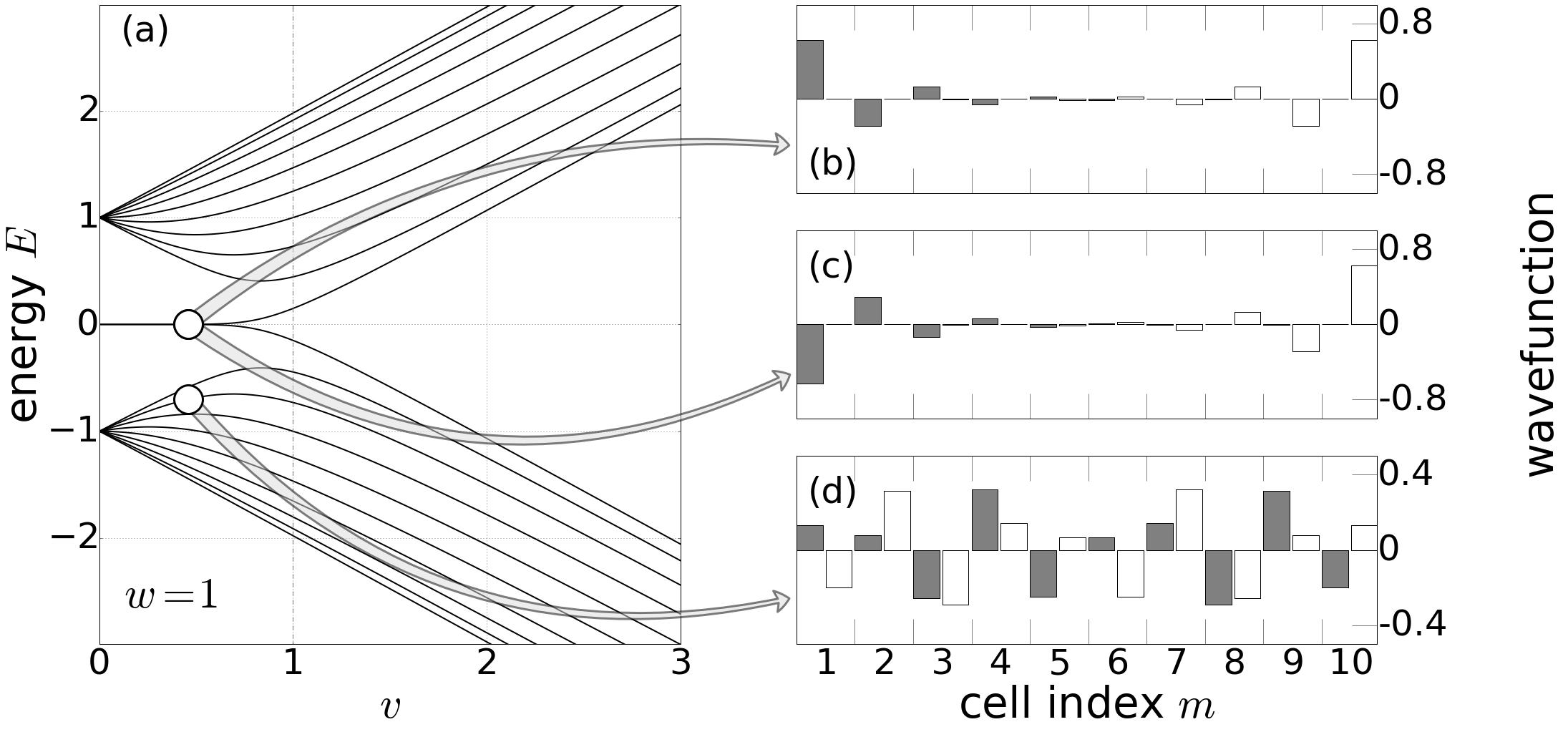}
\caption{
Energy spectrum and wave functions of a finite-sized
SSH model. 
The number of unit cells is $N=10$. 
(a) Energy spectrum of the system for intercell hopping amplitude
$w=1$ as a function the intracell hopping amplitude $v$. 
$v<1$ ($v>1$) corresponds to the topological (trivial) phases. 
(b) and (c) shows the wave functions of the 
hybridized edge states, 
while (d) shows a generic bulk wave function. 
\label{fig:ssh_finite_chain_and_bands}}
\end{figure}

We now examine what happens to the edge states as we move away from
the fully dimerized limit. To be specific, we examine how the spectrum
of an open topological chain, $v=0, w=1$, of $N=10$ unit cells changes,
as we continuously turn on the intracell hopping amplitude $v$.  The
spectra, 
Fig.~\ref{fig:ssh_finite_chain_and_bands}, reveal that the energies of
the edge states remain very close to zero energy.  

The wavefunctions of almost-zero-energy edge states have to be
exponentially localized at the left/right edge, because the zero of
energy is in the bulk band gap.  A plot of the wavefunctions (which
have only real components, since the Hamiltonian is real),
Fig.~\ref{fig:ssh_finite_chain_and_bands}, reveals that the
almost-zero-energy eigenstates are odd and even superpositions of
states localized exponentially on the left and right edge. This is a
result of the exponentially small overlap between the left and the
right edge states. We will later show, in
Sect.~\ref{subsect:exact_edge}, that the edge-state energies are also
controlled by this overlap, and are of the order $E=e^{-N/\xi}$, with
a localization length $\xi = 1/\log(v/w)$.

There is an important property of the right/left edge states, which is
only revealed by the plot of the wavefunctions,
Fig.~\ref{fig:ssh_finite_chain_and_bands}. These states have
nonvanishing components only on the A/B sublattice.

In the following, we show the generality of these properties, and show
the link between the bulk winding number and the presence/absence of
edge states, known as bulk--boundary correspondence. In the case of
the SSH model, all this hinges on a property of the model known as
chiral symmetry.

\section{Chiral symmetry}
\label{sec:ssh-chiral}

In
quantum mechanics, we say that a Hamiltonian $\HH$ has a symmetry
represented by a unitary operator $\hat{U}$ if
\begin{align}
\label{eq:ssh_unitary}
\hat{U} \HH \hat{U}^\dagger &= \HH.
\end{align}
In case of a symmetry, $\hat{U}$ and $\HH$ can be diagonalized
together, and therefore, $\HH$ has no matrix elements between two
eigenstates of $\hat{U}$ with different eigenvalues.  This can be
understood as a superselection rule: if we partition the Hilbert space
into different sectors, i.e., eigenspaces of $\hat{U}$, labeled by the
corresponding eigenvalues, then the dynamics as defined by $\HH$ can be
regarded separately in each sector. 

\subsubsection*{No unitary symmetries}

A unitary symmetry can be simply made to disappear if we restrict
ourselves to one sector of the Hilbert space. This is how we obtained
the bulk momentum-space Hamiltonian, in Sect.~\ref{sec:ssh-bulk},
where the symmetry was the lattice translation operator $\hat{U} =
\ket{\uci+1,A}\bra{\uci,A}+\ket{\uci+1,B}\bra{\uci,B}$, and the labels of
the superselection sectors were the quasimomenta $k$.

\subsubsection*{A different type of symmetry}

The word ``symmetry'' is also used in a different sense in condensed
matter physics.  We say that a system with Hamiltonian $\HH$ 
has \emph{chiral symmetry}, if
\begin{align} 
\label{eq:ssh_def_chiral_symmetry}
\hat{\Gamma} \HH \hat{\Gamma}^\dagger &= -\HH, 
\end{align} 
with an operator $\hat{\Gamma}$ that is not only unitary, but fulfils some
other criteria as well. Notice the extra minus sign on the right hand
side. This has important consequences, which we come to later, but
first discuss the criteria on the symmetry operator. 

First, the chiral symmetry operator has to be unitary and Hermitian, 
$\hat{\Gamma}^\dagger = \hat{\Gamma}$, which can be written succintly
as 
\begin{align}
\label{eq:sigmaz_hermtian_unitary}
\hat{\Gamma}^\dagger \hat{\Gamma} = \hat{\Gamma}^2 &= 1.
\end{align}
The reason for this requirement is that if the operator
$\hat{\Gamma}^2$ was nontrivial, it would represent a unitary
symmetry, since
\begin{align}
\label{eq:sigmaz_hermtian_unitary}
\hat{\Gamma}
\hat{\Gamma}
\HH
\hat{\Gamma}
\hat{\Gamma}
&= -\hat{\Gamma}
\HH
\hat{\Gamma} 
= \HH.
\end{align}
This could still leave room for the chiral symmetry operator to square
to a state-independent phase, $\hat{\Gamma}^2=e^{i\phi}$. However,
this can be got rid of by a redefinition of the chiral symmetry, 
$\Gamma \to e^{-i\phi/2} \Gamma$. 

Second, it is also required that the sublattice operator
$\hat{\Gamma}$ be local. The system is assumed to consist of unit
cells, and matrix elements of $\hat{\Gamma}$ between sites from
different unit cells should vanish. In the SSH chain, this means that
for $\uci\neq \uci'$, we have $\bra{\uci,\alpha}\hat{\Gamma}
\ket{\uci',\alpha'}=0$, for any $\alpha,\alpha'\in(A,B)$. To keep
things simple, we can demand that the sublattice operator act in the
same way in each unit cell (although this is not strictly necessary),
its action represented by a unitary operator $\hat{\gamma}$ acting on
the internal Hilbert space of one unit cell, i.e., 
\begin{align}
\hat{\Gamma} &= \hat{\gamma} \oplus \hat{\gamma} \oplus \ldots \oplus
\hat{\gamma} = \bigoplus_{\uci=1}^N \hat{\gamma},
\label{eq:sigma_sites}
\end{align}
where $N$ is the number of unit cells.  

A third requirement, which is often not explicitly stated, is that the
chiral symmetry has to be \emph{robust}. To understand what we mean by that,
first note that in solid state physics, we often deal with
Hamiltonians with many local parameters that vary in a controlled or
uncontrolled way. An example is the SSH model, where the values of the
hopping amplitudes could be subject to spatial disorder. We gather all
such parameters in a formal vector, and call it $\underline{\xi} \in
\Xi$. Here $\Xi$ is the set of all realizations of disorder that we
investigate.  Instead of talking about the symmetries of a Hamiltonian
$\HH$, we should rather refer to symmetries of a set of
Hamiltonians $\{\HH(\underline{\xi})\}$, for all $\underline{\xi}
\in \Xi$. This set has chiral symmetry represented by $\hat{\Gamma}$
if
\begin{align}
\forall \underline{\xi}  \in \Xi: \quad&\quad 
\hat{\Gamma} \HH (\underline{\xi} ) \hat{\Gamma} = -\HH,
\end{align}
with the symmetry operator $\hat{\Gamma}$ independent of the parameters
$\underline{\xi} $. This is the robustness of the chiral symmetry.

 
\subsection{Consequences of chiral symmetry for energy eigenstates}

We now come to the consequences of chiral symmetry, which are very
different from those of conventional symmetries, due to the extra
minus sign in its definition, Eq.~\eqref{eq:ssh_def_chiral_symmetry}.  

\subsubsection*{Sublattice symmetry}

Chiral symmetry is also called \emph{sublattice symmetry}.
Given the chiral symmetry operator $\hat{\Gamma}$, we can define
orthogonal sublattice projectors $\hat{P_A}$ and $\hat{P_B}$, as
\begin{align}
\hat{P}_A &= \tfrac{1}{2} \left(\mathbb{I}+\hat{\Gamma}\right);&
\hat{P}_B &= \tfrac{1}{2} \left(\mathbb{I}-\hat{\Gamma}\right), 
\label{eq:ssh_def_chiral_sublattice}
\end{align}
where $\II$ represents the unity operator on the Hilbert space of the
system. Note that $\hat{P}_A + \hat{P}_B = \II$, and $\hat{P}_A
\hat{P}_B = 0$.  The defining relation of sublattice symmetry,
Eq.~\eqref{eq:ssh_def_chiral_symmetry}, can be written in an
equivalent form by requiring that the Hamiltonian induces no
transitions from any site on one sublattice to any site on the same
sublattice,
\begin{align}
\hat{P_A} \HH \hat{P_A}&= {P_B} \HH \hat{P_B} = 0;& 
\HH &= \hat{P_A} \HH \hat{P_B} + \hat{P_B} \HH \hat{P_A}.
\label{eq:ssh_def_sublattice_symmetry}
\end{align}
In fact, using the projectors $\hat{P}_A$ and $\hat{P}_B$ is an
alternative and equivalent way of defining chiral symmetry. 

\subsubsection*{Symmetric spectrum}

The spectrum of a chiral symmetric Hamiltonian is symmetric. For any
state with energy $E$, there is a chiral symmetric partner with energy
$-E$.  This is simply seen,
\begin{align}
\HH \ket{\psi_\eii} &= E_\eii \ket{\psi_\eii} \quad \Longrightarrow \quad 
\HH \hat{\Gamma} \ket{\psi_\eii} = -\hat{\Gamma} \HH \ket{\psi_\eii} 
= -\hat{\Gamma} E_\eii
\ket{\psi_\eii} = -E_\eii \hat{\Gamma} \ket{\psi_\eii}.
\end{align}
This carries different implications for nonzero energy eigenstates and
zero energy eigenstates.

For $E_\eii\neq 0$, the states $\ket{\psi_\eii}$ and
$\hat{\Gamma}\ket{\psi_\eii}$ are eigenstates with different energy,
and, therefore, have to be orthogonal.  This implies that every
nonzero energy eigenstate of $\HH$ has equal support on both
sublattices,
\begin{align}
\text{If } E_\eii \neq 0: \quad
0 &= \bra{\psi_\eii} \hat{\Gamma} \ket{\psi_\eii} 
= \bra{\psi_\eii} P_A \ket{\psi_\eii} - 
\bra{\psi_\eii} P_B \ket{\psi_\eii}.
\end{align}

For $E_\eii = 0$, zero energy eigenstates can be chosen to have
support on only one sublattice. This is because
\begin{align}
\text{If } \HH \ket{\psi_\eii} = 0: \quad
\HH \hat{P}_{A/B} \ket{\psi_\eii} = 
\HH \left( \ket{\psi_\eii} \pm \hat{\Gamma} \ket{\psi_\eii} \right) = 0. 
\end{align}
These projected zero-energy eigenstates are eigenstates of
$\hat{\Gamma}$, and therefore are chiral symmetric partners of themselves.

\subsection{Sublattice projectors and chiral symmetry of the SSH model}

The Hamiltonian of the SSH model, Eq.~\eqref{eq:ssh_hamiltonian_def},
is \emph{bipartite}: the Hamiltonian includes no transitions between
sites with the same sublattice index. 
The projectors to the sublattices read
\begin{align}
\hat{P}_A &= \sum_{\uci=1}^{N} \ket{\uci,A}\bra{n,A};& 
\hat{P}_B &= \sum_{\uci=1}^{N} \ket{\uci,B}\bra{n,B}.
\end{align}
Chiral symmetry is represented by the sublattice operator
$\hat{\Sigma}_z$, that multiplies all components of a wavefunction on
sublattice $B$ by (-1),
\begin{align}
\hat{\Sigma}_z &= \hat{P}_A - \hat{P}_B.
\end{align}
Note that this operator has the properties required of the
chiral symmetry operator above: it is unitary, Hermitian, 
and local.   

The chiral symmetry of the SSH model is a restatement of the fact that
the Hamiltonian is bipartite, 
\begin{align} 
\label{eq:def_sigmaz_chiral_symmetry}
\hat{P}_A \HH \hat{P}_A &=
\hat{P}_B \HH \hat{P}_B =0;&\Longrightarrow \quad \quad
\hat{\Sigma}_z \HH \hat{\Sigma}_z &= -\HH. 
\end{align} 
This relation holds because $\HH$ only contains terms that are
multiples of $\ket{\uci,A}\bra{\uci',B}$, or of
$\ket{\uci,B}\bra{\uci',A}$ with $\uci,\uci'\in\mathbb{Z}$. Upon
multiplication from the left and the right by $\hat{\Sigma}_z$, such a
term picks up a single factor of $-1$ (because of the multiplication
from the left or because of the multiplication from the right). Note
that this relation, equivalent to an anticommutation of $\HH$ and
$\hat{\Sigma}_z$, holds whether or not the hopping amplitudes depend
on position: therefore, the chiral symmetry represented by
$\hat{\Sigma}_z$ has the required property of robustness.

\subsection{Consequence of chiral symmetry: Bulk winding number for
  the SSH model}

The path of the endpoint of $\hh(k)$, as the wavenumber goes through
the Brillouin zone, $k=0\to2\pi$, is a closed path on the $d_x,d_y$
plane. This path has to avoid the origin, $\hh=0$: if there was a $k$
at which $\hh(k)=0$, the gap would close at this $k$, and we would not
be talking about an insulator. 
Because of chiral symmetry, the vector $\hh(k)$ is restricted to lie
on the $d_x d_y$ plane, 
\begin{align}
\sigma_z \HH(k) \sigma_z &= 0&\Longrightarrow \quad \quad
d_z &= 0. 
\end{align}
This is then a closed, directed loop on the plane, and thus has a well
defined integer \emph{winding number} about the origin.

\subsubsection*{Winding number as the multiplicity of solutions}

\begin{figure}[b!]
\sidecaption
\includegraphics[width=0.6\linewidth]
{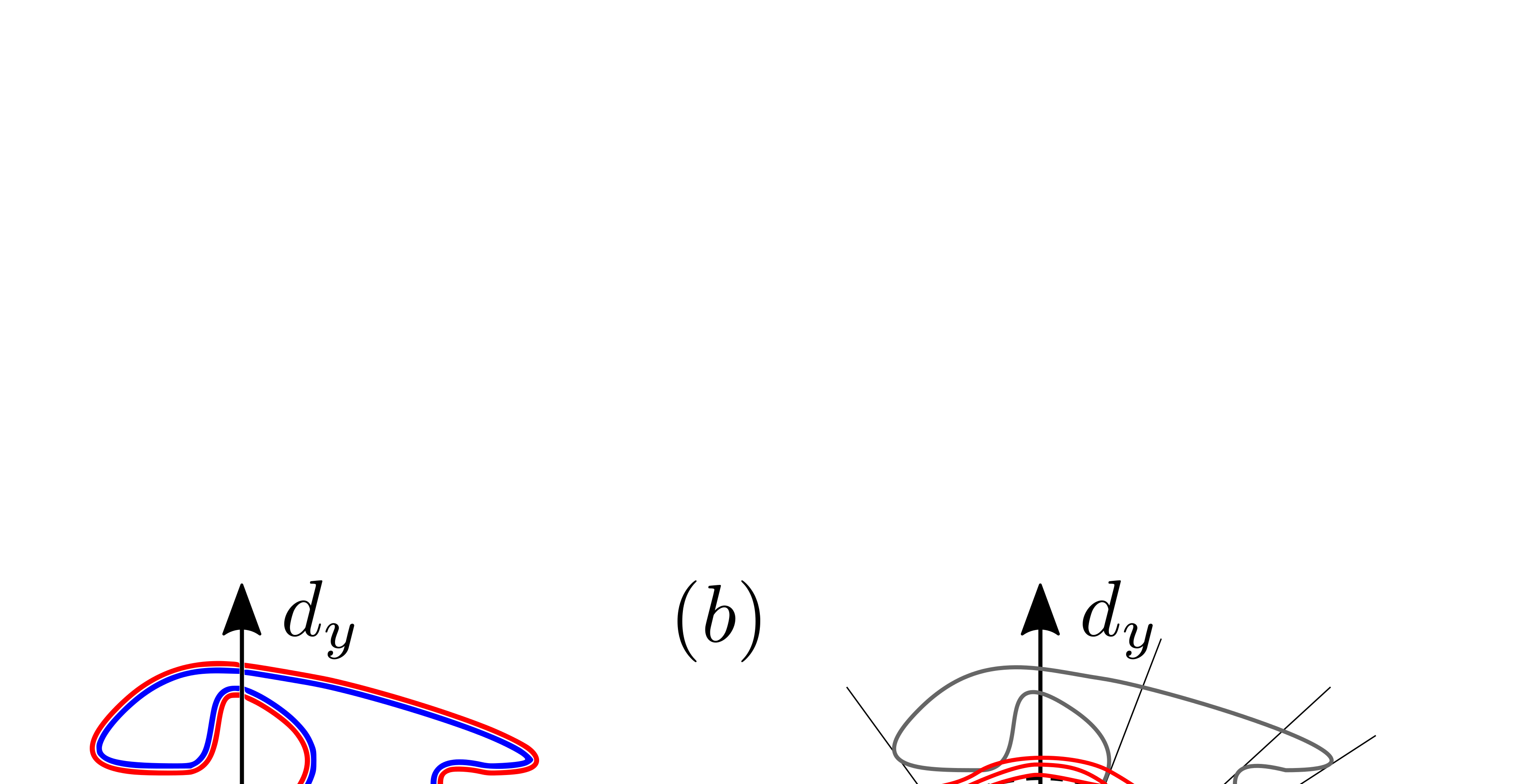}
\caption{The endpoints of the vector $\hh(k)$ as $k$ goes across
  the Brillouin zone (red or blue closed circles), 
\label{fig:ssh-winding-geometric}}
\end{figure}

The simplest way to obtain the winding number graphically is counting
the number of times $\hh(k)$ intersects a curve that goes from the
origin of the $d_x,d_y$ plane to infinity.
\begin{enumerate}
\item Since $\hh(k)$ is a directed curve, it has a left side and a
  right side. Paint the left side blue, the
  right side red, as shown in Fig.~\ref{fig:ssh-winding-geometric} (a). 
\item Take a directed curve $\mathcal{L}$ going from 0 to infinity. We
  can call this the ``line of sight to infinity'', although it need
  not be a straight line. A simple choice is the half-infinite line,
  $d_y=0$, $d_x\ge 0$. Two other choices are shown in
  Fig.~\ref{fig:ssh-winding-geometric}.
\item Identify the intersections of $\hh(k)$ with $\mathcal{L}$.
\item Each intersection has a signature: this is $+1$ if the line of
  sight meets it from the blue side, $-1$ for the red side.
\item The winding number $\nu$ is the sum of the signatures. 
\end{enumerate}

To show that the winding number $\nu$ defined above is a topological
invariant, we need to consider how it can change under continuous
deformations of $\mathcal{L}$ or of $\hh(k)$. Due to the deformations
the intersections of $\mathcal{L}$ and $\hh(k)$ can move, but this
does not change $\nu$. They can also appear or disappear, at points
where $\mathcal{L}$ and $\hh(k)$ touch. However, they can only appear
or disappear pairwise: a red and a blue intersection together, which
does not change $\nu$. As an example, the two choices of the line of
sight $\mathcal{L}$ in Fig.~\ref{fig:ssh-winding-geometric} (a), have
1 or 3 intersections, but the winding number is $+1$, for either of
them.

\subsubsection*{Winding number as an integral}

The winding number can also be written as a compact formula using the 
unit vector $\tilde{\hh}$, defined as
\begin{align}
\tilde{\hh} &= \frac{\hh}{\abs{\hh}}.
\end{align}
This is the result of projecting the curve of $\hh(k)$ to the unit
circle, as shown in Fig.~\ref{fig:ssh-winding-geometric} (b).  The
vector $\tilde{\hh}(k)$ is well defined for all $k$ because $\hh(k)\neq 0$. 

You can check easily 
that the winding number $\nu$ is given by 
\begin{align}
\nu &= \frac{1}{2\pi} \int \left(\tilde{\hh}(k) \times
\frac{d}{dk}\tilde{\hh}(k)\right)_z dk.
\label{eq:ssh_winding_def}
\end{align}

To calculate $\nu$ directly from the bulk momentum-space Hamiltonian,
note that it is off-diagonal (in the basis
of eigenstates of the chiral symmetry operator $\sigma_z$),
\begin{align} 
H(k) &= \begin{pmatrix}
0 & h(k) \\
h^\ast(k) & 0 
\end{pmatrix};& h(k) &= d_x(k)-id_y(k).
\end{align} 
The winding number of $\hh(k)$ 
can be written as an integral, using the
complex logarithm function, $\log (\abs{h} e^{i\, \mathrm{arg}\,h}) =
\log\abs{h} + i \mathrm{arg}\, h$. It is easy to check that 
\begin{align} 
\nu &= \frac{1}{2\pi i} \int_{-\pi}^{\pi} dk \frac{d}{dk} \log h(k). 
\end{align} 
Here during the calculation of the integral, the branch cut for the
logarithm is always shifted so that the derivative is always well
defined. The above integral is always real, since
$\abs{h(k=-\pi)}=\abs{h(k=\pi)}$.  

\subsubsection*{Winding number of the SSH model}

For the SSH model, the winding number is either 0 or 1, depending on
the parameters. In the trivial case, when the intracell hopping
dominates the intercell hopping, $v > w$, the winding number is
$\nu=0$. In the topological case, when $w > v$, we have $\nu=1$.
\begin{figure}[b!]
\includegraphics[width=\linewidth]{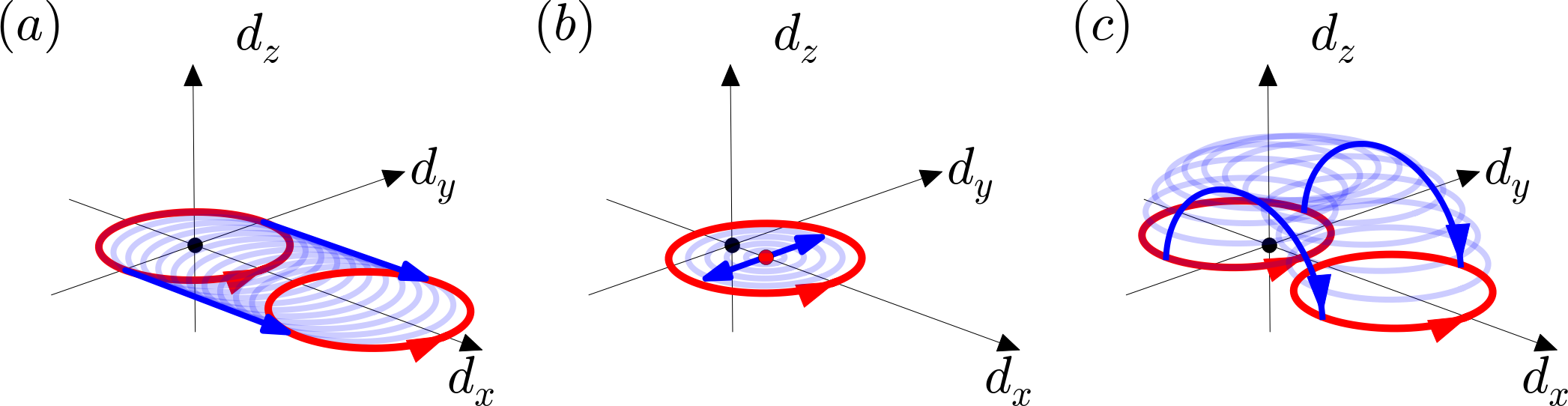}
\caption{The endpoints of the vector $\hh(k)$ as $k$ goes across
  the Brillouin zone (red or blue closed circles), for various
  parameter settings in the SSH model. In (a), intracell hopping $v$ is
  set to 0.5, and intercell hopping is gradually increased from 0 to 1
  (outermost red circle). In the process, the bulk gap was closed and
  reopened, as the origin (black point) is included in some of the
  blue circles. The winding number changed from 0 to 1.  
In (b), we gradually
  decrease intracell hopping $v$ to 0 so that the winding number again
  changes to $1$. In the process, the bulk gap closes and reopens. 
In (c),
  keeping $w=1$, we increase the intracell hopping from 0.5 to $2.5$,
  but avoid closing the bulk gap by introducing a sublattice
  potential, $H_\mathrm{sublattice} = u {\hsigma}_z$. We do this by
  tuning a parameter $\theta$ from $0$ to $\pi$, and setting $v =
  1.5-\cos \theta$, and $u= \sin \theta$.  At the end of the process,
  $\theta = 0$, there is no sublattice potential, so chiral symmetry
  is restored, but the winding number is 0 again. 
\label{fig:ssh_winding_change}}
\end{figure}


To change the winding number $\nu$ of the SSH model, we need to either
a) pull the path of $\hh(k)$ through the origin in the $d_x,d_y$
plane, or (b) lift it out of the plane and put it back on the plane at
a different position. This is illustrated in
Fig.~\ref{fig:ssh_winding_change}. Method (a) requires closing the bulk
gap. Method (b) requires breaking chiral symmetry. 

\section{Number of edge states as topological invariant}
\label{sec:ssh-number_edge}

We now introduce the notion of \emph{adiabatic deformation} of
insulating Hamiltonians.  An insulating Hamiltonian is adiabatically
deformed if 
\begin{itemize}
\item its parameters are changed continuously, 
\item the important symmetries of the system are maintained, 
\item the bulk gap around $E=0$ remains open.  
\end{itemize}
The deformation is a fictitious process, and does not take place in
time. However, if we do think of it as a process in real time, the adiabatic
theorem \cite{griffiths} tells us, that, starting from
the many-body ground state (separated from excited states by the
energy gap), and performing the deformation slowly enough, we end up
in the ground state, at least as far as the bulk of the system is
concerned.  At the edges of a system, changes can occur, and there is
a subtle point to be made about adiabatic deformations being slow, but
not too slow, that the edges should still be considered separately. We
will come back to this point in Chapt.~\ref{chap:ricemele}.

\subsubsection*{Adiabatic equivalence of Hamiltonians}
\begin{figure}[b!]
\sidecaption
\includegraphics[width=0.55\linewidth]{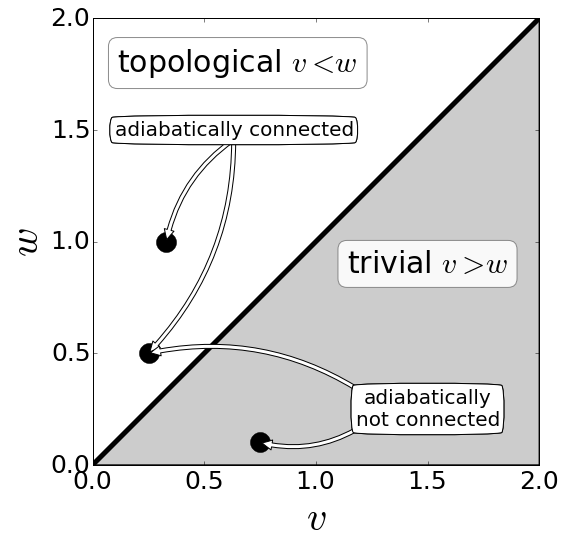}
\caption{Phase diagram of the SSH model.  The winding number of the
  bulk momentum-space Hamiltonian $\ido{H}(k)$ can be $\nu=0$, if
  $v>w$, or $\nu=1$, if $v<w$. This defines the trivial (gray) and the
  topological phase (white).  The 
  boundary separating these phases (black solid line),
  corresponds to $v=w$, where the bulk gap closes at some $k$. 
  Two Hamiltonians in
  the same phase are adiabatically connected.
\label{fig:ssh_phase_diagram}}
\end{figure}

Two insulating Hamiltonians are said to be \emph{adiabatically
  equivalent} or \emph{adiabatically connected} if there is an
adiabatic deformation connecting them, that respects the important
symmetries.  For example, in the phase diagram
Fig.~\ref{fig:ssh_phase_diagram} of the SSH model, the two
Hamiltonians corresponding to the two black points in the topological
phase ($w>v$) are adiabatically connected, as one can draw a path
between them which does not cross the gapless topological-trival phase
boundary $w=v$.


\subsubsection*{Topological invariant}

We call an integer number characterizing an insulating Hamiltonian a
\emph{topological invariant}, or adiabatic invariant, if it cannot
change under adiabatic deformations. Note that the use of adiabatic
deformations implies two properties of the topological invariant: 1)
it is only well defined in the thermodynamic limit, 2) it depends on
the symmetries that need to be respected.  An example for a
topological invariant is the winding number $\nu$ of the SSH model.

We know that two insulating Hamiltonians are not adiabatically
equivalent if their topological invariants differ. Consider as an
example two Hamiltonians corresponding to two points on different
sides of the phase boundary in Fig.~\ref{fig:ssh_phase_diagram} of the
SSH model. One might think that although there is no continous path
connecting them in the phase diagram, continuously modifying the bulk
Hamiltonian by the addition of extra terms can lead to a connection
between them. However, their winding numbers differ, and since winding
numbers cannot change under adiabatic deformation, we know that they
are not adiabatically equivalent.

\subsubsection*{Number of edge states as a topological invariant}

We have seen in Sect.~\ref{subsec:ssh_moving_away}, that the number of
edge states at one end of the SSH model was an integer that did not
change under a specific type of adiabatic deformation. We now
generalize this example. 

Consider energy eigenstates at the left end of a gapped chiral
symmetric one-dimensional Hamiltonian in the thermodynamic limit,
i.e., with length $N\to\infty$, in an energy window from $-\varepsilon
< E < \varepsilon$, with $\varepsilon$ in the bulk gap. There can be
nonzero energy edge states in this energy window, and zero energy edge
states as well. Each nonzero energy state has to have a chiral
symmetric partner, with the state and its partner occupying the same
unit cells (the chiral symmetry operator is a local unitary). The
number of zero energy states is finite (because of the gap in the
bulk), and they can be restricted to a single sublattice each. There
are $N_A$ zero energy states on sublattice $A$, and $N_B$ states on
sublattice $B$.

Consider the effect of an adiabatic deformation of the Hamiltonian,
indexed by some continuous parameter $d:0\to 1$, on the number $N_A-N_B$.
The Hamiltonian respect chiral symmetry, and its bulk energy gap
exceeds $2\varepsilon$, for all values of $d$. 


The deformation can create zero energy states by bringing a nonzero
energy edge state $\ket{\Psi_0(d=0)}$ to zero energy, $E_0(d)=0$ for
$d \ge d'$ but not for $d< d'$. In that case, the chiral symmetric
partner of $\ket{\Psi_0}$, which is $\Gamma \ket{\Psi_0(d)}$ up to a
phase factor, has to move simultaneously to zero energy. The newly
created zero energy edge states are $\hat{P}_A \ket{\Psi_0(d')}$ and
$\hat{P}_B \ket{\Psi_0(d')}$, which occupy sublattice $A$ and $B$,
respectively.  Thus, the number $N_A-N_B$ is unchanged.


The deformation can also bring a zero energy state $\ket{\Psi_0}$ to
energy $E>0$ at some $d=d'$. However, it must also create a chiral
symmetric partner with energy $E<0$ at the same $d'$. This is the time
reverse of the process of the previous paragraph: here, both $N_A$ and $N_B$ must decrease by 1, and, again, $N_A-N_B$ is
unchanged.

The deformation can move nonzero energy states in or out of the
$-\varepsilon < E < \varepsilon$ energy window. This obviously has no
effect on the number $N_A-N_B$.

Due to the deformation, the wavefunction of a zero energy eigenstate
can change so that it extends deeper and deeper into the
bulk. However, because of the gap condition, zero energy states have
to have wavefunctions that decay exponentially towards the bulk, and
so this process cannot move them away from the edge. Thus, $N_A$ and
$N_B$ cannot be changed this way.

The arguments above show that $N_A-N_B$, the net number of edge states
on sublattice $A$ at the left edge, is a topological invariant.

\subsubsection*{Bulk--boundary correspondence in the SSH model}

We have introduced two topological invariants for the SSH model: the
winding number $\nu$, of Eq.~\eqref{eq:ssh_winding_def}, and the net
number of edge states, $N_A-N_B$, of this section. The first one was
obtained from the bulk Hamiltonian only, the second by looking at the
low energy sector of the left edge. In the trivial case of the SSH
model, $v>w$, both are 0; in the topological case, $v<w$, both are
1. This shows that we can use the bulk topological invariant (the
winding number) to make simple robust predictions about the low-energy
physics at the edge. This is a simple example for the
\emph{bulk--boundary correspondence}, a recurrent theme in the theory
of topological insulators, which will reappear in various models in
the forthcoming chapters.

\subsection{Bound states at domain walls}
\label{subsec:ssh_domain}

\begin{figure}
\includegraphics[width=\columnwidth]{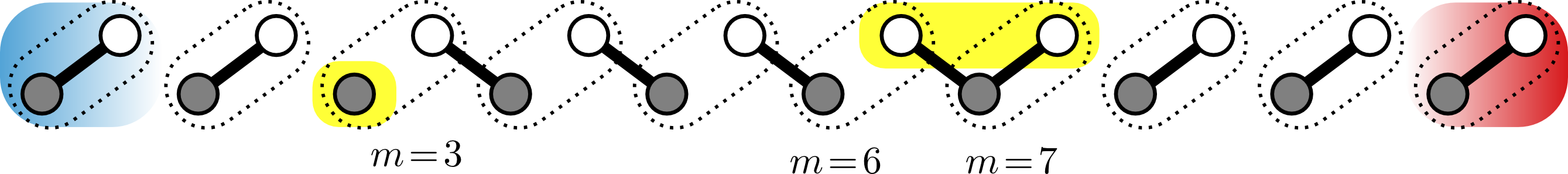}
\caption{A long, fully dimerized SSH chain with 3 domains. The
  boundaries between the domains, the ``domain walls'', host zero
  energy eigenstates (yellow shading). These can be localized on a
  single site, as for the domain wall at $n=3$, or on a superposition
  of sites, as the odd superposition of the ends of the trimer shared
  between the $n=6$ and $n=7$ unit cells.
\label{fig:fully_dimerized_domains}.}
\end{figure}

Edge states do not only occur at the ends of an open chain, but also
at interfaces between different insulating domains of the same
chain. This can be understood via the fully dimerized limit,
Fig.~\ref{fig:fully_dimerized_domains}. There are two types of domain
walls here: those containing single isolated sites, which host 0
energy states on a single sublattice (no onsite potentials are
allowed), and those containing trimers. On a trimer, the odd
superposition of the two end sites form a zero energy eigenstate. In the
the example of Fig.~\ref{fig:fully_dimerized_domains}, this is
\begin{align}
\HH (\ket{6,B}-\ket{7,B}) &=0.
\end{align}
Note that, just as the
edge states at the ends of the chain, these zero energy states at the
interfaces have wavefunctions that take nonzero values on one
sublattice only.

From a perfect dimerized phase without domains it is only possible to
germinate an even number of interfaces.  This means that if one
encounters a domain wall with a localized state on one sublattice then
there will be another domain wall somewhere in the system -- possibly
at the system's edge -- with a localized state on the opposite
sublattice.

Consider a domain wall in an SSH system that is not in the fully
dimerized limit. The wavefunctions of the edge states at the domain
walls will penetrate to some small depth into the bulk, with
exponentially decaying evanescent tails.  
For two domain
walls at a distance of $M$ unit cells, the two edge states on the
walls will hybridize, form ``bonding'' and ``anti-bonding''
states. 
At half filling, of these only the negative energy eigenstate
will be occupied. This state hosts a single electron, however, its
wavefunction is localized with equal weight on the two domain
walls. Hence each domain wall, when well separated from other domain
walls and the ends of the chain, will carry half an electronic
charge. This effect is sometimes referred to as ``fractionalization''
of the charge.

\subsection{Exact calculation of edge states}
\label{subsect:exact_edge}

The zero energy edge states of the SSH model can also be calculated
exactly, even in the absence of translational invariance.  Take an SSH
model on $N$ unit cells, with complex intracell and intercell hopping
amplitudes, 
\begin{align}
\label{eq:ssh_hamiltonian_disorder_def}
\HH &= 
\sum_{\uci=1}^{N} \big( 
v_\uci \ket{\uci,B} \bra{\uci,A} + h.c. \big) +
\sum_{\uci=1}^{N-1} \big( 
w_\uci \ket{\uci+1,A}\bra{\uci,B}  + h.c. \big).
\end{align}
We are looking for a zero energy eigenstate of this Hamiltonian,
\begin{align}
\HH \sum_{\uci=1}^N \big( 
a_\uci \ket{\uci,A} + b_\uci \ket{\uci,B} \big) &=0.
\label{eq:ssh_disorder_0energy}
\end{align}
This gives us $2N$ equations for the amplitudes $a_\uci$ and 
$b_\uci$, which read
\begin{subequations}
\begin{align}
\uci &= 1,\ldots,N-1:& 
\quad v_\uci a_\uci + w_\uci a_{\uci+1} &= 0;&
\quad w_\uci b_\uci + v_{\uci+1} b_{\uci+1} &= 0;\\
\label{eq:ssh_ab_boundary}
\quad&\text{boundaries}:&
 \quad v_{N} a_N &= 0;& 
 \quad v_{1} b_1 &= 0. 
\end{align}
\end{subequations}
The first set of equations is solved by
\begin{align}
\label{eq:ssh_disorder_A}
\uci &= 2,\ldots,N:& 
a_\uci &= \prod_{j=1}^{\uci-1} \frac{-v_j}{w_j}\, a_1;\\
\label{eq:ssh_disorder_B}
\uci &= 1,\ldots,N-1:& 
b_\uci &= \frac{-v_N}{w_\uci} \prod_{j=\uci+1}^{N-1} \frac{-v_j}{w_j}\, b_N. 
\end{align}
However, we also have to fulfil Eqs.~\eqref{eq:ssh_ab_boundary}, which give
\begin{align}
\label{eq:ssh_disorder_boundarycond}
b_1 &= a_N = 0.
\end{align}
These equations together say that, in the generic case, there
is no zero energy eigenstate, $a_\uci=b_\uci=0$. 

Although there is no exactly zero energy state,
Eqs.~\eqref{eq:ssh_disorder_A}, \eqref{eq:ssh_disorder_B} and
\eqref{eq:ssh_disorder_boundarycond} admit two approximate solutions
in the thermodynamic limit, $N\to \infty$, if the average intercell
hopping is stronger than the intracell hopping.  More precisely, we
define the ``bulk average values'',
\begin{align}
\overline{\log{\abs{v}}} &= 
\frac{1}{N-1}\sum_{\uci=1}^{N-1} \log \abs{v_\uci};&
\overline{\log{\abs{w}}} &= 
\frac{1}{N-1}\sum_{\uci=1}^{N-1} \log \abs{w_\uci}.
\label{eq:ssh_vw_average_def}
\end{align}
Eqs.~\eqref{eq:ssh_disorder_A} and
\eqref{eq:ssh_disorder_B} translate to 
\begin{align}
\abs{a_N} &= \abs{a_1} e^{-(N-1)/\xi};&
\abs{b_1} &= \abs{b_N} e^{-(N-1)/\xi} \frac{\abs{v_N}}{\abs{v_1}},
\end{align}
with the localization length 
\begin{align}
\label{eq:ssh-loclength_vw}
\xi &= \frac{1}{\overline{\log{\abs{w}}} -
\overline{\log{\abs{v}}}}.
\end{align}
If in the thermodynamic limit, the bulk average values, 
Eqs.~\eqref{eq:ssh_vw_average_def} make sense, and $\xi>0$, we have 
two approximate zero energy solutions, 
\begin{align}
\ket{L} &= \sum_{\uci=1}^{N} a_\uci \ket{\uci,A};&
\ket{R} &= \sum_{\uci=1}^{N} b_\uci \ket{\uci,B}, 
\end{align}
with the coefficients $a_\uci$ and $b_\uci$ chosen
according to Eqs.~\eqref{eq:ssh_disorder_A} and \eqref{eq:ssh_disorder_B}, 
and $a_1$, respectively, $b_N$, used to fix the norm of $\ket{L}$, respectively, $\ket{R}$. 

\subsubsection*{Hybridization of edge states}

The two states $\ket{L}$ and $\ket{R}$ hybridize under $\HH$ to an
exponentially small amount, and this induces a small energy splitting.
We can obtain an estimate for the splitting, and the energy
eigenstates, to a good approximation using adiabatic elimination
of the other eigenstates. In this approximation, the central
quantity is the overlap
\begin{align}
\bra{R} \HH \ket{L} &= \abs{a_1 e^{-(N-1)/\xi} v_N b_N} e^{i\phi}, 
\end{align}
with some $\phi\in [0,2\pi)$.
The energy eigenstates are approximated as 
\begin{align}
\ket{0+} &= \frac{e^{-i\phi/2}\ket{L}+e^{i\phi/2} \ket{R}}{\sqrt{2}};&
E_+ &= \abs{a_1 e^{-(N-1)/\xi} v_N b_N};\\
\ket{0-} &= \frac{e^{-i\phi/2}\ket{L}-e^{i\phi/2} \ket{R}}{\sqrt{2}};&
E_- &= -\abs{a_1 e^{-(N-1)/\xi} v_N b_N}.
\end{align}
The energy of the hybridized states thus is exponentially small in the
system size. 


\section*{Problems}
\addcontentsline{toc}{section}{Problems}
\begin{prob}
\label{prob:ssh-arbitrary-winding}
\textbf{Higher winding numbers}\\
The SSH model is one-dimensional in space, 
and has a two-dimensional internal Hilbert space. 
Construct a lattice model that has these properties
of the SSH model, but which has a bulk winding number of 2.
Generalize the construction for an arbitrary integer 
bulk winding number.  
\end{prob}

\begin{prob}
\label{prob:ssh-complex-hopping}
\textbf{Complex-valued hopping amplitudes}\\
Generalize the SSH model in the following way. 
Assume that the hopping amplitudes $v = |v| e^{i\phi_v}$ and 
$w= |w| e^{i\phi_w}$ are complex, 
and include a third complex-valued hopping amplitude
$z = |z| e^{i\phi_z}$ between the states
$\ket{\uci,A}$ and $\ket{\uci+1,B}$
for every $\uci$.
Provide a specific example where the tuning of  one of the 
phases changes the bulk winding number. 
\end{prob}

\begin{prob}
\label{prob:ssh-2d-generalization}
\textbf{A possible generalization to two dimensions}\\ 
Consider a two dimensional generalization of the SSH model. Take parallel copies of
the SSH chain and couple them without breaking chiral symmetry. What will happen with the edge states?
\end{prob}

%


%% file: berry_chern.tex

\chapter{Berry phase, Chern number}
\label{chap:berry_chern}

\abstract*{
To describe the theory of topological band insulators we will use the
language of adiabatic phases. In this chapter we review the basic
concepts: 
the Berry phase, the Berry curvature, and
the Chern number.
We further describe the relation between the Berry phase and
adiabatic dynamics in quantum mechanics. 
Finally, we illustrate these concepts using the 
two-level system as a simple example. }

To describe the theory of topological band insulators we will use the
language of adiabatic phases. In this chapter we review the basic
concepts: 
the Berry phase, the Berry curvature, and
the Chern number.
We further describe the relation between the Berry phase and
adiabatic dynamics in quantum mechanics. 
Finally, we illustrate these concepts using the 
two-level system as a simple example. 


For pedagogical introductions, we refer the reader to Berry's original
paper\cite{berry_phase_1984}, and papers from the Americal Journal of
Physics\cite{apj_holstein,apj_berry_degenerate}.  For the application
to solid state physics, we will mostly build on Resta's lecture
note\cite{resta_cycle}, and Niu's review
paper\cite{RevModPhys.82.1959}.

\section{Discrete case}
\label{sec::berry_chern-discrete_berry}

The subject of adiabatic phases is strongly related to 
adiabatic quantum dynamics, when a Hamiltonian is slowly 
changed in time, and the time evolution of the quantum state
follows the instantaneous eigenstate of the Hamiltonian. 
In that context, as time is a continuous variable and the
time-dependent Schr\"odinger equation is a differential equation, 
the adiabatic phase and the related concepts are expressed
using differential operators and integrals. 
We will arrive to that point later during this chapter;
however, we start the discussion using the language
of discrete quantum states. 
Besides the conceptual simplicity, this language also
offers an efficient tool for the numerical evaluation of Chern number,
which is an important topological invariant in the context of 
two-dimensional electron systems.

\subsection{Relative phase of two nonorthogonal quantum states}

In quantum mechanics, the state of a physical system is represented by
an equivalence class of vectors in a Hilbert space: a multiplication
by a complex phase factor does not change the physical content. 
A gauge transformation is precisely
such a multiplication:
\begin{align}
\ket{\Psi} &\to e^{i\alpha} \ket{\Psi},\quad \text{with }\alpha \in [0,2\pi). 
\end{align} 
In that sense, the phase of a vector $\ket{\Psi}$ does not represent
physical information.  We can try to define the relative phase
$\gamma_{12}$ of two nonorthogonal states $\ket{\Psi_1}$ and
$\ket{\Psi_2}$ as 
\begin{align} 
\gamma_{12} &= -\arg \braket{\Psi_1}{\Psi_2} ,
\end{align} 
where $\arg(z)$ denotes the phase of the complex number $z$,
with the specification that $\arg (z) \in (-\pi,\pi]$.  Clearly, the
relative phase $\gamma_{12}$ fulfils
\begin{align}
\label{eq:berry_relativephase}
e^{-i\gamma_{12}} &=
\frac{\braket{\Psi_1}{\Psi_2}}{\abs{\braket{\Psi_1}{\Psi_2}}}.
\end{align}
However, the relative phase is
not invariant under a local gauge transformation,
\begin{align} 
\ket{\Psi_j} &\to e^{i\alpha_j} \ket{\Psi_j}&
e^{-i\gamma_{12}} &\to e^{-i\gamma_{12} + i(\alpha_2-\alpha_1)}. 
\end{align}

\subsection{Berry phase}

\begin{figure}
\centering
\includegraphics[width=0.9\linewidth]{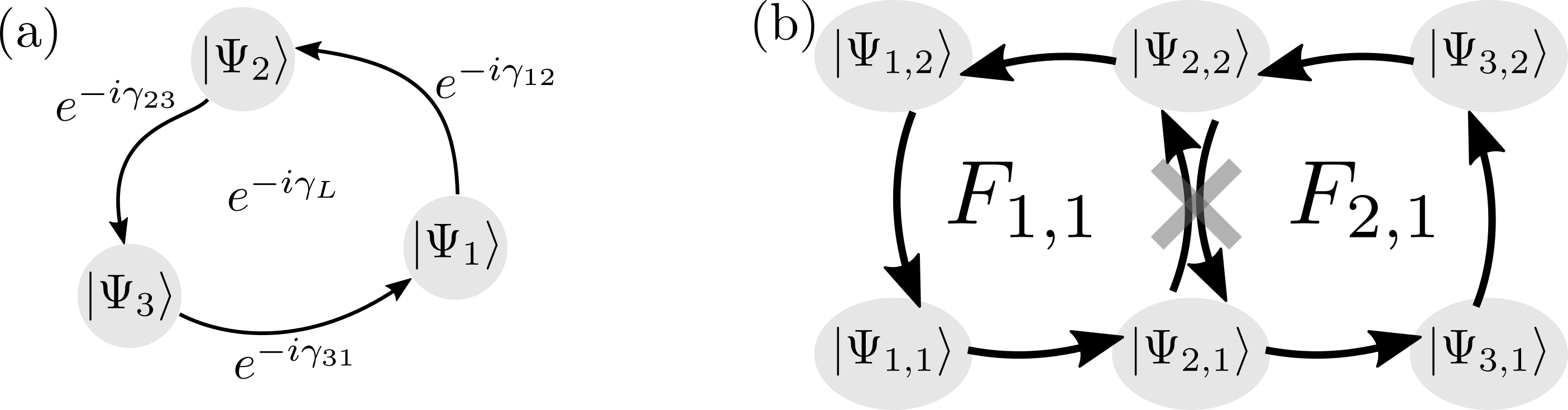}
\caption{
Berry phase,
Berry flux and Berry curvature for discrete quantum states. 
(a) The Berry phase $\gamma_L$ for the loop $L$ 
consisting of $N=3$ states 
is defined from the relative phases $\gamma_{12}$, $\gamma_{23}$,
$\gamma_{31}$. 
(b) The Berry phase of a loop defined on a lattice of states can
be expressed as the sum of the Berry phases 
$F_{1,1}$ and $F_{2,1}$
of the plaquettes
enclosed by the loop. 
The plaquette Berry phase $F_{n,m}$ is also called 
Berry flux. 
\label{fig:berry-discrete}}
\end{figure}

Take $N \ge 3$ states in a Hilbert space, order them in a loop, and
ask about the phase around the loop. As we show below, the answer --
the \emph{Berry phase} -- is gauge invariant. For states
$\ket{\Psi_j}$, with $j=1,2,\ldots,N$, and for the ordered list $L =
(1,2,\ldots,N)$ which define the loop, shown in
Fig.~\ref{fig:berry-discrete}, the Berry phase is defined as
\begin{align}
\label{eq:berry_def_discrete}
\gamma_L &= - \arg
e^{-i
\left(\gamma_{12} + \gamma_{23}
+ \ldots + \gamma_{N1}
\right)} = 
- \arg
\left(
\braket{\Psi_1}{\Psi_2}
\braket{\Psi_2}{\Psi_3}
\ldots
\braket{\Psi_N}{\Psi_1}
\right).
\end{align}
To show the gauge invariance of the Berry phase, it can be rewritten
as
\begin{align}
\label{eq:berry_projector_discrete}
\gamma_L &= - \arg
\Tr \left(
{\ket{\Psi_1}\bra{\Psi_1}}
  \ket{\Psi_2}\bra{\Psi_2}
\ldots
\ket{\Psi_N}\bra{\Psi_N} \right).
\end{align}
Here, we expressed the Berry phase $\gamma_L$ 
using projectors that are themselves gauge invariant. 

Even though the
Berry phase is not the expectation value of some operator, it is a
gauge invariant quantity, and as such, it can have a direct physical
significance. We will  find such a significance, but first,
we want to gain more intuition about its behaviour.

\subsection{Berry flux}

Consider a Hilbert space of quantum states, 
and a finite two-dimensional square lattice with points labelled by 
$n,m \in \mathbb{Z}$, $1 \le n \le N$, and $1 \le m \le M$.
Assign a quantum state $\ket{\Psi_{n,m}}$
from the Hilbert space to
each lattice site. 
Say you want to know the Berry phase
of the loop $L$ around this set,
\begin{multline}
\label{eq:berry_gamma_L_lattice_def}
\gamma_L = - \arg \exp \Bigg[ -i \left( \sum_{n=1}^{N-1}
  \gamma_{(n,1),(n+1,1)} + \sum_{m=1}^{M-1} \gamma_{(N,m),(N,m+1)} \right. \\
 + \left. \sum_{n=1}^{N-1} \gamma_{(n+1,M),(n,M)} +
  \sum_{m=1}^{M-1} \gamma_{(1,m+1),(1,m)} \right) \Bigg]
\end{multline}
as shown in Fig.~\ref{fig:berry-discrete}. Although the Berry phase is
a gauge invariant quantity, calculating it according to the recipe
above involves multiplying together many gauge dependent complex
numbers.  The alternative route, via
Eq.~\eqref{eq:berry_projector_discrete}, involves multliplying gauge
independent matrices, and then taking the trace.

There is a way to break the calculation of the Berry phase of the loop
down to a product of gauge independent complex numbers. To each
plaquette (elementary square) on the grid, with $n,m$ indexing the
lower left corner, we define the \emph{Berry flux} $F_{n,m}$ of the
plaqette using the sum of the relative phases around its boundary,
\begin{multline}
F_{nm} = -\arg 
\exp \left[-i\left(
	\gamma_{(n,m),(n+1,m)} 
	+ 
	\gamma_{(n+1,m),(n+1,m+1)} \right. \right.\\
	\left. \left. +
	\gamma_{(n+1,m+1),(n,m+1)}
	+
	\gamma_{(n,m+1),(n,m)}
\right) \right],
\label{eq:berry-def2-discrete-plaquette}
\end{multline}
for $n=1,\ldots,N$ and $m=1,\ldots,M$.
Note that the Berry flux is itself a Berry phase and is therefore
gauge invariant. 
Alternatively, we can also write
\begin{multline}
F_{nm} = -\arg \big(
\braket{\Psi_{n,m}}{\Psi_{n+1,m}}   
\braket{\Psi_{n+1,m}}{\Psi_{n+1,m+1}} \\ 
\braket{\Psi_{n+1,m+1}}{\Psi_{n,m+1}}   
\braket{\Psi_{n,m+1}}{\Psi_{n,m}} \big),
\label{eq:berry-def-discrete-plaquette}
\end{multline}
Now consider the product of all plaquette phase factors $e^{-iF_{nm}}$, 
\begin{multline}
\label{eq:berry_intermediate1}
\prod_{n=1}^{N-1}\prod_{m=1}^{M-1} e^{-i F_{nm}} =  
\exp \Big[ -i\sum_{n=1}^{N-1} \sum_{m=1}^{M-1} 
F_{nm} \Big]
=
\exp \Big[ -i\sum_{n=1}^{N-1} \sum_{m=1}^{M-1} 
\big(\gamma_{(n,m),(n+1,m)} \\ 
+\gamma_{(n+1,m),(n+1,m+1)}
+\gamma_{(n+1,m+1),(n,m+1)}
+\gamma_{(n,m+1),(n,m)}
\big) \Big]
\end{multline}
Each internal edge of the lattice is shared between two plaquettes,
and therefore occurs twice in the product. However, since we fixed the
orientation of the plaquette phases, these two contributions will
always be complex conjugates of each other, and cancel each other.
Therefore the exponent in the right-hand-side of
Eq.~\eqref{eq:berry_intermediate1} simplifies to the exponent
appearing in Eq.~\eqref{eq:berry_gamma_L_lattice_def}, implying 
\begin{align}
\exp \Big[-i\sum_{n=1}^{N-1} \sum_{m=1}^{M-1} 
F_{nm}\Big]
&= e^{-i\gamma_L}.
\label{eq:berry-discrete-stokes}
\end{align}
This result is reminsicent of the Stokes theorem
connecting the integral of the curl of a vector field on an open
surface and the line integral of the vector field along the
boundary of the surface. 
In Eq. \eqref{eq:berry-discrete-stokes},
the sum of the relative phases, i.e., the 
Berry phase  $\gamma_L$, plays the role of the
line integral, 
whereas the double sum of the Berry fluxes plays the role of 
the surface integral. 
There is an important difference with respect to the Stokes theorem,
namely, the equality of the total Berry flux and the Berry phase is
not guaranteed: Eq.~\eqref{eq:berry-discrete-stokes} only tells that
they are either equal or have a difference of $2\pi$ times an integer.

\subsection{Chern number}
\label{subsect:berry-discrete-chern}

Consider states in a Hilbert space arranged on a grid as above,
$\ket{\Psi_{n,m}}$, with $n,m \in \mathbb{Z}$, $1 \le n \le N$, and $1
\le m \le M$, but now imagine this grid to be on the surface of a
torus. We use the same definition for the Berry flux per plaquette as
in \eqref{eq:berry-def-discrete-plaquette}, but now with $n \text{ mod
} N + 1$ in place of $n+1$ and $m \text{ mod } M + 1$ in place of
$m+1$.

The product of the Berry flux phase factors of all plaquettes is now 1, 
\begin{eqnarray}
\prod_{m=1}^{M}\prod_{n=1}^{N} e^{-i F_{nm}} = 1.
\label{eq:berry-discrete-product1}
\end{eqnarray}
The same derivation can be applied as for
Eq.~\eqref{eq:berry-discrete-stokes} 
above, but now every edge is an
internal edge, and so all contributions to the product cancel.

The Chern number $Q$ associated to
our structure is defined via the sum of the Berry fluxes
of all the plaquettes forming the closed torus surface: 
\begin{equation}
\label{eq:berry_chern_discretecherndef}
Q = \frac{1}{2\pi} \sum_{nm} F_{nm}.
\end{equation}
The fact that the Chern number $Q$ is defined via the gauge invariant
Berry fluxes ensures that $Q$ itself is gauge invariant. 
Furthermore, 
taking the $\arg$ of Eq.~\eqref{eq:berry-discrete-product1} 
proves that the Chern number $Q$ is an integer. 

It is worthwhile to look a little deeper into the discrete formula for
the Chern number. We can define modified Berry fluxes $\tilde{F}_{nm}$ as 
\begin{equation}
\tilde{F}_{nm} = \gamma_{(n,m),(n+1,m)} + \gamma_{(n+1,m),(n+1,m+1)} + 
\gamma_{(n+1,m+1),(n,m+1)} + \gamma_{(n,m+1),(n,m)}. 
\label{eq:berry-tildeF-def}
\end{equation}
Since each edge is shared between
two neighboring plaquettes, the sum of the modified Berry fluxes
over all plaquettes vanishes,  
\begin{equation}
\sum_{m=1}^{M}\sum_{n=1}^{N} \tilde{F}_{nm} = 0.
\end{equation}
If $-\pi \le \tilde{F}_{nm} < \pi$, then we have $\tilde{F}_{nm} =
F_{nm}$. However, $\tilde{F}_{nm}$ can be outside the range
$[-\pi,\pi)$: then as the logarithm is taken in
  Eq.~\eqref{eq:berry-def2-discrete-plaquette}, $F_{nm}$ is taken back
  into $[-\pi,\pi)$ by adding a (positive or negative) integer
    multiple of $2\pi$.  In that case, we say the plaquette $nm$ contains a
    number $Q_{nm} \in \mathbb{Z}$ of vortices, with
\begin{equation}
Q_{nm} = \frac{F_{nm}
- \tilde{F}_{nm}}{2\pi} \in \mathbb{Z}.
\label{eq:berry-vortex-definition}
\end{equation}

We have found a simple picture for the Chern number: \emph{The Chern
  number $Q$, that is, the sum of the Berry fluxes of all the
  plaquettes of a closed surface, is the number of vortices on the
  surface},
\begin{equation}
Q = \frac{1}{2\pi} \sum_{nm} F_{nm} = \sum_{nm} Q_{nm} \in \mathbb{Z}.
\label{eq:berry-discrete-chern}
\end{equation}
Although we proved it here for the special case of a torus, the
derivation is easily generalized to all orientable closed surfaces. We
focused on the torus, because this construction can 
be used as a
very efficient
numerical recipe to discretize and calculate the 
(continuum) Chern
number of a 2-dimensional insulator\cite{hatsugai_chern_discrete},
to be defined in Sect.~\ref{sec::berry_chern-continuum_chern}.

\section{Continuum case}
\label{sec::berry_chern-continuum_berry}


We now assume that instead of a discrete set of states, $\{
\ket{\Psi_j} \}$, we have a continuum, $\ket{\Psi(\RR)}$, where the
$\RR$'s are elements of some $D$-dimensional parameter space
$\mathcal P$. 

\subsection{Berry connection}

We take a
smooth directed path $\CC$, i.e., a \emph{curve} in 
the parameter  space $\mathcal P$,
\begin{align}
\CC:  [0,1)\to \mathcal P,\quad t \mapsto \RR(t). 
\end{align}
We assume that all components of $\ket{\Psi(\RR)}$ are smooth, at
least in an open neighborhood of the the curve $\CC$.
The relative phase between two neighbouring states on the curve
$\CC$, 
corresponding to the parameters
$\RR$ and $\RR+d\RR$, is
\begin{align}
e^{-i\Delta\gamma} &= \frac{\braket{\Psi(\RR)}{\Psi(\RR+d\RR)}}
{\abs{\braket{\Psi(\RR)}{\Psi(\RR+d\RR)}}};&
\Delta \gamma &= i \bra{\Psi(\RR)} \nab_\RR \ket{\Psi(\RR)} \cdot d\RR,
\label{eq:berry_relativephase_continuum}
\end{align}
obtained to first order in $d \RR\to 0$. The quantity multiplying
$d\RR$ on the right-hand side defines the \emph{Berry connection},
\begin{align}  
\AAA (\RR) &= i \braket{\Psi(\RR)}{\nabr \Psi(\RR)} 
= -\im  \braket{\Psi(\RR)}{\nabr \Psi(\RR)}.
\label{eq:berry_vectorpot_def}
\end{align}
Here $\ket{\nabr \Psi(\RR)}$ is defined by requiring for every Hilbert
space vector $\ket{\Phi}$, that
\begin{align}
\braket{\Phi}{\nabr \Psi(\RR)} &= \nabr \braket{\Phi}{\Psi(\RR)}.
\end{align}
The second equality in Eq.~\eqref{eq:berry_vectorpot_def} follows from
the conservation of the norm, $\nabr\braket{\Psi(\RR)}{\Psi(\RR)}=0$.

We have seen in the discrete case that
the relative phase of two states is not gauge invariant;
neither is the Berry connection.
Under a gauge transformation, it changes as
\begin{align}   
\ket{\Psi(\RR)} &\to e^{i\alpha(\RR)}\ket{\Psi(\RR)}: \quad \quad
\AAA(\RR) \to \AAA(\RR) - \nabr \alpha(\RR).
\end{align}




\subsection{Berry phase}

Consider a closed 
directed curve $\CC$ in parameter space. 
The Berry phase along
the curve is defined as 
\begin{align} 
\gamma(\CC) &= - \arg \exp \left[ -i
\underset{\CC}{\oint} 
\AAA \cdot d \RR \right]
\label{eq:Berry_phase_continuous} 
\end{align} 
The Berry phase of a closed directed curve is
gauge invariant, 
since it can be interpreted as a limiting case of the
discrete Berry phase, via Eqs. 
\eqref{eq:berry_vectorpot_def},
\eqref{eq:berry_relativephase_continuum},
and \eqref{eq:berry_def_discrete},
and the latter has been shown 
to be gauge invariant.

\subsection{Berry curvature}

As in the discrete case above, we would like to express the gauge
invariant Berry phase as a surface integral of a gauge invariant
quantity. This quantity is the \emph{Berry curvature}.
Similarly to the discrete case, 
we consider a two-dimensional 
parameter space, and for simplicity denote
the parameters as $x$ and $y$.
We take a simply connected region $\mathcal{F}$ in this
two-dimensional parameter space, with the 
oriented boundary curve of
this surface denoted by $\partial \mathcal{F}$, and consider
the continuum Berry phase corresponding to the boundary.

\subsubsection*{Smoothness of the manifold of states}

Before relating the Berry phase to the Berry curvature, 
an important note on the manifold $\ket{\Psi(\RR)}$ 
of considered states 
is in order. 
From now on, we consider a manifold of states, living in our 
two-dimensional parameter space, that is smooth,
in the sense that the map
$\vec R \mapsto \ketbra{\Psi(\RR)}$
is smooth. 
Importantly, this condition does 
not necessarily imply that 
that the function $\RR \mapsto \ket{\Psi(\RR)}$,
also referred to as a \emph{gauge}
describing our manifold, is smooth. 
(For further discussion and examples, see 
Sect.~\ref{sec:berry_nocontinuous}.)
Nevertheless, even if the gauge
$\RR \mapsto \ket{\Psi(\RR)}$ is not smooth in 
a point $\RR_0$ of the parameter space, one can always find an 
alternative gauge $\ket{\Psi'(\RR)}$ which is
(i) locally smooth, that is, smooth in  
 the point $\RR_0$, 
and (ii) locally 
generates the same map as $\ket{\Psi(\RR)}$, 
that is, for which $\ketbra{\Psi'(\RR)} = \ketbra{\Psi(\RR)}$
in an infinitesimal neighborhood of $\RR_0$. 
Let us formulate an intuitive argument supporting the latter 
claim using quantum-mechanical perturbation theory.
Take the Hamiltonian $\hat{H}(\RR) = - \ketbra{\Psi(\RR)}$,
which can be substituted in the infinitesimal neighborhood
of $\RR_0$ with $\hat{H}(\RR_0+\Delta \RR) = \hat{H}(\RR_0) + 
\Delta \RR \cdot (\nabla \hat{H})(\RR_0)$.
According to first-order perturbation theory, the ground state of the
latter is given by
\begin{equation}
\label{eq:berry_perturbation}
\ket{\Psi'(\RR_0 + \Delta \RR)} = 
\ket{\Psi(\RR_0)} - \sum_{n=2}^{D} \ketbra{\Psi_n(\RR_0)} 
\Delta \RR \cdot (\nabla \hat{H})(\RR_0)
\ket{\Psi(\RR_0)},
\end{equation}
where the states
$\ket{\Psi_n(\RR_0)}$ ($n=2,3,\dots,D$),
together with $\ket{\Psi(\RR_0)}$, 
form a basis of the Hilbert space. 
On the one hand, Eq. \eqref{eq:berry_perturbation}
defines a function that is smooth in $\RR_0$, hence
the condition (i) above is satisfied. 
On the other hand, as $\ket{\Psi'(\RR_0 + \Delta \RR)}$ is 
the ground state of $\hat{H}(\RR_0 + \Delta \RR)$,
condition (ii) is also satisfied.

\subsubsection*{Berry phase and Berry curvature}

Now return to our original goal and
try to express the Berry phase as a surface integral of a
gauge invariant quantity;
to this end, we start by relating the Berry phase to its discrete
counterpart: 
\bean
\label{eq:berry_connection_disccont}
\oint_{\partial \mathcal F} \AAA \cdot d \RR =
\lim_{\Delta x,\Delta y \to 0} \gamma_{\partial \mathcal F},
\eean 
where we discretize the parameter space using a square grid
of steps $\Delta x$, $\Delta y$, 
and express the integral as the discrete Berry phase  
$\gamma_{\partial \mathcal F}$ of a loop approximating 
$\partial \mathcal F$, in the limit of an infinitesimally fine
grid. 
Then, from Eq. \eqref{eq:berry_connection_disccont}
and the Stokes-type theorem in 
Eq. \eqref{eq:berry-discrete-stokes},
we obtain
\begin{align} 
\exp \left[ -i\oint_{\partial \mathcal F} \AAA \cdot d \RR \right]&= 
\lim_{\Delta x,\Delta y \to 0}
e^{-i \sum_{nm} F_{nm}},
\label{eq:berry_phase_curvature_step} 
\end{align} 
where the $nm$ sum goes for the plaquettes forming
the open surface $\mathcal F$.
Furthermore, let us take 
a guage  $\ket{\Psi'(\RR)}$ 
and the corresponding Berry connection $\AAA'$ 
that is smooth in the plaquette
$nm$;
this could be $\ket{\Psi(\RR)}$ and $\AAA$ if
that was already smooth. 
Then, due to the gauge invariance of the 
Berry flux we have 
\bean
\label{eq:berry_F_Fprime}
e^{-i F_{nm}} 
=
e^{-i F'_{nm}},
\eean
where $F'_{nm}$ is the Berry flux corresponding to the locally smooth
gauge. 
Furthermore, in the limit of an infinitely fine grid it holds that 
\bean
\label{eq:berry_F_vs_A}
F'_{nm}
&=&  
A'_x\left(x_n+\frac{\Delta x}{2},y_m\right) \Delta x 
+ 
A'_y\left(x_{n+1},y_m+\frac{\Delta y}{2}\right) \Delta y
\nonumber \\
&-& 
A'_x\left(x_{n}+\frac{\Delta x}{2},y_{m+1}\right) \Delta x
- 
A'_y \left(x_{n},y_{m}+\frac{\Delta y}{2}\right) \Delta y.
\eean
Taylor expansion of the Berry connection around
$\RR_{nm}=\left(x_n+\frac{\Delta x}{2}, y_n+\frac{\Delta y}{2}\right)$
to first order yields
\bean
\label{eq:berry-flux-curvature}
F'_{nm} &=&
\left[\partial_x A'_y(\RR_{nm}) - \partial_y A'_x(\RR_{nm})\right]
\Delta x \Delta y.
\eean
Thereby, with the definition of the \emph{Berry curvature} as
\bean
B = \lim_{\Delta x,\Delta y \to 0} \frac{F'_{nm}}{\Delta x \Delta y},
\eean
we obtain a quantity that is gauge invariant, 
as it is defined via the gauge invariant
Berry flux, 
and is related to the Berry connection via
\bean
\label{eq:berry_curvature_vs_connection}
B = \partial_x A'_y(\RR_{nm}) - \partial_y A'_x(\RR_{nm}).
\eean
We can rephrase Eq. \eqref{eq:berry-flux-curvature}
as follows: the Berry flux for the $nm$ plaquette is expressed
as the product of the Berry curvature on the plaquette
and the surface area of the plaquette.

Substituting Eqs. 
\eqref{eq:berry_F_Fprime} and
\eqref{eq:berry-flux-curvature} 
into Eq. \eqref{eq:berry_phase_curvature_step} yields
\bean
\label{eq:berry_continuum_phasevscurvature}
\exp\left[ -i\oint_{\partial \mathcal F} \AAA \cdot d \RR \right]
=
\exp\left[ -i\int_{ \mathcal F} B(x,y) dx dy \right],
\eean
which is the continuum version of the result 
\eqref{eq:berry-discrete-stokes}.
Equation \eqref{eq:berry_continuum_phasevscurvature} 
can also be rephrased as 
\bean
\gamma(\partial \mathcal F) = - \arg e^{-i\int_{ \mathcal F} B(x,y) dx dy}.
\eean


\subsubsection*{A special case where the usual Stokes theorem works}

A shortcut towards a stronger result than
Eq. \eqref{eq:berry_continuum_phasevscurvature}
is offered in the special case when $\ket{\Psi(\RR)}$
is smooth in the neighborhood of the open surface $\mathcal F$.
Then, a direct application of the two-dimensional Stokes theorem 
implies 
\bean
\label{eq:berry_continuum_phasevscurvature_smooth}
\oint_{\partial \mathcal F} \AAA \cdot d \RR
= \int_{\mathcal F} (\partial_x A_y - \partial_y A_x)dx dy
= \int_{\mathcal F} B dx dy
\eean
Summarizing Eqs. \eqref{eq:berry_continuum_phasevscurvature}
and 
\eqref{eq:berry_continuum_phasevscurvature_smooth},
we can say that line integral of the Berry connection
equals the surface integral of the Berry curvature if the
set of states $\ket{\Psi(\RR)}$ is smooth in the neighborhood
of $\mathcal F$, but they might differ with an integer multiple
of $2\pi$ otherwise.

\subsubsection*{The case of the three-dimensional parameter space}

Let us briefly discuss also the case of a 3D parameter space. 
This will be particularly useful in the context of
two-level systems. 
Starting with the case when the gauge
 $\ket{\Psi(\RR)}$
on the two-dimensional open surface $\mathcal F$
embedded in the 3D parameter space is smooth
in the neighborhood of $\mathcal F$,
we can directly apply the 3D Stokes theorem to 
convert the line integral of $\AAA$ to the
surface integral of the curl of $\AAA$ to 
obtain
\bean
\oint_{\partial \mathcal F} \AAA \cdot d\RR = 
\int_{\mathcal F}\BB \cdot d \vec S,
\eean
where
the Berry curvature is defined as the vector
field $\BB(\RR)$ via
\begin{align}
\BB(\RR) &= \nabr \times \AAA(\RR),
\label{eq:berry_B_def}
\end{align} 
which is gauge invariant as in the two-dimensional case.  Even if $\ket{\Psi(\RR)}$
is not smooth on $\mathcal F$, the relation 
\bean 
\gamma(\partial
\mathcal F)= - \arg e^{-i \oint_{\partial \mathcal F} \AAA \cdot d\RR}
= - \arg e^{-i \int_{\mathcal F}\BB \cdot d \vec S} 
\eean 
holds,
similarly to the two-dimensional result
Eq. \eqref{eq:berry_continuum_phasevscurvature}.


Note furthermore that the Berry phase $\gamma(\partial \mathcal F)$
is not only gauge invariant, but also invariant
against continuous deformations of the two-dimensional surface $\mathcal F$
embedded in 3D, 
as long as the
Berry curvature is smooth everywhere along the way.  

We also remark that although we used the
three-dimensional notation here, but the above results can be
generalized for any dimensionality of the parameter space. 

The notation $\AAA$ and $\BB$ for the Berry connection and Berry
curvature suggest that they are much like the vector potential and the
magnetic field. This is a useful analogy, for instance, $\nabr \BB =
0$, from the definition \eqref{eq:berry_B_def}. Nevertheless, it is
not true that in every problem where the Berry curvature is nonzero,
there is a physical magnetic field.


\subsection{Chern number}
\label{sec::berry_chern-continuum_chern}

In the discrete case, we defined the Chern number as
a sum of Berry fluxes for a square lattice living on a torus. 
Here, we take a continuum parameter space that
has the topology of a torus. 
The motivation is that 
certain physical parameter spaces 
in fact have this torus topology, and the corresponding 
Chern number does have physical significance. 
One example will be the Brillouin zone of a two-dimensional
lattice representing a solid crystalline material.
The Brillouin zone has a torus topology, as
the momentum vectors $(k_x,k_y)$, $(k_x+2\pi,ky)$,
and $(k_x,k_y+2\pi)$ are equivalent. 

Quite naturally, in the continuum definition of the Chern number, 
the sum of Berry fluxes is replaced by the surface integral
of the Berry curvature: 
\bean
\label{eq:berry_continuumchern_def}
Q = -\frac{1}{2\pi} \int_{\mathcal P} B dx dy.
\eean
As this can be interpreted as a continuum limit of the 
discrete Chern number, it inherits the properties of the latter: 
the continuum Chern number is a gauge invariant integer.

\label{sec:chernoflattice}

For future reference, let us lay down the notation to be used
for calculating the Chern numbers of electronic energy bands
in two-dimensional crystals. 
Consider a square lattice for simplicity, which has
a square-shaped Brillouin zone as well. 
Our parameter space $\mathcal P$ is the two-dimensional Brillouin zone now,
which has a torus topology as discussed above. 
The parameters are the Cartesian components 
$k_x,k_y \in [-\pi,\pi)$ of the momentum
vector $\vec k$. 
The electronic energy bands and the corresponding 
electron wavefunctions can be obtained from the 
bulk momentum-space Hamiltonian $\ido{H}(k_x,k_y)$.
The latter defines the Schr\"odinger equation
\begin{align}
\ido{H}(\kk)\ket{u_n(\kk)} &= E_n(\kk) \ket{u_n(\kk)},
\end{align}
where $n=1,2,\dots$ is the band index, which has as many possible
values as the dimension of the Hilbert space of the
internal degree freedom of our lattice model. 
Note that defining the Berry connection, 
the Berry curvature and the Chern number for the $n$th band is possible
only if that band is separated from the other bands
by finite energy gaps. 
The Berry connection of the $n$th band, in line 
with the general definition \eqref{eq:berry_vectorpot_def}, reads 
\begin{align}
  A^{(n)}_{j}(\kk) = i \bra{u_n(\kk)} \partial_{k_j} \ket{u_n(\kk)},
  \quad \mathrm{for}\quad j=x,y  .
\end{align}
The Chern number of the $n$th band,
in correspondence with Eqs. \eqref{eq:berry_continuumchern_def} and 
\eqref{eq:berry_curvature_vs_connection},
reads
\begin{align}  
Q^{(n)} &=  -\frac{1}{2\pi} \int_{BZ} dk_x dk_y \left(\frac{\partial
  A^{(n)}_y}{\partial k_x}- \frac{\partial A^{(n)}_x}{\partial k_y}\right).
\label{eq:chern_torus_def}
\end{align}

Certain approximations of the  band-structure theory 
of electrons provide low-dimensional momentum-space Hamiltonians
that can be diagonalized analytically, allowing for 
an analytical derivation of the Chern numbers 
of the electronic bands. 
More often, however, the electronic wave functions are
obtained from numerical techniques on a finite-resolution grid 
of $(k_x,k_y)$ points in the Brillouin zone. 
In that case, the Chern number of a chosen band 
can still be effectively evaluated using the discrete version of its
definition \eqref{eq:berry_chern_discretecherndef}.

The Chern number of a band of an insulator is a 
topological invariant in the following sense.
One can imagine that the Hamiltonian describing the electrons
on the lattice is deformed adiabatically, that is, continuously and 
with the energy gaps separating the $n$th band from the 
other bands kept open. 
In this case, the Berry curvature varies continuously, and
therefore its integral for the Brillouin zone, which is the Chern number, cannot 
change as the value of the latter is restricted to integers. 
If the deformation of the crystal Hamiltonian is such that
some energy gaps separating the $n$th band from 
a neighboring band is closed and 
reopened, that is, the deformation of the Hamiltonian is 
not adiabatic, then the Chern number might change. 
In this sense, the Chern number is a similar topological invariant 
for two-dimensional lattice models as the winding number is 
for the one-dimensional SSH model.

\section{Berry phase and adiabatic dynamics}
\label{sec::berry_chern-adiabatic_phases}

In most physical situations of interest, the set of states whose
geometric features (Berry phases) we are interested in are eigenstates
of some Hamiltonian $\hat{H}$.  Take a physical system with $D$ real
parameters that are gathered into a formal vector $\RR = (R_1, R_2,
\ldots, R_D)$.  
The Hamiltonian is a smooth function $\hat{H}(\RR)$ of
the parameters, at least in the region of interest. We order the
eigenstates of the Hamiltonian according to the energies
$E_n(\RR)$, 
\begin{align}
\hat{H}(\RR) \ket{n(\RR)} = E_n(\RR) \ket{n(\RR)} .
\label{eq:berry_Ham_general}
\end{align}
We call the set of eigenstates $\ket{n(\RR)}$ the \emph{snapshot
  basis}.  

The definition of the snapshot basis involves \emph{gauge fixing},
i.e., specifying the otherwise arbitrary phase prefactor for every
$\ket{n(\RR)}$. This can be a tricky issue: even if in theory a gauge
exists where all elements of the snapshot basis are smooth functions
of the parameters, this gauge might be very challenging to construct. 

We consider the following problem. 
We assume that the system is
initialized with $\RR=\RR_0$ and in an eigenstate $\ket{n(\RR_0)}$
that is in the discrete part of the spectrum, i.e.,
$E_n(\RR)-E_{n-1}(\RR)$ and $E_{n+1}(\RR)-E_{n}(\RR)$ are nonzero.  At
time $t=0$ we thus have
\begin{align}
\RR(t=0) &= \RR_0;&
\ket{\psi(t=0)}&=\ket{n(\RR_0)}.
\end{align} 
Now assume that during the time $t=0 \to T$ the parameters $\RR$ are
slowly changed: $\RR$ becomes $\RR(t)$, and the values of $\RR(t)$
define a continuous directed curve $\mathcal{C}$. 
Also, assume that $\ket{n(\RR)}$ is smooth along the
curve $\mathcal C$.
The state of the
system evolves according to the time-dependent Schr\"odinger equation:
\begin{align}
i \frac{d}{dt}\ket{\psi(t)}&= \hat{H}(\RR(t)) \ket{\psi(t)}.
\label{eq:schr_e}
\end{align}

Further, assume that $\RR$ is varied in such a way that at all times
the energy gaps around the state $\ket{n(\RR(t))}$ remain finite.  We
can then choose the rate of variation of $\RR(t)$ along the path
$\mathcal{C}$ to be slow enough compared to the frequencies
corresponding to the energy gap, so the \emph{adiabatic approximation}
holds 
In that case, the system remains in the energy
eigenstate $\ket{n(\RR(t))}$, only picking up a phase. 
We are now
going to find this phase.


By virtue of the adiabatic approximation, we take as Ansatz
\begin{align}
\ket{\psi(t)} &= e^{i\gamma_n(t)} e^{-i \int_0^t E_n(\RR(t'))dt'}
\ket{n(\RR(t))}.
\label{eq:ansatz}
\end{align} 
For better readability, in the following we often drop the $t$
argument where this leads to no confusion. The time derivative of 
Eq.~\eqref{eq:ansatz} reads
\begin{align}
i \frac{d}{dt} \ket{\psi(t)} &= e^{i\gamma_n} e^{-i \int_0^t E_n(\RR(t'))dt'} 
\left(-\frac{d\gamma_n}{dt} \ket{n(\RR)}+
  E_n(\RR) \ket{n(\RR)} + i \ket{\tfrac{d}{dt}n(\RR)} \right).  
\end{align}
To show what we mean by $\ket{\tfrac{d}{dt}n(\RR(t))}$, we write it
out explicitly in terms of a fixed basis, that of the eigenstates at
$\RR=\RR_0$:
\begin{align}
\ket{n(\RR)} &= \sum_m c_m(\RR) \ket{m(\RR_0)};\\
\ket{\tfrac{d}{dt}  n(\RR(t))} &= \frac{d\RR}{dt} \cdot \ket{\nabr n(\RR)} = 
\frac{d\RR}{dt}\sum_m  \nabr c_m(\RR) \ket{m(\RR_0)}.
\label{eq:expression_dndt_ket}
\end{align} 
We insert the Ansatz \eqref{eq:ansatz} into the right hand side of the
Schr\"odinger equation \eqref{eq:schr_e}, use
the snapshot eigenvalue relation \eqref{eq:berry_Ham_general},
simplify and reorder the Schr\"odinger equation, and obtain
\begin{align}
-\frac{d\gamma_n}{dt}  \ket{n(\RR)} + i\ket{\tfrac{d}{dt} n(\RR)} &= 0.
\label{eq:gamma_dot1}
\end{align}
Multiplying from the left by $\bra{n(\RR)}$, and using
Eq.~\eqref{eq:expression_dndt_ket}, we obtain
\begin{align} 
\frac{d}{dt} \gamma_n(t) &= 
i \braket{n(\RR(t))}{\tfrac{d}{dt}n(\RR(t))} = 
\frac{d\RR}{dt} i \braket{n(\RR)}{\nabr n(\RR)}. 
\label{eq:gamma_dot}
\end{align} 

We have found that for the directed curve $\mathcal{C}$ in parameter
space, traced out by $\RR(t)$, there is an adiabatic phase
$\gamma_n(\CC)$, which reads
\begin{align} 
\gamma_n(\CC) &= \underset{\CC}{\int} 
i \braket{n(\RR)}{\nabr n(\RR)} d\RR.
\label{eq:Berry_phase_integral} 
\end{align} 
A related result is obtained after a similar derivation, 
if the parameter space of the $\RR$ points
is omitted and the snapshot basis 
$\ket{n(t)}$ is parametrized directly by 
the time variable.
Then, the adiabatic phase is 
\bean
\gamma_n(t) = \int_0^t i \braket{n(t')}{\partial_{t'} n(t')} dt'. 
\eean

Equation \eqref{eq:Berry_phase_integral} allows
us to formulate the key message of this section as the following. 
Consider the case of an adiabatic and \emph{cyclic}
change of the Hamiltonian, that is, 
when the curve $\mathcal C$ is closed, 
implying $\RR(T) = \RR_0$. 
In this case, the adiabatic phase reads
\begin{align}
\label{eq:berry-gamma-def2} 
\gamma_n(\CC) &= \underset{\mathcal C}{\oint}
i \braket{n(\RR)}{\nabr n(\RR)} d\RR.
\end{align} 
Therefore, the adiabatic phase picked up by the 
state during a cyclic adiabatic change of the Hamiltonian 
is equivalent to the Berry phase corresponding to the
closed oriented curve representing the Hamiltonian's
path in the parameter space. 

Two further remarks are in order.
First,
on the face of it, our derivation seems to do too much. 
It seems that we have
produced an exact solution of the Schr\"odinger equation. 
Where did we
use the adiabatic approximation? 
In fact, Eq.~\eqref{eq:gamma_dot} does not imply
Eq.~\eqref{eq:gamma_dot1}. For the more complete derivation, showing how
the nonadiabatic terms appear, see \cite{griffiths}. 


The second remark concerns the measurability of the 
Berry phase. 
The usual way to experimentally detect phases is by an interferometric
setup. This means coherently splitting the wavefunction of the system
into two parts, taking them through two adiabatic trips in parameter
space, via $\RR(t)$ and $\RR'(t)$, and bringing the parts back
together. The interference only comes from the overlap between the
states: it is maximal if $\ket{n(\RR(T))}=\ket{n(\RR'(T))}$, which is
typically ensured if $\RR(T)=\RR'(T)$.  The difference in the
adiabatic phases $\gamma_n$ and $\gamma_n'$ is the adiabatic phase
associated with the closed loop $\CC$, which is the path obtained by
going forward along $t=0\to T: \RR(t) $, then coming back along $ t =
T \to 0: \RR'(t)$.

\section{Berry's formulas for the Berry curvature}

Berry provided\cite{berry_phase_1984} two practical formulas for the
Berry curvature. Here we present them in a form corresponding to a
three-dimensional parameter space.
To obtain the two-dimensional case, where the Berry curvature
$B$ is a scalar, one can identify the latter with the 
component $B_z$ of the three-dimensional case treated below; 
for generalization to higher than 3 
dimensions, see the discussion in Berry's
paper\cite{berry_phase_1984}. First,
\begin{align} 
B_j = -\im \,\epsilon_{jkl} \,\partial_k \braket{ n}{\partial_l n} = 
-\im\, \epsilon_{jkl} \braket{\partial_k n}{\partial_l n} + 0,
\end{align} 
where the second term is 0 because $\partial_k\partial_l =
\partial_l\partial_k$ but $\epsilon_{jkl}=-\epsilon_{jlk}$.

To obtain Berry's second formula, inserting a resolution of
identity in the snapshot basis in the above equation, we obtain
\begin{align}
\BB^{(n)} = -\im \sum_{n'\neq n} \braket{ \nabla n}{n'} \times 
\braket{n'}{\nabla n},
\label{eq:berry_1}
\end{align}
where the parameter set $\RR$ is suppressed for brevity. 
The term with $n'=n$ is omitted from the sum, as it is zero, since
because of the conservation of the norm, 
$\braket{\nabla n}{n} = -\braket{n}{\nabla n}$. To calculate
$\braket{n'}{\nabla n}$, start from the definition of the eigenstate
$\ket{n}$, act on both sides with $\nabla$, and then project unto $\ket{n'}$:
\begin{align}
\hat{H} \ket{n} &= E_n \ket{n};\\
(\nabla \hat{H}) \ket{n} + \hat{H} \ket{\nabla n} &= (\nabla E_n)
\ket{n} + E_n \ket{\nabla n};\\
\bra{n'} \nabla \hat{H} \ket{n} + \bra{n'} \hat{H} \ket{\nabla n} &= 
0 + E_n \braket{n'}{\nabla n}.
\label{eq:gradn}
\end{align}
Act with $H$ towards the left in Eq.~\eqref{eq:gradn}, rearrange,
substitute into \eqref{eq:berry_1}, and you should obtain the second
form of the Berry curvature, which is manifestly gauge invariant:
\begin{align}
\BB^{(n)} &= -\im \sum_{n'\neq n} \frac{\bra{n} \nabla \hat{H} \ket{n'}\times
\bra{n'}\nabla \hat{H} \ket{n}}{(E_n-E_{n'})^2}.
\label{eq:Berry_curvature_2}
\end{align}
This shows that the monopole sources of the Berry curvature, if they
exist, are the points of degeneracy.

A direct consequence of Eq.~\eqref{eq:Berry_curvature_2}, is
that the sum of the Berry curvatures of all eigenstates of a
Hamiltonian is zero. If all the spectrum of $\hat{H}(\RR)$ is discrete
along a closed curve $\CC$, then one can add up the Berry phases of
all the energy eigenstates. 
\begin{align}
\sum_n\BB^{(n)} &= -\im \sum_n \sum_{n'\neq n} \frac{\bra{n} \nab_\RR \hat{H} \ket{n'}\times
\bra{n'}\nab_\RR \hat{H} \ket{n}}{(E_n-E_{n'})^2} \nonumber\\
\quad &= -\im \sum_n \sum_{n'<n} \frac{1}{(E_n-E_{n'})^2} \Big( \bra{n} \nab_\RR \hat{H} \ket{n'}\times
\bra{n'}\nab_\RR \hat{H} \ket{n} \nonumber\\
\quad &\quad + \bra{n'} \nab_\RR \hat{H} \ket{n}\times
\bra{n}\nab_\RR \hat{H} \ket{n'} \Big)= 0. 
\end{align}
The last equation holds because $\vec{a}\times\vec{b} = -
\vec{b}\times\vec{a}$ for any two vectors $\vec{a},\vec{b}$.

\section{Example: the two-level system}
\label{sec:berry-example-spinhalf}


So far, most of the discussion on the Berry phase and the
related concepts have been kept rather general. 
In this section, we illustrate these concepts
via the simplest nontrivial example, that is, the two-level system.

\subsection{No continuous global gauge}
\label{sec:berry_nocontinuous}

Consider a Hamiltonian describing 
a two-level system:
\begin{align}
\hat{H}(\hh)= d_x \ido{\sigma}_x + d_y \ido{\sigma}_y +
d_z \ido{\sigma}_z = \hh \cdot \hat{\ssigma},
\label{eq:def_hh_reminder}
\end{align}
with $\hh = (d_x,d_y,d_z)\in \mathbb{R}^3 \backslash \{0\}$.
Here, the vector $\hh$ plays the role of the parameter $\RR$ in
of preceding sections, and the parameter space is
the punctured 3D Euclidean space $\mathbb{R}^3\backslash\{0\}$,
to avoid the degenerate case of the energy spectrum. 
Note the absence of a
term proportional to $\sigma_0$: this would play no role in adiabatic
phases. Because of the anticommutation relations of the Pauli
matrices, the Hamiltonian above squares to a multiple of the identity
operator, $\hat{H}(\hh)^2 = \hh^2 \sigma_0$. 
Thus, the eigenvalues of
$\hat{H}(\hh)$ have to have absolute value $\abs{\hh}$.

\begin{figure}
\centering
\includegraphics[width=0.9\columnwidth]{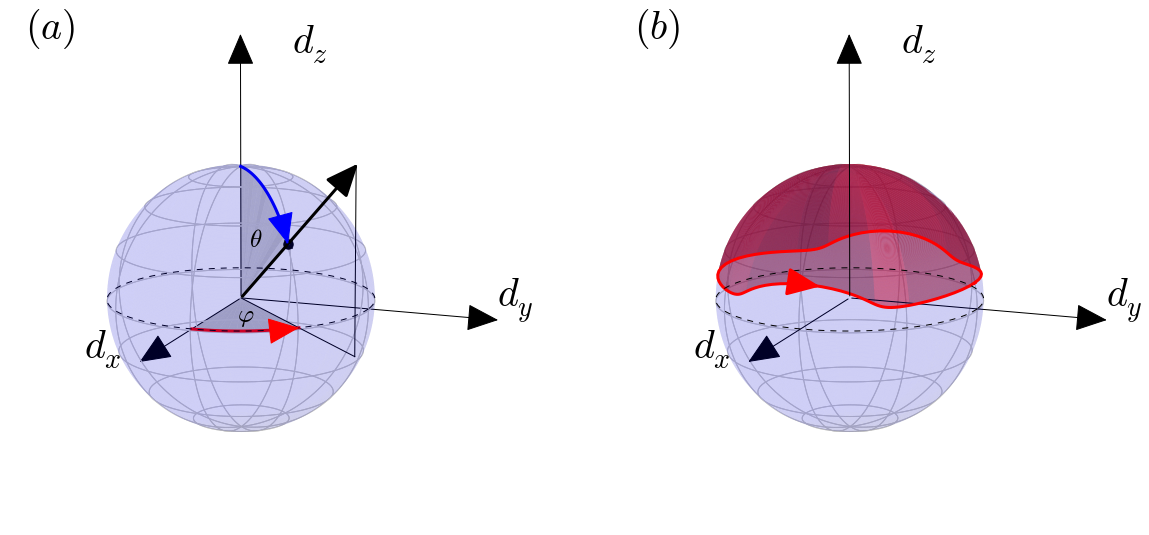}
\caption{
The Bloch sphere. A generic traceless gapped two-level Hamiltonian
  is a linear combination of Pauli matrices, $\hat{H}(\hh)= \hh \cdot \hat{\ssigma}$. This
  can be identified with a point in $\mathbb{R}^3 \backslash \{0\}$. 
  The eigenenergies
  are given by the distance of the point from the origin, the
  eigenstates depend only on the direction of the vector $\hh$, i.e.,
  on the angles $\theta$ and $\varphi$, as defined in subfigure (a)
  and in  Eq.~\eqref{eq:def_theta_phi}
The Berry phase of a closed curve $\CC$ is half the area enclosed
by the curve when it is projected onto the surface of the Bloch sphere.  
  \label{fig:bloch_sphere_12}
}
\end{figure}

A practical graphical representation of $\hat{H}(\hh)$ is the Bloch sphere,
shown in Fig.~\ref{fig:bloch_sphere_12}.  The spherical angles
$\theta\in[0,\pi)$ and $\varphi\in[0,2\pi)$ are defined as
\begin{align} 
\cos \theta &= \frac{d_z}{\abs{\hh}};&
e^{i\varphi} &= \frac{d_x+i
  d_y}{\sqrt{d_x^2+d_y^2}}.  
\label{eq:def_theta_phi}
\end{align} 

We denote the two eigenstates of the Hamiltonian $\hat{H}(\hh)$ by
$\ket{+_\hh}$ and $\ket{-_\hh}$, with
\begin{align}
\hat{H}(\hh) \ket{\pm_\hh} &= \pm | \hh |
\ket{\pm_\hh}.
\end{align}
These eigenstates depend on the direction of the 3-dimensional vector
$\hh$, but not on its length.  
The eigenstate with $E=+\abs{\hh}$ of the corresponding Hamiltonian is:
\begin{align} 
\label{eq:def_kethplus}
\ket{+_\hh} =   
e^{i\alpha(\theta,\varphi)}
\begin{pmatrix} 
e^{-i\varphi/2} \cos{\theta/2}\\ 
e^{i\varphi/2} \sin{\theta/2}
\end{pmatrix}, 
\end{align} 
while the eigenstate with $E=-\abs{\hh}$ is $\ket{-_\hh} =
e^{i\beta(\hh)} \ket{+_{-\hh}}$.  The choice of the phase factors
$\alpha$ and $\beta$ above corresponds to fixing a gauge. We will now
review a few gauge choices.

Consider fixing $\alpha(\theta, \varphi)=0$ for all
$\theta,\varphi$. This is a very symmetric choice, in this way in
formula \eqref{eq:def_kethplus}, we find $\theta/2$ and $\varphi/2$.
There is problem, however, as you can see if you consider a full
circle in parameter space: at any fixed value of $\theta$, let
$\varphi=0 \to 2\pi$. We should come back to the same Hilbert space
vector, and we do, but we also pick up a phase of $-1$. We can either
say that this choice of gauge led to a discontinuity at $\varphi=0$,
or that our representation is not single-valued. We now look at some
attempts at fixing these problems, to find a gauge that is both
continuous and single valued. 

As a first attempt, let us fix $\alpha = \varphi/2$; denoting this
gauge by subscript $S$, we have
\begin{align} 
\label{eq:berry_gauge_S}
\ket{+_\hh}_S = 
\begin{pmatrix}
\cos{\theta/2}\\ 
e^{i\varphi} \sin{\theta/2}
\end{pmatrix}. 
\end{align} 
The phase prefactor now gives an additional factor of $-1$ as we make
the circle in $\varphi$ at fixed $\theta$, and so it seems we have a
continuous, single valued representation.  There are two tricky
points, however: the North Pole, $\theta=0$, and the South Pole,
$\theta=\pi$. At the North Pole, $\ket{(0,0,1)}_S=(1,0)$ no
problems. This gauge is problematic at the South Pole, however (which
explains the choice of subscript): there,
$\ket{(0,0,-1)}_S=(0,e^{i\varphi})$, the value of the wavefunction
depends on which direction we approach the South Pole from.

We can try to solve the problem at the South Pole by choosing $\alpha
= -\varphi/2$, which gives us
\begin{align} 
\label{eq:berry_gauge_N}
\ket{+_\hh}_N = 
\begin{pmatrix} 
e^{-i\varphi}\cos{\theta/2}\\
\sin{\theta/2}
\end{pmatrix}. 
\end{align} 
As you can probably already see, this representation runs into trouble at
the North Pole: $\ket{(0,0,1)}_N=(e^{-i\varphi},0)$. 

We can try to overcome the problems at the poles by taking linear
combinations of $\ket{+_\hh}_S$ and $\ket{+_\hh}_N$, with prefactors that
vanish at the South and North Poles, respectively. A family of options
is: 
\begin{align} 
\ket{+_\hh}_\chi &= 
e^{i\chi} 
\sin\frac{\theta}{2} \ket{+_\hh}_S +
\cos\frac{\theta}{2} \ket{+_\hh}_N 
\\
\quad &=
\begin{pmatrix}
\cos\frac{\theta}{2} ( \cos\frac{\theta}{2} + \sin\frac{\theta}{2}
e^{i\chi} e^{-i\varphi})\\ \sin\frac{\theta}{2} e^{i\varphi}(
\cos\frac{\theta}{2} + \sin\frac{\theta}{2} e^{i\chi}e^{-i\varphi})
\end{pmatrix}. 
\end{align}
This is single valued everywhere, solves the problems at the
Poles. However, it has its own problems: somewhere on the Equator, at
$\theta=\pi/2$, $\varphi = \chi \pm \pi$, its norm disappears.

It is not all that surprising that we could not find a
well-behaved gauge: there is none. By the end of this chapter, it
should be clear, why. 


\subsection{Calculating the Berry curvature and the Berry phase}

Consider the two-level system as defined in the previous section. 
Take a closed curve $\CC$ in the parameter space
$\mathbb{R}^3 \backslash \{0\}$. 
We are going to calculate the Berry phase $\gamma_-$
of the $\ket{-_\hh}$ eigenstate on this curve: 
\begin{align} 
\gamma_-(\CC) &= \underset{\CC}{\oint} 
\AAA(\hh) d\hh, 
\end{align} 
with the Berry vector potential defined as 
\begin{align}
\AAA(\hh) &= i \bra{-_\hh}\nab_\hh \ket{-_\hh}.
\end{align} 

The calculation
becomes straightforward if we use the Berry curvature,
\begin{align}
\BB(\hh) &= \nab_\hh \times \AAA(\hh);\\
\gamma_-(\CC) &= \int_\SSS \BB(\hh) d\mathrm{\SSS},
\label{eq:gamma_berry}
\end{align}
where $\SSS$ is any surface whose boundary is the loop $\CC$.
(Alternatively, it is a worthwhile exercise to calculate the Berry phase
directly in a fixed gauge, e.g., one of the three gauges of the
previous chapter.)

Specifically, we make use of Berry's gauge invariant formulation 
\eqref{eq:Berry_curvature_2} of
the Berry curvature, derived in the last chapter.
In the case of the generic two-level  Hamiltonian 
\eqref{eq:def_hh_reminder}, 
Eq. \eqref{eq:Berry_curvature_2} gives
\begin{align}
\BB^{\pm}(\hh) = -\im \frac{\bra{\pm} \nab_\hh \hat{H} \ket{\mp}\times
\bra{\mp}\nab_\hh \hat{H} \ket{\pm}} {4 \hh^2},
\label{eq:Berry_curvature_2x2}
\end{align}
with 
\begin{align} 
\nab_\hh \hat{H} &= \hat{\ssigma}.
\end{align} 
To evaluate \eqref{eq:Berry_curvature_2x2}, we choose the quantization axis parallel
to $\hh$, thus the eigenstates simply read
\begin{align}
\ket{+_\hh} &= \begin{pmatrix} 1\\0\end{pmatrix};&
\ket{-_\hh} &= \begin{pmatrix} 0\\1\end{pmatrix}.
\end{align}
The matrix elements can now be computed as
\begin{align}
\bra{-} \hat{\sigma}_x \ket{+} &= \begin{pmatrix} 0 & 1\end{pmatrix}
\begin{pmatrix} 0& 1 \\   
1 &0\end{pmatrix}
\begin{pmatrix} 1\\0\end{pmatrix} =1,
\end{align}  
and similarly, 
\begin{align}
\bra{-} \sigma_y \ket{+} &= i;\\
\bra{-} \sigma_z \ket{+} &= 0.
\end{align}
So the cross product of the vectors reads 
\begin{align}
\bra{-} \hat{\ssigma} \ket{+} \times \bra{+} \hat{\ssigma} \ket{-}
&= 
\begin{pmatrix} 1\\i\\0 \end{pmatrix}
\times
\begin{pmatrix} 1\\-i\\0 \end{pmatrix}
=
\begin{pmatrix} 0\\ 0\\ 2i \end{pmatrix}.
\end{align} 
This gives us for the Berry curvature, 
\begin{align}
\BB^{\pm} (\hh) &= \pm\frac{\hh}{\abs{\hh}} \frac{1}{2 \hh^2}. 
\end{align} 
We can recognize in this the field of a pointlike monopole source in
the origin. Alluding to the analog between the Berry curvature and the
magnetic field of electrodynamics (both are derived from a ``vector
potential'') we can refer to this field, as a ``magnetic
monopole''. Note however that this monopole exists in the abstract
space of the vectors $\hh$ and not in real space.

The Berry phase of the closed loop $\CC$ in parameter space, according
to Eq.~\eqref{eq:gamma_berry}, is the flux of the monopole field
through a surface $\SSS$ whose boundary is $\CC$. It is easy to
convince yourself that this is half of the solid angle subtended by
the curve,
\begin{align}
\gamma_-(\CC) &= \frac{1}{2} \Omega_\CC.
\end{align}
In other words, 
the Berry phase is half of the area enclosed by the image of $\CC$,
projected onto the surface of the unit sphere, as illustrated in
Fig.~\ref{fig:bloch_sphere_12}.


What about the Berry phase of the other energy eigenstate? From
Eq.~\eqref{eq:Berry_curvature_2x2}, the corresponding Berry curvature
$\BB_+$ is obtained by inverting the order of the factors in the cross
product: this flips the sign of the cross product. Therefore the Berry
phases of the ground and excited state fulfil the relation
\begin{align}
\gamma_+(\CC) &= - \gamma_-(\CC).
\end{align}
One can see the same result on the Bloch sphere. Since
$\braket{+}{-}=0$, the point corresponding to $\ket{-}$ is antipodal
to the point corresponding to $\ket{+}$. Therefore, the curve
traced by the $\ket{-}$ on the Bloch sphere is the inverted image of the curve
traced by $\ket{+}$. These two curves have the same orientation,
therefore the same area, with opposite signs.

\subsection{Two-band lattice models and their Chern numbers}

The simplest case where a Chern number can arise is a two-band
system. Consider a particle with two internal states, hopping on a two-dimensional
lattice. The two internal states can be the spin of the conduction
electron, but can also be some sublattice index of a spin polarized
electron. In the translation invariant bulk, the 
wave vector $\kk = (k_x,k_y)$
is a good quantum number, and the Hamiltonian reads
\begin{align}
\hat{H}(\kk) &= \hh(\kk) \hat{\ssigma}, 
\end{align}
with the function $\hh(\kk)$ mapping the each point of the Brillouin
Zone to a 3D vector.  Since the Brillouin zone is a torus, the
endpoints of the vectors $\hh(\kk)$ map out a deformed torus in
$\mathbb{R}^3\backslash\{0\}$.  This torus is a directed surface: its
inside can be painted red, its outside, blue.

The Chern number of $\ket{-}$ (using the notation of
Sect.~\ref{sec:ssh-bulk}, of $\ket{u_{1}(\kk)}$) is the flux of
$\BB_-(\hh)$ through this torus. We have seen above that $\BB_-(\hh)$
is the magnetic field of a monopole at the origin $\hh=0$. If the
origin is on the inside of the torus, this flux is +1. If it is
outside of the torus, it is 0. If the torus is turned inside out, and
contains the origin, the flux is -1. The torus can also intersect
itself, and therefore contain the origin any number of times.

One way to count the number of times is to
take any line from the origin to infinity, and count the number of
times it intersects the torus, with a +1 for intersecting from the
inside, and a -1 for intersecting from the outside. The sum is
independent of the shape of the line, as long as it goes all the way
from the origin to infinity.

\section*{Problems}
\addcontentsline{toc}{section}{Problems}
%
\begin{prob}
\label{prob:berry-spinhalf}
\textbf{Discrete Berry phase and Bloch vectors}\\
Take an ordered set
of three arbitrary, normalized states of a two-level system.
Evaluate the corresponding discrete Berry phase.
Each state is represented by a vector on the Bloch sphere. 
Show analytically that if two of the vectors coincide,
then the discrete Berry phase vanishes. 
\end{prob}

\begin{prob}
\label{prob:gauges}
\textbf{Two-level system and the Berry connection}\\
Consider the two-level system defined in 
Eq. \eqref{eq:def_hh_reminder},
and describe the excited energy eigenstates using the 
 gauge $\ket{+_{\hh}}_S$ defined in Eq. \eqref{eq:berry_gauge_S}.
Using this gauge, evaluate and visualize the
corresponding Berry connection 
vector field $\vec A(\vec d)$.
Is it well-defined in every point of the parameter space? 
Complete the same tasks using the gauge
 $\ket{+_{\hh}}_N$ defined in Eq. \eqref{eq:berry_gauge_N}.
\end{prob}

\begin{prob}
\label{prob:berry_infinite}
\textbf{Massive Dirac Hamiltonian}\\
Consider the two-dimensional \emph{massive Dirac Hamiltonian}
$\hat{H}(k_x,k_y) = m \hat{\sigma}_z + 
k_x \hat{\sigma_x} + k_y \hat{\sigma_y}$, 
where $m \in \mathbb{R}$ is a constant
and the parameter space is $\mathbb{R}^2 \ni (k_x,k_y)$. 
(a) Take a circular loop with radius $\kappa$ 
in the parameter space, centered around the origin. 
Calculate the Berry phase associated to this loop
and the ground-state manifold of the Hamiltonian:
$\gamma_-(m,\kappa) = ?$. 
(b) Calculate the Berry connection $B_-(k_x,k_y)$ 
for the ground-state manifold. 
(c) Integrate the Berry connection for the whole parameter space. 
How does the result depend on $m$? 
\end{prob}

\begin{prob}
\label{prob:berry_nogloballysmoothgauge}
\textbf{Absence of a continuous global gauge}\\
In Sect.~\ref{sec:berry_nocontinuous}, we
have shown example gauges for the two-level system
that were not globally smooth on the parameter space. 
Prove that such globally smooth gauge does not exist. 
\end{prob}

\begin{prob}
\label{prob:chern-d}
\textbf{Chern number of two-band models}\\
Consider a two-band lattice model 
with the Hamiltonian $\hat{H}(\kk) = \hh(\kk) \cdot \hat{\ssigma}$.
Express the Chern number of the lower-energy band
in terms of $\hh(\kk)/|\hh(\kk)|$.
\end{prob}


%% file: polarization.tex

\chapter{Polarization and Berry phase}
\label{chap:polarization}

\abstract*{What is the bulk polarization of a band insulator? We review
  the answer to this question provided by the modern theory of
  polarization. This defines a set of orthonormal, localized Wannier
  states for the electrons in the occupied bands, whose centers of
  masses can be identified with the electron positions. We show that
  the Wannier centers of states from a band are calculated using the
  Berry phase of the band, for the path of the wavenumber $k$ across
  the Brillouin zone. 
In later chapters we will corroborate the identification of the Berry
phase with the bulk polarization by calculating the current in the
bulk. We will use the concepts of this chapter to obtain the
bulk--boundary correspondence for two-dimensional topological
insulators.  As an illustration of these concepts, we use the
Rice-Mele model, obtained by adding a sublattice potential to the
Su-Schrieffer-Heeger model of polyacetylene.}

The bulk polarization of a band insulator is a tricky concept.
Polarization of a neutral molecule is easily defined using the
difference in centers of the negative and positive charges
constituting the system. 
When we try to apply this simple concept to the periodic bulk of a
band insulator (assuming for simplicity that the positive atom cores
are immobile and localized), we meet complications. The center of the
negative charges should be calculated from electrons in the fully
occupied valence bands. However, all energy eigenstates in the valence
band are delocalized over the bulk, and so the center of charge of each
electron in such a state is ill defined. Nevertheless, insulators are
polarizable, and respond to an external electric field by a
rearrangement of charges, which corresponds to a (tiny) current in the
bulk. Thus, there should be a way to define a bulk polarization.

In this chapter we show how a bulk polarization can be defined for
band insulators using the so-called modern theory of
polarization\cite{resta1999macroscopic,resta2000makes,rmp_wannier_vanderbilt}.
The contribution of the electrons to the polarization is a property of
the many-body electron state, a Slater determinant of the energy
eigenstates from the fully occupied valence bands. The central idea is
to rewrite the same Slater determinant using a different orthonormal
basis, one composed of localized states, the so-called Wannier
states. The contribution of each electron in a Wannier state to the
center of charge can then be easily assessed, and then added up.

We discuss the simplest interesting case, that of a one-dimensional
two-band insulator with one occupied and one empty band, and leave the
multiband case for later. We show that the center of charge of the
Wannier states can be identified with the Berry phase of the occupied
band over the Brillouin zone, also known as the Zak
phase\cite{zak_phase}.

For a more complete and very pedagogical introduction to the Berry
phase in electron wavefunctions, we refer the reader to a set of lecture
notes by Resta\cite{resta_cycle}.

\subsubsection*{The Rice-Mele model}

The toy model we use in this chapter is the Rice-Mele model, which is
the SSH model of Chapt.~\ref{chap:ssh} with an extra staggered onsite
potential. 
The Hamiltonian for the Rice-Mele model on a chain of $N$ unit cells
reads
\begin{multline}
\hat{H} = 
v \sum_{\uci=1}^{N} \big( \ket{\uci,B}\bra{\uci,A} + h.c. \big) +
w \sum_{\uci=1}^{N-1} \big( \ket{\uci+1,A}\bra{\uci,B}  + h.c. \big) \\
+ u \sum_{\uci=1}^{N} \big( \ket{\uci,A}\bra{\uci,A} - \ket{\uci,B}\bra{\uci,B} \big),
\label{eq:ricemele_Hamiltonian_def}
\end{multline}
with the staggered onsite potential $u$, the intracell hopping
amplitude $v$, and intercell hopping amplitude $w$ all assumed to be
real. 
%
The matrix of the Hamiltonian for the Rice-Mele model on a chain of
$N=4$ sites reads
\begin{align}
H&=\begin{pmatrix}
u & v & 0 & 0 & 0 & 0 & 0 & 0 \\
v & -u & w & 0 & 0 & 0 & 0 & 0\\
0 & w & u & v & 0 & 0 & 0 & 0\\
0& 0 & v & -u & w & 0 & 0 & 0 \\
0 & 0 & 0 & w& u & v & 0 & 0\\
0 & 0 & 0 & 0 & v & -u & v & 0\\
0 & 0 & 0 & 0 & 0 & w& u & v \\
0 & 0 & 0 & 0 & 0 & 0 & v & -u \\
\end{pmatrix},
\end{align}


\section{Wannier states in the Rice-Mele model}
\label{sec:polarization-wannier}


The bulk energy eigenstates of a band insulator are delocalized over
the whole system.  
We use as an example the bulk Hamiltonian of the Rice-Mele model,
i.e., the model on a ring of $N$ unit cells.  As in the case of the SSH
model, Sect.~\ref{sec:ssh-bulk}, the energy eigenstates are the plane
wave Bloch states,
\begin{align}
\ket{\Psi(k)} &= \ket{k} \otimes \ket{u(k)},
\end{align}
with 
\begin{align}
\label{eq:polarization-planewave_def}
\ket{k} &= \frac{1}{\sqrt{N}} \sum_{\uci=1}^N 
e^{i\uci k} \ket{\uci},& \quad \quad 
\text{for }
k&\in \{ \delta_k, 2\delta_k,\ldots, N \delta_k\}\quad \text{with }\,\,
\delta_k = \frac{2\pi}{N}.
\end{align}
We omit the index $1$ from the eigenstate for simplicity. 
The $\ket{u(k)}$ are eigenstates of the bulk 
momentum-space Hamiltonian, 
\begin{align}
\label{eq:polarization-bulk_ham_uk_def}
H(k) &= 
\begin{pmatrix} u & v+we^{-ik}\\ v+we^{ik}& -u \end{pmatrix}, 
\end{align}
with eigenvalue $E(k)$.  

The Bloch states $\ket{\Psi(k)}$ are spread
over the whole chain. They span the occupied subspace, defined by the
projector
\begin{align}
\hat{P} &= \sum_{k\in BZ} \ket{\Psi(k)}\bra{\Psi(k)} .
\label{eq:polarization_P_def}
\end{align} 

The phase of each Bloch eigenstate $\ket{\Psi(k)}$ can be set at
will. A change of these phases, a gauge transformation,
\begin{align}
\ket{u(k)} &\to e^{i\alpha(k)} \ket{u(k)};& 
\ket{\Psi(k)} &\to e^{i\alpha(k)} \ket{\Psi(k)}, 
\end{align}
gives an equally good set of Bloch states, with an arbitrary set of
phases $\alpha(k)\in\mathbb{R}$ for $k= \delta_k,2\delta_k,\ldots,
2\pi$.  Using this freedom it is in principle possible to ensure that
in the thermodynamic limit of $N\to \infty$, the components of
$\ket{\Psi(k)}$ are smooth, continuous functions of $k$. However, this
gauge might not be easy to obtain by numerical methods. 
We therefore prefer, if possible, to work with gauge-independent
quantities, like the projector to the occupied subspace defined in
Eq.~\eqref{eq:polarization_P_def}.

\subsubsection*{Defining properties of Wannier states}

The Wannier states
$\ket{w(\jjj)}\in\mathcal{H}_\text{external}\otimes\mathcal{H}_\text{internal}$,
with $\jjj=1,\ldots,N$, are defined by the following
properties:
\begin{subequations}
\label{eq:polarization_wannier_requirements}
\begin{align} 
\braket{w(\jjj')}{w(\jjj)} &= \delta_{\jjj' \jjj}&  &\text{Orthonormal set}\\
\sum_{\jjj=1}^N \ket{w(\jjj)}\bra{w(\jjj)} &= \hat{P} &  &\text{Span the occupied subspace}
\\
\phantom{\sum_{\jjj=1}^N} 
\braket{\uci+1}{w(\jjj+1)} &= \braket{\uci}{w(\jjj)}& &\text{Related by translation} 
\label{eq:polarization-wannier_translation_requirement}
\\
\lim_{N\to\infty} \bra{w(N/2)}
(\hat{x}-N/2)^2 &\ket{w(N/2)} < \infty&   & \text{Localization} 
\label{eq:polarization-wannier_localization_requirement}
\end{align}
\end{subequations}
with the addition in
Eq.~\eqref{eq:polarization-wannier_translation_requirement} defined
modulo $N$.  Requirement
\eqref{eq:polarization-wannier_localization_requirement}, that of
localization, uses the position operator, 
\begin{align}
\label{eq:polarization_xhat_def}
\hat{x} &= \sum_{\uci=1}^N \uci \left( 
\ket{\uci, A}\bra{\uci,A} + 
\ket{\uci, B}\bra{\uci,B} \right),
\end{align}
and refers to a property of $\ket{w(\jjj)}$ in the
thermodynamic limit of $N\to\infty$ that is not easy to define
precisely. In this one-dimensional case it can be turned into an even
stricter requirement of exponential localization,
$\braket{w(\jjj)}{\uci} \braket{\uci}{w(\jjj)} <
e^{-\abs{\jjj-\uci}/\xi}$ for some finite localization length
$\xi\in\mathbb{R}$.

\subsubsection*{Wannier states are inverse Fourier transforms 
of the Bloch eigenstates}

Because of Bloch's theorem, all energy eigenstates have a plane wave
form not only in the canonical basis $\ket{\uci,\beta}$, with
$\beta=A,B$, but in the Wannier basis as well,
\begin{align}
\ket{\Psi(k)} &= e^{-i \alpha(k)} \frac{1}{\sqrt{N}} 
\sum_{\jjj=1}^N e^{i k\jjj} \ket{w(\jjj)},
\label{eq:polarization_wannier_1}
\end{align}
with some phase factors $\alpha(k)$. To convince yourself of this,
consider the components of the right-hand-side in the basis of Bloch
eigenstates. The right-hand-side is an eigenstate of the lattice
translation operator $S$, with eigenvalue $e^{-ik}$, and therefore,
orthogonal to all of the Bloch eigenstates $\ket{\Psi(k')}$ with
$k'\neq k$. It is also orthogonal to all positive energy eigenstates,
since it is in the occupied subspace. Thus, the only state left is
$\ket{\Psi(k)}$. 

From Eq.~\eqref{eq:polarization_wannier_1}, an inverse Fourier
transformation gives us a practical Ansatz for Wannier states, 
\begin{align}
\ket{w(\jjj)} &= \frac{1}{\sqrt{N}} \sum_{k=\delta_k}^{N \delta_k} 
e^{-i\jjj k} e^{i\alpha(k)}
\ket{\Psi(k)}.
\label{eq:polarization_wn_fourier}
\end{align}
There is still a large amount of freedom left by this form, since the
gauge function $\alpha(k)$ is unconstrained. This freedom can be used
to construct Wannier states as localized as possible. If, e.g., a
smooth gauge is found, where in the $N\to\infty$ limit, the components
of $e^{i\alpha(k)} \ket{\Psi(k)}$ are analytic functions of $k$, we
have exponential localization of the Wannier states due to
properties of the Fourier transform. (More generally, if a
discontinuity appears first in the $l$th derivative of
$\ket{\Psi(k)}$, the components of the Wannier state $\ket{w(\jjj)}$
will decay as $K/n^{l+1}$.)


\subsubsection*{Wannier centers can be identified
  with the Berry phase}

We first assume that we have found a continuous gauge.  The center of
the Wannier state $\ket{w(0)}$ can be calculated, using
\begin{multline}
\hat{x} \ket{w(0)} = 
\frac{1}{2\pi}\int_{-\pi}^\pi dk \sum_\uci 
\uci e^{ik\uci} \ket{\uci}\otimes \ket{u(k)} \\
= -\frac{i}{2\pi} 
\left[ \sum_\uci e^{ik\uci} \ket{\uci} \otimes \ket{u(k)} \right]_{-\pi}^{\pi}
+ \frac{i}{2\pi} \int_{-\pi}^\pi dk \sum_\uci e^{ik\uci} \ket{\uci} \otimes \ket{\partial_k u(k)} \\
= \frac{i}{2\pi} \int_{-\pi}^\pi dk \sum_\uci e^{ik\uci} \ket{\uci} \otimes \ket{\partial_k u(k)}.
\end{multline}
We find that the center of the Wannier state $\ket{w(\jjj)}$ is
\begin{equation}
\bra{w(\jjj)} \hat{x} \ket{w(\jjj)} = \frac{i}{2\pi} 
\int_{-\pi}^\pi dk \braket{u(k)}{\partial_k u(k)} + \jjj.
\label{eq:polarization-wannier_centers}
\end{equation}
The second term in this equation shows that the centers of the Wannier
states are equally spaced, at a distance of one unit cell from each
other. The first term, which is the Berry phase (divided by $2\pi$) of
the occupied band across the Brillouin zone,
cf.~Eq.~\eqref{eq:berry-gamma-def2}, corresponds to a uniform
displacement of each Wannier state by the same amount.

We define the bulk electric polarization to be the Berry phase of the
occupied band across the Brillouin zone, the first term in
Eq.~\eqref{eq:polarization-wannier_centers}, 
\begin{equation}
P_\text{electric} = \frac{i}{2\pi} \int_{-\pi}^\pi dk 
\braket{u(k)}{\partial_k u(k)}.
\label{eq:polarization-polarization_def}
\end{equation}
Although the way we derived this above is intuitive, it remains to be
shown that this is a consistent definition. From
Chapt.~\ref{chap:berry_chern}, it is clear that a gauge transformation
can only change the bulk electric polarization by an integer. We will
show explicitly in Chapt.~\ref{chap:current} that the change of this
polarization in a quasi-adiabatic process correctly reproduces the
bulk current.

\subsection{Wannier states using the projected position operator}
\label{sec:polarization_wannierfromprojected}

A numerically stable, gauge invariant way to find a tightly localized
set of Wannier states is using the unitary position
operator\cite{resta2000makes}, 
\begin{align}
\hat{X} &= e^{i \delta_k \hat{x}}.
\end{align}
This operator is useful, because it fully respects the periodic
boundary conditions of the ring. The eigensystem of $\hat{X}$ consists
of eigenstates localized in cell $\uci$ with eigenvalue $e^{i \delta_k
  \uci}$. Thus, we can associate the expectation value of the position
in state $\ket{\Psi}$ with the phase of the expectation value of
$\hat{X}$,
\begin{align}
\expect{x} = \frac{N}{2\pi} \im \log \bra{\Psi} \hat{X} \ket{\Psi}. 
\end{align}
The real part of the logarithm carries information about the degree
of localization\cite{resta2000makes,ortiz_localization}. 

In order to obtain the Wannier states, we restrict the unitary
position operator to the filled bands, defining 
\begin{align}
\hat{X}_P &= \hat{P} \hat{X} \hat{P}. 
\end{align}
We will show below that in the thermodynamic limit of $N\to\infty$,
the eigenstates of the projected position operator $\hat{X}_P$ form
Wannier states.

To simplify the operator $\hat{X}_P$, consider
\begin{multline}
\bra{\Psi(k')} \hat{X} \ket{\Psi(k)} = \frac{1}{\nuc}
\sum_{\uci'=1}^\nuc e^{-i\uci'k'} \bra{\uci'} \otimes \bra{u(k')}
\sum_{\uci=1}^\nuc e^{i\delta_k \uci} e^{i \uci k} \ket{\uci}
\otimes\ket{u(k)} \\ 
= \frac{1}{\nuc} \braket{u(k')}{u(k)}
\sum_{\uci=0}^{\nuc-1}e^{i \uci (k+\delta_k-k')} =
\delta_{k+\delta_k,k'} \,\braket{u(k+\delta_k)}{u(k)}
\end{multline}
where $\delta_{k+\delta_k,k'}=1$ if $k'=k+\delta_k$, and 0 otherwise.
Using this, we have
\begin{multline}
\hat{X}_P = 
\sum_{k' k} \ket{\Psi(k')} \bra{\Psi(k')} \hat{X} 
\ket{\Psi(k)}\bra{\Psi(k)} \\ 
= \sum_{k} 
\braket{u(k+\delta_k)}{u(k)} \cdot 
\ket{\Psi(k+\delta_k)}\bra{\Psi(k)}.
\label{eq:polarization-xp2}
\end{multline}

We can find the eigenvalues of $\hat{X}_P$ using a direct consequence
of Eq.~\eqref{eq:polarization-xp2}, namely, that raising $\hat{X}_P$
to the $N$th power gives an operator proportional to the unity in the
occupied subspace,
\begin{align}
\left(\hat{X}_P\right)^N &= W \, \hat{P}.
\end{align}
We will refer to the constant of proportionality, $W\in\mathbb{C}$,  
given by 
\begin{align}
W &= \braket{u(2\pi)}{u(2\pi-\delta_k)}
\cdot \ldots \cdot \braket{u(2 \delta_k)}{u(\delta_k)} 
\braket{u(\delta_k)}{u(2\pi)},
\end{align}
as the Wilson loop.  Note that $W$ is very similar to a discrete Berry
phase, apart from the fact that $\abs{W} \le 1$ (although
$\lim_{N\to\infty} \abs{W}=1$).  The spectrum of eigenvalues of
$\hat{X}_P$ is therefore composed of the $N$th roots of $W$,
\begin{align}
\lambda_n &= e^{i n \delta_k + \log (W)/N}, \quad \text{with} \quad n = 1,\ldots,N;
& \Longrightarrow \quad \lambda_n^N &= W.
\label{eq:polarization-Nthroot} 
\end{align}
These eigenvalues have the same magnitude
$\abs{\lambda_n}=\sqrt[\leftroot{-2}\uproot{2}N]{\abs{W}}<1$, and
phases in the interval $[0,2\pi)$, spaced by $\delta_k$.  Because
  $\bra{w(\jjj)} \hat{X}_P \ket{w(\jjj)} = \bra{w(\jjj)} \hat{X}
  \ket{w(\jjj)}$, the magnitude tells us about the localization
  properties of the Wannier states, and the phases can be
  interpreted as position expectation values.

We now check whether eigenstates of $\hat{X}_P$ fulfil the properties
required of Wannier states,
Eq.~\eqref{eq:polarization_wannier_requirements}.  The
relation $(\hat{X}_P)^N=W \hat{P}$ above shows that eigenvectors of
$\hat{X}_P$ span the occupied subspace.  The eigenstates are related
by translation, since
\begin{align}
\hat{S}^\dagger \hat{X}_P \hat{S} &= e^{i\delta_k} \hat{X}_P;\\
\hat{X}_P \ket{\Psi} &= \abs{W}^{1/N} e^{i\alpha} \ket{\Psi};\\ 
\hat{X}_P \hat{S} \ket{\Psi} &= \abs{W}^{1/N} e^{i\alpha+\delta_k} 
\hat{S} \ket{\Psi}. 
\end{align}
There is a problem with the orthogonality of the eigenstates though.
We leave the proof of localization as an exercise for the reader.  

The projected unitary position operator $\hat{X}_P$ is a normal
operator only in the thermodynamic limit of $N\to\infty$. For finite
$N$, it is not normal, i.e., it does not commute with its adjoint, and
as a result, its eigenstates do not form an orthonormal
basis. This can be seen as a discretization error. 


%

\section{Inversion symmetry and polarization}
\label{sec:polarization-inversion}

For single-component, continuous-variable wavefunctions $\Psi(r)$,
inversion about the origin (also known as parity) has the effect $
\Psi(r) \to \Inv \Psi(r) = \Psi(-r)$.
Two important properties of the unitary operator
$\Inv$ representing inversion follow: 
$\Inv^2 = 1$, and $\Inv e^{ikr} = e^{-ikr}$.
A Hamiltonian has inversion symmetry if 
$\Inv \hat{H} \Inv^\dagger = \hat{H}$. 

When generalizing the inversion operator to lattice models of solid
state physics with internal degrees of freedom, we have to keep two
things in mind.

First, in a finite sample, the edges are bound to break inversion
symmetry about the origin (except for very fine-tuned sample
preparation). We therefore only care about inversion symmetry in the
bulk, and require that it take $\ket{k} \to \ket{-k}$.

Second, each unit cell of the lattice models we consider also has its
internal Hilbert space, which can be affected by inversion. This
includes spin components (untouched by inversion) and orbital type
variables (affected by inversion) as well. In general, we represent
the action of inversion on the internal Hilbert space by a unitary
operator $\Invi$ independent of the unit cell. 

The inversion operator is represented on the bulk Hamiltonian of a
lattice model by an operator $\Inv$, which acts on
$\mathcal{H}_\text{internal}$ as $\Invi$, 
\begin{align}
\Inv \ket{k}\otimes\ket{u} &= \ket{-k} \otimes \Invi \ket{u};\\
\Invi^2 &= \Invi^\dagger \Invi = \II_\text{internal}. 
\end{align}
The action of the inversion operator on the bulk momentum-space Hamiltonian
can be read off using its definition, 
\begin{align}
\Inv \HH(k) \Inv^{-1} &= \Inv \bra{k} \Hbulk \ket{k} \Inv^{-1} = 
\bra{-k} \Inv \Hbulk \Inv^{-1} \ket{-k} = \Invi \HH(-k) \Invi^\dagger.  
\end{align}

A lattice model has inversion symmetry in the bulk, if there exists a
unitary and Hermitian $\Invi$ acting on the internal space, such that
\begin{align}
\Invi \HH(-k) \Invi &= \HH(k). 
\end{align}

If all occupied bands can be adiabatically separated in
energy, so we can focus on one band, with wavefunction $\ket{u(k)}$,
inversion symmetry has a simple consequence.  
The eigenstates at $-k$ and $k$ are related by 
\begin{align}
\HH(k) \ket{u(k)} &= E(k) \ket{u(k)} \quad \Longrightarrow 
\HH(-k) \Invi \ket{u(k)} = 
E(k) \Invi \ket{u(k)};\\
\Longrightarrow & \quad \ket{u(-k)} = e^{i\phi(k)}\Invi \ket{u(k)}. 
\label{eq:polarization-ukumk}
\end{align}
For the wavenumbers $k=0$ and $k=\pi$, the so-called time-reversal
invariant momenta, this says that they have states with a definite
parity,  
\begin{align}
\ket{u(0)} &= p_0  \ket{u(0)};&  \ket{u(\pi)} &= p_\pi \ket{u(\pi)}, \\
\text{with} \quad p_0&=\pm 1;& p_\pi&=\pm 1.
\end{align}

\subsection{Quantization of the Wilson loop due to inversion symmetry}

We now rewrite the Wilson loop $W$ of a band of an inversion-symmetric
one-dimensional Hamiltonian, assuming we have a discretization into a number $2M$
of $k$-states, labeled by $j=-M+1,\ldots,M$, as
\begin{align}
\ket{u_j} &= 
\begin{cases}
    \ket{u(2\pi+j\delta_k)} & \text{if $j \le 0$};\\
    \ket{u(j\delta_k)}, & \text{otherwise}.
\end{cases}
\end{align}
We use Eq.~\eqref{eq:polarization-ukumk}, which takes the form 
\begin{align}
\ket{u_{-j}} &= e^{i\phi_j}\Invi \ket{u_j}. 
\label{eq:polarization-ukumkj}
\end{align}

The Wilson loop $W$ of a band of an inversion symmetric one-dimensional insulator
can only take on the values $\pm1$. We show this, using $M=3$ as an
example,
\begin{multline}
W = 
\braket{u_M}{u_2} \braket{u_2}{u_1} \braket{u_1}{u_0} 
\braket{u_0}{u_{-1}} \braket{u_{-1}}{u_{-2}} \braket{u_{-2}}{u_M} \\
= \braket{u_M}{u_{2}} \braket{u_2}{u_1} \braket{u_1}{u_0} 
\bra{u_0}  e^{i\phi_1} \Invi \ket{u_1} \\
\bra{u_1} \Invi e^{-i\phi_1} e^{i\phi_2} \Invi   \ket{u_2} 
\bra{u_2} \Invi e^{-i\phi_{2}} \ket{u_M} \\ 
= \braket{u_1}{u_0} \bra{u_0}\Invi \ket{u_1}
\bra{u_{2}} \Invi \ket{u_M} \braket{u_M}{u_2}= p_0 p_\pi 
\end{multline}

\begin{multline}
W = 
\braket{u_M}{u_{M-1}} \ldots
\braket{u_1}{u_0} \braket{u_0}{u_{-1}} \ldots
\braket{u_{-M+1}}{u_M} \\
= \braket{u_M}{u_{M-1}} \ldots 
\braket{u_1}{u_0} 
\bra{u_0}  e^{i\phi_1} \Invi \ket{u_1} \\
\bra{u_1} \Invi e^{-i\phi_1} e^{i\phi_2} \Invi   \ket{u_2} 
\bra{u_2} \Invi e^{-i\phi_2} e^{i\phi_3} \Invi   \ket{u_3} \ldots
\bra{u_{M+1}} \Invi e^{-i\phi_{M+1}} \ket{u_M} \\ 
= \braket{u_1}{u_0} \bra{u_0}\Invi \ket{u_1}
\bra{u_{M-1}} \Invi \ket{u_M} 
\braket{u_M}{u_{M-1}} = \pm 1 
\end{multline}

The statement about the Wilson loop can be translated to the bulk
polarization, using Eq.~\eqref{eq:polarization-polarization_def}.
\emph{Each band of an inversion symmetric one-dimensional insulator
  contributes to the bulk polarization 0 or 1/2.}

\section*{Problems}
\addcontentsline{toc}{section}{Problems}
%

\begin{prob}
\label{prob:polarization_sshinversion}
\textbf{Inversion symmetry of the SSH model}\\
Does the SSH model have inversion symmetry? 
If it has, then provide the corresponding local
operator acting in the internal Hilbert space. 
\end{prob}

\begin{prob}
\label{prob:polarization_sshinversion}
\textbf{Eigenstates of the projected position operator are
localized}\\
In Sect.~\ref{sec:polarization_wannierfromprojected}, 
it is claimed that the eigenstates of the
projected position operator $\hat{X}_P$
form a Wannier set. 
One necessary condition for that statement to be true is 
that the eigenstates are localized, see 
Eq. \eqref{eq:polarization-wannier_localization_requirement}. 
Prove this. 
\end{prob}

%% file: thouless_pump.tex

\chapter{Adiabatic charge pumping, Rice-Mele model}
\label{chap:ricemele}

\abstract*{
We apply the Berry phase and the Chern number to show that the
periodically and slowly changing the parameters of a one-dimensional
solid, it is possible to pump particles in it. The number of particles
(charge) pumped is an integer per cycle, that is given by a Chern
number. Along the way we will introduce important concepts of edge
state branches of the dispersion relation, and bulk--boundary
correspondence.
Since we are working towards understanding time-independent
topological insulators, this chapter might seem like a detour.
However, bulk--boundary correspondence of 2-dimensional Chern
insulators, at the heart of the theory of topological insulators, is
best understood via a mapping to an adiabatic charge pump. 
The concrete system we use in this chapter is the simplest adiabatic
charge pump, obtained by adding staggered onsite potential to the SSH
model. It is known as the Rice-Mele (RM) model. 
}

We now apply the Berry phase and the Chern number to show that the
periodically and slowly changing the parameters of a one-dimensional
solid, it is possible to pump particles in it. The number of particles
(charge) pumped is an integer per cycle, that is given by a Chern
number. Along the way we will introduce important concepts of edge
state branches of the dispersion relation, and bulk--boundary
correspondence.
Since we are working towards understanding time-independent
topological insulators, this Chapter might seem like a detour.
However, bulk--boundary correspondence of 2-dimensional Chern
insulators, at the heart of the theory of topological insulators, is
best understood via a mapping to an adiabatic charge pump. 
The concrete system we use in this Chapter is the simplest adiabatic
charge pump, 
the time-dependent version of the Rice-Mele (RM) model,
\begin{multline}
\HH(t) = 
v(t) \sum_{\uci=1}^{N} \big( \ket{\uci,B}\bra{\uci,A} + h.c. \big) +
w(t) \sum_{\uci=1}^{N-1} \big( \ket{\uci+1,A}\bra{\uci,B}  + h.c. \big) \\
+ u(t) \sum_{\uci=1}^{N} \big( \ket{\uci,A}\bra{\uci,A} - \ket{\uci,B}\bra{\uci,B} \big),
\label{eq:thouless-ricemele_Hamiltonian_def}
\end{multline}
with the staggered onsite potential $u$, intracell hopping amplitude
$v$, and intercell hopping amplitude $w$ all assumed to be real and
periodic functions of time $t$. In this Chapter, we are going to see
how, by properly choosing the time sequences, we can ensure that
particles are pumped along the chain.



\section{Charge pumping in a control freak way}

The most straightforward way to operate a charge pump in the Rice-Mele
model is to make sure that the system falls apart at all times to
disconnected dimers. This will happen if at any time either the
intercell hopping amplitude $w$, or the intracell hopping amplitude
$v$ vanishes. We can then use the staggered onsite potential to nudge
the lower energy eigenstate to the right.  If during the whole cycle
we keep a finite energy difference between the two eigenstates, we can
do the cycle slowly enough to prevent excitation.

\subsubsection*{Adiabatic shifting of charge on a dimer}

As a first step towards the charge pumping protocol, consider a single
dimer, i.e., $N=1$. Using the adiabatic limit introduced in the last
chapter, we can shift charge from one site to the other. The
Hamiltonian reads
\begin{align}
\HH(t) &= u(t) \hat{\sigma}_z + v(t)\hat{\sigma}_x, 
\end{align}
with no hopping allowed at the beginning and end of the cycle,
at $t =0$, we have $u=1; \quad v=0;$ and 
at $t =T$, $u=-1; \quad v=0;$.

We initialize the system in the ground state, which at time $t=0$
corresponds to $\ket{A}$, a particle on site $A$.  Then we switch on
the hopping, which allows the particle to spill over to site $B$, and
once it has done that, we switch the hopping off.  To ensure that the
particle spills over, we raise the onsite potential at $A$ and lower
it at $B$. A practical choice is
\begin{align} u(t) &= \cos (\pi t/T);& v(t) &= \sin
  (\pi t/T),
\end{align}
whereby the energy gap is at any time $2$.  According to the adiabatic
theorem, if $H(t)$ is varied \emph{slowly enough}, we will have
shifted the charge to $B$ at the end of the cycle. 

\subsubsection*{Putting together the control freak sequence}

\begin{figure}
\centering
\includegraphics[width=0.85\columnwidth]
{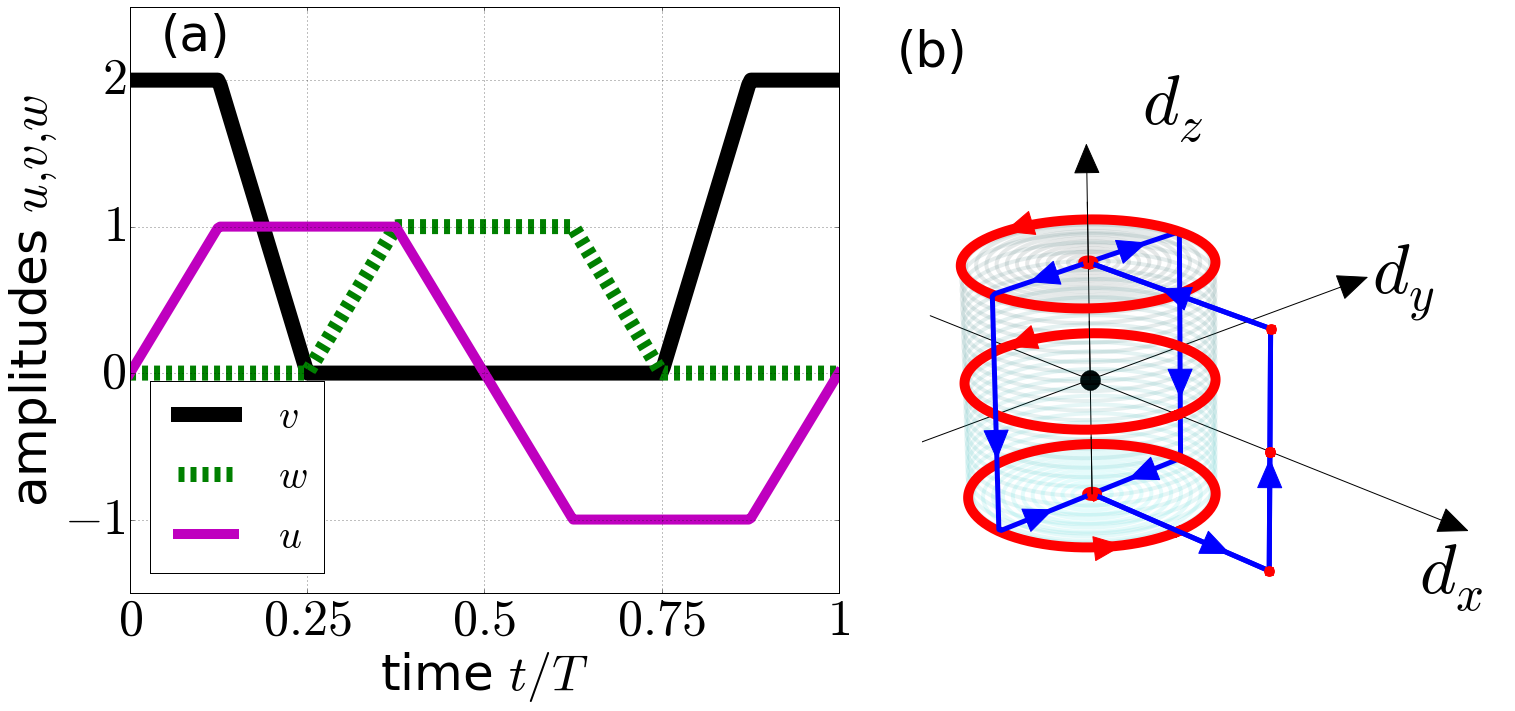}
\includegraphics[width=0.85\columnwidth]
{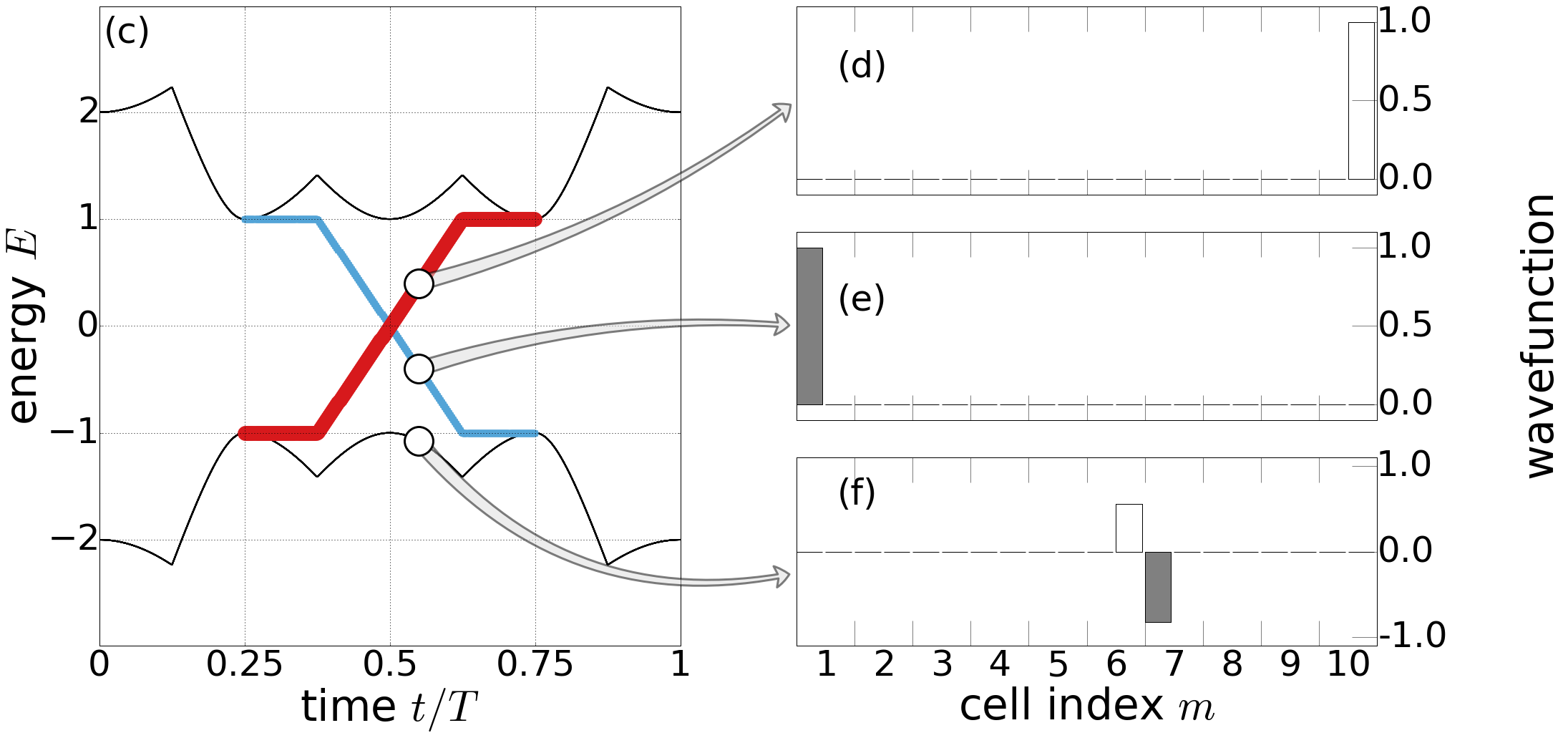}
\caption{The control freak pump sequence in the Rice-Mele model.
The sequence is defined via Eqs. 
\eqref{eq:ricemele_controlfreak_pump_sequence} and
\eqref{eq:thouless-general-d}.
(a) Time dependence of the hopping amplitudes $v$, $w$
and the sublattice potential $u$.
(b) The surface formed by the vector $\hh(k,t)$ 
corresponding to the bulk momentum-space Hamiltonian.
The topology of the surface is a torus,
but its parts corresponding to $t \in [0,0.25]T$ and
$t \in [0.75,1]T$ are infinitely thin and appear as a line
due to the vanishing value of $w$ in these time intervals.
(c) Instantaneous spectrum of the Hamiltonian
$\hat{H}(t)$ of an open chain of $N=10$ sites.
Red (blue) points represent states that are localized in the 
rightmost (leftmost) unit cells and have energies
between -1 and 1. 
\label{fig:ricemele-controlfreak-sequence}}
\end{figure}

Once we know how to shift a particle from $\ket{\uci,A}$ to $\ket{\uci,B}$,
we can use that to shift the particle further from $\ket{\uci,B}$ to
$\ket{\uci+1,A}$. For simplicity, we take a sequence constructed from
linear ramps of the amplitudes, using the function $f:
[0,1)\to\mathbb{R}$:
\begin{align}
   f(x)&= 
\begin{cases}
    8 x, & \text{if } x \le 1/8\\ 
    1, & \text{if } 1/8 \le x < 3/8\\ 
    1 - 8 (x-3/8), & \text{if } 3/8 \le x < 1/2\\ 
    0, & \text{otherwise }
\end{cases}
\end{align}
One period of the pump sequence, for $0 \le t < T$, reads  
\begin{subequations}
\label{eq:ricemele_controlfreak_pump_sequence}
\begin{align}
u(t) &= f(t/T) - f(t/T+\tfrac{1}{2}) ;\\
v(t) &= 2 f(t/T + \tfrac{1}{4}) ;\\
w(t) &= f(t/T - \tfrac{1}{4}). 
\end{align}
\end{subequations}
This period, shown in Fig.~\ref{fig:ricemele-controlfreak-sequence}
a), is assumed to then be repeated. 
Note that we shifted the beginning
time of the sequence: now at times $t/T=n\in\mathbb{Z}$, the
Hamiltonian is the trivial SSH model, $t/T = n+1/4$, disconnected
monomers, at times $t/T=n+1/2$, it is the nontrivial SSH model.

The time-dependent bulk momentum-space Hamiltonian reads 
\begin{align}
\label{eq:thouless-general-d}
\HH(k,t) &= \hh(k,t) \hsigma = (v(t) + w(t) \cos k ) \hsigma_x + w(t) \sin
k \hsigma_y + u(t) \hsigma_z,
\end{align}
which can be represented graphically as the path of the vector
$\hh(k,t)$ as the quasimomentum goes through the Brillouin zone, $k:
0\to 2\pi$, for various fixed values of time $t$, as in
Fig.~\ref{fig:ricemele-controlfreak-sequence} (b).

\subsubsection*{Visualizing the motion of energy eigenstates}

We can visualize the effects of the control freak pumps sequence in
the Rice-Mele model by tracing the trajectories of the energy
eigenstates. At any time $t$, each instantaneous energy eigenstate can
be chosen confined to a single dimer: either on a single unit cell, or
shared between two cells. In both cases, we can associate a position
with the energy eigenstates: the expectation value of the position
operator $\hat{x}$ defined as per Eq.~\eqref{eq:polarization_xhat_def}.

\begin{figure}
\includegraphics[width=0.98\columnwidth]{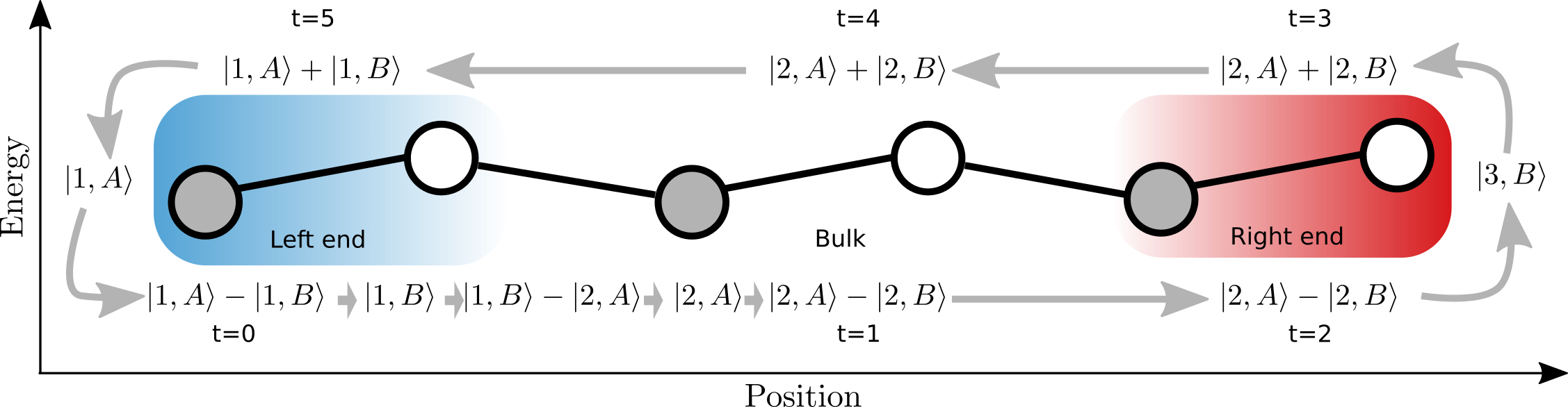}
\caption{An instantaneous energy eigenstate of the control freak pump
  sequence, as it is pumped through the system. At negative (positive)
  energy, it is pumped towards the right (left), upon reaching the
  right (left) end, it is pumped into the upper (lower) band.  
\label{fig:ricemele-controlfreak-sequence-eigenstate}.}
\end{figure}

The trajectories of energy eigenstates in the position-energy space,
Fig.~\ref{fig:ricemele-controlfreak-sequence-eigenstate}, show that
the charge pump sequence works rather like a conveyor belt for the
eigenstates.
%
%
We engineered the sequence as a unitary operation that pushes all
negative energy states in the bulk to the right at the rate of one
unit cell per cycle (a current of one particle per cycle).  These
orthogonal states, one by one, are pushed into the right end region,
which has only room for one energy eigenstate.  Eigenstates cannot
pile up in the right end region: if they did, this would violate
unitarity of the time evolution operator $U(t) = \mathbb{T}
e^{-i\int_0^t H(t') dt'}$, where $\mathbb{T}$ stands for time
ordering, since initially orthogonal states would
acquire finite overlap. So, states pumped to the right edge have to go
somewhere, and the only direction they can go is back towards the
bulk.  This on the other hand is only possible, if they acquire enough
energy to be in the upper band, since all states in the lower band in
the bulk are pushed towards the right. Moreover, in order to carry
these states away from the right edge, and make room for those coming
from the bulk, the pump sequence has to push upper band bulk states
towards the left.

To summarize, the control freak pump sequence is characterized by
three statements.  The protocol
\begin{itemize}
\item in the bulk, pushes all $E<0$ eigenstates rightwards, by 1
  unit cell per cycle,
\item at the right end, pushes 1 eigenstate per cycle with $E<0$ to $E>0$,
\item in the bulk, pushes all $E>0$ eigenstates leftwards, by 1 unit
  cell per cycle.
\end{itemize}
If any one of these statements holds, the other two must also hold as
a consequence.

\subsubsection*{Edge states in the instantaneous spectrum}

We can see the charge pump at work indirectly -- via its effect on the
edge states -- using the instantaneous spectrum, the eigenvalues of
$H(t)$ of the open chain. An example is shown for the control freak
pump sequence of the Rice-Mele model on a chain of 10 unit cells (20
sites) in Fig.~\ref{fig:ricemele-controlfreak-sequence}. Due to the
special choice of the control freak sequence, the bulk consists of
$N-1$-fold degenerate states (the bands are flat).  More importantly,
there is an energy gap separating the bands, which is open around
$E=0$ at all times. However, there are branches of the spectrum
crossing this energy gap, which must represent edge states.

To assign ``right'' or ``left'' labels to edge states in the
instantaneous spectrum, it is necessary to examine the corresponding
wavefunctions.  In case of the control freak pump sequence, right
(left) edge state wavefunctions are localized on the $\uci=N$ (
$\uci=1$) unit cells, and the corresponding energy values are
highlighted in green (red).  The edge state branches in the dispersion
in Fig.~\ref{fig:ricemele-controlfreak-sequence}. clearly show that 1
state per cycle is pushed up in energy at the right edge.

\section{Moving away from the control freak limit} 

We will now argue that the number of particles pumped by a cycle of a
periodic adiabatic modulation of an insulating chain is an integer,
even if the control freak attitude is relaxed. In the generic case,
the energy eigenstates are delocalized over the whole bulk, and so we
will need new tools to keep track the charge pumping process. The
robust quantization of charge pumping was shown by Thouless,
who calculated the bulk current directly: we defer this calculation to
the next Chapter, and here argue using adiabatic deformations.

\begin{figure}
\centering
\includegraphics[width=0.85\columnwidth]{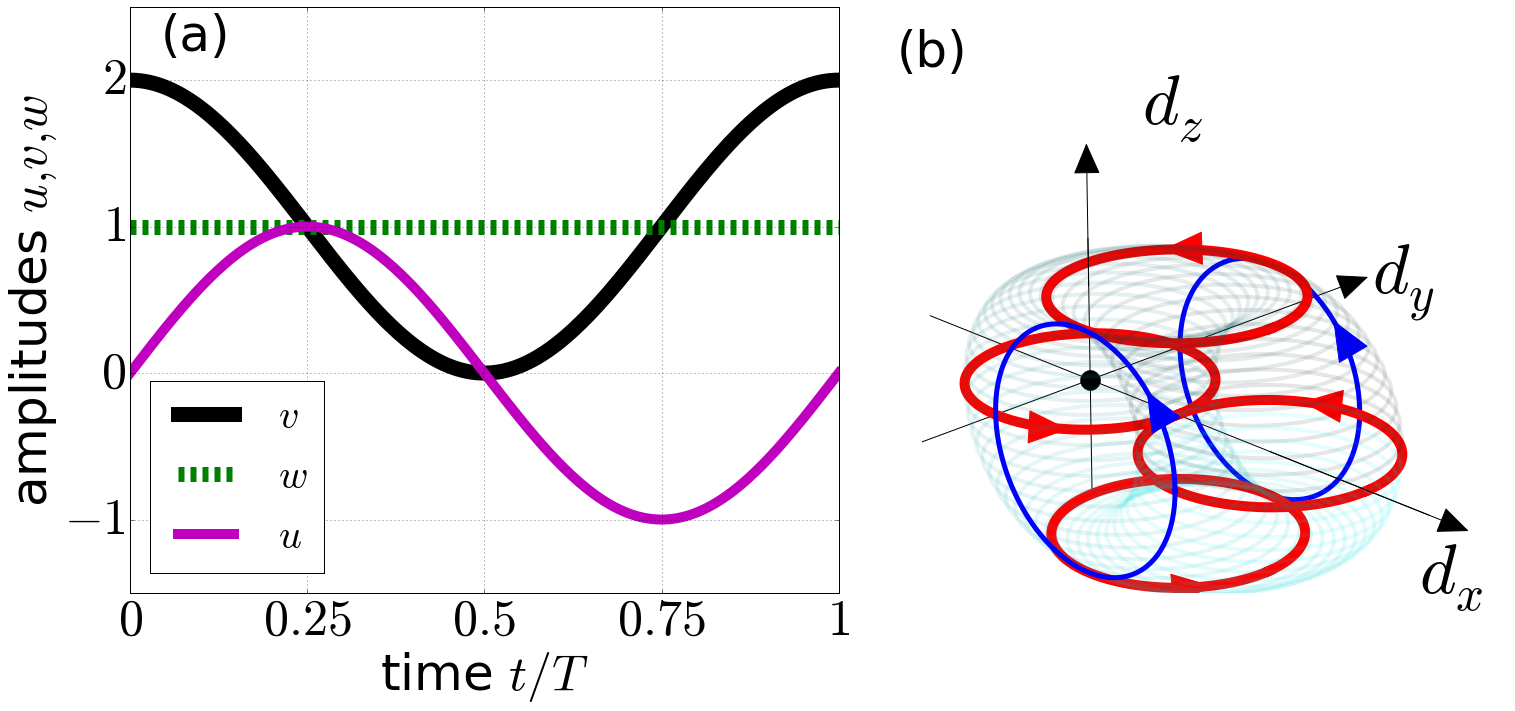}
\par\bigskip
{\centering
\includegraphics[width=0.85\columnwidth]
{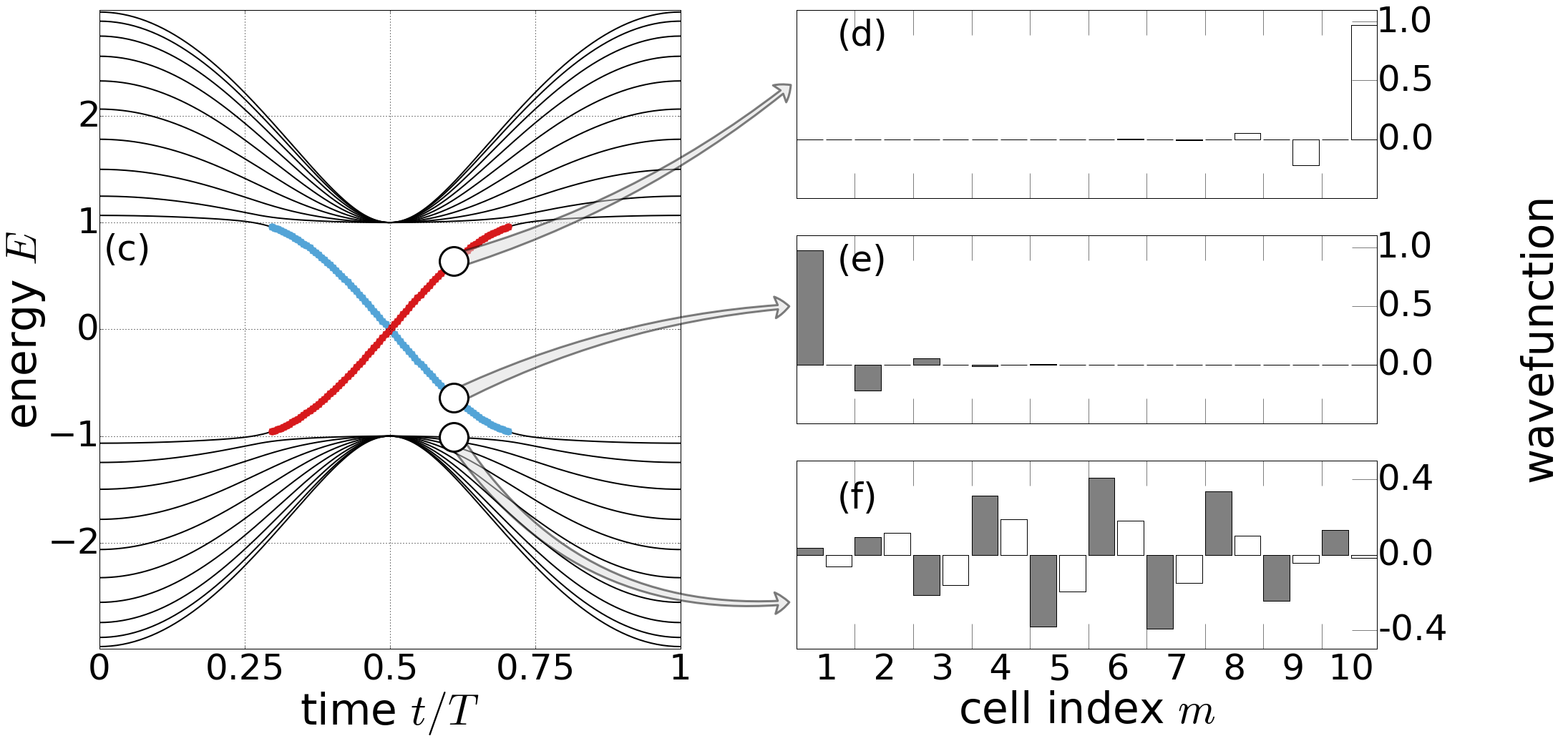}\par
}
\caption{The smooth pump sequence of the Rice-Mele model
for $\bar v = 1$. The hopping
  amplitudes and the sublattice potential (a) are varied smoothly as a
  function of time. The vector $\hh(k,t)$ corresponding to the bulk
  momentum-space Hamiltonian (b) traces out a torus in the
  3-dimensional space. Instantaneous spectrum of the Hamiltonian
  $\hat{H}(t)$ on an open chain of $N=10$ sites (c) reveals that
  during a cycle, one state crosses over to the upper band on the
  right edge, and one to the lower band on the left edge (dark
  red/light blue highlights energies of edge states, whose
  wavefunctions have than 60\% weight on the rightmost/leftmost 2 unit
  cells). The wavefunctions of the edge states (d,e) are exponentially
  localized to one edge and have support overwhelmingly on one
  sublattice each. In contrast a typical bulk state (f) has a delocalized
  wavefunction with support on both sublattices. 
\label{fig:ricemele-smooth_sequence}.}
\end{figure}

As an example for a generic periodically modulated insulator, we take
the Rice-Mele model, but we relax the control freak attitude. We consider a smooth modulation sequence, 
\begin{subequations}
\label{eq:ricemele_smooth_pump_sequence}
\begin{align}
\label{eq:ricemele_smooth_pump_sequence_u}
u(t) &= \sin \Omega t;\\
\label{eq:ricemele_smooth_pump_sequence_v}
v(t) &= \overline{v} + \cos \Omega t;\\
\label{eq:ricemele_smooth_pump_sequence_w}
w(t) &= 1,
\end{align}
\end{subequations}
where the sequence is fixed by choosing the average value of the
intracell hopping, $\overline{v}$. With $\overline{v}=1$, this
sequence can be obtained by an adiabatic deformation of the control
freak sequence. We show the smooth pump sequence and its representation in
the $\hh$ space for $\overline{v}=1$ in Fig.~\ref{fig:ricemele-smooth_sequence}.

\subsubsection*{Edge states in the instantaneous spectrum}



Consider the spectrum of the instantaneous energies on an open chain,
with an example for $N=20$ unit cells shown in
Fig.~\ref{fig:ricemele-smooth_sequence}. Since this charge sequence
was obtained by adiabatic deformation of the control freak sequence
above, each branch in the dispersion relation is deformed continuously
from a branch in Fig.~\ref{fig:ricemele-controlfreak-sequence}.

The edge states are no longer confined to a single unit cell, as in
the control freak case. However, as long as their energy lies deep in
the bulk band gap, they have wavefunctions that decay exponentially
towards the bulk, and so they can be unambiguously assigned to the
left or the right end. (In case of a degeneracy between edge states at
the right and left end, we might find a wavefunction with components
on both ends. In that case, however, restriction of that state to the
left/right end results in two seperate eigenstates, to a precision
that is exponentially high in the bulk length). In Fig.~ we used the
same simple criterion as in Chapt.~\ref{chap:ssh} to define edge
states:
\begin{align}
\ket{\Psi} \text{ is on the right edge } &\Leftrightarrow
\sum_{\uci=N-1}^N \left(\abs{\braket{\Psi}{\uci,A}}^2 
+ \abs{\braket{\Psi}{\uci,B}}^2 \right) > 0.6; \\
\ket{\Psi} \text{ is on the left edge } &\Leftrightarrow
\sum_{\uci=1}^2 \left(\abs{\braket{\Psi}{\uci,A}}^2 
+ \abs{\braket{\Psi}{\uci,B}}^2 \right) > 0.6.
\end{align}
As in the control freak case, there is a branch of energy eigenstates
crossing over from $E<0$ to $E>0$ at the right edge, and from $E>0$ to
$E<0$, at the left.

We can define the \emph{edge spectrum} to consist of edge state
branches of the dispersion relation, that are clearly assigned to the
right end. More precisely, we take two limiting energies,
$\varepsilon_-$ and $\varepsilon_+$, deep in the bulk gap, and only
consider energy eigenstates of the open chain with eigenvalues
$E_\eii(t)$ between these limits, $\varepsilon_- < E_\eii(t) <
\varepsilon_+$, with eigenstates localized at the right edge.  Each
edge state branch can begin (1) at $t=0$, as a continuation of another
(or the same) edge state branch ending at $t=T$, or (2) at
$E=\varepsilon_-$, or (3) at $E=\varepsilon_+$.  Each edge state
branch can end (1) at $t=T$, to then continue in another (or the same)
edge state branch at $t=0$, or (2) at $E=\varepsilon_-$, or (3) at
$E=\varepsilon_+$. That is 9 possible types of edge state
branches. Taking into account that the edge state spectrum, like the
total spectrum, has to be periodic in $t$, \emph{the number of edge
  state branches entering the energy range $\varepsilon_- < E <
  \varepsilon_+$ during a cycle is equal to the number of branches
  leaving it}. 

\subsubsection*{The net number of edge states pumped in energy is 
a topological invariant}

We now define an integer $Q$, that counts the number of edge states
pumped up in energy across at the right edge. Although this quantity
is not easily represented by a closed formula, it is straightforward
to read it off from the dispersion relation of an open system.  We
restrict our attention to the neighbourhood of an energy $\varepsilon$
deep in the bulk gap around $E=0$, such that at all points where
$E_n=\varepsilon$, the derivative $d E_n/dt$ does not vanish. Then
every edge state energy branch entering this neighborhood crosses
$E=\varepsilon$ either towards $E>\varepsilon$ or towards
$E<\varepsilon$. During one cycle, we define for the states at the
right edge
\begin{align}
N_+ &= \text{number of times $E=\varepsilon$ is crossed from $E<\varepsilon$ to
  $E>\varepsilon$};\\ 
N_- &= \text{number of times $E=\varepsilon$ is crossed from
  $E>\varepsilon$ to $E<\varepsilon$};\\ 
Q &= N_+ - N_- = \text{net number of edge states 
  pumped up in energy }.
\label{eq:ricemele_Q_def0}
\end{align}
Note that within the gap, $Q$ is independent of the choice of
$\varepsilon$. If we found a value $Q_0$ at $E=\varepsilon_0$, but a
different $Q_1 \neq Q_0$ at $E=\varepsilon_1>\varepsilon_0$, this
would require a net number $Q_0-Q_1$ of edge state branches at the
right edge to enter the energy region $\varepsilon_0<E<\varepsilon_1$
during a cycle but never exit it. Since both $\varepsilon_0$ and
$\varepsilon_1$ are deep in the bulk gap, away from the bulk bands,
this is not possible.

\begin{figure}
\includegraphics[width=0.9\linewidth]{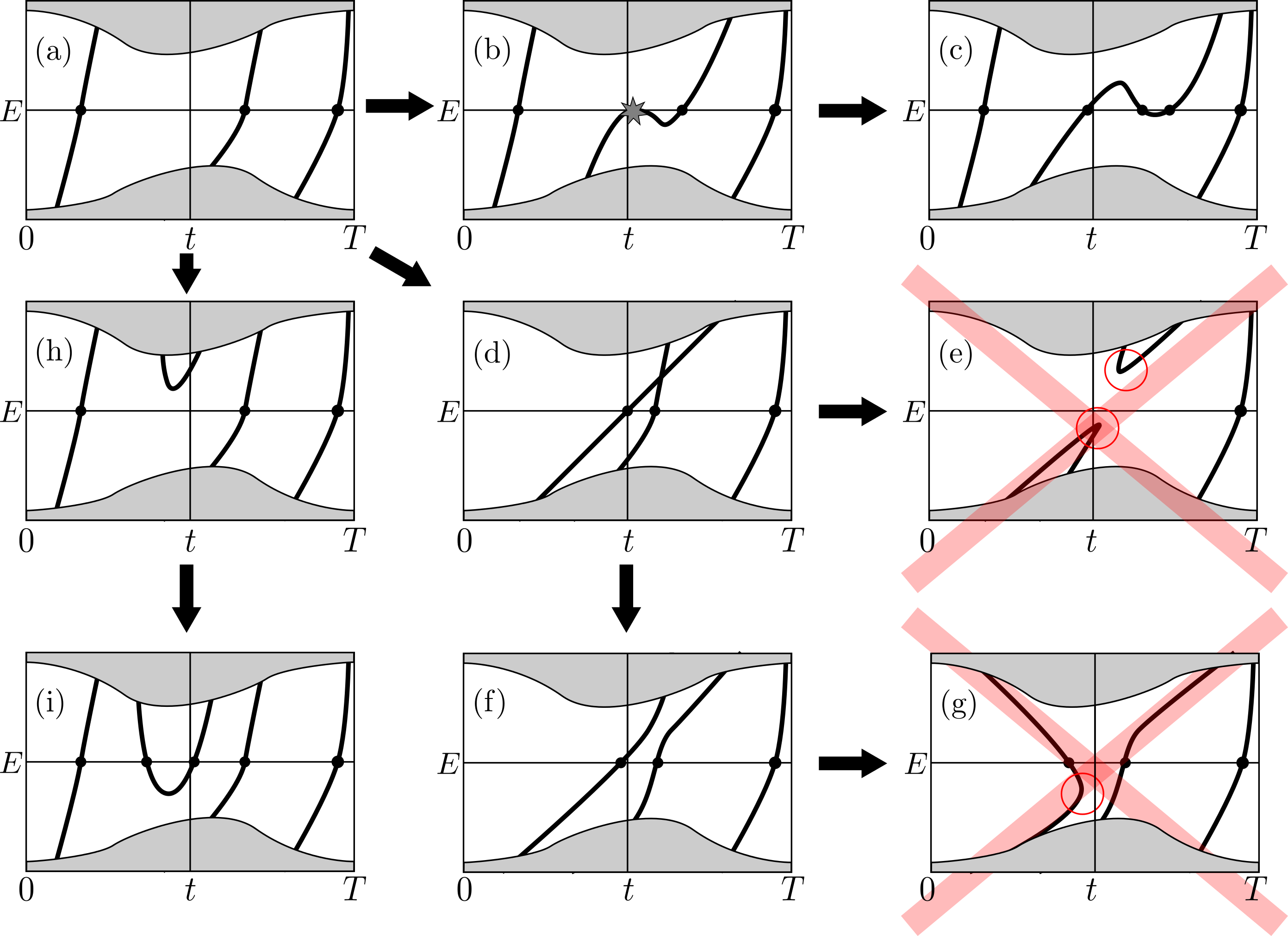}
\caption{ Adiabatic deformations of dispersion relations of edge
  states on one edge, in an energy window that is deep inside the bulk
  gap. Starting from a system with 3 copropagating edge states (a), an
  edge state's dispersion relation can develop a ``bump'',
  (b)-(c). This can change the number of edge states at a given energy
  (intersections of the branches with the horizontal line
  corresponding to the energy), but always by introducing new edge
  states pairwise, with opposite directions of propagation. Thus the
  signed sum of edge states remains unchanged. Alternatively, two edge
  states can develop a crossing, that because of possible coupling
  between the edge states turns into an avoided crossing (d),(f). This
  cannot open a gap between branches of the dispersion relation (e),
  as this would mean that the branches become multivalued functions of
  the wavenumber $k_x$ (indicating a discontinuity in $E(k_x)$, which
  is not possible for a system with short-range hoppings). Therefore,
  the signed sum of edge states is also unchanged by this process. One
  might think the signed sum of edge states can change if an edge
  state's direction of propagation changes under the adiabatic
  deformation, as in (g). However, this is also not possible, as it
  would also make a branch of the dispersion relation
  multivalued. Deformation of the Hamiltonian can also form a new edge
  state dispersion branch, as in (a)-(h)-(i), but because of periodic
  boundary conditions along $k_x$, this cannot change the signed sum
  of the number of edge states.
 \label{fig:chern_edge_states}}
\end{figure}

The net number of edge states pumped up inside the gap on the right
edge, $Q$, is a topological invariant: its value cannot change under
continuous deformations of the Hamiltonian $H(t)$ that preserve the
bulk gap.  This so-called topological protection is straightforward to
prove, by considering processes that might change this number. We do
this using Fig.~\ref{fig:chern_edge_states}.

The number of times edge state branches intersect $E=0$ can change
because new intersection points appear. These can form because an edge
state branch is deformed, and as a result, it gradually develops a
``bump'', local maximum, and the local maximum gets displaced from
$E<0$ to $E>0$. For a schematic example, see
Fig.~\ref{fig:chern_edge_states} 
(a)-(c).  Alternatively, the
dispersion relation branch of the edge state can also form a local
minimum, gradually displaced from $E>0$ to $E<0$. In both cases, the
number of intersections of the edge band with the $E=0$ line grows by
2, but the two new intersections must have opposite pump directions.
Therefore, both $N_+$ and $N_-$ increase by 1, but their difference,
$Q = N_+-N_-$, stays the same.

New intersection points can also arise because a new edge state branch
forms. As long as the bulk gap stays open, though, this new edge state
band has to be a deformed version of one of the bulk bands, as shown
in Fig.~\ref{fig:chern_edge_states} (a)-(h)-(i).  Because the periodic
boundary conditions must hold in the Brillouin zone, the dispersion
relation of the new edge state has to ``come from'' a bulk band and go
back to the same bulk band, or it can be detached from the bulk band,
and be entirely inside the gap. In both cases, the above argument
applies, and it has to intersect the $E=0$ line an even number of
times, with no change of $Q$.

The number of times edge state branches intersect $E=0$ can also
decrease if two edge state branches develop an energy gap. However, to
open an energy gap, the edge states have to be pumped in opposite
directions. For states pumped in the same direction, energy crossing
between them can become an avoided crossing, but no gap can be opened,
as this would violate the single-valuedness of a dispersion relation
branch, as illustrated in Fig.~\ref{fig:chern_edge_states} (d)-(f).
This same argument shows why it is not possible for an edge state to
change its direction of propagation under an adiabatic deformation
without developing a local maximum or minimum (which cases we already
considered above). As shown in Fig.~\ref{fig:chern_edge_states} (g),
this would entail that at some stage during the deformation the edge
state branch was not single valued.

\section{Tracking the charges with Wannier states}

Electrons in a solid are often described via Bloch states delocalized
over the whole lattice. 
As we have seen in Sect.~\ref{sec:polarization-wannier} though, 
one can represent a certain energy band with a set of Wannier states,
which inherit the spatial structure (discrete translational invariance)
of the lattice, and are well localized. 
Therefore, it seems possible to visualize the adiabatic pumping process
by following the adiabatic motion of the Wannier functions
as the parameters of the lattice Hamiltonian are varied in time. 

In fact, we will describe the adiabatic evolution of both the
position and energy expectation values of the Wannier functions.
By this, the toolbox for analyzing the adiabatic pumping procedure
for control-freak-type pumping  is extended to 
arbitrary pumping sequences.

\subsection{Plot the Wannier centers}

According to the result \eqref{eq:polarization-wannier_centers},
the Wannier center positions of a certain band, in units of the lattice
constant, are given by the Berry phase of that band
divided by $2\pi$. 
Hence, to follow the motion of the Wannier centers during the
pumping procedure, 
we calculate the Berry phase of the given band
for each moment of time. 

\begin{figure}
\centering
\includegraphics[width=0.47\columnwidth]
{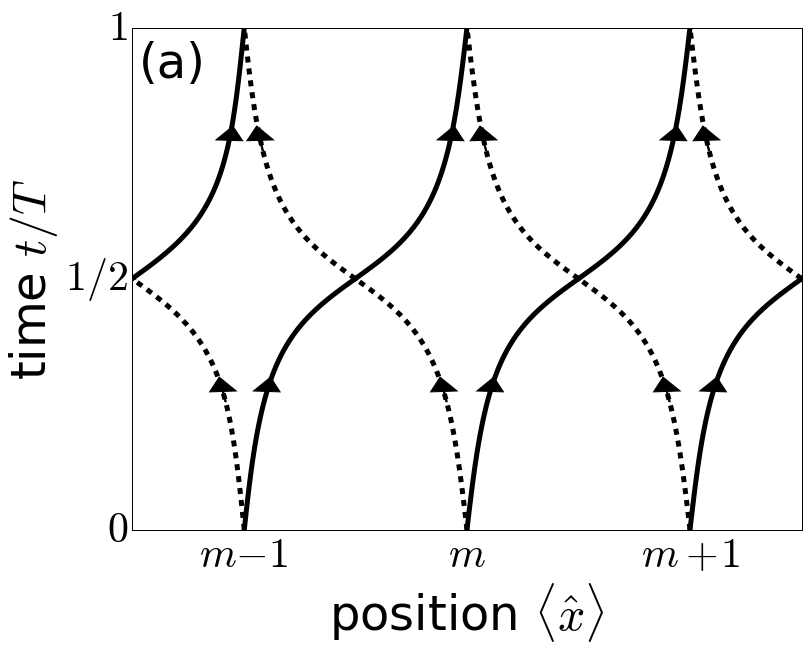} \hspace{0.2cm}
\includegraphics[width=0.47\columnwidth]
{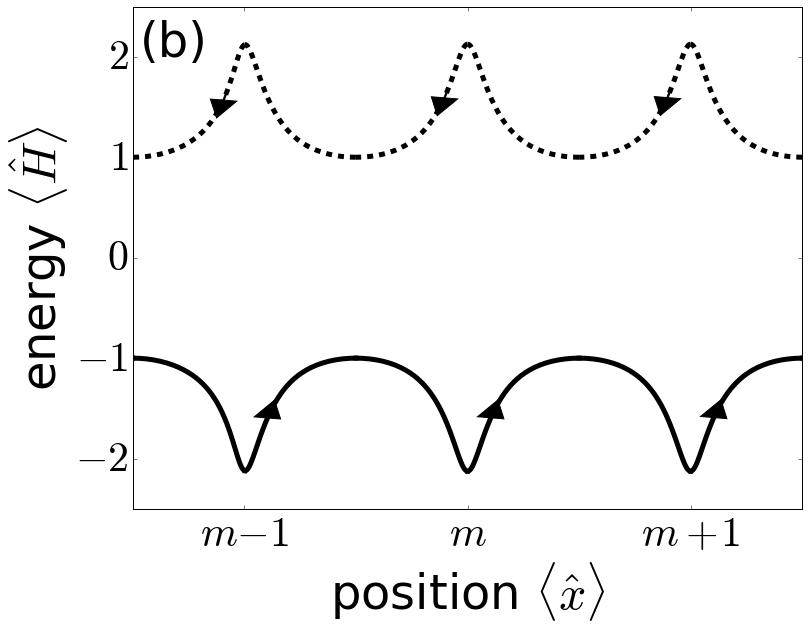}
\caption{Time evolution of Wannier centers and Wannier energies
in a smoothly modulated topological Rice-Mele pumping sequence
[defined via Eqs. \eqref{eq:thouless-general-d} and 
\eqref{eq:ricemele_controlfreak_pump_sequence}]. 
The parameter of the sequence is $\bar v =1 $, corresponding
to a Chern number of 1.
In both subfigures, a solid (dotted) line corresponds to the valence
(conduction) band. 
(a) Time evolution of the Wannier centers of the bands.
During a cycle, each Wannier center of the valence (conduction) band
moves to the right (left) with a single lattice constant. 
(b) Time evolution of the position and energy expectation 
value of a single Wannier center in the valence/conduction
band, over a few complete cycles. 
\label{fig:ricemele-smooth-sequence-motion}}
\end{figure}

The numerically computed Wannier-center positions
obtained for the smoothly modulated Rice-Mele
sequence
[defined via Eqs. \eqref{eq:thouless-general-d} and 
\eqref{eq:ricemele_smooth_pump_sequence}]. 
, with $\bar v = 1$, on a finite lattice, 
are shown in Fig. \ref{fig:ricemele-smooth-sequence-motion}a.
Solid (dotted) lines correspond to Wannier states of the 
valence (conduction) band. 
The results show that 
during a complete cycle, 
each Wannier center of the valence (conduction) band
moves to the right (left) with a single lattice constant. 
Figure \ref{fig:ricemele-smooth-sequence-motion}b shows
time evolution of the position and energy expectation 
value of a single Wannier center in the valence/conduction
band, over a few complete cycles. 
In complete analogy with  control-freak pumping, 
these results suggest that the considered pumping sequence
operates as a conveyor belt: it transports valence-band
electrons from left to right, with a speed of 
one lattice constant per cycle, 
and would transport conduction-band 
electrons, if they were present, from right to left, with the same pace. 

An important result, which is not specific to the considered
pumping cycle, arises from the above considerations.
As pumping is cyclic, the Berry phase of a given band
at $t=0$ is equivalent to that at $t=T$. 
Therefore, the displacement of the Wannier center
during a complete cycle is an integer.

\subsection{Number of pumped particles is the Chern number}

Our above results for the smoothly modulated Rice-Mele cycle
suggest the interpretation that during a complete cycle, 
each electron in the filled valence band is displaced to the right
by a single lattice constant.
From this interpetation, it follows that the number particles
pumped from left to right,
through an arbitrary cross section of the lattice, during a complete
cycle, is one.
Generalizing this consideration for arbitary
one-dimensional lattice models and pumping cycles, it
suggests that the number of pumped particles is an integer. 

We now shown that this integer is the Chern number
associated to the valence band, that is, 
to the ground-state manifold of the time-dependent
bulk momentum-space Hamiltonian $\ido{H}(k,t)$.
To prove this is, we first write the Wannier-center
displacement $\Delta x_{0,T}$
for the complete cycle by splitting 
up the cycle $[0,T]$ to small pieces $\Delta t$: 
\bean
\label{eq:thouless_deltax0T}
\Delta x_{0,T} = 
\lim_{n\to \infty } \sum_{i=0}^{n-1} \Delta x_{t_i,t_i+\Delta t},
\eean
where $\Delta t = T/n$ and $t_i = i \Delta t$. 
Then, we express the infinitesimal displacements
with the Berry phases, 
\bean
\Delta x(t_i,t_i+\Delta t) = \frac{i}{2\pi} 
\int_{-\pi}^{\pi} dk
\left[
	\braket{u_n(t_i+\Delta t)}{\partial_k u_n(t_i+\Delta t)} - 
	\braket{u_n(t_i)}{\partial_k u_n(t_i)}
\right],
\eean
where the $k$ argument is suppressed for brevity. 
Since the $k=-\pi$ and $k=\pi$ values are equivalent, 
the above integral can be considered as a
line integral of the Berry connection 
to the closed boundary line $\partial R_i$ of the infinitesimally
narrow rectangle 
 $(k,t) \in R_i = [-\pi,\pi) \times [t_i, t_i+\Delta t]$;
 that is, 
\bean
\Delta x(t_i,t_i+\Delta t) = \frac{1}{2\pi} 
\oint_{\partial R_i}
\AAA^{(n)} \cdot d\RR.
\eean
 Using the fact that we can choose a gauge that is 
 locally smooth 
on that  rectangle $R_i$,
we obtain 
\bean
\Delta x(t_i,t_i+\Delta t) = \frac{1}{2\pi} 
\int_{\partial R_i}
B^{(n)} dk dt, 
\eean
where $B^{(n)}$ is the Berry curvature 
associated to the $n$th eigenstate manifold of $\ido{H}(k,t)$.
Together with Eq. \eqref{eq:thouless_deltax0T}, this result ensures that 
the Wannier-center displacement is the Chern number:
\bean
\Delta x_{0,t} = 
\frac{1}{2\pi} 
\oint_{-\pi}^\pi
B^{(n)} dk dt.
\eean 

Note that even though we have not performed an explicit calculation
of the valence-band Chern number of the smooth Rice-Mele
pump cycle with $\bar v= 1$, by looking at the motion of the 
corresponding Wannier centers we can conclude that the 
Chern number is 1.

\subsection{Tuning the pump using the average intracell hopping amplitude $\overline{v}$}

\begin{figure}
\centering \includegraphics[width=0.85\columnwidth]
           {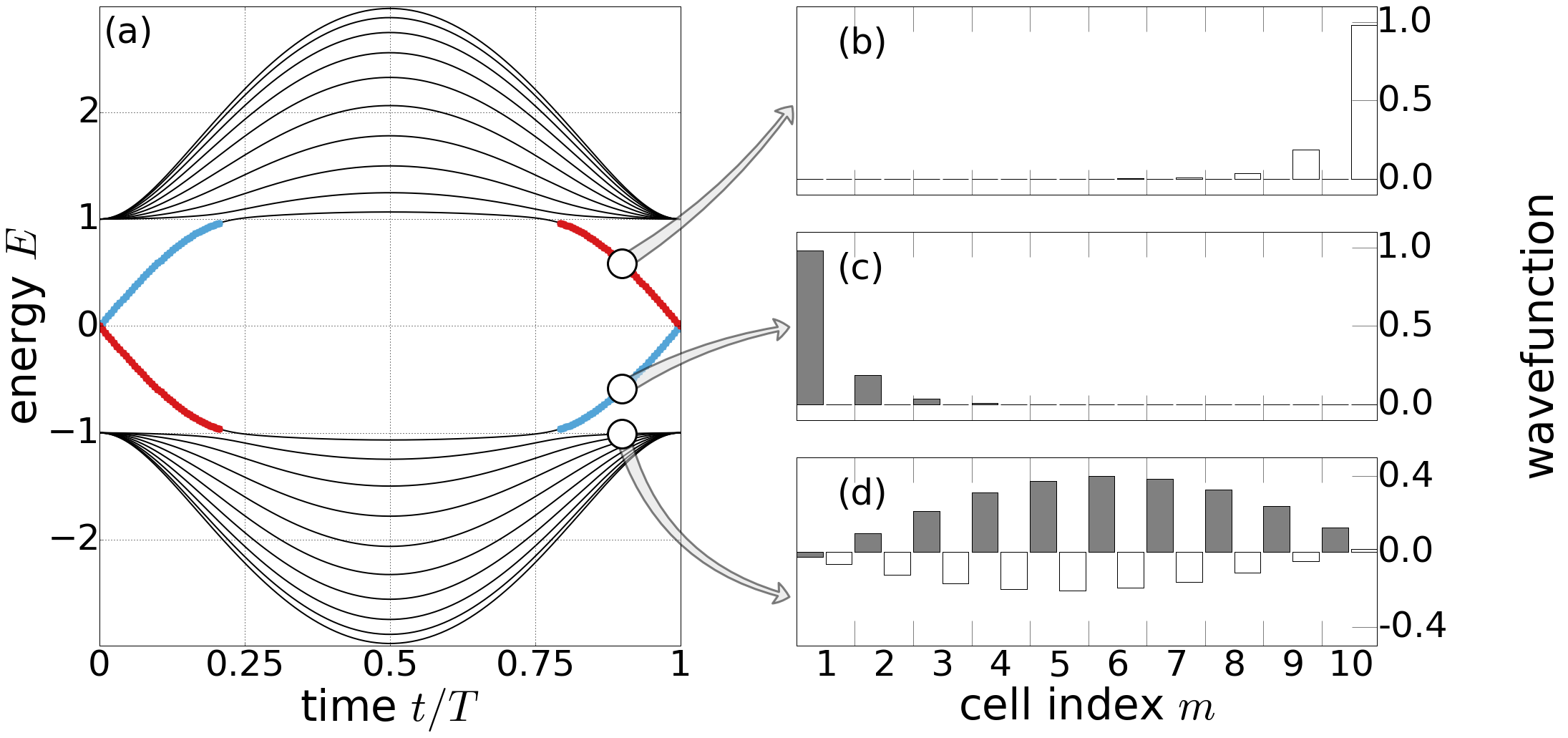}
\caption{
The smooth pump sequence of the Rice-Mele model for
$\bar \nu = -1$, revealing reversed pumping with respect to the
$\bar \nu =1$ case. 
  (a) Instantaneous spectrum of the Hamiltonian
  $\hat{H}(t)$ on an open chain of $N=10$ sites. 
  During a cycle, one edge state on the left (light blue)
  crosses over from the valence to the conduction 
  band, revealing that there is a single particle per cycle is pumped
  from right to left in the valence band. 
  Wavefunctions of the edge states (b,c) as well as a typical
  bulk state (d) are also shown. 
\label{fig:ricemele-smooth_dispersions_-1}}
\end{figure}

So far, the discussed results were obtained for the $\bar v=1$ special
case of the smoothly modulated Rice-Mele pumping cycle. 
Now we ask the question: 
can the number of pumped particles be changed
by tuning the parameter $\bar v$?
To show that the answer is affirmative and the pump
has such a tunability, 
on Fig. \ref{fig:ricemele-smooth_dispersions_-1}a
we plot the instantaneous energy spectrum corresponding 
to $\bar v = -1$. 
The spectrum reveals that this sequence, 
similarly to the $\bar v = 1$ case,  does pump a single particle
per cycle.
However, the direction of pumping is opposite in the two cases: 
Fig. \ref{fig:ricemele-smooth_dispersions_-1}a shows that
during a cycle, one edge state on the left (light blue)
crosses over from the valence band to the conduction 
band, revealing that the particles are
pumped from right to left in the valence band.

\subsection{Robustness against disorder}

So far, we have the following picture
of an adiabatic pump in a long open chain.
If we take a cross section at the middle of the chain, 
a single particle will be pumped through that, from left to right, 
during a complete cycle. 
This implies that at the end of the cycle, the number of particles
on the right side of the cross section has grown by one, therefore
the final state reached by the electron system is different from 
the original, ground state. 
This implies that during the course of the cycle,
a valence-band energy eigenstate 
deformed into a conduction-band state,
and that occured on the right edge of the chain;
the opposite happens on the left edge. 

Does that qualitative behavior change if we introduce
disorder in the edge regions of the open chain? 
No: as long as the bulk of the chain remains regular, 
the pump works at the middle of the chain, and therefore the above
conclusion about the exchange of a pair of states between
the valence and conduction bands still holds. 
On the other hand, introducing disorder in the bulk seems to
complicate the above-developed 
description of pumping in terms of Wannier-center motion,
and therefore might change number of edge states
and the qualitative nature of the instantaneous energy spectrum. 

\section*{Problems}
\addcontentsline{toc}{section}{Problems}
%
\begin{prob}
\label{prob:thouless-flat-controlfreak}
\textbf{Further control-freak pump cycles}\\
Construct a control-freak pump cycle 
where the spectrum of the bulk does not change during the entire 
cycle, and the pumped charge is (a) zero (b) one. 

\end{prob}



%% file: adiabatic_pumping_current.tex

\chapter{Current operator and particle pumping}
\label{chap:current}
 
\abstract*{
In the previous chapter, we described quantized adiabatic pumping 
of particles in a one-dimensional lattice in an intuitive and visual fashion,
using the concepts of the control-freak pumping cycle 
and the time evolution of the Wannier centers. 
Here, we provide a more formal description of the same
effect.
Based on Ehrenfest's theorem, we identify the current operator
describing the flow of probability density through a cross
section of the one-dimensional lattice, and find 
that the momentum- and time-resolved current in a given filled
band of the lattice is proportional to the Berry curvature associated
to that band. 
Naturally, this leads to the same conclusion as we have seen before: 
that the number of particles adiabatically pumped through
a cross section of the crystal is given by the Chern number of the
corresponding filled band, and therefore it is an integer. 
}


In the previous chapter, we described quantized adiabatic pumping 
of particles in a one-dimensional  lattice
in an intuitive and visual fashion,
using the concepts of the control-freak pumping cycle 
and the time evolution of the Wannier centers. 
Here, we provide a more formal description of the same
effect.
For simplicity, we consider 
two-band insulator lattice models with a completely filled 
lower band, 
which are  described by a periodically
 time-dependent bulk momentum-space
Hamiltonian of the form
\bean
\label{eq:current:twolevelh}
\ido{H}(k,t) = \hh(k,t) \cdot \ido{\ssigma},
\eean
where
$\hh(k,t)$ is a dimensionless three-dimensional vector fulfilling
$\hh(k,t) \geq 1$, and 
$\ido{\ssigma}$ is the vector of Pauli matrices. 
This Hamiltonian is periodic both in momentum  and in time, 
$\ido{H}(k+2\pi,t) = \ido{H}(k,t+T)=\ido{H}(k,t)$, where 
$T$ is the  period of the time dependence of the Hamiltonian.
The minimal energy gap 
between the two eigenstates of the Hamiltonian is $2$.
Furthermore, the frequency characterising the periodicity of the Hamiltonian
 is $\Omega \equiv 2\pi / T$. 
We call the periodically time-dependent Hamiltonian \emph{quasi-adiabatic}, 
if $\Omega \ll 1$, and 
the \emph{adiabatic limit} is defined as $\Omega \to 0$, that is, $T\to \infty$. 

For example, $\hh$ can be chosen as
\bean
\label{eq:current:dexample}
\hh (k,t) = \left(\bna{c}
	\bar v +\cos \Omega t +  \cos k \\
	\sin k \\
	\sin \Omega t
\eda\right),
\eean
corresponding to the smoothly modulated Rice-Mele model, see Eq. \eqref{eq:ricemele_smooth_pump_sequence}
and Eq. \eqref{eq:thouless-general-d}.

We will denote the eigenstate of $\ido{H}(k,t)$
with a lower (higher)  energy eigenvalue 
as $\ket{u_\ground (k,t)}$ ($\ket{u_\excited (k,t)}$).
With this notation, we can express the 
central result of this chapter:
the momentum- and time-resolved current 
carried by the electrons of the filled band 
equals the Berry curvature associated to that band. 
As a consequence, 
the number $\mathcal Q$
of particles pumped through an arbitrary cross section of an 
infinite one-dimensional crystal during a complete adiabatic cycle is
the momentum- and time integral of the Berry curvature, that is, 
\bean
\label{eq:centralresult}
\mathcal Q =  -i \frac{1}{2\pi}\int_0^T dt \int_{\rm BZ} dk 
\left(\partial_k \bra{u_\ground (k,t)} \partial_t u_\ground (k,t) \rangle - 
\partial_t \bra{u_\ground (k,t)} \partial_k u_\ground (k,t) \rangle\right).
\eean
This is the Chern number
associated to the ground-state manifold of $\ido{H}(k,t)$.
As the latter is an integer, the number of pumped particles is quantized.
This result was discovered by David Thouless \cite{Thouless}.

We derive Eq. \eqref{eq:centralresult} via the following steps.
In Sect.~\ref{sec:current:operator}, we consider
a generic time-dependent lattice
Hamiltonian, we express the number of particles 
moving through a cross section 
of the lattice using the current operator and the time-evolving 
 energy eigenstates of the lattice. 
Then, in Sect.~\ref{sec:current:quasiadiabatic}, 
we provide a description of the 
time-evolving energy eigenstates of the lattice  
in the case of periodic and quasi-adiabatic time dependence of the lattice 
Hamiltonian. 
This allows us to express the number of pumped particles 
for quasi-adiabatic time dependence.
Finally, in Sect.~\ref{sec:current:chern}, 
building on the latter result for quasi-adiabatic pumping,
we take the adiabatic limit and thereby establish
the connection between the current, the Berry curvature, 
 the number of pumped particles, and the Chern number.

\section{Particle current at a cross section of the lattice}
\label{sec:current:operator}

Our aim here is to express the number of particles
pumped through a cross section of the
lattice, assuming that the time evolution of the Bloch states due to the
time-dependence of the Hamiltonian is known. 
As intermediate steps toward this end, 
we derive the real-space current operator
and the diagonal matrix elements of the 
momentum-space current operator, 
and establish an important relation between those diagonal
matrix elements and the momentum-space Hamiltonian. 
For concreteness, we first discuss these using the example of 
the Rice-Mele model introduced in the preceding chapter.
It is straightforward to generalize the results for lattice models with 
an generic internal degree of freedom;
the generalized results are also given below. 
Finally, we use the relation between the current and the 
Hamiltonian to express the 
number of pumped particles
with the time-evolving states and the Hamiltonian.

\subsection{Current operator in the Rice-Mele model}
\label{sec:current:ricemele}

We consider the Rice-Mele model with $N\gg 1 $ unit cells and
periodic boundary conditions.
The real-space bulk Hamiltonian $\rso{H}_\textrm{bulk}$ 
has the almost the same form as Eq. \eqref{eq:ricemele_Hamiltonian_def},
with the difference that the sum corresponding to intercell
hopping runs up to $N$, 
and in accordance with the periodic boundary condition,  
the unit cell index $\uci$ should be understood 
as $(\uci \mod N)$.
The bulk momentum-space Hamiltonian $\ido{H}(k)$ of the model 
is given in Eq. \eqref{eq:thouless-general-d}.

\begin{figure}[!ht]
\begin{center}
\includegraphics[width=0.8\textwidth]{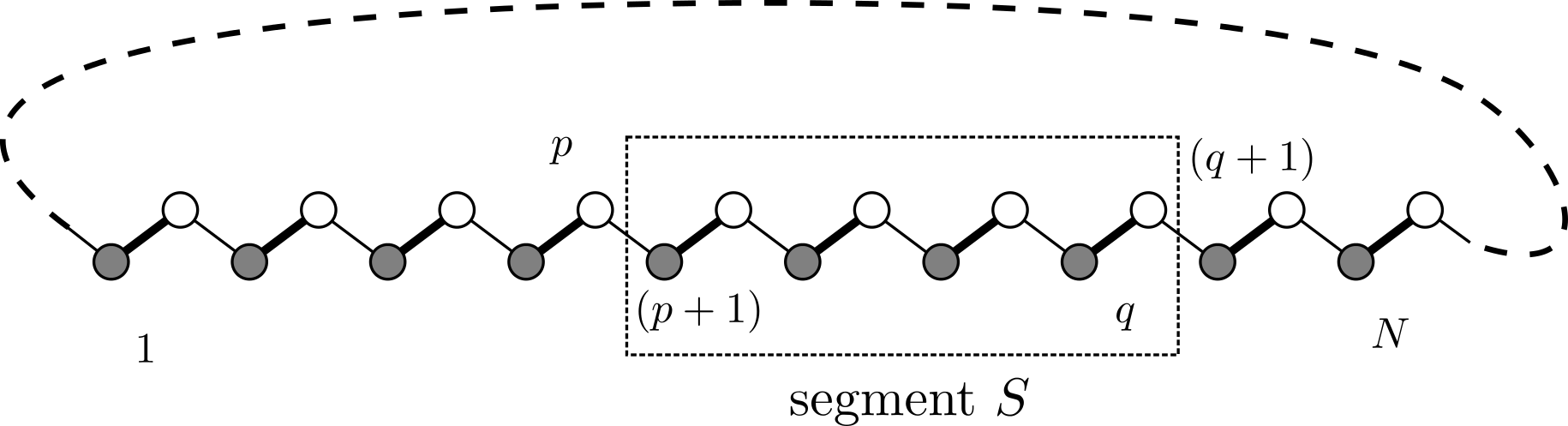}
\caption{\label{fig:current_segment}
A segment enclosed by two 
cross sections in the one-dimensional SSH model.
The segment $S$ is defined as the part of the chain 
between the $p$th and $(q+1)$th unit cells.
The current operators corresponding to the two 
cross sections can be established by considering
the temporal change of the number of particles
in the segment. 
The dashed line represents the periodic boundary condition.
} 
\end{center}
\end{figure}

\subsubsection*{Influx of particles into a segment of the crystal}

We aim at establishing the operator representing the 
particle current flowing through 
a cross section of the one-dimensional crystal. 
We take a cross section between the $p$th and $(p+1)$th unit cells,
and denote the corresponding current operator
as $\rso{j}_{p+1/2}$.
To find the current operator, we first consider a segment $S$ 
of the crystal, stretching between (and including) the 
$(p+1)$th and $q$th unit cells, where
$q\geq p+1$.
The number of particles in that segment $S$
is represented by the operator 
\bean
\rso{\mathcal Q}_S \equiv \sum_{\uci \in S}  \sum_{\idi \in \{A,B\}} \ket{\uci, \idi} \bra{\uci, \idi}.
\eean
Now, Ehrenfest's theorem ensures that 
the time-evolution of the number of particles embedded in the segment 
$S$
follows 
\bean
\partial_t \avg{\rso{\mathcal Q}_S}_t 
= -i \avg{[\rso{\mathcal Q}_S,\rso{H}(t)]}_t
\eean
Hence, we identify the operator describing 
the influx of particles into the  segment
as
\bean
\label{eq:current:jqh}
\rso{j}_S(t) = - i  [\rso{\mathcal Q}_S,\rso{H}(t)].
\eean
A straigthforward calculation shows that Eq. \eqref{eq:current:jqh}
implies 
\bean
\label{eq:jsegmentsimple}
\rso{j}_S(t) &=& 
-i  w(t) 
\left(
	\ket{p+1,A}\bra{p,B} - \ket{p,B}\bra{p+1,A}
	\right. \nonumber \\
	&+&
	\left. \ket{q,B}\bra{q+1,A} - \ket{q+1,A}\bra{q,B}
\right) 
\eean
Remarkably, the operator $\rso{j}_S(t)$ is  
time dependent, if the hopping amplitude $w(t)$ is time dependent.

\subsubsection*{Current operator at a cross section of the crystal}

Clearly, the terms in Eq. \eqref{eq:jsegmentsimple} can be separated 
into two groups: the first two terms are hopping operators
bridging the cross section $p+1/2$, and 
the last two terms are bridging the cross section $q+1/2$.
Thereby we define
\bean
\label{eq:current:1dmccurrentoperator}
\rso{j}_{\uci+1/2}(t) \equiv 
-i  w(t) 
\left(
	\ket{\uci+1,A}\bra{\uci,B} - \ket{\uci,B}\bra{\uci+1,A}
\right),
\eean
and use this definition to express $\rso{j}_S$ as 
\bean
\label{eq:current:jsegment}
\rso{j}_S = \rso{j}_{p+1/2} - \rso{j}_{q+1/2}.
\eean 
This relation allows us to interpret $\rso{j}_{\uci+1/2}(t)$ as the current operator
describing particle flow, from left to right, across the cross section $\uci+1/2$.

\subsubsection*{Relation of the current operator to the Hamiltonian
and to the group velocity}
  
Later we will need the momentum-diagonal matrix elements of 
the current operator, which are defined as 
\bean
\label{eq:current:momentumdiagonal}
\ido{j}_{\uci+1/2}(k,t) = \bra{k} \rso{j}_{\uci+1/2} (t) \ket{k}.
\eean
For the Rice-Mele model under consideration, 
these can be expressed using 
 Eqs. \eqref{eq:ssh_planewave_def} and
\eqref{eq:current:1dmccurrentoperator} as
\bean
\ido{j}_{\uci+1/2}(k,t)
 =  
\frac{1}{N} \ttmatrix{0}{-i w(t) e^{-ik}}{i w(t) e^{ik}}{0}.
\eean
From a comparison of this result and Eq. \eqref{eq:thouless-general-d},
we see that the momentum-diagonal matrix  elements of the
current operator are related to the momentum-space
Hamiltonian as 
\bean
\label{eq:currentvshamiltonian}
\ido{j}_{\uci+1/2}(k,t) = \frac{1}{N} \partial_k \ido{H}(k,t).
\eean
This is the central result of this section. 
Even though we have derived it only for the case of the 
Rice-Mele model, it is a generic result. 
A generalization is outlined in the next section. 

In the case of a lattice without an internal degree of freedom,
it is easy to see that 
 Eq. \eqref{eq:currentvshamiltonian}  establishes the
equivalence
$j_{m+1/2} (k,t) = v_k(t)/N$
between the current 
and the (instantaneous) group velocity $v_k(t)$:
in this case, the bulk momentum-space Hamiltonian $\ido{H}(k,t)$
equals the dispersion relation $E(k,t)$, 
and its momentum derivative $v_k(t) = \frac{\partial E(k,t)}{\partial k}$ 
is defined as  
the group velocity of the energy eigenstates.
This correspondence generalizes to lattices with an internal degree of
freedom as well.
The current carried by an instantaneous energy eigenstate 
$\ket{\Psi_n(k,t)} = \ket{k} \otimes \ket{u_n(k,t)}$
of such a lattice is
\bean
\bra{u_n} \ido{j}_{\uci+1/2} \ket{u_n} &=& 
\frac{1}{N}
\bra{u_n} \left[\partial_k \ido{H} \right] \ket{u_n} \nonumber \\
&=&
\frac 1 N \bra{u_n} \left[\partial_k \sum_{n'} E_{n'} \ket{u_{n'}} \bra{u_{n'}} \right] \ket{u_n} \nonumber \\
 &=&
\frac 1 N\left\{ \bra{u_n} \left[
\left(\partial_k  E_{n}\right) \ket{u_{n}} \bra{u_{n}} 
+
 E_{n}   \ket{ \partial_k u_{n}}   \bra{u_{n}} 
 +
  E_{n} \ket{u_{n}}   \bra{\partial_k u_{n}} 
\right] \ket{u_n} \right\} \nonumber \\ 
&=& 
\frac 1 N \left(
\partial_k E_n + E_n \partial_k \bra{u_n} u_n \rangle \right)
= \frac{\partial_k E_n} N 
 = \frac{v_{n,k}}{N} \label{eq:current:currentvsgroupvelocity},
\eean
where the arguments $(k,t)$ are suppressed for brevity.


\subsection{Current operator in a generic one-dimensional lattice model}
\label{sec:currentgeneric}

Here we prove Eq. \eqref{eq:currentvshamiltonian}
in a more general setting. 
Previously, we focused on the Rice-Mele model that
has only two bands and hopping only between nearest-neighbor cells. 
Consider now a  general one-dimensional lattice model with 
$N_b$ bands and finite-range hopping with range $1 \leq r \ll N$;
here, $r=1$ corresponds to hopping between neighbouring 
unit cells only.
As before, we take a long chain with $N \gg 1$ unit cells, and assume 
periodic boundary conditions. 

The real-space Hamiltonian has the form
\bean
\rso{H}(t) = \sum_{\uci,\uci' =1}^N \sum_{\idi,\idi' =  1}^{N_b} H_{\uci \idi, \uci' \idi'}(t) \ket{\uci,\idi}\bra{\uci' \idi'},
\eean
where $\uci$ and $\uci'$ are unit cell indices, and
$\idi, \idi'  \in \{1,2,\dots,N_b\}$ correspond to the
internal degree of freedom within the unit cell. 
Due to the finite-range-hopping assumption, the Hamiltonian
can also be written as
\bean
\rso{H}(t) = \sum_{\uci =1}^N \sum_{i=-r}^r \sum_{\alpha,\alpha'= 1}^{N_b} H_{\uci+i,\idi;\uci,\idi'}(t) \ket{\uci+i,\idi}\bra{\uci \idi'}.
\eean
Again, a unit cell index $\uci$ should be understood 
as $(\uci \mod N)$.
Also, due to the discrete translational invariance, we have
$H_{\uci+i,\idi;\uci,\idi'} = H_{i \idi; 0, \idi'}$, 
implying
\bean
\rso{H}(t) = \sum_{\uci =1}^N \sum_{i =-r}^r \sum_{\idi,\idi' = 1}^{N_b} H_{i, \idi; 0\idi'}(t) \ket{\uci+i,\idi}\bra{\uci \idi'}
\eean
The bulk momentum-space Hamiltonian then 
reads 
\bean
\label{eq:current:generalizedHk}
\ido{H}(k,t) \equiv
\bra{k } \rso{H}(t) \ket{k} =
\sum_{\uci=-r}^r H_{\uci,\idi;0,\idi'}(t) e^{-ikm} \ket{\alpha}\bra{\alpha'}.
\eean

Next, we establish the operator representing the 
particle current flowing through 
a cross section of the one-dimensional crystal,
the same way we did in Sect.~\ref{sec:current:ricemele}. 
We take a cross section between the $p$th and $(p+1)$th unit cells,
and denote the corresponding current operator
as $\rso{j}_{p+1/2}$.
To find the current operator, we first consider a segment $S$
of the crystal, stretching between (and including) the 
$(p+1)$th and $q$th unit cells, where $q-p \geq r$.
The number of particles embedded in that long segment 
is represented by the operator 
\bean
\rso{\mathcal Q}_S \equiv  \sum_{\uci \in S} \sum_{\idi=1}^{N_b} \ket{\uci \idi} 
\bra{\uci \idi}.
\eean
As discussed in the preceding section, we identify the operator describing 
the influx of particles into the wire segment
as
\bean
\rso{j}_S(t) = - i  [\rso{\mathcal Q}_S,\rso{H}(t)].
\eean
From this, a straigthforward calculation shows that
\bean
&\rso{j}_S(t) = 
-{i} \sum_{\uci \in S} \sum_{\uci' \notin S} \sum_{\idi,\idi'=1}^{N_b}
 &\left[ H_{\uci \idi,\uci' \idi'}(t) \ket{\uci \idi}\bra{\uci' \idi'} \right. 
 \nonumber
 \\
&&-
\left.
H_{\uci' \idi',\uci \idi}(t) \ket{\uci' \idi'}\bra{\uci \idi}\right].
\label{eq:jsegment}
\eean

Note that Eq. \eqref{eq:jsegment} testifies that 
the operator $\rso{j}_S(t)$ is 
constructed only from those hopping matrix elements of the Hamiltonian 
that bridge either the $p+1/2$ or the $q+1/2$ cross sections of the
crystal, i.e., one of the two cross sections that terminate the 
segment under
consideration.
This is ensured by the condition that the segment is at least
as long as the range $r$ of  hopping. 
A further consequence of this is that
the terms in Eq. \eqref{eq:jsegment} can be separated 
into two groups: one containing the hopping matrix elements
bridging the cross section $p+1/2$, and 
one with those bridging the cross section $q+1/2$.
The former reads
\bean
\label{eq:current:generalized}
&\rso{j}_{p+1/2}(t) = 
- i
\sum_{\uci=p+1}^{p+r} 
\sum_{\uci'= p+1-r}^p  \sum_{\idi,\idi' =1}^{N_b}&
\left[
H_{\uci \idi,\uci' \idi'}(t) \ket{\uci \idi}\bra{\uci' \idi'} \right.\\
&&- \left.
H_{\uci' \idi',\uci \idi}(t) \ket{\uci' \idi'}\bra{\uci \idi}\right].
\nonumber
\eean
Using this as a definition for any cross section
$\uci+1/2$, 
we conclude that Eq. \eqref{eq:current:jsegment}
holds without any change in this generalized case as well.
This conclusion 
allows us to interpret $\rso{j}_{\uci+1/2}$ as the current operator
describing particle flow, from left to right, across the cross section $n+1/2$.

After defining the momentum-diagonal matrix elements
of the current operator exactly the same way 
as in Eq. \eqref{eq:current:momentumdiagonal},
the relation between the current operator and 
the Hamiltonian has exactly the same form as
in Eq. \eqref{eq:currentvshamiltonian}.
This can be proven straightforwardly 
using
Eqs. \eqref{eq:current:momentumdiagonal},
\eqref{eq:current:generalized},
and the $k$-derivative of Eq. \eqref{eq:current:generalizedHk}.

\newcommand{\mestate}{\Phi} 
\newcommand{\sestate}{\Psi} 

\subsection{Number of pumped particles}

Let us return to particle pumping in 
insulating two-band models. 
As the electrons are assumed to be non-interacting, 
and the time-dependent 
lattice Hamiltonian has a discrete translational invariance for all
times, 
the many-electron state $\mestate(t)$ is a Slater determinant of
Bloch-type single-particle states 
$\ket{\tilde{\sestate}_{1}(k,t)} = \ket{k} \otimes \ket{\tilde{u}_1(k,t)}$.
Here, we adopted the notation introduced in Sect.~\ref{sec:ssh-bulk},
with $n=1$ referring to the filled band. 
There are also two additions with respect to the notation of 
Sect.~\ref{sec:ssh-bulk}: 
first, we explicitly denote the time dependence of the state;
second, we added a tilde here to denote that the state 
$\ket{\tilde{\sestate}_1(k,t)}$ is not an
instantaneous lower-band eigenstate 
$\ket{\sestate_{1}(k,t)}$ of the Hamiltonian, but is
slightly different from that due to the quasi-adiabatic driving.

The number of particles pumped through the cross section $\uci+1/2$
within the time interval $t \in [0,T]$ is 
the time-integrated current, that is, 
\bean
\label{eq:current_Q1}
\mathcal Q = \int_0^T dt \bra{\mestate(t) } j^M_{\uci+1/2}(t) \ket{\mestate(t)},
\eean
where $j^M_{\uci+1/2}(t)$ is the many-particle generalisation 
of the current operator defined in 
Eq. \eqref{eq:current:generalized},
or, for the special case of the Rice-Mele model, in 
Eq. \eqref{eq:current:1dmccurrentoperator}),
and
$\mestate(t)$ is the many-electron Slater determinant
formed by the filled Bloch-type single-particle
states, introduced in the preceding paragraph. 
Equation \eqref{eq:current_Q1} can be converted to an
expression with single-particle states:
\bean
\label{eq:current:qrealspacej}
\mathcal Q = \int_0^T dt \sum_{k \in {\rm BZ}} \bra{\tilde{\sestate}_1(k,t)} \rso{j}_{\uci+1/2}(t) \ket{\tilde{\sestate}_1(k,t)},
\eean
which is related to the momentum-diagonal 
matrix elements of the current operator 
as
\bean
\mathcal Q = \int_0^T dt \sum_{k \in {\rm BZ}} 
\bra{\tilde{u}_1(k,t)} \ido{j}_{\uci+1/2}(k,t) \ket{\tilde{u}_1(k,t)}.
\eean
Finally, this is rewritten using the
current-Hamiltonian relation 
Eq. \eqref{eq:currentvshamiltonian} as
\bean
\label{eq:Qquasiadiabatic}
\mathcal Q =  \frac{1}{N}
\int_0^T dt \sum_{k \in {\rm BZ}} \bra{\tilde{u}_1(k,t)} \partial_k \ido{H}(k,t)
\ket{\tilde{u}_1(k,t)}.
\eean
To evaluate this in the case of adiabatic, periodically time-dependent
Hamiltonian, we first need to understand how the 
two-level wave functions $\ket{\tilde{u}_1(k,t)}$ 
evolve in time in the quasi-adiabatic case;
then we can insert those in Eq. \eqref{eq:Qquasiadiabatic},
and take the adiabatic limit.

\section{Time evolution governed by a quasi-adiabatic Hamiltonian}
\label{sec:current:quasiadiabatic}

Our goal here is to describe the time evolution 
of Bloch-type electronic energy eigenstates.
Nevertheless, as the pumping dynamics preserves the 
wavenumber $k$, the task simplifies to describe 
the dynamics of distinct two-level systems,
labelled by the wavenumber $k$. 
Therefore, in this section we discuss the dynamics 
of a single two-level system, hence the wavenumber $k$
does not appear in the formulas. 
We will restore $k$ when evaluating the number of pumped particles
in the next section.  
To describe the time evolution of the electronic states
subject to quasi-adiabatic driving, 
it is convenient to use the so-called \emph{parallel-transport gauge}
or \emph{parallel-transport time parametrization},
which we introduce below. 
Then, our goal 
is reached by 
performing perturbation theory in the small frequency $\Omega\ll 1$
characterizing the quasi-adiabatic driving.

\subsection{The parallel-transport time parametrization}
\label{sec:current:paralleltransport}

As mentioned earlier, the instantaneous energy eigenstates
of the bulk momentum-space Hamiltonian $\ido{H}(k,t)$ 
are denoted as $\ket{u_n(k,t)}$.
Here, in order to simplify the derivations,
we will use a special time parametrization (gauge)
for these eigenstates,
which is called the 
\emph{parallel-transport time parametrization}
or
\emph{parallel-transport gauge}.
As mentioned above, we suppress the 
momentum $k$.

We will call the smooth time parametrization $\ket{u_n(t)}$ 
of the instantaneous $n$th eigenstate of the Hamiltonian
$\ido{H}(t)$ a \emph{parallel-transport time parametrization}, 
if 
for any time point $t$ and any band $n$, 
it holds that
\bean
\label{eq:current:ptdef}
\bra{u_n(t)} \partial_t \ket{u_n(t)}=0.
\eean
Using a time parametrization with this property will simplify the
upcoming calculations of this section. 

In Sect.~\ref{sec::berry_chern-adiabatic_phases},
we used smooth parametrizations that were defined 
via a parameter space, and therefore were cyclic.
For any time parametrization $\ket{u'_n(t)}$ having those properties, 
we can construct a parallel-transport
time parametrizaton $\ket{u_n(t)}$ via the definition
\bean
\label{eq:current:ptconstruction}
\ket{u_n(t)} =
e^{i\gamma_n(t)}  \ket{u'_n(t)},
\eean
where $\gamma_n(t)$ is the adiabatic phase associated to 
the adiabatic time evolution of  
the initial state $\ket{u'_n(t=0)}$, governed by our 
adiabatically varying Hamiltonian
$\ido{H}(t)$: 
\bean
\gamma_n(t) = i \int_0^{t} dt' \bra{u'_n(t')} \partial_{t'} \ket{u'_n(t')}.
\eean 
The fact that $\ket{u_n(t)}$ indeed fulfils Eq. \eqref{eq:current:ptdef}
can be checked by performing the time derivation and the
scalar product on the left hand side of the latter. 

As an interpretation of Eq. \eqref{eq:current:ptconstruction},
we can  say that a parallel-transport time parametrization is
an adiabatically time-evolving state without the dynamical phase factor.
Furthermore, as the Berry phase factor
$e^{i\gamma_n(T)}$ is, in general, different from $1$, 
the parallel-transport time parametrization \eqref{eq:current:ptconstruction}
is, in general, not cyclic.

\subsection{Quasi-adiabatic evolution}
\label{sec:current:quasiadiabatic}

Here, following Thouless \cite{Thouless}, 
we describe the quasi-adiabatic time evolution
using \emph{stationary states}, also known as \emph{Floquet states},
that are characteristic of periodically driven quantum systems. 
Again, we focus on two-level systems as introduced in 
Eq. \eqref{eq:current:twolevelh}, 
and suppress the wave number $k$ in our notation. 
The central result of this section is Eq. \eqref{eq:psitresult}, 
which expresses how the instantaneous ground state
mixes weakly with the instantaneous excited state 
due to the quasi-adiabatic time dependence of the Hamiltonian.
In the next section, this result is used to evaluate the 
particle current and the number of pumped particles.

As an example, we can consider the state
corresponding to the wavenumber $k=0$ in the smoothly
modulated Rice-Mele model with $\bar v = 1$, 
see Eq. \eqref{eq:current:dexample}:
\bean
\label{eq:ricemeleexample}
\hh(t) = \left(\bna{c}
2+\cos \Omega t\\
0 \\
\sin (\Omega t)
\eda\right).
\eean

\subsubsection*{Stationary states of periodically driven dynamics}

The stationary states
are special solutions of the periodically 
time-dependent Schr\"odinger equation, which 
 are essentially periodic with period $T$;
that is, which fulfill $\ket{\psi(t+T)} = e^{-i \phi}\ket{\psi(t)}$ for any $t$,
with $\phi$ being a $t$-independent real number. 
The number of such nonequivalent solutions equals the dimension of the 
Hilbert space of the quantum system, i.e., there are two of them for
the case we consider. 
Here we describe stationary states in the quasi-adiabatic case, 
when the time evolution of the Hamiltonian is slow compared
to the energy gap between the instantaneous energy eigenvalues:
$\Omega \ll 1$. 
This condition suggest that the deviation from the adiabatic
dynamics is small, and therefore each stationary state is
in the close vicinity of either the instantaneous ground state
or the instantaneous excited state. 
Thereby, we will label the stationary states with the band index $n$, 
and denote them as $\ket{\tilde{u}_n(t)}$. 

Since, after all, we wish to describe pumping in a lattice
with a filled lower band and an empty upper band,
we mostly care about the stationary state corresponding to the 
lower band, 
$\ket{\tilde{u}_1(t)}$, and therefore want to solve 
the time-dependent Schr\"odinger equation
\bean
\label{eq:tdse}
-i \partial_t \ket{\tilde{u}_1(t)} + \ido{H}(t) \ket{\tilde{u}_1(t)} = 0.
\eean

%

\subsubsection*{Making use of the parallel-transport gauge}

We characterize the time evolution of the wave function $\ket{\tilde{u}_1(t)}$ 
by a time-dependent linear combination of the instantaneous energy 
eigenstates:
\bnen
\label{eq:psit}
\ket{\tilde{u}_1(t)} = 
a_1(t) e^{- i \int_0^t dt' E_{\ground}(t')} \ket{u_{\ground}(t)}
+ a_{2}(t) e^{-i \int_0^t dt' E_{\excited}(t')} \ket{u_\excited(t)},
\eden
Recall that we are using the parallel-transport time parametrization, 
having the properties \eqref{eq:current:ptdef} and 
\eqref{eq:current:ptconstruction}.
Therefore, in the adiabatic limit $\Omega \to 0$, we already now that 
$a_1(t) = 1$ and $a_2(t) = 0$. 
Here, we are mostly interested in the quasi-adiabatic 
case defined via $\Omega \ll 1$,
and then it is expected that 
$a_1(t) \sim 1$ and $a_{2}(t) \sim \Omega  \ll 1$.

Before making use of that consideration in the form of 
perturbation theory in $\Omega$, we 
convert the time-dependent Schr\"odinger equation \eqref{eq:tdse}
to two differential equations for the two unknown functions
$a_1(t)$ and $a_2(t)$. 
We insert $\ket{\tilde{u}_1(t)}$ 
of Eq. \eqref{eq:psit} to the time-dependent Schrodinger equation
\eqref{eq:tdse},
yielding
\bean
-i \partial_t \left[ a_1(t)
e^{-i\int_0^t dt' E_{1}(t')} \ket{u_{\ground}(t')}
+a_{2}(t) e^{-i\int_0^t dt' E_{2}(t')} \ket{u_{\excited}(t')}\right] &&
\\
\nonumber
+
\ido{H}(t) 
\left[ a_1(t)
e^{-i\int_0^t dt' E_{1}(t')} \ket{u_{\ground}(t)}
+a_{2}(t) e^{-i\int_0^t dt' E_{2}(t')} \ket{u_{\excited}(t)}\right] &=& 0
\eean
After evaluating the time derivatives, the left hand side 
consists of 8 terms.
Using the instantaneous eigenvalue relations
$\ido{H}(t) \ket{u_n(t)} = E_n(t) \ket{u_n(t)}$, two pairs 
of terms annihilate each other, and
only 4 terms remain:
\begin{multline}
\label{eq:tdse2}
\dot a_1(t)  \ket{u_{\ground}(t)}
+
a_1(t) \partial_t \ket{u_{\ground}(t)}
+
\dot a_{2}(t) e^{-i\int_0^t dt' E(t')} \ket{u_{\excited}(t)} \\
+
 a_{2}(t) e^{-i\int_0^t dt' E(t')} \partial_t \ket{u_{\excited}(t)}
=0,
\end{multline}
where $E(t) = E_2(t)-E_1(t)=2 \hhabs(t)$.

Projecting Eq. \eqref{eq:tdse2} onto 
$\bra{u_{\ground}(t)}$ and $\bra{u_{\excited}(t)}$, respectively, 
and making use of the parallel-transport gauge, yields
\bean
\label{eq:dota1}
\dot a_1(t)
+
a_{2}(t) e^{-i\int_0^t dt' E(t')} 
\bra{u_{\ground}(t)} \partial_t \ket{ u_{\excited}(t) }
 &=&0, \\
a_1(t) \bra{u_{\excited}(t)} \partial_t \ket{u_{\ground}(t)}
+
\dot a_{2}(t) e^{-i\int_0^t dt' E(t')} 
&=&0.
\eean
The latter result can be rewritten as
\bean
\label{eq:dota2}
\dot a_2(t) = -a_1(t) 
\bra{u_{\excited}(t)} \partial_t \ket{u_{\ground}(t)} e^{i\int_0^t dt' E(t')} 
\eean

\subsubsection*{Making use of the quasi-adiabatic condition}

Note that the quasi-adiabatic condition has not been invoked so far;
this is the next step.
As mentioned above, the quasi-adiabatic nature of the Hamiltonian
suggests that one of the two stationary states will be close to the instantaneous
ground state, suggesting $a_1(t) \sim 1$ and 
$a_2(t)\sim \Omega$. 
Furthermore, we know that 
$\bra{u_{\ground}(t)} \partial_t \ket{u_{\excited}(t)} \sim \Omega$, since variations in $\ket{u_\eii(t)}$ become slower as the adiabatic limit is approached. 
The latter relation is explicitly demonstrated by the example in
Eq. \eqref{eq:ricemeleexample}:
if we use $\ket{u_{\ground}(t)} = \left(-\sin (\theta/2), \cos(\theta/2)\right)^T$
and 
$\ket{u_{\excited}(t)} = \left(\cos (\theta/2), \sin(\theta/2)\right)^T$
with $\theta = \arctan\left(\frac{2+\cos \Omega t}{\sin \Omega t }\right)$, 
fulfilling the parallel-gauge criterion,
we find
$\bra{u_{\ground}(t)} \partial_t \ket{u_{\excited}(t)}=
- \Omega \frac{1+2\cos \Omega t}{10+8 \cos \Omega t}$.

As we are interested in the quasi-adiabatic case $\Omega \ll 1$,  
we drop those terms from 
\eqref{eq:dota1} and \eqref{eq:dota2} 
that are at least second order in $\Omega$. 
This results in 
\bean
\dot a_1(t) &=& 0, \\
\dot a_2(t) &=& -a_1(t) \bra{u_{\excited}(t)} \partial_t \ket{u_\ground(t)}
 e^{i\int_0^t dt' E(t')}.
\label{eq:a2}
\eean
If  we assume $a_1(t=0) = 1 + o(\Omega)$, 
then the first equation guarantees that $a_1(t) = 1 + o(\Omega)$.
Then this allows for a further simplification of Eq. \eqref{eq:a2}:
\bean
\label{eq:quasiadiabaticdota2}
\dot a_2(t) &=& -\bra{u_{\excited}(t)} \partial_t \ket{ u_{\ground}(t) } e^{i\int_0^t dt' E(t')}.
\eean

\subsubsection*{Solution of the equation of motion}

The remaining task is to solve Eq. \eqref{eq:quasiadiabaticdota2} for $a_2(t)$. 
Instead of doing this in a constructive fashion, we give the solution $a_2(t)$
and prove that it indeed fulfills 
Eq. \eqref{eq:quasiadiabaticdota2} up to the desired order. 
The solution reads
\bean
\label{eq:a2solution}
a_2(t) =  i \frac{\bra{u_{\excited}(t)}\partial_t \ket{u_{\ground}(t) }}{E_t} 
e^{i\int_0^t dt' E(t')}.
\eean

First, let us check if it  solves the differential equation \eqref{eq:quasiadiabaticdota2}:
\bean
\nonumber
\partial_t a_2(t) &=& 
i \frac{(\partial_t \bra{u_{\excited}(t)} \partial_t \ket{ u_{\ground}(t) })}{E_t} 
	e^{i\int_0^t dt' E(t')}
- i
\frac{(\partial_t E_t) \bra{u_{\excited}(t)} \partial_t \ket{ u_{\ground}(t) }}{E_t^2}
	e^{i\int_0^t dt' E(t')}
\\
&-&
\bra{u_{\excited}(t)} \partial_t \ket{ u_{\ground}(t) } e^{i\int_0^t dt' E(t')}.
\eean
The first two terms on the right hand side scale as $\Omega^2$, 
whereas the third one scales as $\Omega$. 
Hence we conclude that 
in the quasi-adiabatic case, Eq. \eqref{eq:a2solution} is the solution of 
Eq. \eqref{eq:quasiadiabaticdota2}
we were after. 
The corresponding solution of the time-dependent Schr\"odinger equation
\eqref{eq:tdse} is constructed using Eqs. \eqref{eq:psit}, 
$a_1(t) = 1$ and \eqref{eq:a2}, and reads
\bean
\label{eq:psitresult}
\ket{\tilde{u}_{\ground}(t)} = 
e^{- i \int_0^t dt' E_{1}(t')}\left[
	 \ket{u_{\ground}(t)} 
	+ i \frac{\bra{u_{\excited}(t)}\partial_t \ket{ u_{\ground}(t) }}{E_t} \ket{u_{\excited}(t)} 
\right].
\eean
In words, Eq. \eqref{eq:psitresult} assures that
the stationary state has most of its weight
in the instantaneous ground state $\ket{u_{\ground}(t)}$,
with a small, $\sim \Omega \ll 1$ admixture of the instantaneous
excited state $\ket{u_{\excited}(t)}$. 
Interestingly, even though this small admixture vanishes in the 
adiabatic limit $\Omega \to 0$, 
the corresponding contribution to the number of pumped particles
can give a finite contribution, as the cycle period $T$ goes to 
infinity in the adiabatic limit. 
This will be shown explicitly in the next section.

Finally, we show that this state $\ket{\tilde{u}_{\ground}(t)}$ is 
indeed stationary. 
That is proven if we can prove that $\ket{\tilde{u}_{\ground}(T)}$ is 
equal to $\ket{\tilde{u}_{\ground}(0)}$ up to a phase factor. 
This arises as the consequence of the following fact.
If the Berry phase associated to the state $\ket{u_{\ground}}$ 
is $\gamma$, that is, 
if $\ket{u_{\ground}(T)} = e^{i\gamma} \ket{u_{\ground}(0)}$,
then 
\bean
\nonumber
\left[\partial_t \ket{u_\ground(t) }\right]_T &=&
\lim_{\epsilon \to 0} \frac{\ket{u_\ground (T+\epsilon)} - \ket{u_\ground (T)}}{\epsilon}
=
\lim_{\epsilon \to 0} 
\frac{ e^{i\gamma} \ket{u_\ground(\epsilon)} - e^{i\gamma}\ket{u_\ground(0)}}{\epsilon}\\
&=&
 e^{i\gamma} \partial_t \ket{u_\ground (0)}.
\eean
Therefore, the two terms in the square bracket of Eq. \eqref{eq:psitresult}
acquire the same phase factor $e^{i\gamma}$
at the end of the cycle, hence the obtained $\ket{\tilde u_\ground(t)}$
solution is stationary.

\section{The pumped current is the Berry curvature}
\label{sec:current:chern}

The number of particles pumped through an arbitrary cross section
of the one-dimensional lattice, in the duration $T$ of a quasi-adiabatic cycle, 
is evaluated combining Eqs. \eqref{eq:Qquasiadiabatic}
and \eqref{eq:psitresult}.
We define the \emph{momentum- and time-resolved current
of the filled band} as
\bean
\label{eq:current:momentumtime}
j^{(1)}_{m+1/2}(k,t) = \frac{1}{N}
\bra{\tilde{u}_1(k,t) } \partial_k \ido{H}(k,t) \ket{\tilde u_1(k,t)},
\eean
and perform the usual substitution
$\frac{1}{N} \sum_{k \in {\rm BZ}} \dots = \int_{\rm BZ}\frac{dk}{2\pi} \dots$,
yielding the following formula for the number of pumped particles:
\bean
\label{eq:current_number_vs_momentumtimeresolved}
\mathcal Q = \int_0^T dt \int_{\textrm{BZ}} \frac{dk}{2\pi}
j^{(1)}_{m+1/2}(k,t).
\eean
In the rest of this section, we show that 
the momentum- and time-resolved current 
is the Berry curvature associated to the filled band, 
and therefore the number of pumped particles
is the Chern number, which in turn is indeed an integer.

To this end, we insert the result \eqref{eq:psitresult} to 
the definition \eqref{eq:current:momentumtime}. 
The contribution 
that incorporates two lower-band wave functions $\ket{u_1(k,t)}$,
is finite; however, its integral over the BZ vanishes, 
and therefore we disregard it as it does not contribute to particle pumping.
Hence the leading relevant contribution is the one
incorporating one filled-band $\ket{u_1(k,t)}$ 
and one empty-band $\ket{u_2(k,t)}$
wavefunction:
\bean
j^{(1)}_{m+1/2}(k,t) = 
i
\frac{\bra{u_\ground} [\partial_k \ido{H}] 
\ket{u_\excited } \bra{u_\excited } \partial_t \ket{u_\ground }}{E}
+ c.c.
\eean
where the $k$ and $t$ arguments are suppressed for brevity.

Now we use
\bean
\bra{u_\ground } [\partial_k \ido{H}] \ket{u_\excited} =
(E_1- E_2) \bra{\partial_k u_\ground } u_\excited  \rangle
=
- E \bra{\partial_k u_\ground } u_\excited \rangle,
\eean
which has a straightforward proof using the spectral decomposition 
$\ido{H} = E_1 \ket{u_1} \bra{u_1} + E_2 \ket{u_2} \bra{u_2}$
of the Hamiltonian and the fact that $\partial_k \bra{u_1} u_2 \rangle = 0$.
Therefore, 
\bean
j^{(1)}_{m+1/2} =
- i \bra{\partial_k u_\ground }  u_\excited  \rangle \bra{u_\excited } \partial_t \ket{u_\ground }
	+ c.c..
\eean
Since we use the parallel-transport gauge, 
we can replace
the projector $\ket{u_\excited }\bra{u_\excited}$ with unity in 
 the preceding formula, hence the latter can be simplified as
\bean
\label{eq:current_berrycurvature}
j^{(1)}_{m+1/2} & = &
- 
i \bra{\partial_k u_\ground }  \partial_t  u_\ground  \rangle
+ c.c.
=
- i 
\left(
	\bra{\partial_k u_\ground }  \partial_t  u_\ground  \rangle
	- \bra{\partial_t u_\ground }  \partial_k u_\ground  \rangle
\right) \nonumber
\\
&=&
- i 
\left(
	\partial_k \bra{ u_\ground }  \partial_t  u_\ground  \rangle
	- \partial_t \bra{u_\ground }  \partial_k u_\ground  \rangle
\right).
\eean
This testifies that the momentum- and time-resolved 
current is indeed the Berry curvature corresponding to the 
filled band, and thereby 
confirms the result promised in Eq. \eqref{eq:centralresult}.

\begin{figure}[!ht]
\begin{center}
\includegraphics[width=0.9\textwidth]{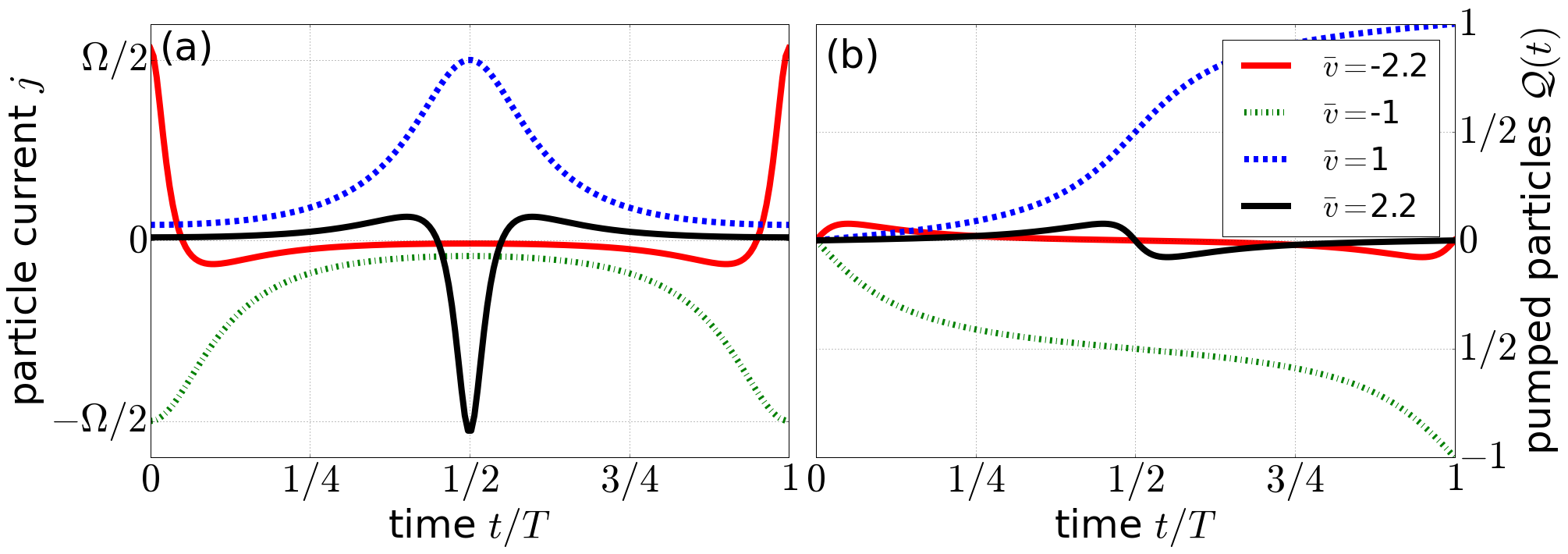}
\caption{\label{fig:current_results}
Time dependence of the current and the number of 
pumped particles in an adiabatic cycle.} 
\end{center}
\end{figure}

As a straightforward application of our result, 
we calculate the time dependence of the
current and the number of pumped particles
through an arbitrary cross section in the 
smoothly modulated Rice-Mele model, see
Eq. \eqref{eq:current:dexample}.
The results corresponding to
four different values of the parameter $\bar v$ 
are shown in Fig. \ref{fig:current_results}.
The momentum- and time-resolved current $j^{(1)}_{m+1/2}(k,t)$
of the filled band
can be obtained analytically from Eq. \eqref{eq:current_berrycurvature}.
Then, the time-resolved current $j$ is defined as 
the integrand of the $t$ integral in 
Eq. \eqref{eq:current_number_vs_momentumtimeresolved}.
We evaluate $j$ via a numerical $k$ integration, 
and plot the result in Fig. \eqref{fig:current_results}a. 
The number of pumped particles as a function of time
is then evaluated numerically 
via $\mathcal Q(t) = \int_0^t dt j(t)$;
the results are shown in Fig. \eqref{fig:current_results}b. 
These results confirm that the number of particles
pumped through the cross section during the complete
cycle is an integer, and is given by the Chern number
associated to the pumping cycle.

In this chapter, we have provided a formal description of 
adiabatic pumping in one-dimensional lattices. 
After identifying the current operator describing particle
flow at a cross section of the lattice, 
we discussed the quasi-adiabatic time evolution
of the lower-band states in a two-band model, 
and combined these results 
to express the number of pumped particles
in the limit of adiabatic pumping. 
The central result is that the momentum- 
and time-resolved current carried by the lower-band electrons
is the Berry curvature associated to their band.

\section*{Problems}
\addcontentsline{toc}{section}{Problems}
%
\begin{prob}
\label{prob:current-scf}
\textbf{The smooth pump sequence of the Rice-Mele model.}\\
For the smoothly modulated Rice-Mele pumping
cycle, see \eqref{eq:current:dexample}, 
evaluate the momentum- and time-dependent current
density, and the number of particles pumped through 
an arbitrary unit cell boundary as the 
function of time; that is, reproduce Fig. \ref{fig:current_results}.
\end{prob}

\begin{prob}
\label{prob:current-scf}
\textbf{Parallel-transport time parametrization.}\\
Specify a parallel-transport time parametrization 
for the ground state of the two-level Hamiltonian 
defined by Eqs. \eqref{eq:current:twolevelh}, \eqref{eq:current:dexample}, 
and 
(a) $k=0$ (b) $k=\pi$.
\end{prob}

\begin{prob}
\label{prob:current-initialcondition}
\textbf{Quasi-adiabatic dynamics with a different boundary condition.}\\
In Sect.~\ref{sec:current:quasiadiabatic}, 
we described a \emph{stationary state}
of a quasi-adiabatically driven two-level system,
and used the result to express the number of 
particles pumped during a complete cycle.
How does the derivation and the result change,
if we describe the dynamics not via the stationary state,
but by specifying that the initial state is the instantaneous
ground state of the Hamiltonian at $t=0$?
Is the final result for the number of pumped particles different 
in this case? 
\end{prob}

\begin{prob}
\label{prob:current-generalization}
\textbf{Adiabatic pumping in multiband models.}\\
Generalize the central result of this chapter in the following 
sense. 
Consider adiabatic charge pumping in a one-dimensional multi-band system 
$(n=1,2,\dots,N)$, 
where the energies of the first $N_-$ bands $(n=1,2,\dots N_-)$ 
are below the Fermi energy and 
the energies of the remaining bands $(n=N_-+1, \dots, N$)
are above the Fermi energy, 
and the bands do not cross each other. 
Show that the number of particles adiabatically 
pumped through an arbitrary cross section 
of the crystal is the sum of the Chern numbers of the filled bands. 
\end{prob}


%% file: qiwuzhang.tex

\chapter{Two-dimensional Chern insulators -- the Qi-Wu-Zhang model}
\label{chap:qiwuzhang}


\abstract*{In this Chapter we construct two-dimensional insulators with
nonvanishing Chern numbers, by \emph{promoting} the time $t$ in a an
adiabatic pump cycle on a one-dimensional insulator, to a wavenumber
$k_y$. This construction not only gives us control over the Chern
number, but also brings with it edge states. These edge states are
more than just bound states at the edge: they form continuous bands
across the bulk energy gap, and combine into a discrete number of
channels that conduct particles in one way only. As we will show, the
conduction along the edge in these channels is unimpeded even by
arbitrarily strong disorder at the edge.
}

The unique physical feature of topological insulators is the
guaranteed existence of low-energy states at their boundaries. We have
seen an example of this for a one-dimensional topological insulator, the
SSH model: A finite, open, topologically \emph{nontrivial} SSH chain
hosts 0 energy bound states at both ends. The \emph{bulk--boundary
  correspondence} was the way in which the topological invariant of
the bulk -- in the case of the SSH chain, the winding number of the
bulk Hamiltonian -- can be used to predict the number of edge states.



We will show that the connection between the Chern number and the
number of edge-state channels is valid in general for two-dimensional
insulators. This is the statement of \emph{bulk--boundary
  correspondence} for Chern insulators. The way we will show this
inverts the argument above: taking any two-dimensional insulator, we
can map it to an adiabatic pump sequence in a one-dimensional
insulator by \emph{demoting} one of the wavenumbers to time. The
connection between the Chern number and the number of edge states in
the higher dimensional Hamiltonian is a direct consequence of the
connection between Chern number and charge pumping in the lower
dimensional system. 

Chern insulators (two-dimensional band insulators with
nonvanishing Chern number) were first used to explain the Quantum Hall
Effect. 
There an external magnetic field, included in lattice models via a
Peierls substitution, is responsible for the nonzero value of the
Chern number. Peierls substitution, however, breaks the lattice
translation invariance. This neccessitates extra care, including the
use of magnetic Brillouin zones whose size depends on the magnetic
field. 

The models we construct in this Chapter describe the so-called Quantum
Anomalous Hall Effect. Here we have the same
connection between edge states and bulk Chern number as in the Quantum
Hall Effect, however, there is no external magnetic field, and thus no
complications with magnetic Brillouin zones. The Quantum Anomalous
Hall Effect has recently been observed in thin films of chromium-doped
(Bi,Sb)$_2$Te$_3$ \cite{qah_experiment2013}.

To illustrate the concepts of Chern insulators, we will use a toy
model introduced by Qi, Wu and Zhang\cite{qi_wu_zhang2006}, which we
call the QWZ model. This model is also important because it forms the
basic building block of the Bernevig-Hughes-Zhang model for the
quantum spin Hall effect (Chapt.~\ref{chap:BHZ}), and thus it is also
sometimes called ``half BHZ''.


\section{Dimensional extension: from an adiabatic pump to a
  Chern insulator}

We want to construct a two-dimensional lattice Hamiltonian $\hat{H}$
with a nonvanishing bulk Chern number. We will do this by first
constructing the bulk momentum-space Hamiltonian $\hat{H}(k_x,k_y)$,
from which the real-space Hamiltonian can be obtained by Fourier
transformation. For the construction we simply take an adiabatic pump
sequence on a one-dimensional insulator, $\hat{H}(k,t)$, and
reinterpret the cyclic time variable $t$ as a new momentum variable
$k_y$. This way of gaining an extra dimension by \emph{promotion} of a
cyclic parameter in a continuous ensemble to a momentum is known as
\emph{dimensional extension}. This, and the reverse process of
\emph{dimensional reduction}, are key tools to construct the general
classification of topological insulators\cite{tenfold_way}.

\subsubsection*{From the Rice-Mele model to the Qi-Wu-Zhang model}

To see how the construction of a Chern insulator works, we take the
example of the smooth pump sequence on the Rice-Mele
model from the previous Chapter,
Eqs.~\eqref{eq:ricemele_smooth_pump_sequence}.  In addition to the
promotion of time to an extra wavenumber, $\Omega t \to k_y$, we also
do an extra unitary rotation in the internal Hilbert space, to arrive
at the Qi-Wu-Zhang model, 
\begin{align}
\HH(k) &= \sin k_x \hsigma_x +\sin k_y \hsigma_y + [\uu + \cos k_x
  + \cos k_y ] \hsigma_z.
\label{eq:qiwuzhang-bulk_Hamiltonian_def}
\end{align}
The mapping is summarized in Table~\ref{tab:qiwuzhang-mapping}.

\begin{table}[ht]
\centering
\begin{tabular}{|c|c|}
\hline
Adiabatic pump in the RM model & QWZ model (Chern Insulator) \\
\hline
average intracell hopping $\overline{v}$ & staggered onsite potential $u$ \\
wavenumber $k$ & wavenumber $k_x$\\
time $t$ [in units of $T/(2\pi)$] & wavenumber $k_y$ \\
$\sigma_x, \sigma_y, \sigma_z$ & $\sigma_y, \sigma_z,\sigma_x$ \\
\hline
\end{tabular}
\caption{Mapping of an adiabatic pump sequence of the Rice-Mele model, 
$\hat{H}(k,t)$
  to the QWZ model for the Anomalous Hall Effect, $\hat{H}(k_x,k_y)$.}
\label{tab:qiwuzhang-mapping}
\end{table}

The corresponding $\hh(\kk)$ vector reads,
\begin{align}
\hh(k_x,k_y) &= 
\begin{pmatrix}
\sin k_x\\
\sin k_y\\
\uu + \cos k_x + \cos k_y
\end{pmatrix}. 
\label{eq:qiwuzhang-dvector}
\end{align}


\subsection{Bulk dispersion relation}

\begin{figure}
\centering
\includegraphics[width=0.8\linewidth]{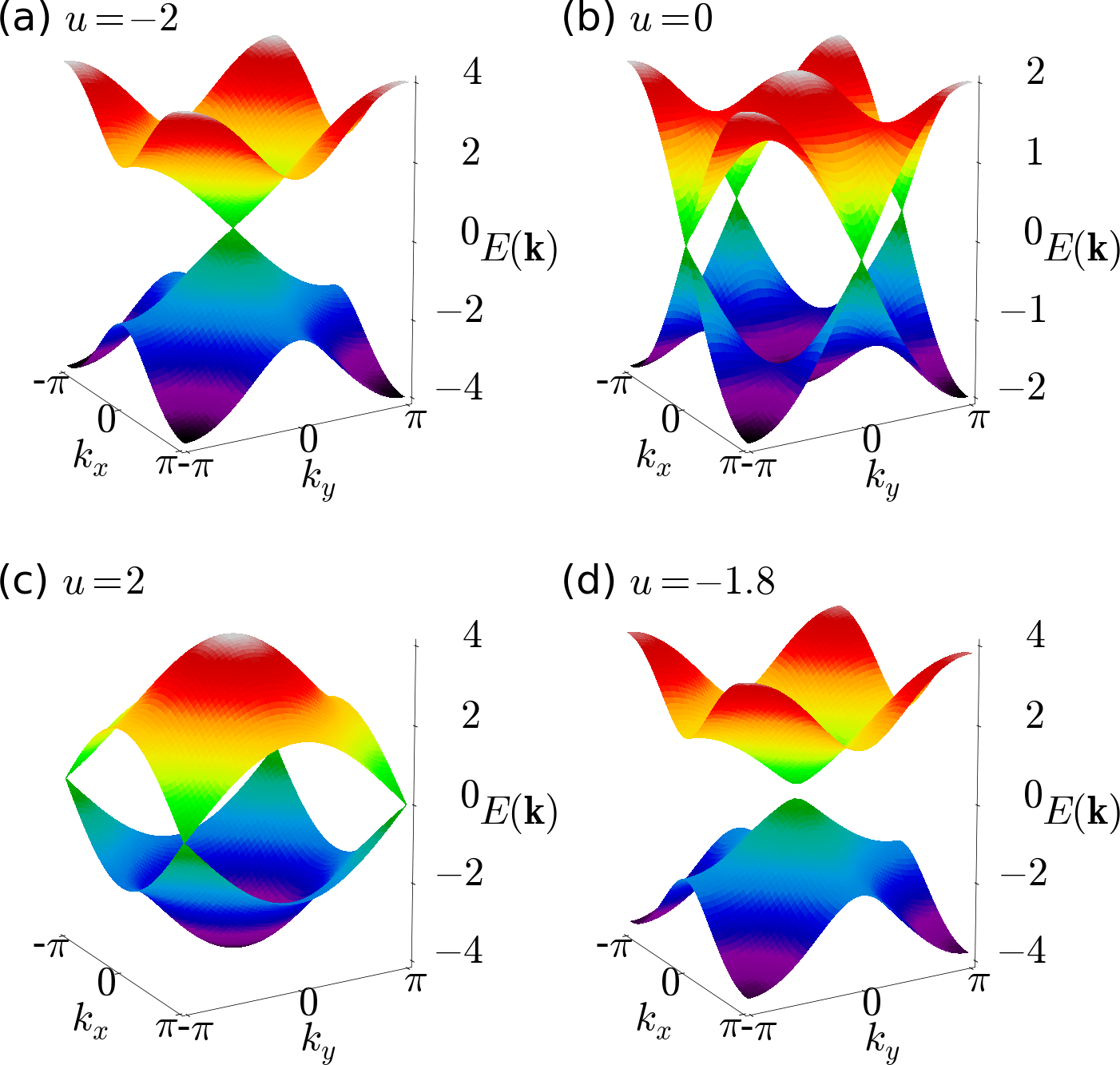}
\caption{ The bulk dispersion relation of the QWZ model, for various values
  of $\uu$, as indicated in the plots. In (a)-(c), the gapless cases
  are shown, where the bulk gap closes at so-called Dirac points.  In
  (d), a generic value $\uu=-1.8$, the system is insulating.
\label{fig:qiwuzhang-dispersion}}
\end{figure}

We can find the dispersion relation of the QWZ model using the
algebraic properties of the Pauli matrices, whereby $\hat{H}^2 =
E(\kk) \II_2$, with $\II_2$ the unit matrix. 
Thus, the spectrum of the QWZ model has two bands, the two eigenstates
of $\HH(\kk)$, with energies 
\begin{align}
E_\pm (k_x,k_y)&=\pm|\hh (k_x,k_y)|\\
&=\pm\sqrt{\sin^2(k_x)+\sin^2(k_y)+ (\uu +\cos(k_x)+\cos(k_y))^2}.
\label{eq:qiwuzhang-bulk_spectrum}
\end{align}
The spectrum of the QWZ model is depicted in
Fig.\ref{fig:qiwuzhang-dispersion}.  

There is an energy gap in the spectrum of the QWZ model, which closes
at finetuned values of $\uu=+2,0,-2$. This is simple to show,
since the gap closing requires $\hh(\kk)=0$ at some $\kk$. From
Eq.~\eqref{eq:qiwuzhang-dvector}, $d_x(\kk)=d_y(\kk)=0$ restricts us
to four inequivalent points in the Brillouin zone: 
\begin{itemize} 
\item if $\uu=-2$: at $k_x=ky=0$, the $\Gamma$ point;
\item if $\uu = 0$: at $k_x=0,k_y=\pi$ and $k_x=\pi,k_y=0$, two
  inequivalent $X$ points
\item if $\uu=+2$: at $k_x=\pi,k_y=\pi$, the $M$ point; note that
  $k_x=\pm\pi$, $k_y=\pm \pi$ are all equivalent
\end{itemize}  
In the vicinity of a gap closing point, called \emph{Dirac point}, the
dispersion relation has the shape of a \emph{Dirac cone}, as seen in
Fig.~\ref{fig:qiwuzhang-dispersion}.  For all other values of $\uu
\neq -2,0,2$, the spectrum is gapped, and thus it makes sense to
investigate the topological properties of the system.

\subsection{Chern number of the QWZ model}

\begin{figure}
\centering
\includegraphics[width=0.99\linewidth]{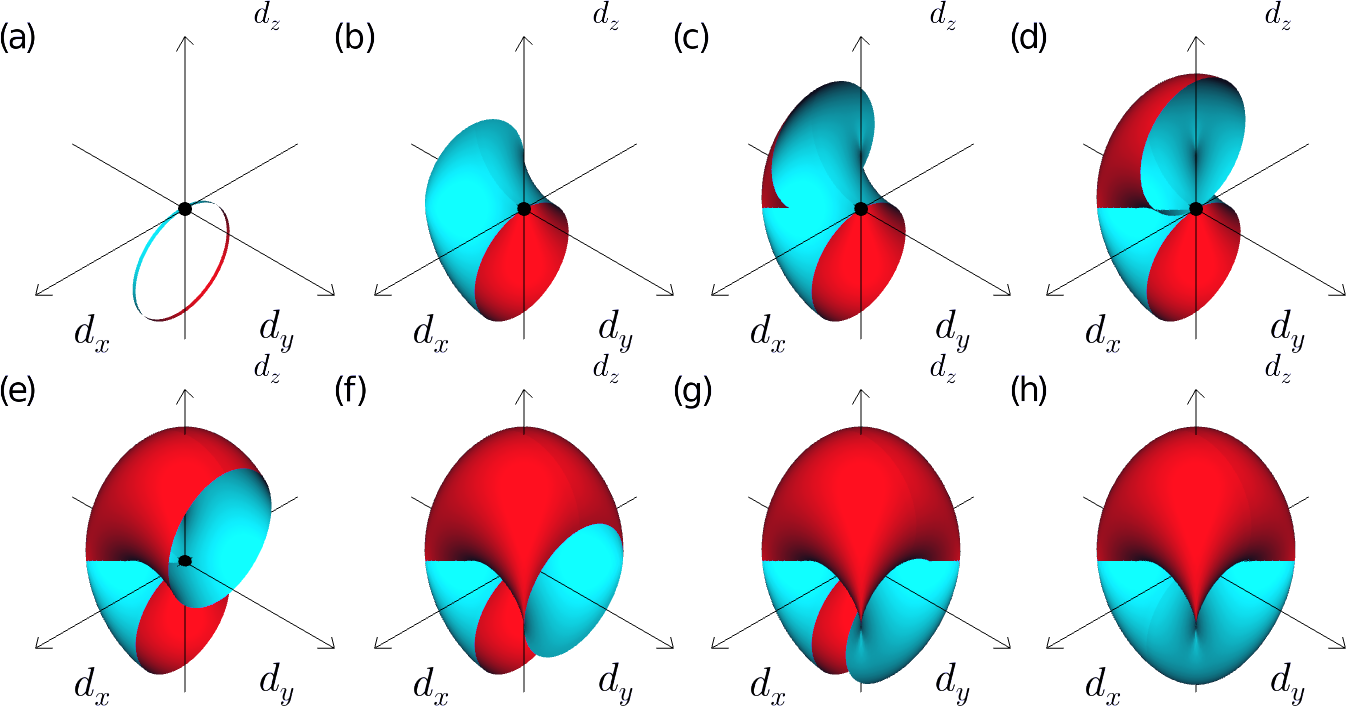}
\caption{The surface $\hh(\kk)$ for the QWZ model as $\kk$ sweeps
  through the whole Brillouin zone.  To illustrate how this surface is
  a torus the sweeping is done gradually with $\uu=0$.  In (a) the
  image of the $k_y=-\pi$ line is depicted. In (b) the image for the
  region $k_y=-\pi\cdots -0.5\pi$, in (c) $k_y=-\pi\cdots -0.25\pi$, in
  (d) $k_y=-\pi\cdots -0$, in (e) $k_y=-\pi\cdots 0.25\pi$, in (f)
  $k_y=-\pi\cdots 0.5\pi$, in (g) $k_y=-\pi\cdots 0.75\pi$ and finally
  in (h) the image of the whole Brillouin zone is depicted and the
  torus is closed.
\label{fig:half_bhz-torus}}
\end{figure}

\begin{figure}
\centering
\includegraphics[width=0.99\linewidth]{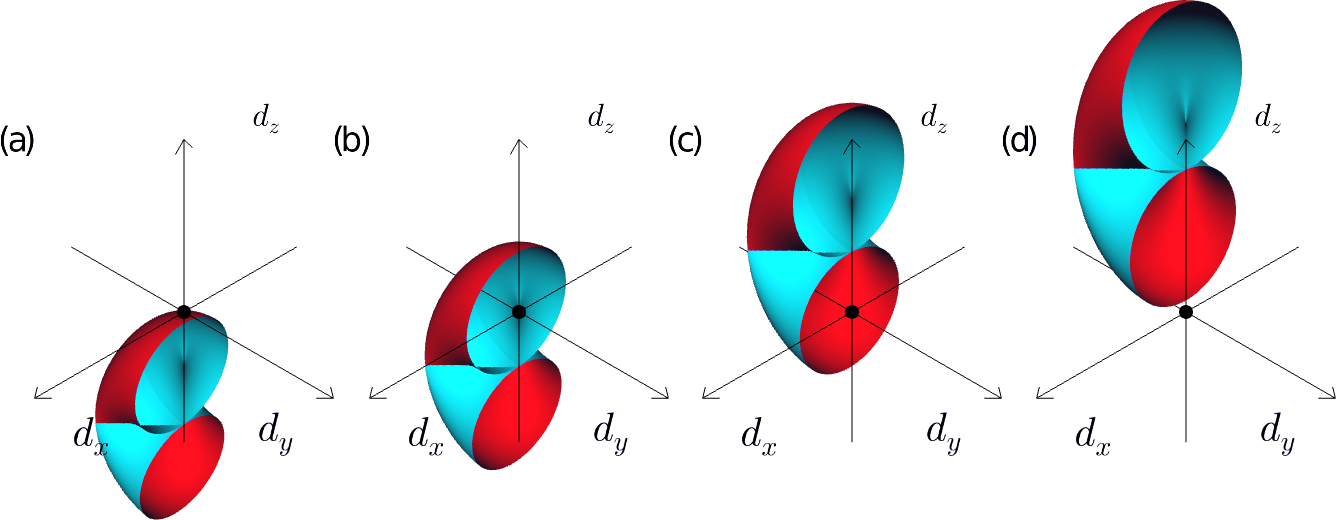}
\caption{ The torus $\hh(\kk)$ of the QWZ model for different values
  of $\uu$ keeping $A=1$.  For clarity only the image of half of the
  Brillouin zone is shown.  In (a) and (d) $\uu=\mp2.2$ and the torus
  does not contain the origin hence $Q=0$. In (b) $\uu=-1$, taking an
  infinite line from the origin along the positive z axis we hit the
  blue side of the torus once hence $Q=-1$.  In (c) $\uu=1$, taking the
  infinite line in the negative z direction we hit the red side of the
  torus thus $Q=1$.
\label{fig:half_bhz-torus-chern}}
\end{figure}

Although we calculated the Chern number of the corresponding pump
sequence in the previous chapter, we show the graphical way to
calculate the Chern number of the QWZ model.  We simply count how many
times the torus of the image of the Brillouin zone in the space of
$\hh$ contains the origin. To get some feeling about the not so
trivial geometry of the torus, it is instructive to follow a gradual
sweep of the Brillouin zone in Fig. \ref{fig:half_bhz-torus}. The
parameter $\uu$ shifts the whole torus along the $d_z$ direction, thus
as we tune it we also control whether the origin is contained inside
it or not.  Generally three situations can occur as it is also
depicted in Fig.~\ref{fig:half_bhz-torus-chern}.  The torus either
does not contain the origin as in (a) and (d) and the Chern number is
$Q=0$ for $|\uu|>2$, or we can take a straight line to infinity from
the origin that pierces the torus first from the blue side (outside)
of the surface as in (b) with $Q=-1$ for $-2<\uu<0$, or piercing the
torus from the red side (inside) as in (c) with $Q=1$ for $0<\uu<2$.

To summarize, the Chern number $Q$ of the QWZ model is
\begin{subequations}
\label{eq:qiwuzhang_chern_of_u}
\begin{align}
u<-2 \quad &: \quad Q = 0; \\
-2 < u< 0 \quad &: \quad Q = -1; \\
0 < u< 2 \quad &: \quad Q = +1; \\
2 < u  \quad &: \quad Q = 0.
\end{align}
\end{subequations}

%
%

\subsection{The real-space Hamiltonian }

\begin{figure}
\centering
\includegraphics[width=0.9\linewidth]{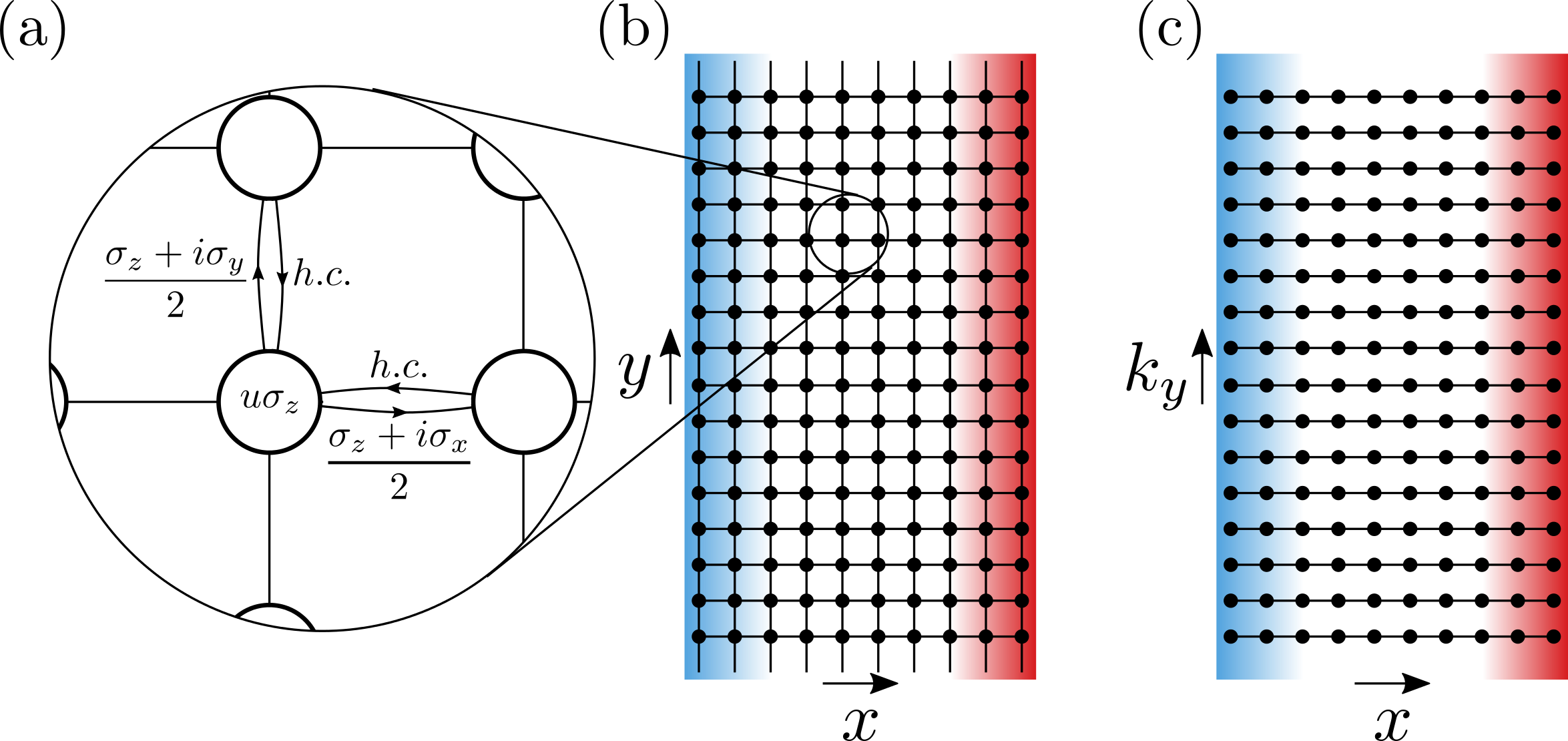}
\caption{ Sketch of the QWZ model: a particle with two internal states
  hopping on a square lattice. (a): The onsite potential and the
  hopping amplitudes are operators acting on the internal states. 
  (b): A strip, with periodic boundary conditions along $y$,
  open boundaries along $x$ (hopping amplitudes set to
  zero). Light blue / dark red highlights left/right edge region. 
  (c): Upon Fourier transformation along $y$, the strip falls apart 
  to an ensemble of one-dimensional Hamiltonians, indexed by $k_y$. 
 \label{fig:qiwuzhang-strip_to_pump}}
\end{figure}

We obtain the full Hamiltonian of the Qi-Wu-Zhang model by
inverse Fourier transform of the bulk momentum-space Hamiltonian,
Eq.~\eqref{eq:qiwuzhang-bulk_Hamiltonian_def}, as
\begin{multline}
\label{eq:qiwuzhang-realspace_Hamiltonian}
\HH = \sum_{\uci_x = 1}^{\nuc_x-1} \sum_{\uci_y = 1}^{\nuc_y} 
\left( \ket{\uci_x+1,\uci_y}\bra{\uci_x,\uci_y} 
\otimes \frac{\hsigma_z + i \hsigma_x}{2} + h.c. \right)\\
+ \sum_{\uci_x = 1}^{\nuc_x} \sum_{\uci_y = 1}^{\nuc_y-1} 
\left( \ket{\uci_x,\uci_y+1}\bra{\uci_x,\uci_y} 
\otimes \frac{\hsigma_z + i \hsigma_y}{2} + h.c. \right)\\
+ \uu \sum_{\uci_x = 1}^{\nuc_x} 
 \sum_{\uci_y = 1}^{\nuc_y} \ket{\uci_x,\uci_y}\bra{\uci_x,\uci_y} 
\otimes \hsigma_z. 
\end{multline}
As sketched in  Fig.~\ref{fig:qiwuzhang-strip_to_pump}), 
the model describes a
particle with two internal states hopping on a lattice 
where the nearest neighbour hopping is accompanied by an operation on the
internal degree of freedom, and this operation is different for the
hoppings along the $x$ and $y$ directions. In addition, there is a
staggered onsite potential of strength $\uu$.
Unlike in the case of the SSH model (Chapt.~\ref{chap:ssh}), the
real-space form of the QWZ Hamiltonian is not intuitive. 



%

\section{Edge states}
\label{sec:qiwuzhang_edgestates}

We constructed a Chern insulator using an adiabatic charge pump. As we
saw in Chapt.~\ref{chap:ricemele}, charge pumps also induce energy
eigenstates at the edge regions that cross from negative energy to
positive energy bands, or vice versa. What do these energy eigenstates
correspond to for the Chern insulator?  

%

\subsubsection*{Dispersion relation of a strip shows the edge states}

To see edge states, consider a strip of a two-dimensional insulator
depicted in
Fig.~\ref{fig:qiwuzhang-strip_to_pump}. Along $y$, we take periodic
boundary conditions (close the strip to a cylinder), and go to the
limit $N_y\to\infty$. Along $x$, the strip is terminated by setting
the hopping amplitudes to 0 (open boundary condition), and it consists
of $N$ sites. Translation invariance holds along $y$, so we can
partially Fourier transform -- only along $y$. After the Fourier
transformation, the original Hamiltonian has falls apart to a set of
one-dimensional lattice Hamiltonians indexed by a continuous parameter $k_y$, 
the wavenumber along $y$. For the QWZ model,
Eq.~\eqref{eq:qiwuzhang-realspace_Hamiltonian}, the $k_y$ -dependent
Hamiltonian reads
\begin{multline}
\HH(k_y) = \sum_{\uci_x = 1}^{\nuc_x-1} 
\left( \ket{\uci_x+1}\bra{\uci_x} 
\otimes \frac{\hsigma_z + i \hsigma_x}{2} + h.c. \right)+\\
\sum_{\uci_x = 1}^{\nuc_x} 
\ket{\uci_x}\bra{\uci_x} 
\otimes (\cos k_y \hsigma_z +  \sin k_y  \hsigma_y
\uu \otimes \hsigma_z ).
\label{eq:qiwuzhang-strip}
\end{multline}
Note that this is the same dimensional reduction
argument as before, but for a system with edges.
Energy eigenstates $\ket{\Psi(k_y)}$ of the strip fall into the
categories of bulk states and edge states, much as in the
one-dimensional case. All states are delocalized along $y$, but bulk
states are also delocalized along $x$, while edge states are
exponentially confined to the left ($x=0$) or the right ($x=N$)
edge. If we find energy eigenstates with energy deep in the bulk gap,
they have to be edge states, and can be assigned to the left or the
right edge.

An example for the dispersion relation of a strip is shown in
Fig. \ref{fig:qiwuzhang-spectrum1}, edge states on the left/right edge
highlighted using dark red/light blue. We used the QWZ model, strip
width $N=10$, sublattice potential parameter $\uu = -1.5$, and the
same practical definition of edge states as in the Rice-Mele model,
\begin{align}
\ket{\Psi(k_y)} \text{ is on the right edge } &\Leftrightarrow
\sum_{\uci_x=N-1}^N \sum_{\alpha \in \{A,B\}} 
\abs{\braket{\Psi(k_y)}{\uci_x,\alpha}}^2  > 0.6; \\
\ket{\Psi(k_y)} \text{ is on the left edge } &\Leftrightarrow
\sum_{\uci_x=1}^{2} \sum_{\alpha \in \{A,B\}} 
\abs{\braket{\Psi(k_y)}{\uci_x,\alpha}}^2  > 0.6.
\end{align}

\subsubsection*{Edge states conduct unidirectionally}

\begin{figure}
\centering
\includegraphics[width=0.8\linewidth]{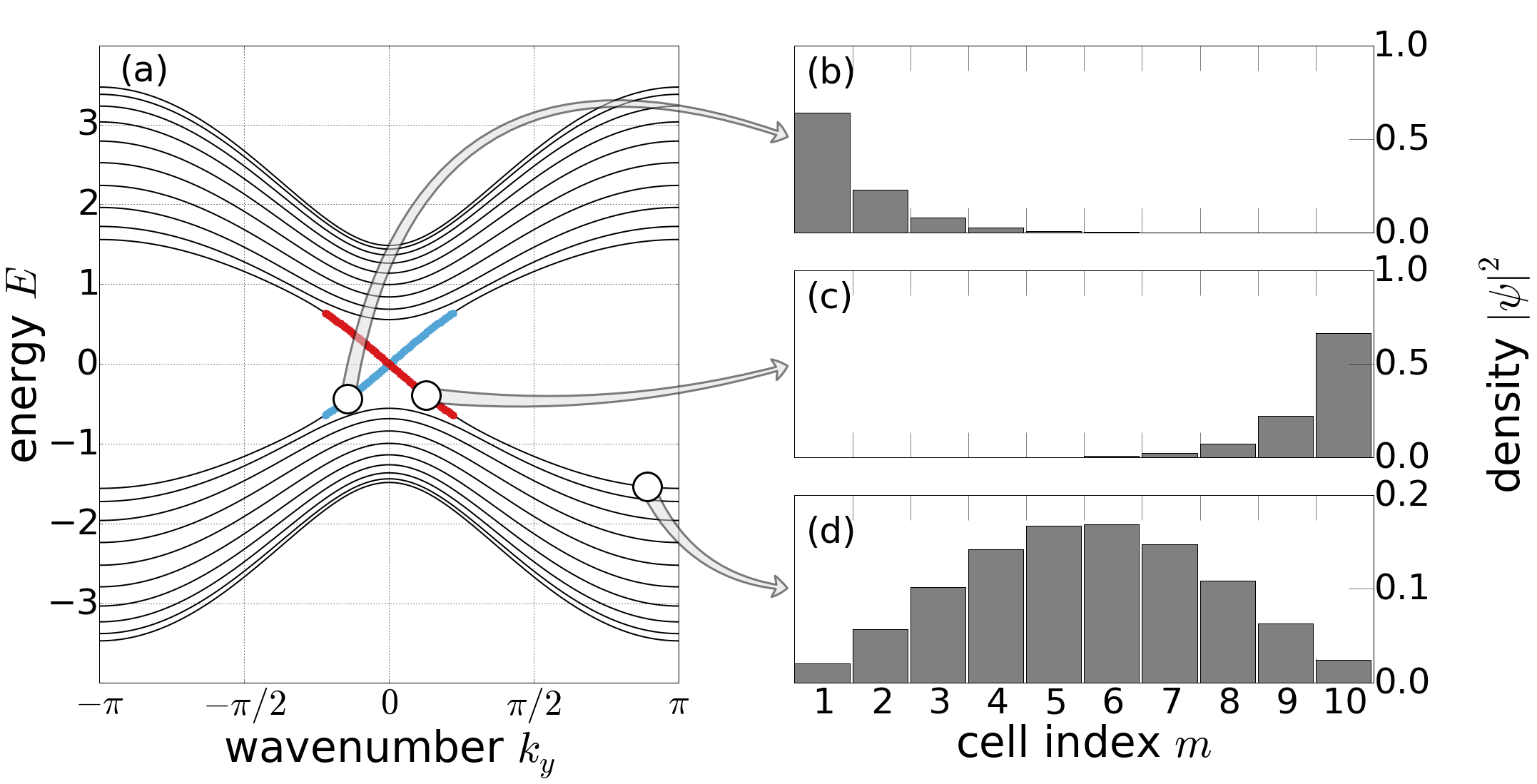}%
\caption{Dispersion relation of a strip of QWZ model, of width $N=10$,
  and sublattice potential parameter $\uu = -1.5$.  Because the
  strip is translation invariant along the edge, the wavenumber $k_y$
  is a good quantum number, and the energy eigenvalues can be plotted
  (a) as a function of $k_y$, forming branches of a dispersion
  relation. Light blue/dark red highlights energies of edge states,
  whose wavefunction has over 60\% weight on unit cells with
  $\uci_x\le 2$/ $\uci_x \ge N-1$. These are parts of the $N$th and
  $N+1$th branch, which split off from the bulk around
  $-\pi/4<k_y<\pi/4$, and have an avoided crossing with an
  exponentially small gap near $k_y=0$. We show the marginal position
  probability distribution of the $N$th energy eigenstate, $P_N(\uci_x)
  = \sum_{\alpha \in \{ A,B \} } \sum_{\uci_y} \abs{ \left\langle
    \uci_x,\alpha|\Psi_N\right\rangle }^2$, for three values of
  $k_y$. Depending on $k_y$, this state can be an edge state on the
  right edge (b), on the left edge (c), or a bulk state (d).
  \label{fig:qiwuzhang-spectrum1}}
\end{figure}

Notice the edge state branches of the dispersion relation of the QWZ
strip, Fig.~\ref{fig:qiwuzhang-spectrum1}, which connect the lower and
upper band across the bulk gap. They are the
edge states of the pumped Rice-Mele model,
Fig.~\ref{fig:ricemele-smooth_dispersions_-1}, but we now look at them
with a new eye. For the edge states in the QWZ model, $dE/dk_y$
corresponds to the group velocity along the edge. Thus, the dispersion
relation tells us that 
\emph{particles in the QWZ model at low energy are 
confined either to the to left edge and propagate towards the right,
or to the right edge and propagate towards the left}. 

The presence of one-way conducting edge state branches implies that
the QWZ model is no longer, strictly speaking, an insulator. Because
of the bulk energy gap, it cannot conduct (at low energies) between
the left and right edges. However, it will conduct along the edges,
but only unidirectionally.

\subsection{Edge states and edge perturbation}

\begin{figure}
\centering
\includegraphics[width=0.7\linewidth]
{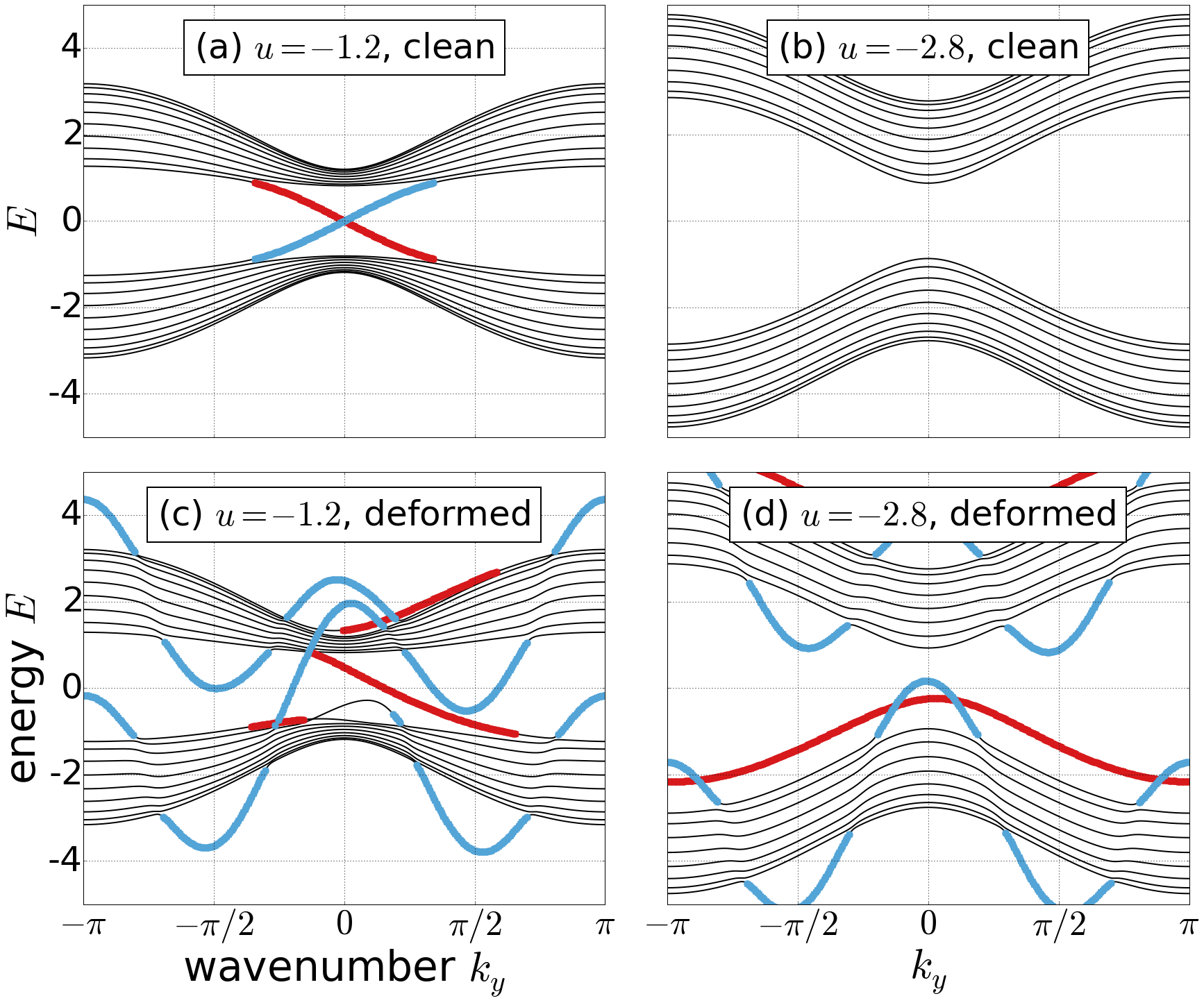}
\includegraphics[width=0.28\linewidth]
{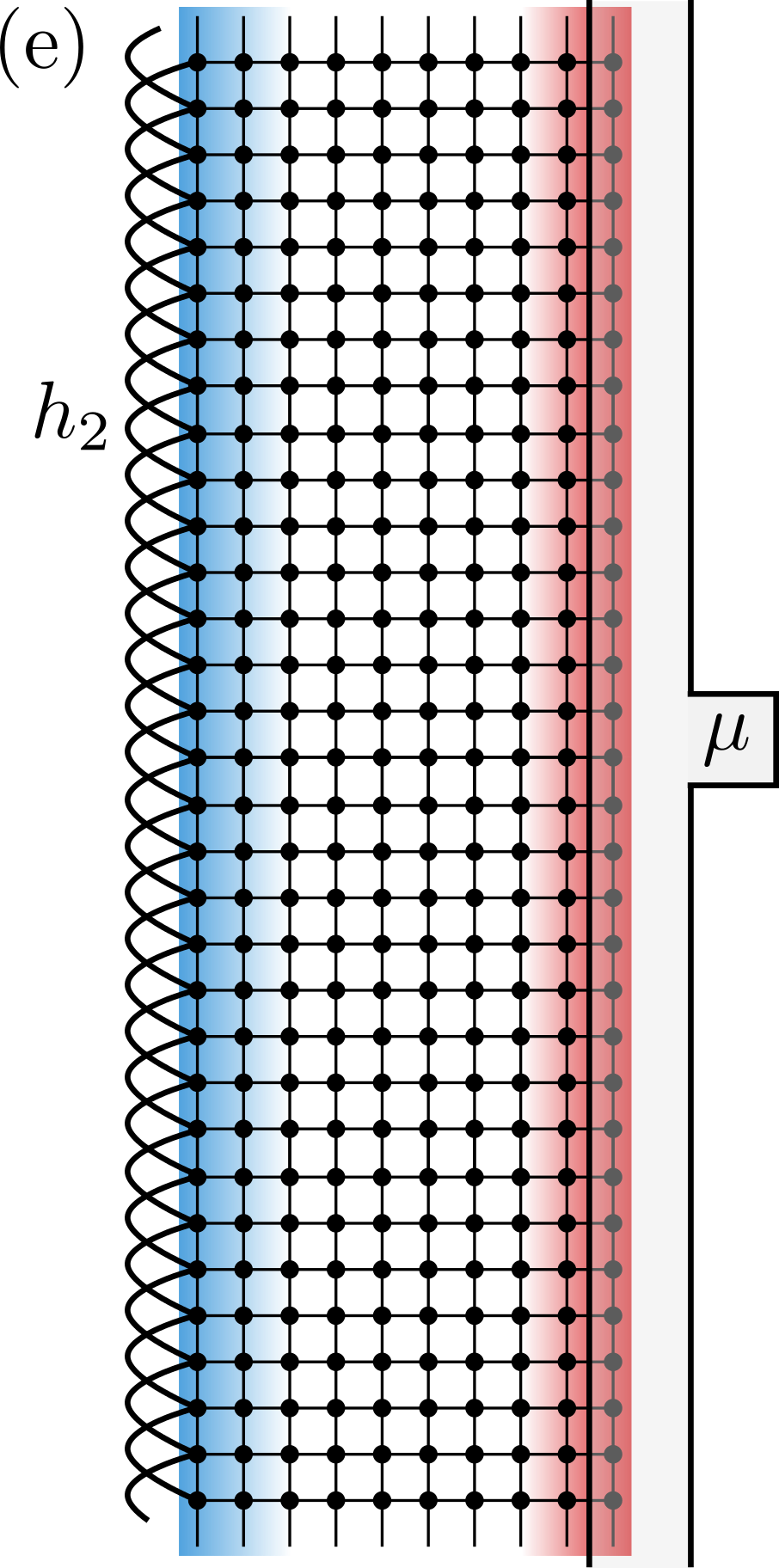}
\caption{ Dispersion relation of a strip of the QWZ model, with edge
  states on the left/right edge in highlighted in dark red / light
  blue.  Top row: clean system. Bottom row: with extra next nearest
  neighbor hopping along $y$, only the left edge, with
  amplitude $h^{(1)}_2=2$ and an onsite potential $\mu^{(N)}$ at the right edge.
  The value of the potential in (c) is $\mu^{(N)}=0.5$ and in
  (d) it is $\mu^{(N)}=1.5$.  In (e) the schematics of the considered
  perturbations is shown.  The additional potential terms do not
  affect the bulk states but distort the edge modes and bring in new
  edge modes at energies that are in the bulk gap or above/below all
  bulk energies. \label{fig:qiwuzhang_edge_deform}}
\end{figure}

We can use dimensional reduction and translate the discussion about
the robustness of edge states from Chapt.~\ref{chap:ricemele} to the
edge states of the QWZ model. This treats the case where the
Hamiltonian is modified in a way that only acts in the edge regions,
and is translation invariant along the edges. As an example, we
introduce an extra, state-independent next-nearest neighbor hopping, 
and onsite potentials at the left and right edge of the sample. 
As Fig.\ref{fig:qiwuzhang_edge_deform} shows, this can modify the
existing edge state branches, as well as create new edge state
branches by deforming bulk branches. Including the new local terms the Hamiltonian 
of Eq.~\eqref{eq:qiwuzhang-strip} is augmented to read

\begin{multline}
\HH(k_y) = \sum_{\uci_x = 1}^{\nuc_x-1} 
\left( \ket{\uci_x+1}\bra{\uci_x} 
\otimes \frac{\hsigma_z + i \hsigma_x}{2} + h.c. \right)+\\
\sum_{\uci_x = 1}^{\nuc_x} 
\ket{\uci_x}\bra{\uci_x} 
\otimes (\cos k_y \hsigma_z +  \sin k_y  \hsigma_y
\uu \otimes \hsigma_z )
+\\\sum_{\uci_x\in\{1,N\}}
\ket{\uci_x}\bra{\uci_x}\otimes\hat{\II}_2 
\left(\mu^{(\uci_x)}+h^{(\uci_x)}_2\cos 2k_y\right). 
\label{eq:qiwuzhang-strip-extra-terms}
\end{multline}
Where $\mu^{(1)/(N)}$ is the onsite potential on the left/right edge 
and $h^{(1)/(N)}_2$ describes a second nearest neighbor hopping on the left/right edge.
If not stated otherwise these terms are considered to be zero.

In the top row of Fig.\ref{fig:qiwuzhang_edge_deform} the spectrum of
strips without edge perturbations are depicted for Chern number $Q=-1$
(a) and $Q=0$ (b) respectively. As we expected, a nonzero Chern number
results in edge states, one on each edge.  In (c) and (d) switching on
perturbations, we see new edge states moving in to the gap.  The
onsite potential acts as an overall shift in energy on the
states around $\uci_x=N$, the second nearest neighbor hopping adds a
considerable warping to the states localized around $\uci_x=1$ The
deformations can change the number of edge states at a specific
energy, but only by adding a pair of edge states with opposite
propagation directions. This leaves the topological invariant
unchanged.

\begin{figure}
\centering
\includegraphics[width=0.6\linewidth]{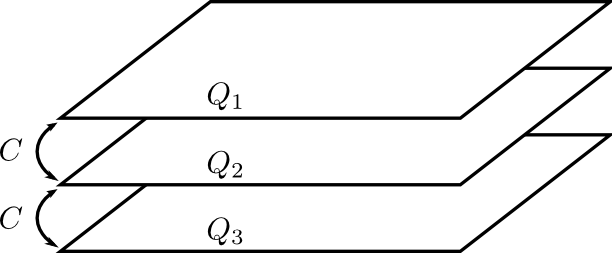}
\caption{Layering sheets of 2-dimensional insulators on top of each
  other is a way to construct a two-dimensional insulator with higher Chern
  numbers. For uncoupled layers, the Chern numbers can simply be
  summed to give the total Chern number of the 3-layer structure,
  $Q_1+Q_2+Q_3$. Switching on coupling (hopping terms) between the
  sheets cannot change the Chern number as long as the bulk gap is not
  closed.
 \label{fig:chern_layer}}
\end{figure}

\subsection{Constructing models with higher Chern numbers}
\label{subsec:qwz-constructing-layers}

A systematic way to construct models with higher Chern numbers is
to layer sheets of Chern insulators onto each other, as illustrated in
Fig.~\ref{fig:chern_layer}. The single-particle Hilbert space of the
composite system of $D$ layers is a direct sum of the Hilbert spaces
of the layers. 
\begin{align}
\mathcal{H}_D &= \mathcal{H}_{L1}\oplus\mathcal{H}_{L2}\oplus\ldots
\oplus\mathcal{H}_{LD}.
\end{align}


The Hamiltonian, including a state-independent interlayer coupling with
amplitude $C$, is  
\begin{align}
\hat{H}_D &= \sum_{d=1}^{D} \ket{d}\bra{d} \otimes \hat{H}_{Ld} + 
\sum_{d=1}^{D-1} (\ket{d+1}\bra{d} + \ket{d}\bra{d+1} \otimes C\,
\hat{\II}_{2\nuc_x\nuc_y}, 
\end{align}
with $\hat{\II}_{2\nuc_x\nuc_y}$ the unit operator on the Hilbert
space of a single layer.  The operators $\hat{H}_{Ld}$ we consider
below are of the form of
Eq.~\eqref{eq:qiwuzhang-strip-extra-terms} with different values
of $\uu$, and can have an overall real prefactor. In layer $d$ the 
strength of the local edge potential is denoted as $\mu^{(1)/(N)_d}$.
As an example, the matrix of the Hamiltonian of a system with 
three coupled layers reads
\begin{align}
H_3 = 
\begin{bmatrix}
H_{L1} & C\, \II & 0 \\
C\, \II & H_{L2} & C\, \II \\
0 & C\, \II & H_{L3} \\
\end{bmatrix}.
\end{align}

\begin{figure}
\centering
\includegraphics[width=0.9\linewidth]
{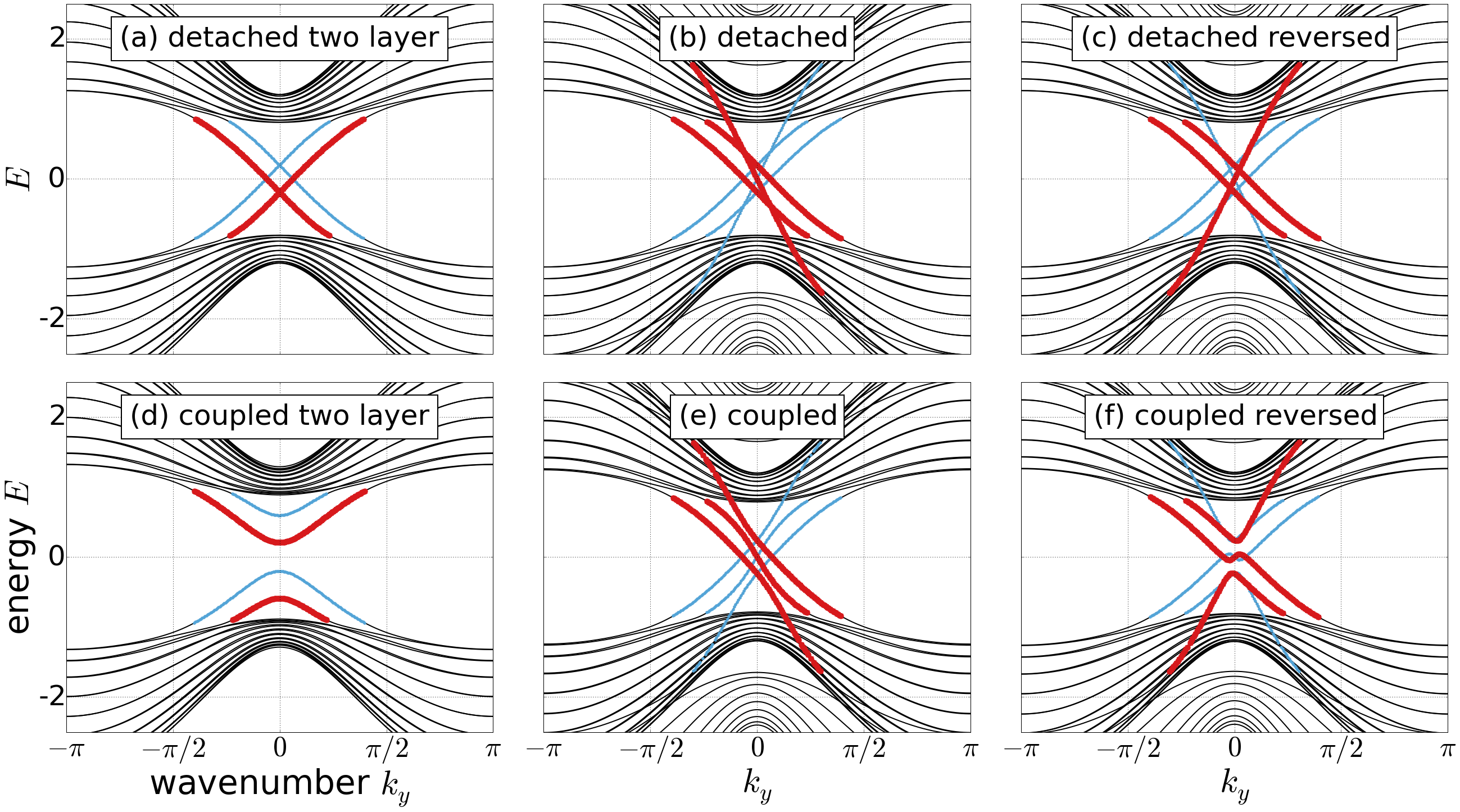}
\caption{Dispersion relations of strips of multilayered QWZ model,
  $N=10$ unit cells wide. In all cases the bulk Hamiltonian of the first layer $\HH_{L1}$ is
  the QWZ Hamiltonian with $\uu=-1.2$. To elucidate the interplay of 
  the edge states we consider a finite potential acting on the edge $\mu^{(1)/(N)}_{L1}=0.2$/$-0.2$.  
  In the left column ((a) and  (d)), we have two layers, with bulk $\HH_{L2}=-\HH_{L1}$ 
  and edge onsite potential $\mu^{(1)/(N)}_{L2}=\mu^{(1)/(N)}_{L1}$.  In the middle
  ((b), (e)) and right ((c), (f)) columns, we have three layers. The third layer is 
  characterized in the bulk by  $\HH_{L3}=\HH_{L1}$ and on the edge by $\mu^{(1)/(N)}_{L3}=-\mu^{(1)/(N)}_{L1}$. 
  In (b) and (e)  $\HH_{L2}=2 \HH_{L1}$, while in (d) and (e) $\HH_{L2}=-2 \HH_{L1}$.
  For all four cases $\mu^{(1)/(N)}_{L2}=0$. Coupling in (d), (e) and  (f) is uniform with magnitude $C=0.4$.
\label{fig:qiwuzhang-spectrum_23_layers}}
\end{figure}

Numerical results for two and three coupled layers, with different
Chern numbers in the layers, are shown in
Figs.~\ref{fig:qiwuzhang-spectrum_23_layers}. The coupling of
copropagating edge modes lifts the degeneracies, but cannot open gaps
in the spectrum, except in the case of strong coupling.  This is
a simple consequence of the fact that an energy eigenstate has to be a
single valued function of momentum. To open a gap, counterpropagating
edge states have to be coupled. We achieve this by coupling layers of
the QWZ model with opposite sign of the Chern number $Q$. For the
case of two layers (Fig.~\ref{fig:qiwuzhang-spectrum_23_layers},
second row), this opens a gap in the spectrum. If there are three
layers, there is a majority direction for the edge states, and so one
edge state survives the coupling.

\section{Robustness of edge states}
\label{sect:qiwuzhang-edge_robustness}

Up to now, we have considered clean edges, i.e., two-dimensional Chern Insulators
that were terminated by an edge (at $m_x=1$ and $m_x=N$), but
translationally invariant along the edge, along $y$. This
translational invariance, and the resulting fact that the wavenumber
$k_y$ is a good quantum number, was used for the definition of the
topological invariant $N_+-N_-$, which was the net number of edge
bands propagating along the edge, equal to the bulk Chern number,
$Q$. With disorder in the edge region that breaks translational
invariance along $y$, we no longer have a good quantum number $k_y$,
and edge state bands are not straightforward to define. However, as we
show in this section, the edge states must still be there in the
presence of disorder, since disorder at the edges cannot close the
bulk gap. 

\begin{figure}
\sidecaption
\includegraphics[scale=.4]{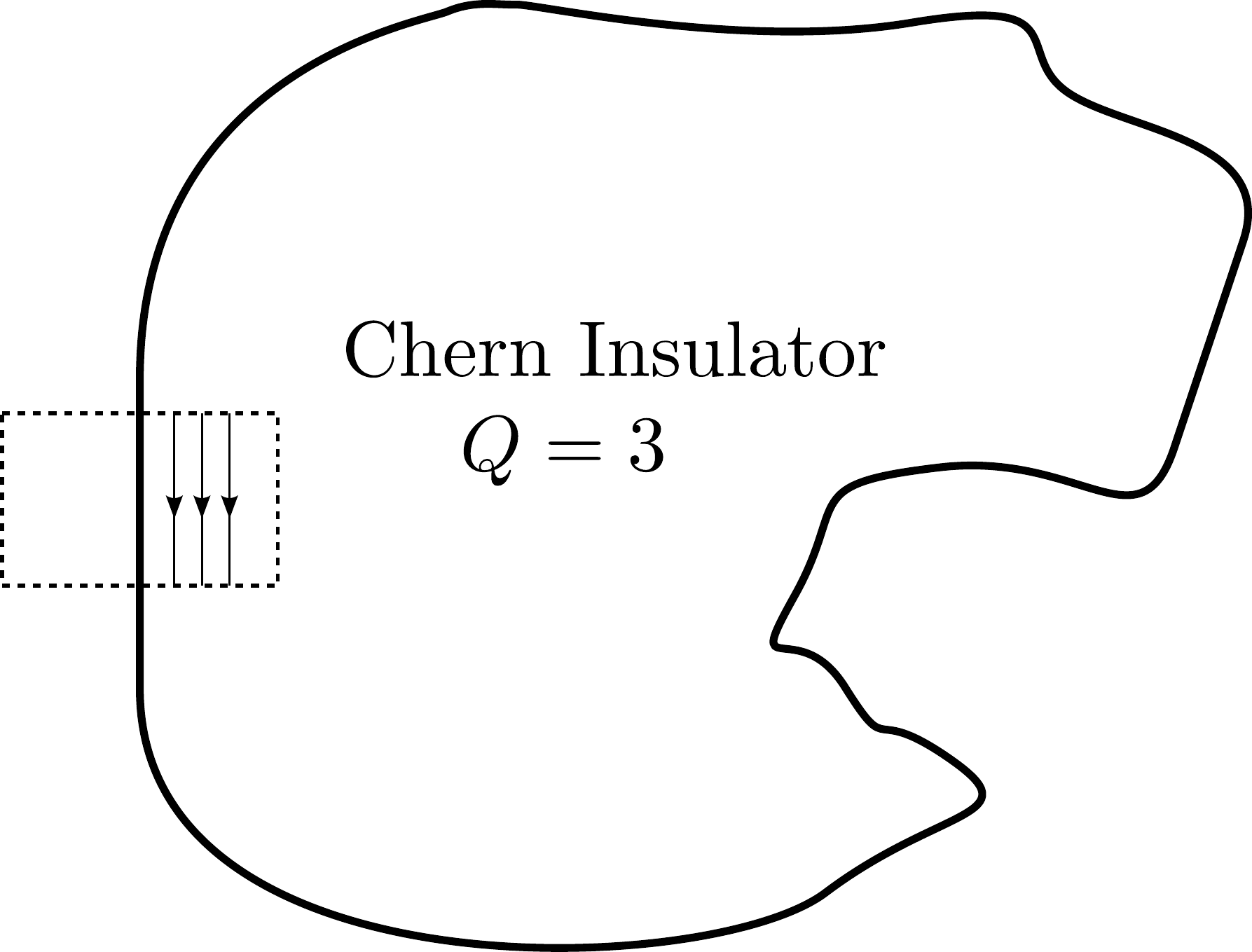}
\caption{ A disordered sample of Chern insulator. The dotted lines
  indicate rectangular parts of the sample, where disorder can be
  turned off adiabatically to reveal edge states (indicated in
  black). 
Since
particles cannot turn back (unidirectional, or
  \emph{chiral} channels), and cannot go into the bulk (in the gap),
  they have to travel all the way on the perimeter of the disordered
  sample, coming back to the rectangular, clean part. 
 \label{fig:qiwuzhang_2contacts}}
\end{figure}

\subsubsection*{Smoothly removing disorder}

Consider a finite, sample of a Chern insulator, with a clean bulk part
but disordered edge region, as depicted in
Fig.~\ref{fig:qiwuzhang_2contacts}. The bulk gap of the sample
decreases due to disorder, but we suppose that it is not closed
completely (just renormalized). Consider now a small part of the
sample, containing some of the edge, indicated by the dotted rectangle
on the right of Fig.~\ref{fig:qiwuzhang_2contacts}. Although this is
much smaller than the whole sample, it is big enough so that part of
it can be considered as translation invariant ``bulk''.  Now in this
small part of the sample, we adiabatically (and smoothly) deform the
Hamiltonian in such a way that we set the disorder gradually to
0. This includes straightening the part of the open boundary of the
sample that falls into the dotted rectangle, to a straight line. The
deformation is adiabatic in the sense that the bulk gap is not closed
in the process. Since this small part is a clean Chern insulator, with
a bulk Chern number of $Q$, it can be deformed in such a way that the
only edge states it contains are $\abs{Q}$ states propagating
counterclockwise (if $Q>0$, say).

\subsubsection*{Unitarity: particles crossing the clean part 
have to go somewhere}

Consider a particle in an edge state in the small clean part of the
sample, with energy deep inside the bulk gap ($E\approx 0$). What can
its future be, as its time evolution follows the Schr\"odinger
equation appropriate for this closed system? Since the edge state is a
chiral mode, the particle has to propagate along the edge until it
leaves the clean region.  Because of unitarity, the particle cannot
``stop'' at the edge: that would mean that at the ``stopping point'',
the divergence of the particle current in the energy eigenstate is
nonzero. In other words, the particle current that flows in the edge
state has to flow somewhere. (Put differently, if the mode describing
an edge state particle ``stopped at the interface'', two particles,
initially orthogonal, following each other in the mode, would after
some time arrive to the same final state. This would break unitarity.)
After leaving the clean part of the sample, the particle cannot
propagate into the bulk, since its energy is deep in the bulk gap. The
disorder in the clean part was removed adiabatically, and thus there
are no edge states at the interface of the clean part and the
disordered part of the sample, along the dashed line. The particle
cannot turn back, as there are no edge states running ``down'' along
the edge in the clean part.  The only thing the particle can do is
propagate along the edge, doing a full loop around the sample until it
comes back to the clean part from below again.

The argument of the previous paragraph shows that even though the sample is
disordered, there has to be a low energy mode that conducts perfectly
(reflectionless) along the edge. Since at 0 energy there are $Q$
orthogonal states a particle can be in at the edge of the clean part
of the sample, unitarity of the dynamics of the particles requires
that all along the edge of the disordered sample there are $Q$
orthogonal modes that conduct counterclockwise. There can be
additional low energy states, representing trapped particles, or an
equal number of extra edge states for particles propagating
counterclockwise and clockwise. However, the total number of
counterclockwise propagating edge modes at any part of the edge always
has to be larger by $Q$ than the number of clockwise propagating edge
modes.  Because the Hamiltonian is short range, our conclusions
regarding the number of edge states at any point far from the deformed
region have to hold independent of the deformation.

To be precise, in the argument above we have shown the existence of
$Q$ edge states all along the edge of the sample, except for the small
part that was adiabatically cleaned from disorder. 
One way to finish the argument is by
considering another part of the sample. If we now remove the
disorder adiabatically only in this part, we obtain the existence of
the edge modes in parts of the sample including the original dotted
rectangle, which was not covered by the argument of the previous
paragraph.

\section*{Problems}
\addcontentsline{toc}{section}{Problems}

\begin{prob}
\label{prob:qwz-phasediagram}
\textbf{Phase diagram of the anisotropic QWZ model }
\\
The lattice Hamiltonian of the QWZ model is provided 
in Eq. \eqref{eq:qiwuzhang-realspace_Hamiltonian}.
Consider the anisotropic modification of the Hamiltonian
when the first term of Eq. \eqref{eq:qiwuzhang-realspace_Hamiltonian},
describing the hopping along the $x$ axis, 
is multiplied by a real number $A$. 
Plot the phase diagram of this model, 
that is, evaluate the Chern number 
as a function of two parameters $u$ and $A$. 
\end{prob}

%


%% file: 2D_Dirac.tex
\chapter{Continuum model of localized states at a domain wall}
\label{ch:a_2DDirac}

\abstract*{So far, we have discussed edge states in lattice models, in which 
the states live on discrete lattice sites, and
the Hamiltonian governing the physics is a matrix. 
Here we argue that in certain cases, the interesting edge states arising in lattice models, discussed in earlier chapters of the book, 
can also be described by continuum models, 
in which the states live in continuous space,
and the Hamiltonian is a differential operator.
The method applied to derive the continuum models 
is known as the envelope-function approximation.
We obtain continuum Hamiltonians for three basic lattice 
models: the one-dimensional monatomic chain, the one-dimensional SSH model, and the 
two-dimensional QWZ model.  
In the cases of the
SSH and QWZ models, we use the resulting effective
Schr\"odinger equations to 
analytically characterize the localized states appearing at boundaries
between regions with different topological invariants.}

So far, we have discussed edge states in lattice models, in which 
the states live on discrete lattice sites, and
the Hamiltonian governing the physics is a matrix. 
In this chapter, we argue that in certain cases, 
it is also possible  to describe these states 
via a continuum model,
in which the states live in continuous space,
and the Hamiltonian is a differential operator.
One benefit of such a continuum description is that it
allows one to use the vast 
available toolkit of differential equations for solid-state problems in general,
including the description of topologically protected states in particular.
Another interesting aspect of these continuum models is their strong
similarity with the Dirac equation describing relativistic fermions. 
A limitation of the continuum models is that their validity
is restricted to narrow windows in momentum and energy;
typically they are applied in the vicinities of band edges. 
Here, we obtain the continuum differential equations for three basic lattice 
models: the one-dimensional monatomic chain, the one-dimensional SSH model, and the 
two-dimensional QWZ model.  
In the cases of the
SSH and QWZ models, the resulting equations will be used to
analytically characterize the localized states appearing at boundaries
between regions with different topological invariants.
Even though in the entire chapter we build our discussion on the three
specific lattice models, the applicability of the technique introduced
here, called \emph{envelope-function approximation}, is more
general and widely used to describe electronic states in various
crystalline solids.

\section{one-dimensional monatomic chain in an electric potential}
\label{sec:dirac_1dmc}

We use this minimal lattice model 
to illustrate the basic concepts
of the envelope-function approximation (EFA) \cite{Bastard-book},
the technique that allows us to find the continuum versions
of our lattice models.
Of course, the one-dimensional monatomic chain does not host topologically protected
states.

\subsection{The model}

\begin{figure}[!ht]
\begin{center}
\includegraphics[width=0.8\textwidth]{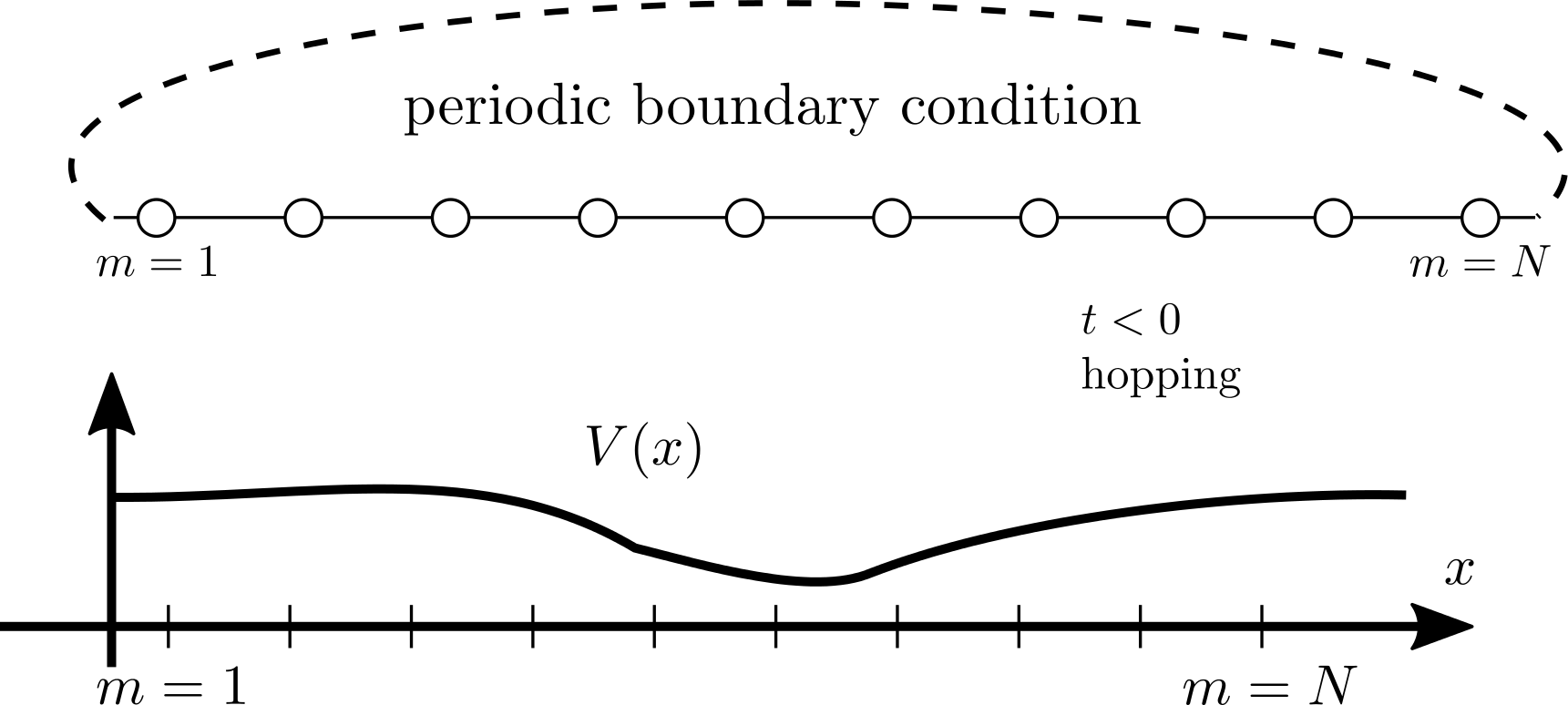}
\caption{\label{fig:dirac_1dmc}
one-dimensional monatomic chain in an inhomogeneous potential $V(x)$.}
\end{center}
\end{figure}

We take a long lattice with $N\gg 1$ unit cells (or sites) without
an internal degree of freedom,
with periodic boundary conditions, and a negative hopping amplitude
 $t<0$.
We consider the situation when
the electrons are subject to an inhomogeneous electric
potential $V(x)$.
This setup is pictured in Fig. \ref{fig:dirac_1dmc}.
The lattice Hamiltonian describing this 
inhomogeneous system reads
\bean
H_\textrm{i} = H + V,
\eean
where
\bean
H &=& t \sum_{\uci=1}^N \ket{\uci} \bra{\uci+1} + h.c., \\
V &=& \sum_{\uci=1}^N V_\uci \ket{\uci} \bra{\uci},  \label{eq:dirac_V}
\eean
with $V_\uci = V(x=n)$.
Our aim is to construct a continuum model that
accurately describes the low-energy eigenstates of this lattice
Hamiltonian.

Before discussing the inhomogeneous case incorporating $V(x)$, 
focus first on the homogeneous system.
The bulk momentum-space Hamiltonian 
is a scalar ($1\times 1$ matrix) 
in this model, since there is no integral degree of freedom
associated to the unit cell;
it reads 
\bean
H(k) = \epsilon(k) \ket{k} \bra{k},
\eean
where $\epsilon(k)$ is the electronic dispersion relation:
\bean
\epsilon(k) = -2|t| \cos k.
\eean
The low-energy part of the dispersion relation is located around
zero momentum, and is approximated by a parabola: 
\bean
\label{eq:dirac:1dmclowenergy}
\epsilon(k) \approx -2 |t| +  |t| k^2
= 
\epsilon_0 + \frac{k^2}{2m^*},
\eean
where we introduced the minimum energy of the band 
$\epsilon_0 = -2|t|$,  and the effective mass
$m^* = \frac{1}{2|t|}$
characterizing the 
low-energy part of the dispersion relation.
(Using proper physical units, the effective mass would have
the form $m^* = \hbar^2 / (2  |t| a^2)$,
with $a$ being the lattice constant.)
For simplicity, we suppress $\epsilon_0$ in what follows; 
that is, we measure energies with respect to $\epsilon_0$.

\subsection{Envelope-function approximation}

Our goal is to find the low-energy eigenstates
of the inhomogeneous lattice Hamiltonian $H_\textrm{i}$. 
The central proposition of the EFA, applied to 
our specific example of the one-dimensional monatomic chain, 
says that it is possible to complete this goal 
by solving the simple continuum Schr\"odinger equation
\bean
H_\textrm{EFA} \varphi(x) = E \varphi(x), 
\eean
where the \emph{envelope-function Hamiltonian} $H_\textrm{EFA}$
has a very similar form to the free-electron Hamiltonian 
with the electric potential $V$: 
\bean
\label{eq:dirac:1dmchefa}
H_\textrm{EFA} = \frac{\hat{p}^2}{2m^*} + V(x).
\eean
Here $\hat{p} = -i \partial_x$ is the usual real-space 
representation of the momentum operator,
and the  function $\varphi(x)$ is usually called
\emph{envelope function}. 
Note the very simple relation between the low-energy dispersion 
in Eq. \eqref{eq:dirac:1dmclowenergy} and the kinetic term in the 
EFA Hamiltonian \eqref{eq:dirac:1dmchefa}:
the latter can be obtained from the former by
substituting the momentum operator $\hat{p}$
in place of the momentum $k$.

Before formulating the EFA proposition more precisely, we
introduce the concept of a
\emph{spatially slowly varying} envelope function. 
We say that $\varphi(x)$ is spatially slowly varying, if its Fourier transform 
\bean
\tilde \varphi(q) = \int_0^{N} dx \frac{e^{-iqx}}{\sqrt{N}} \varphi(x)
\eean
is localized to the $|q| \ll \pi$ region, i.e., 
to the vicinity of the center of the BZ.

With this definition at hand, we can formulate the EFA proposition.
Consider the inhomogeneous one-dimensional monatomic chain 
described by $H_\textrm{i}$. 
Assume that  $\varphi(x)$
is a spatially slowly varying eigenfunction of $H_\textrm{EFA}$
with eigenvalue $E$. 
Then, the state $\ket{\psi}$ defined on the lattice via
\bean
\label{eq:dirac_tbef}
\ket{\psi} =  \sum_{\uci=1}^N \varphi(x=\uci) \ket{\uci}.
\eean
is approximately an eigenstate of the lattice Hamiltonian $H_\textrm{i}$
with eigenvalue $E$.
The proof follows below in Sect.~\ref{sec:dirac:proof}.

Note that if  
the envelope function $\varphi(x)$ 
fulfils the normalization condition 
\bean
\int_0^{N} dx |\varphi(x)|^2 &=& 1,
\eean
then, for a long lattice $N \gg 1$, 
the lattice state $\ket{\psi}$ will also be normalized
to a good accuracy: 
\bean
\bra{\psi} \psi \rangle &\approx& 1.
\eean

It is important to point out two possible interpretations
of the lattice state $\ket{\psi}$ introduced
in Eq. \eqref{eq:dirac_tbef}.
(1) The  state $\ket{\psi}$ 
can be interpreted
as the zero-momentum band-edge eigenstate 
$\ket{k=0} = \frac{1}{\sqrt{N}} \sum_{\uci=1}^{N} \ket{\uci}$
of the homogeneous system, 
modulated by the envelope function $\varphi(x)$ restricted
to the lattice-site positions $x=\uci$. 
(2) The state $\ket{\psi}$ can also be interpreted as
a wave packet, composed of 
those eigenstates $\ket{k}$ of the homogeneous lattice Hamiltonian
$H$
that have wave numbers $k$ close to the band-edge wave number,
the latter being zero in this case.
To see this, we first Fourier-decompose $\varphi(x)$:
\bean
\varphi(x) = \sum_{k \in \textrm{BZ}} \tilde \varphi(k) \frac{e^{ikx}}{\sqrt{N}}
\approx
\sum'_{k} \tilde \varphi(k) \frac{e^{ikx}}{\sqrt{N}}
.
\label{eq:effourier}
\eean
In the approximate equality, 
we used the fact that $\varphi(x)$ is spatially slowly varying, 
i.e., its Fourier transform $\tilde \varphi(k)$ is localized to the 
central part of the BZ,
and introduced the notation $\sum'_k$ 
for a wave-number sum that goes only for the central part of the BZ. 
By inserting Eq. \eqref{eq:effourier} to Eq. \eqref{eq:dirac_tbef},
we find
\bean
\label{eq:dirac_wavepacket}
\ket{\psi} =\sum_{\uci=1}^{N} 
\left(\sum'_k \tilde \varphi(k) \frac{e^{ik \uci}}{\sqrt{N}} \right) \ket{\uci}
= \sum'_k \tilde \varphi(k) \ket{k}.
\eean
That is, $\ket{\psi}$ is indeed a packet of plane waves with
small wave numbers.

\subsection{Envelope-function approximation: the proof}
\label{sec:dirac:proof}

To prove the EFA proposition, 
we calculate $H_\textrm{i} \ket{\psi}$ and
utilize the spatially-slowly-varying condition on $\varphi(x)$. 
Start with the contribution of the bulk Hamiltonian $H$:
\bean
\label{eq:dirac_1dproofstart}
H \ket{\psi} = 
\left[
	\sum_{k \in \textrm{BZ}} \epsilon(k) \ket{k} \bra{k}
\right]
\left[
	\sum_{\uci=1}^N \varphi(\uci) \ket{\uci}
\right].
\eean
Express the envelope function via its Fourier transform: 
\bean
H \ket{\psi} = 
\left[
	\sum_{k \in \textrm{BZ}} \epsilon(k) \ket{k} \bra{k}
\right]
\left[
	\sum_{\uci=1}^N 
	\left(
		\sum_{q\in \textrm{BZ}} \tilde \varphi(q) \frac{e^{i q \uci}}{\sqrt{N}}
	\right) \ket{\uci}
\right].
\eean
Now use the fact that $\varphi(x)$ is spatially slowly varying;
that implies that the $q$ sum can be restricted to the central part 
of the BZ:
\bean
H \ket{\psi} = 
\left[
	\sum_{k \in \textrm{BZ}} \epsilon(k) \ket{k} \bra{k}
\right]
\left[
	\sum_{\uci=1}^N 
	\left(
		\sum'_{q} \tilde \varphi(q) \frac{e^{i q \uci}}{\sqrt{N}}
	\right) \ket{\uci}
\right].
\eean
Performing the sum for $m$, we find
 \bean
H \ket{\psi} = 
\left[
	\sum_{k \in \textrm{BZ}} \epsilon(k) \ket{k} \bra{k}
\right]
\left[
		\sum'_{q} \tilde \varphi(q) \ket{q}
\right].
\eean
Performing the scalar product yields
 \bean
H \ket{\psi} = 
	\sum'_{q} \epsilon(q) 
	\tilde \varphi(q)\ket{q}. 
\eean
Using the fact that the $q$ sum goes for the central part of the BZ, 
where the dispersion relation $\epsilon(q)$ is
well approximated by a parabola, 
we find
\bean
H \ket{\psi} \approx 
	\sum'_{q} \frac{ q^2}{2m^*} 
		\tilde \varphi(q)\ket{q} .
\eean
Utilizing the definition of the plane wave $\ket{q}$, we obtain
\bean
H \ket{\psi} \approx 
		\sum'_{q}  \frac{ q^2}{2m^*}
	\tilde \varphi(q) \sum_{\uci=1}^N \frac{e^{iq \uci}}{\sqrt N} 
	\ket{\uci},
\eean
which can be rewritten as 
\bean
H \ket{\psi} 
&\approx&
\sum_{\uci=1}^N 
	\left[
	\sum'_{q} \frac{ q^2}{2m^*} 
	\tilde \varphi(q)\frac{e^{iq x}}{\sqrt N} 
	\right]_{x=\uci}\ket{\uci}
\\
&=&
	\sum_{\uci=1}^N 
	\left[
	 - \frac{1 }{2m^*} \partial_x^2 
	\sum'_{q} \tilde \varphi(q)\frac{e^{iq x}}{\sqrt N} 
	\right]_{x=\uci}\ket{\uci}
\\
&=&
	\sum_{\uci=1}^N 
	\left[
	\frac{\hat{p}^2}{2m^*} 
	 \varphi(x)
	\right]_{x=\uci}\ket{\uci} .
	\label{eq:dirac_Hpsi}
\eean

Continue with the contribution of the potential $V$.
Using Eqs. \eqref{eq:dirac_V} and \eqref{eq:dirac_tbef}, we find
\bean
\nonumber
V \ket{\psi} 
&=& \left[ \sum_{\uci=1}^N V_\uci \ket{\uci} \bra{\uci} \right]
\left[\sum_{\uci'=1}^N  \varphi({\uci'}) \ket{\uci'} \right]
=   \sum_{\uci=1}^N V_\uci  \varphi(\uci) \ket{\uci} \\
&=& \sum_{\uci=1}^N [V(x) \varphi(x)]_{x=\uci} \ket{\uci}.
\label{eq:dirac_Vpsi}
\eean

Summing up the contribution \eqref{eq:dirac_Hpsi} of $H$ 
and the contribution 
\eqref{eq:dirac_Vpsi} of $V$, 
we find 
\bean
(H+V)\ket{\psi} &\approx &
 \sum_{\uci=1}^N
	\left[
		H_\textrm{EFA} \varphi (x) 
	\right]_{x=\uci} \ket{\uci}
	=
\sum_{\uci=1}^N
	\left[
		E \varphi (x) 
	\right]_{x=\uci} \ket{\uci}	
\\
&=& E
 \sum_{\uci=1}^N
		\varphi (\uci)  \ket{\uci}	
=
	E \ket{\psi},
\label{eq:dirac_1dproofend}
\eean
which concludes the proof.

\section{The SSH model and the one-dimensional Dirac equation}
\label{sec:dirac:ssh}

To illustrate how the EFA captures topologically protected bound
states in one-dimensional, 
we use the SSH model described in detail 
in Chapt.~\ref{chap:ssh}.
The model
is visualized in Fig. \ref{fig:Geometry-of-the-SSH}.
The bulk Hamiltonian is characterized by two parameters,
the intra-cell and inter-cell hopping amplitudes $v,w>0$, respectively.
The bulk lattice Hamiltonian of the SSH model 
is given in Eq. \eqref{eq:ssh_hamiltonian_def}, whereas
the bulk momentum-space Hamiltonian is
given in Eq. \eqref{eq:ssh_bulk_H_def}.
We have seen that the $(v,w)$ parameter space
is separated to two adiabatically connected
partitions by the $v = w$ line. 
In Sect.~\ref{subsec:ssh_domain}, we have also seen that 
localized zero-energy states appear at 
a domain wall between two half-infinite homogeneous regions,
if the two regions have different bulk topological invariants;
that is, if the sign of $v-w$ is different at the two sides of the
domain wall. 

\begin{figure}[!ht]
\begin{center}
\includegraphics[width=0.8\textwidth]{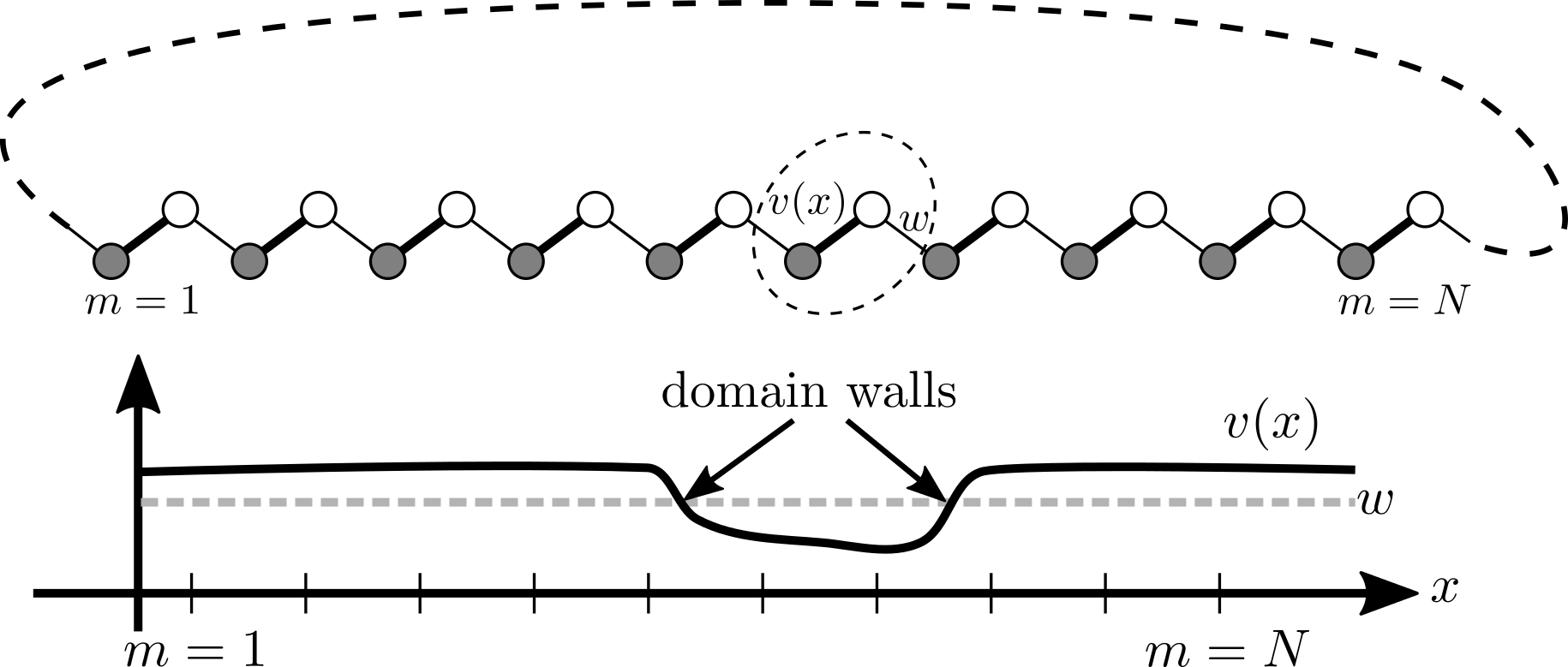}
\caption{\label{fig:dirac_1dssh}
Inhomogeneous intracell hopping and domain walls in the 
SSH model.
The dashed ellipse denotes the unit cell. 
The dashed line connecting the edges of the
chain denotes the periodic boundary condition.
}
\end{center}
\end{figure}

This is the phenomenon that we address in this section: 
we show that an analytical description of such localized
states can be given using the EFA. 
First, we discuss the electronic dispersion 
relation of the metallic ($v=w$)
and nearly metallic ($|v-w| \ll |v+w|$) homogeneous 
SSH model.
Second, 
we obtain  the Dirac-type differential equation
providing a continuum description for the inhomogeneous
 SSH model (see Fig. \ref{fig:dirac_1dssh})
 for the energy range in the
vicinity of the bulk band gap. 
Finally, we solve that differential
equation to find the localized zero-energy states at a domain wall. 
Remarkably, the analytical treatment remains
useful even if the spatial structure  of the 
domain wall is rather irregular.

\subsection{The metallic case}

First, consider the metallic homogeneous SSH model,
where $v=w$. 
The dispersion relation is shown as the 
blue solid line in Fig. \ref{fig:ssh_5_dispersions}c.
The filled and empty bands touch at the 
end of the BZ, at $k=k_0 \equiv \pi$.
Figure \ref{fig:ssh_5_dispersions}c
shows that in the vicinity of that touching point, commonly referred to as
a \emph{Dirac point}, the dispersion relations are linear
functions of the relative wave vector $q= k-k_0$.
The slope of these linear functions, corresponding 
to the group velocity of the electrons, 
can be determined, e.g. by Taylor-expanding the bulk 
momentum-space Hamiltonian 
$\ido{H}(k) = (v+w\cos k) \ido{\sigma}_x + w\sin k \ido{\sigma}_y$,
see Eq. \eqref{eq:ssh_bulk_H_matrixelements_def},
to first order in $q$: 
\bean
\label{eq:dirac_masslesslin}
\ido{H}(k_0+q) \approx - w q \ido{\sigma}_y, \ \ \ \ (v=w).
\eean 
which indeed has a linear dispersion relation, 
\bean
\label{eq:dirac_metallicsshdispersion}
E_\pm(q) = \pm w q.
\eean
The eigenstates of the linearized 
Hamiltonian \eqref{eq:dirac_masslesslin} are
\begin{equation}
\psi_{\pm} (q) = 
\frac{1}{\sqrt 2}
\left(\begin{array}{c}
1 \\ \mp i
\end{array}\right).
\end{equation}


Note that the dispersion relation 
of the Dirac equation of fermions with zero mass is
\begin{equation}
\label{eq:dirac_zeromass}
E_\pm(k ) = 
\pm \hbar k c,
\end{equation}
where $\hbar$ is the reduced Planck's constant and $c$ is the speed of light.
Comparing Eqs. \eqref{eq:dirac_metallicsshdispersion}
and \eqref{eq:dirac_zeromass}, 
we conclude that the dispersion of the metallic SSH model 
is analogous to that of massless Dirac fermions, 
and the hopping amplitude of the metallic 
SSH model plays the role
of $\hbar c$.
Because of the similarity of the 
dispersions
\eqref{eq:dirac_metallicsshdispersion} 
and
\eqref{eq:dirac_zeromass},
the linearized Hamiltonian 
\eqref{eq:dirac_masslesslin} is often called a \emph{massless 
Dirac Hamiltonian}.

At this point, the linearization of the bulk momentum-space Hamiltonian 
of the SSH model 
does not seem to be a particularly
 fruitful simplification: to obtain the dispersion relation and the
eigenstates, a $2\times2$ matrix has to be diagonalized, no matter 
if the linearization has been done or not.
However, linearizing the Hamiltonian is the first step
towards the EFA, as discussed below.

\subsection{The nearly metallic case}

Now consider a homogeneous, insulating SSH model
that is \emph{nearly metallic};
that is, the scale of the energy gap $|v-w|$ opened at $k_0$ 
is significantly smaller than the scale of the band width $v+w$. 
An example is are shown in 
\ref{fig:ssh_5_dispersions}b, where the dispersion 
relation is plotted for the
parameter values $v=1$ and $w=0.6$.

We wish to describe the states close to the band gap located
around zero energy.
Hence, again, we can  use
the approximate bulk momentum-space Hamiltonian
obtained via linearization in the relative momentum $q$:
\begin{equation}
\label{eq:dirac_homlin}
\ido{H}(k_0 + q) \approx
M \ido{\sigma}_x - w q \ido{\sigma}_y,
\end{equation}
where we defined $M = v-w$.
The dispersion relation reads
\begin{equation}
E_{\pm} ( q) = \pm \sqrt{M^2+ w^2 q^2}.
\end{equation}
The (unnormalized) eigenstates of the linearized Hamiltonian
\eqref{eq:dirac_homlin} have the 
form
\begin{equation}
\psi_{\pm}( q) = \left(\begin{array}{c}
M+ i w q \\ E_\pm(q)
\end{array}\right).
\end{equation}

Note that the dispersion relation 
of the Dirac equation for fermions with finite mass $\mu \neq 0$ reads
\begin{equation}
\label{eq:dirac_massivedispersion}
E_\pm( k) = 
\pm \sqrt{\mu^2 c^4 + \hbar^2 k^2 c^2}.
\end{equation}
Therefore, the parameter $M=v-w$ of the SSH model plays the role
of the mass-related term $\mu c^2$ of the relativistic dispersion relation
\eqref{eq:dirac_massivedispersion},
and the linearized Hamiltonian \eqref{eq:dirac_homlin} is
often called a \emph{massive Dirac Hamiltonian}.

\subsection{Continuum description of the nearly metallic case}
\label{sec:dirac_continuumssh}

We are mostly interested in a continuum description of 
the zero-energy localized states formed at a domain wall
between two topologically distinct regions. 
For simplicity and concreteness,  consider the case 
when the domain wall is created so that
the intra-cell hopping amplitude $v$ varies in space while
the inter-cell one is  constant,
as shown in Fig. \ref{fig:dirac_1dssh}.

In what follows, we will focus on one of the two domain walls
shown in Fig. \ref{fig:dirac_1dssh}.
The inhomogeneous Hamiltonian has the form
\bean
H_\textrm{i} = \sum_{\uci=1}^N
v_\uci \left( \ket{\uci,B} \bra{\uci,A} + h.c. \right)
+ 
w 
\sum_{\uci=1}^N
\left(
	\ket{\uci,B} \bra{\uci+1,A}
	+ h.c.
\right),
\eean
where $v_\uci = v(x=\uci)$ and $v(x) \geq 0 $ is a continuously varying
function of position, which takes the constant value
$v_-$ ($v_+$) far on the left (right) from the domain wall.

We also assume that the local Hamiltonian is a nearly metallic SSH 
Hamiltonian everywhere in space. 
That is, $|v(x) - w | \ll v(x)+w$.
This ensures that the local band gaps
$|v_\pm - w|$
on the two sides of the domain wall
are much smaller than the local band widths
$v_\pm+w$.

Based on our experience with the EFA in the 
inhomogeneous one-dimensional monatomic chain
(see Sect.~\ref{sec:dirac_1dmc}), 
now we construct the EFA proposition corresponding to 
this inhomogeneous, nearly metallic SSH model. 
Recall that in the former case, 
we obtained the EFA Hamiltonian 
by (i) Taylor-expanding the bulk momentum-space
Hamiltonian around the wave vector corresponding to the
band extremum (that was $k_0 = 0$ in Sect.~\ref{sec:dirac_1dmc}),
(ii) replacing the relative wave vector $q$ with the
momentum operator $\hat{p} = - i \partial_x$, 
and
(iii) incorporating the inhomogeneity of the respective 
parameter, which was the on-site potential $V(x)$ in that case. 
The same procedure, applied now for the 
SSH model with a first-order Taylor expansion, yields
the following EFA Hamiltonian: 
\bean
\label{eq:dirac_hefassh}
H_\textrm{EFA} = M(x) \ido{\sigma}_x - w \hat p  \ido{\sigma}_y. 
\eean
The EFA proposition is then formulated as follows. 
Assume that $\varphi(x) = (\varphi_A(x), \varphi_B(x))$ 
is a spatially slowly varying eigenfunction 
of $H_\textrm{EFA}$ in Eq. \eqref{eq:dirac_hefassh}, with eigenvalue E. 
Then, the state $\ket{\psi}$ defined on the
lattice  via
\bean
\ket{\psi} =  \sum_{\uci = 1}^N \sum_{\idi = A,B} \varphi_\idi(\uci) 
	e^{i k_0 \uci}\ket{\uci,\idi}
\eean
is approximately an eigenstate of the lattice Hamiltonian $H_\textrm{i}$
with energy $E$.

Note that, in analogy with Eq. \eqref{eq:dirac_wavepacket},
the lattice state $\ket{\psi}$ 
can be reformulated as a wave packet
formed by those eigenfunctions of the homogeneous ($v(x)=w$)
system that are in the vicinity of the band-edge momentum $k_0$:
\bean
\ket{\psi} 
\approx 
\sum'_q \sum_{\idi=A,B} \tilde{\varphi}_\idi(q) \ket{k_0 + q} \otimes \ket{\idi}.
\eean

\subsection{Localized states at a domain wall}
\label{sec:dirac:sshdomain}

\begin{figure}[!ht]
\begin{center}
\includegraphics[width=0.8\textwidth]{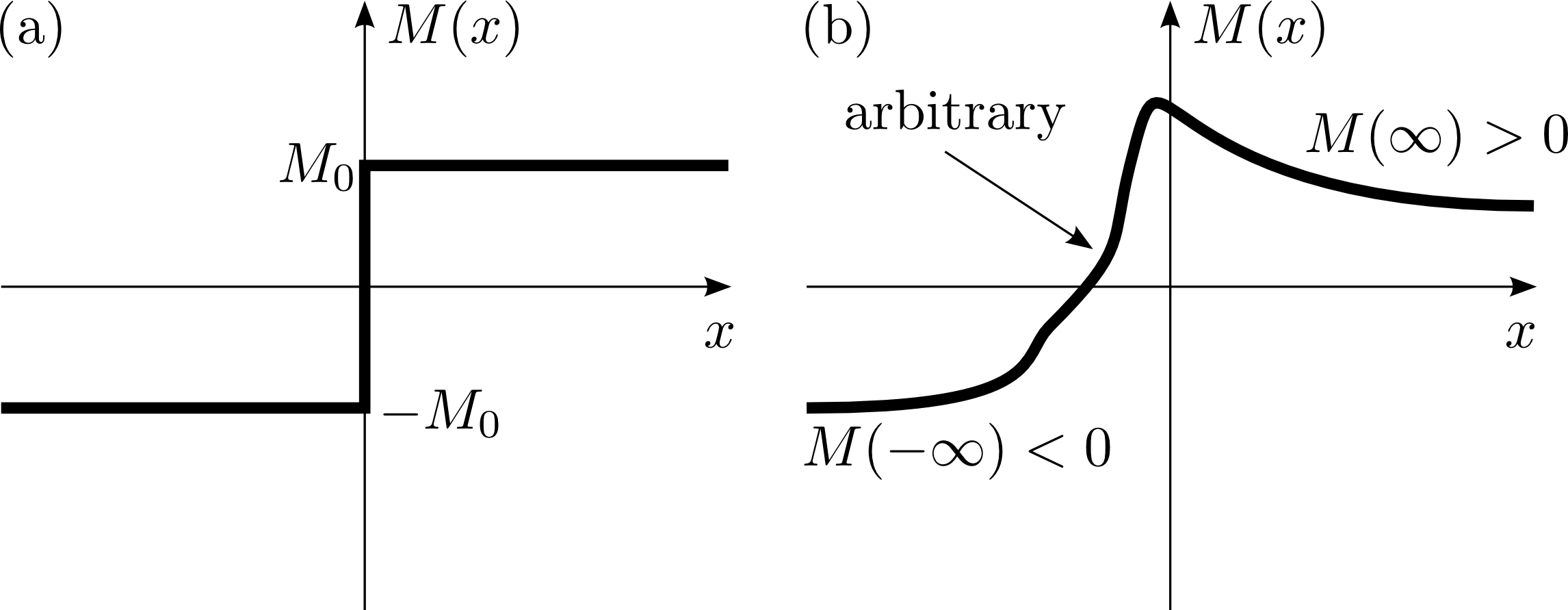}
\caption{\label{fig:a_1Dmassprofile}
(a) Step-like and (b) irregular spatial dependence of the mass
parameter $M(x)$ of the one-dimensional Dirac equation.}
\end{center}
\end{figure}

Having the envelope-function Schr\"odinger equation
\bean
\label{eq:dirac_sshefaschrodinger}
\left[
	M(x) \ido{\sigma}_x - w \hat p \ido{\sigma}_y 
\right] \varphi(x) = E \varphi(x)
\eean
at hand, we can study the domain wall between the
two topologically distinct regions.
First, we consider a step-type domain wall, defined via
\begin{equation}
\label{eq:dirac_massstep}
M(x) = \left\{ \begin{array}{lrl} 
M_0 & {\rm if} & x>0, \\
-M_0 & {\rm if} & x<0
\end{array}\right. ,
\end{equation}
and $M_0 > 0$, 
as shown in Fig.~\ref{fig:a_1Dmassprofile}a. 

We wish to use the EFA Schr\"odinger equation 
\eqref{eq:dirac_sshefaschrodinger}
to establish the zero-energy
states localized to the domain wall, 
which were revealed earlier in the lattice SSH model. 
That is, we look for evanescent solutions of 
Eq. \eqref{eq:dirac_sshefaschrodinger} on both sides
of the domain wall, and try to match them at the interface $x=0$. 
For the $x>0$ region, our evanescent-wave Ansatz reads
\bean
\varphi_{x>0}(x) = \spinor{a}{b} e^{-\kappa x}
\eean
with $\kappa > 0 $. 
Substituting this to Eq. \eqref{eq:dirac_sshefaschrodinger} yields
a quadratic characteristic equation for the energy $E$, 
having two solutions
\bean
\label{eq:dirac_evanescentenergies}
E_\pm = \pm \sqrt{M_0^2 - w^2  \kappa^2}. 
\eean
The corresponding unnormalized 
spinors read 
\bean
\spinor{a_\pm}{b_\pm} = \spinor{\frac{M_0 - w \kappa}{E_\pm}}{1}.
\eean
An analogous Ansatz for the $x<0$ region is
\bean
\varphi_{x<0}(x) = \spinor{c}{d} e^{\kappa x}
\eean
with $\kappa>0$, yielding the same energies as in 
Eq. \eqref{eq:dirac_evanescentenergies}, 
and the spinors
\bean
\spinor{c_\pm}{d_\pm} = \spinor{\frac{-M_0 + w \kappa}{E_\pm}}{1}.
\eean

Now consider an energy eigenstate 
with a given energy $E$. 
For clarity, set $M_0 > E \geq 0$.
The (unnormalized) envelope function of the energy 
eigenstate has the form
\begin{equation}
\label{eq:dirac_completewfn}
\varphi(x) = \varphi_{x>0}(x) \Theta(x) + C \varphi_{x<0}(x) \Theta(-x),
\end{equation}
where
$\varphi_{x<0}$ and $\varphi_{x>0}$
should be evaluated by replacing $E_\pm \mapsto E$ and
$\kappa \mapsto \frac{\sqrt{M_0^2- E^2}}{w}$, 
and $C$ is a yet unknown  parameter to be determined
from the boundary conditions at the domain wall.
The envelope  function \eqref{eq:dirac_completewfn} 
is an eigenstate of the EFA Hamiltonian with 
energy $E$ if the boundary condition that 
the wave function is continuous at $x=0$, that is,
\begin{equation}
\label{eq:dirac_bc}
\varphi_{x<0}(0) = C\varphi_{x>0}(0),
\end{equation}
is fulfilled.
Note that in our case, the Dirac equation
is a first-order differential equation and therefore there is no 
boundary condition imposed on the derivative of the wave function.
From the second component of Eq. \eqref{eq:dirac_bc}, 
we have $C=1$.
From the first component, we have $M_0 - w \kappa = -M_0 + w \kappa$, 
implying $M_0 = w \kappa$ and thereby $E=0$.  
The same result is obtained if the range $-M_0 < E \leq 0$ of 
negative energies is considered. 
Hence we conclude that the zero-energy state
at the domain wall does appear in the continuum model of the
inhomogeneous SSH chain, as expected. 


Let us also determine the coefficients $a$ and $c$
describing this localized state:
\bean
a = \lim_{E\to 0} \frac{M_0 - \sqrt{M_0^2-E^2}}{E} = 0,
\eean
and similarly, $c=0$. 
These imply that the localized state is completely sublattice-polarized,
i.e., it lives on the $B$ sublattice, and therefore it is its own
chiral partner.
Considering the a similar mass profile with negative $M_0$, 
we would have found that the localized state lives
on the $A$ sublattice. 
These properties are in line with our expectations drawn from the
lattice SSH model.  

A further characteristic property of the 
localized state is its localization length;
from our continuum model, we have an analytical result for that:
\bean
\label{eq:dirac_loclength}
\frac{1}{\kappa} = \frac{w}{ M_0 }.
\eean 
(In physical units, 
that is $\frac{1}{\kappa} = \frac{w}{ M_0 } a $.)
Recall that we are constrained to the nearly metallic regime
$w \gg M_0 = v_+ - w$;
together with Eq. \eqref{eq:dirac_loclength},
this implies that the localization length 
is much larger than one (that is, the lattice constant).
This is reassuring: it means that the envelope function
$\varphi(x)$ is spatially slowly varying, 
hence is within the range of validity of the EFA.

The result \eqref{eq:dirac_loclength} can be compared
to the corresponding result for the SSH lattice model.
Eq. \eqref{eq:ssh-loclength_vw}
 provides the localization length $\xi$ of an edge state
in a disordered SSH model, which 
corresponds to the localization length $\frac{1}{\kappa}$ obtained above.
Taking the disorder-free special case of Eq. \eqref{eq:ssh-loclength_vw},
we have
\bean
\label{eq:dirac:xi}
\xi = \frac{1}{\log \frac{w}{v}} = \frac{1}{\log \frac{w}{w+(v-w)}}
\approx \frac{w}{w-v}.
\eean
As we are making a comparison to the 
nearly metallic $v\approx w$ case considered in this section, 
we could approximate $\xi$ in Eq. \eqref{eq:dirac:xi} using
a leading-order Taylor expansion in the small quantity $(w-v)/w$.
The approximate result \eqref{eq:dirac:xi}
is in line with Eq. \eqref{eq:dirac_loclength}
obtained from the continuum model.



A further interesting fact is that the existence of the
localized state 
is not constrained to the case of  a  sharp, step-like
domain wall described above. 
The simple spinor structure
found above also generalizes for less regular domain walls. 
To see this, consider an almost arbitrary one-dimensional spatial dependence $M(x)$
of the mass, illustrated in Fig. \ref{fig:a_1Dmassprofile}b,
with the only condition that $M$ changes sign
between the half-planes $x<0$ and $x>0$, i.e.,
$M(x\to -\infty) < 0 $ and $M(x \to \infty) > 0$. 
We claim that there exists a zero-energy solution of 
the corresponding one-dimensional Dirac equation that 
is localized to the domain wall and 
has the envelope function
\begin{equation}
\label{eq:dirac_generalsolution}
\varphi(x) = \left(\begin{array}{c}
	0 \\ 1 
\end{array}\right)
f(x).
\end{equation}
To prove this claim, insert this wave function $\varphi(x)$ to
the one-dimensional Dirac equation and substitute $E = 0$ therein.
This procedure results in the 
single differential equation $\partial_x f(x) = - \frac{M(x)}{w}$, implying that 
Eq. \eqref{eq:dirac_generalsolution} is indeed a zero-energy
eigenstate of the envelope-function Hamiltonian if
the  function $f$ has the form
\begin{equation}
f(x) = \textrm{const} \times e^{-\frac{1}{w} \int_0^x dx' M(x')}.
\end{equation}
Furthermore, the asymptotic conditions of the mass $M(x)$ 
guarantee that this envelope function decays
as $x\to \pm \infty$, and therefore is localized 
at the domain wall.

\section{The QWZ model and the two-dimensional Dirac equation}

We have introduced the QWZ model as an example for a two-dimensional
Chern insulator in Chapt.~\ref{chap:qiwuzhang}.
The lattice Hamiltonian of the model is given 
in Eq. \eqref{eq:qiwuzhang-realspace_Hamiltonian},
whereas the bulk momentum-space Hamiltonian 
is given in Eq.
\eqref{eq:qiwuzhang-bulk_Hamiltonian_def}.
The dispersion relation is calculated
in Eq. \eqref{eq:qiwuzhang-bulk_spectrum}, 
and examples of it are shown in Fig. \ref{fig:qiwuzhang-dispersion}.

Recall that the model has a single parameter $u$, 
and the Chern number of the model is determined by 
the value of $u$ via Eq. \eqref{eq:qiwuzhang_chern_of_u}.
Similarly to the case of the SSH model in one dimension, 
one can consider a domain wall between locally homogeneous 
regions of the QWZ model that have different Chern numbers. 
Just as the edge of a strip, such a domain wall can support
topologically protected states that propagate along the domain wall 
but are localized at the domain wall in the transverse direction.
The number and propagation direction of those
states is determined by the magnitude and sign of the
difference of the Chern numbers in the two domains, 
respectively.
In this section, we use the EFA to provide a continuum description
of such states.


\subsection{The metallic case}

First, consider the metallic cases of the QWZ model; that
is, when the band structure has no energy gap. 
In particular, we will focus on 
the $u = -2$ case. 
The corresponding band structure
is shown in Fig.~\ref{fig:qiwuzhang-dispersion}(a).
The two bands touch at $\nvec k = (0,0)$,
and form a Dirac cone at that Dirac point. 

To describe  excitations in the vicinity of the Dirac point of
such a metal,
it is sufficient to use a linearized approximation of the 
QWZ Hamiltonian $\ido{H}(\nvec k)$
that is obtained via a Taylor expansion of $\ido{H}(\nvec k)$ up to
first order 
in the $k$-space location $\nvec q = \nvec k - \nvec k_0$ measured
from the Dirac point $\nvec k_0$.
In the case $u = -2$, the Dirac point is
$\nvec k_0 = (0,0)$, and the linearized Hamiltonian reads
\begin{equation}
\hat{H}(\nvec k_0  + \nvec q) \approx
q_x \ido{\sigma}_x + q_y  \ido{\sigma}_y.
\label{eq:a_2ddirac00}
\end{equation}
The dispersion relation is $E_{\pm }(\nvec q) = \pm q$;
again, this is analogous to that of the massless Dirac equation
Eq. \eqref{eq:dirac_zeromass}.

\subsection{The nearly metallic case}

Now consider a QWZ insulator that is nearly metallic:
$u \approx -2$.
The dispersion relation for $u = -1.8$ is shown
in Fig.~\ref{fig:qiwuzhang-dispersion}(d).
In the vicinity of the metallic state, 
as seen in the figure, 
a small gap opens in the band structure
at the Dirac point $\nvec k_0 = (0,0)$.

The states and the band structure around $\nvec k_0$ 
can again be described by 
a linearized approximation of the 
QWZ Hamiltonian $\ido{H}(\nvec k_0+\nvec q)$
in $\nvec q$:
\begin{equation}
\label{eq:a_homlin}
\ido{H}(\nvec k_0 + \nvec q) \approx
M \ido{\sigma}_z 
+ q_x \ido{\sigma}_x + q_y \ido{\sigma}_y,
\end{equation}
where we defined the parameter $M = u +2$.
The dispersion relation reads
\begin{equation}
E_{\pm} (\nvec q) = \pm \sqrt{M^2+q^2}.
\end{equation}
A comparison with the relativistic dispersion relation 
\eqref{eq:dirac_massivedispersion}
reveals that 
the parameter $M$ of the QWZ model plays the role
of $\mu c^2$;
hence $M$ can be called the \emph{mass parameter}.

\subsection{Continuum description of the nearly metallic case}
\label{sec:continuumQWZ}

We have discussed that 
the QWZ lattice with an inhomogeneous $u$ parameter
might support topologically protected  
states at boundaries separating 
locally homogeneous regions with different Chern numbers.
Similarly to the one-dimensional SSH model treated above, 
these localized states can be described analytically, using the 
envelope function approximation (EFA),
also in the two-dimensional QWZ model. 
In the rest of this Chapter, we focus on the 
nearly metallic case where
the inhomogeneous $u(x,y)$ 
is in the vicinity of $-2$ (i.e., $|M(x,y)| = |u(x,y)+2| \ll 1$), 
in which case the the low-energy excitations are
expected to localize in Fourier space around the band extremum point 
$\nvec k_0=(0,0)$ 
(see Figs.~\ref{fig:qiwuzhang-dispersion}a and d).
Here we obtain the EFA Schr\"odinger-type equation,
which resembles the two-dimensional Dirac equation,
and in the next subsection 
we provide its localized solutions 
for simple domain-wall arrangements. 

The considered lattice is inhomogeneous due to the spatial dependence
of the parameter $u(x,y)$. 
In the tight-binding lattice model, 
we denote the value of $u$ in unit cell 
$\nvec{\uci}=(\uci_x,\uci_y)$ as 
$u_{\nvec \uci} = u(x=\uci_x, y = \uci_y)$, 
and correspondingly, we introduce
the \emph{local mass parameter} 
via $M(x,y) = u(x,y) + 2$ and 
$M_{\nvec \uci} = M(x=\uci_x,y=\uci_y)$.

The EFA Hamiltonian can be constructed the same way 
as in sections \ref{sec:dirac_1dmc} and \ref{sec:dirac_continuumssh}.
The bulk momentum-space Hamiltonian $H(\nvec k_0+\nvec q)$
is Taylor-expanded 
around the band-edge wave vector $\nvec k_0 = (0,0)$,
the wave-number components
$q_x$ and $q_y$ are replaced by the differential 
operators $\hat{p}_x$ and $\hat{p}_y$, respectively,
and the inhomogeneous mass 
parameter $M(x,y)$  is incorporated.
This yields the following result in our present case: 
\begin{equation}
\label{eq:a_efahamiltonian}
\hat{H}_{\rm EFA} = M(x,y) \ido{\sigma}_z + 
	\hat p_x \ido{\sigma}_x + \hat p_y \ido{\sigma}_y.
\end{equation}
Then, the familiar EFA proposition is as follows. 
Assume that the two-component envelope function $\varphi(x,y)$
is a spatially slowly varying solution of the EFA 
Schr\"odinger equation
\begin{equation}
\label{eq:a_efaschrodinger}
\hat{H}_{\rm EFA} \varphi(x,y) = E \varphi(x,y).
\end{equation}
Then, the lattice state $\ket{\psi}$
associated to the envelope function $\varphi(x,y)$ 
is defined as 
\bean
\ket{\psi} = \sum_{\nvec \uci,\idi}
\varphi_\idi(\nvec \uci) \ket{\nvec \uci, \idi}.
\eean
It is claimed that the lattice state $\ket{\psi}$
is approximately an eigenstate of the inhomogeneous
lattice Hamiltonian with the eigenvalue $E$.

\subsection{Chiral states at a domain wall}

We can now use the EFA Hamiltonian \eqref{eq:a_efahamiltonian}
to describe the chiral states at a domain wall 
between two topologically distinct regions of the 
QWZ model. 
The Dirac-type EFA Schr\"odinger equation reads:
\begin{equation}
\label{eq:a_dirac}
\left[M(x,y)\ido{\sigma}_z + \hat{p}_x \ido{\sigma}_x  + \hat{p} \ido{\sigma}_y \right] \varphi(x,y)
= E \varphi(x,y).
\end{equation}

Consider the homogeneous case first:
$M(x,y) = M_0$, where $M_0$ might be positive or negative.
What is the dispersion relation for propagating waves? 
What are the energy eigenstates?
The answers follow from the plane-wave Ansatz
\begin{equation}
\label{eq:homogeneousdirac}
\varphi(x,y) = \left(\begin{array}{c}
	a \\ b 
\end{array}\right) e^{i q_x x }e^{i q_y y}
\end{equation}
with $q_x,q_y \in \mathbb{R}$ and $a,b \in \mathbb{C}$.
With this trial wave function, Eq.~\eqref{eq:a_dirac} yields
two solutions: 
\begin{equation}
E_\pm = \pm \sqrt{M_0^2+q_x^2+q_y^2},
\end{equation}
and
\begin{equation}
\frac{a_{\pm}}{b_{\pm}} = \frac{q_x-iq_y}{E_\pm - M_0}.
\end{equation}

Describe now the
states at a domain wall
between two locally homogeneous regions where
the sign of the mass parameter is different. 
Remember that the sign of the mass parameter in the EFA Hamiltonian
is related to the Chern number of the corresponding 
homogeneous half-BHZ lattice:
in our case, a positive (negative) mass implies a Chern number $-1$ ($0$).

To be specific, we will consider the case when the
two domains are defined as the $y<0$ and the $y>0$ half-planes,
i.e., the mass profile in Eq.~\eqref{eq:a_dirac} are
\begin{equation}
\label{eq:massstep}
M(x,y) = \left\{ \begin{array}{lrl} 
M_0 & {\rm if} & y>0, \\
-M_0 & {\rm if} & y<0
\end{array}\right. .
\end{equation}
Let $M_0$ be positive; the corresponding mass profile is the
same as shown in
Fig.~\ref{fig:a_1Dmassprofile}a, with $x$ replaced by $y$.


Now we look for solutions of Eq.~\eqref{eq:a_dirac}
that reside in the energy range 
$-M_0<E<M_0$, i.e., in the bulk gap of the two domains,
and which propagate along, but decay 
perpendicular to, the domain wall at $y=0$.
Our wave-function Ansatz for the upper half plane $y>0$ is
\begin{equation}
\varphi_u(x,y) = \left(\begin{array}{c}
	a_u \\ b_u 
\end{array}\right) e^{i q_x x }e^{i q_y^{(u)} y}
\end{equation}
with $q_x\in \mathbb{R}$, $q_y^{(u)} \in i\mathbb{R}^+$ 
and $a,b \in \mathbb{C}$.
For the lower half plane, $\varphi_l(x,y)$ is defined as
$\varphi_u(x,y)$ but with $u\leftrightarrow l$ interchanged and
$q_y^{(l)} \in i \mathbb{R}^-$.
The wave function $\varphi_u(x,y)$ does solve the two-dimensional Dirac
equation defined by
Eqs. \eqref{eq:a_dirac} and \eqref{eq:massstep}
in the upper half plane $y>0$ provided 
\begin{equation}
\label{eq:a_qyu}
q_y^{(u)} = i \kappa \equiv i \sqrt{M_0^2+q_x^2-E^2}
\end{equation}
and
\begin{equation}
\frac{a_u}{b_u} = \frac{q_x + \kappa}{E-M_0}
\end{equation}
Similar conditions apply for the ansatz $\varphi_l(x,y)$ 
for the lower half plane, with the
substitutions $u \mapsto l$,
$\kappa \mapsto -\kappa$ and
$M_0\mapsto -M_0$.
The complete (unnormalized) envelope function has the form
\begin{equation}
\label{eq:a_completewfn}
\varphi(x,y) = \varphi_u(x,y) \Theta(y) + c \varphi_l(x,y) \Theta(-y),
\end{equation}
where $c$ is a yet unknown complex parameter to be determined
from the boundary conditions at the domain wall.

The wave function \eqref{eq:a_completewfn} 
is an eigenstate of the EFA Hamiltonian with 
energy $E$ if the boundary condition that 
the wave function is continuous on the line $y=0$, that is,
\begin{equation}
\label{eq:diracbc}
\varphi_u(x,0) = \varphi_l(x,0),
\end{equation}
 is fulfilled for every $x$.

\begin{figure}[!ht]
\begin{center}
\includegraphics[scale=0.5]{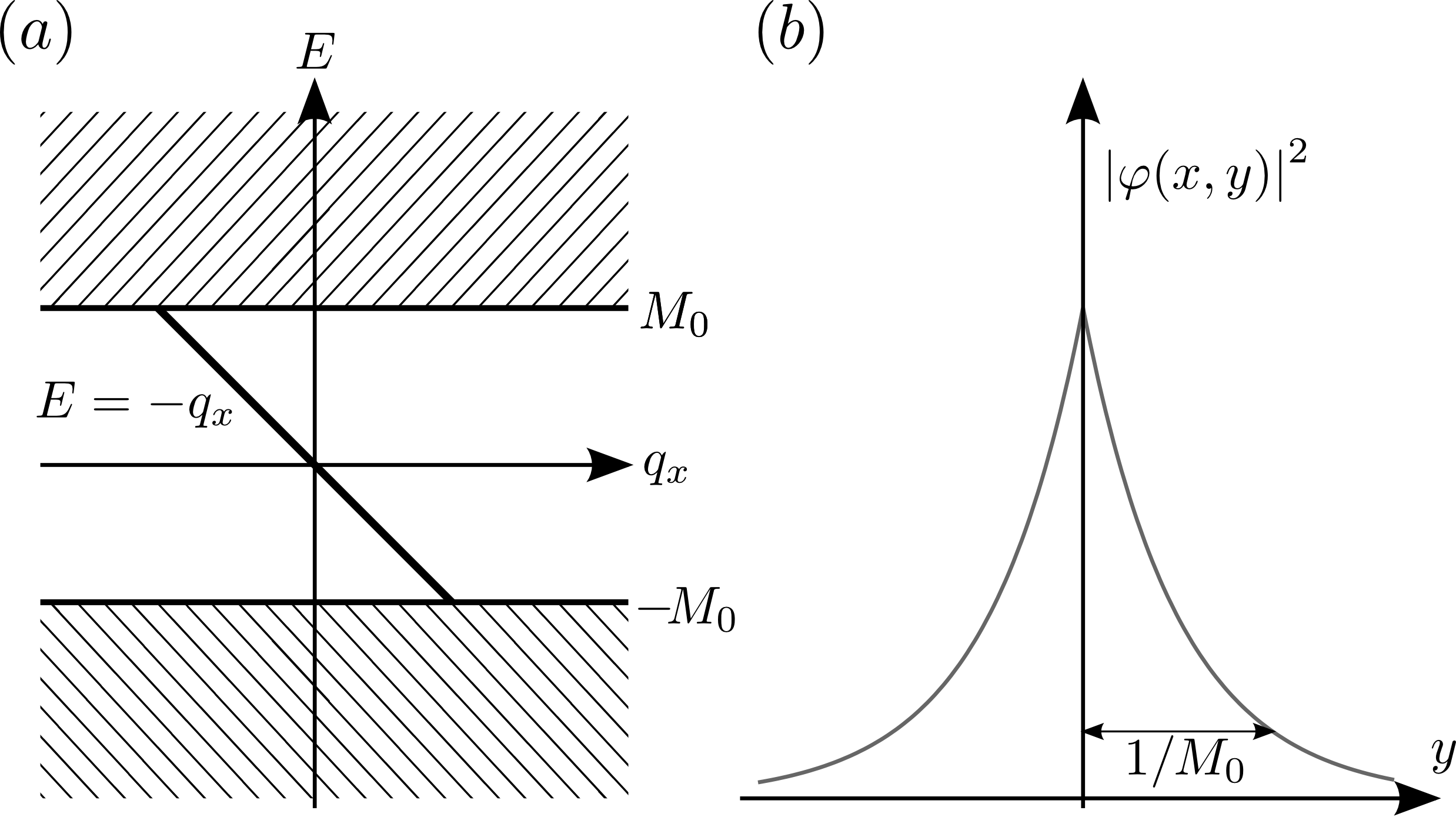}
\caption{\label{fig:a_2ddirac-results}
Chiral state obtained from the two-dimensional Dirac equation.
(a) Dispersion relation and (b) squared wave function
of a chiral state confined to, and propagating along, 
a mass domain wall.
}
\end{center}
\end{figure}

The boundary condition \eqref{eq:diracbc} determines the value
of the parameter $c$ as well as the dispersion relation $E(q_x)$
of the edge states. 
First, \eqref{eq:diracbc} implies
\begin{eqnarray}
q_x - \kappa = c(q_x + \kappa)
&\Rightarrow& c = \frac{q_x-\kappa}{q_x+\kappa}, \\
E+M_0 = c(E-M_0) &\Rightarrow& 
-q_x M_0 = \kappa E.
\end{eqnarray}
Note that $\kappa$ depends on $E$ according to 
Eq. \eqref{eq:a_qyu}.
It is straigthforward to find the dispersion relation of the
edge states by solving $-q_x M_0 = \kappa(E) E$
for $E$ with the condition $-M_0<E<M_0$: 
\begin{equation}
E = -q_x.
\end{equation}
This simple dispersion relation is shown in Fig.~\ref{fig:a_2ddirac-results}a.
Together with Eq.~\eqref{eq:a_qyu}, this dispersion implies that
the localization length of edge states is governed by $M_0$ only, i.e.,
is independent of $q_x$. 
The squared wave function of an edge state 
is shown in Fig.~\ref{fig:a_2ddirac-results}b.

A remarkable consequence of this simple dispersion relation is that the
spinor
components of the envelope function also have a simple form:
\begin{equation}
\left(\begin{array}{c}
	a_u \\ b_u
\end{array}\right) =
\left(\begin{array}{c}
	a_l \\ b_l
\end{array}\right)
=
\left(\begin{array}{c}
	1 \\ -1
\end{array}\right).
\end{equation}

Edge states at similar mass domain walls at $u(y) \approx 0$
and $u(y) \approx 2$ can be derived analogously. 
Note that at $u(y) \approx 0$, the low-energy states can
reside in two different Dirac valleys, around
$\nvec k_0 = (0,\pi)$ or $\nvec k_0=(\pi,0)$, 
and there is one edge state in each valley. 
The number of edge states obtained in the continuum model,
as well as their directions of propagation, 
are in correspondence with those obtained in the
lattice model; as we have seen for the latter case, the number and
direction are given by the magnitude and the sign of Chern-number 
difference across the domain wall, respectively. 

An interesting fact is that the existence of the
edge state 
is not constrained to case of  a  sharp, step-like
domain wall described above. 
Moreover, the simple dispersion relation and spinor structure
found above generalize for more irregular domain walls. 
This generalization is proven in a similar fashion as in 
the case of the SSH model, see Sect.~\ref{sec:dirac:sshdomain}. 
To see this, consider an almost arbitrary one-dimensional spatial dependence of the
mass, similar to the one in Fig. \ref{fig:a_1Dmassprofile}b: 
$M(x,y) = M(y)$ with the only condition that $M$ changes sign
between the half-planes $y<0$ and $y>0$, i.e.,
$M(y\to -\infty) < 0 $ and $M(y \to \infty) > 0$. 
We claim that there exists a solution of 
the corresponding two-dimensional Dirac equation that propagates
along the domain wall, has the 
dispersion relation $E = -q_x$, 
is confined in the direction perpendicular to the domain wall, and
has the wave function
\begin{equation}
\label{eq:a_generalsolution}
\varphi(x,y) = \left(\begin{array}{c}
	1 \\ -1 
\end{array}\right)
e^{iq_x x} f(y).
\end{equation}
To prove this proposition, insert this wave function $\varphi(x,y)$ to
the two-dimensional Dirac equation and substitute $E$ with $-q_x$ therein.
This procedure results in two equivalent equations that are
fulfilled if $\partial_y f(y) = - M(y) f(y)$, implying that 
Eq. \eqref{eq:a_generalsolution} is indeed a normalizable 
solution with $E = -q_x$
provided that the  function $f$ has the form
\begin{equation}
f(y) = e^{-\int_0^y dy' M(y')}.
\end{equation}

To summarize: 
In the preceding chapters, we introduced the topological 
characterization of lattice models and the corresponding edge states
and states bound to domain walls between regions of different
topological character. 
In this chapter, we demonstrated that 
a low-energy continuum description (the EFA Schr\"odinger equation)
can be derived from a lattice model, and can be used to 
describe those electronic states.
Besides being a convenient analytical tool to describe 
inhomogeneous lattices, the 
envelope-function approximation also demonstrates that the emergence
of topologically protected states is not restricted to lattice models.

\section*{Problems}
\addcontentsline{toc}{section}{Problems}
%

\begin{prob}
\label{prob:dirac-ssh-v}
\textbf{SSH model with spatially dependent intracell hopping.}\\
In Sect.~\ref{sec:dirac_continuumssh},
we provide the EFA proposition for the SSH model
with spatially dependent intracell hopping. 
Prove this proposition,  following the procedure 
detailed in Sect.~\ref{sec:dirac:proof}
for the one-dimensional monatomic chain. 
\end{prob}

\begin{prob}
\label{prob:dirac-ssh-w}
\textbf{SSH model with spatially dependent intercell hopping.}\\
Derive the EFA Hamiltonian for an inhomogeneous SSH model, 
where $w$ varies in space and $v$ is constant.
Assume a nearly metallic scenario, $w(x) \approx v$. 
\end{prob}

\begin{prob}
\label{prob:dirac-qwz}
\textbf{QWZ model.}\\
Prove the EFA proposition for the QWZ model. 
The proposition is outlined in Sect.~\ref{sec:continuumQWZ}.
The proof is analogous to that used for the SSH model. 
\end{prob}

\begin{prob}
\label{prob:dirac-qwz-Delta2}
\textbf{QWZ model at $u \approx 2$.}\\
The bulk momentum-space Hamiltonian $\ido{H}(k)$ 
of the QWZ model
is given in Eq. \eqref{eq:qiwuzhang-bulk_Hamiltonian_def}.
Starting from this $\ido{H}(k)$,
derive the EFA Hamiltonian describing low-energy
excitations in the case of 
an inhomogeneous $u$ parameter 
for which $u \approx 2$. 
\end{prob}


%% file: bhz_Smatrix.tex

\chapter{Time-reversal symmetric two-dimensional topological insulators -- the Bernevig--Hughes--Zhang model}
\label{chap:BHZ}

\chaptermark{Time-reversal symmetric two-dimensional topological insulators}

\abstract*{Time-reversal symmetry ensures that for every edge state in
  the spectrum, there is another edge state at the same edge, at the
  same energy, hosting particles that propagate in the opposite
  direction. Time-reversal symmetric insulators must thus have a Chern
  number of zero.  In this chapter we show that if the time reversal
  squares to -1 (fermionic time reversal), then these systems can host
  edge states that are protected against localization by time-reversal
  invariant disorder. There is at most one pair of protected edge
  states, and thus time-reversal symmetric two-dimensional insulators
  come in two flavors: topological or trivial. }

In the previous chapters, we have seen how two-dimensional insulators
can host one-way propagating (a.k.a.~chiral) edge states, which
ensures reflectionless transport along the edge. 
The existence of
chiral edge states precludes time-reversal symmetry. Indeed,
time-reversed edge states would describe particles propagating
backwards along the edge. In Chern Insulators (two-dimensional
insulators with nonvanishing Chern number), the absence of these
counterpropagating states from the spectrum is what ensures the
reflectionless propagation of particles along the edges.

What about time-reversal symmetric or ti\-me-rever\-sal invariant
two dimen\-sional insulators? According to the above, they cannot be
Chern insulators. Interestingly though, the same time-reversal
symmetry that ensures that for every edge state mode there is a
counterpropagating time reversed partner,
can also ensure that no scattering between these two modes
occurs. This means that it is possible for time-reversal invariant two-dimensional
insulators to host edge states with reflectionless propagation,
in both directions, at both edges. The details of why and how this
happens are discussed in this and the following chapters.

We will find that all two-dimensional time-reversal invariant
insulators fall into two classes: the trivial class, with an even
number of pairs of edge states at a single edge, and the topological
class, with an odd number of pairs of edge states at a single edge. We
then subsequently show that disorder that breaks translational
invariance along the edge can destroy edge state conduction in the
trivial class, but not in the topological class.

The bulk--boundary correspondence for Chern Insulators stated that the
the net number of edge states on the edge, $N_+-N_-$, is the same as
the Chern number of the bulk, $Q$. We showed this by mapping the
2-dimensional system to a periodically, adiabatically pumped
one-dimensional chain. After the mapping, the unit of charge pumped
through the chain during a period could be identified with the net
number of chiral edge states. 

Unfortunately, identifying and calculating the bulk topological
invariant of a time-reversal invariant two-dimensional insulator is
much more cumbersome than for a Chern insulator. We therefore come
back to this problem in the next Chapter.

\section{Time-Reversal Symmetry} 
\label{sec:bhz_smatrix-timereversal}

Before we discuss time-reversal symmetric topological insulators, we
first need to understand what we mean by time reversal symmetry, and
how it leads to Kramers' degeneracy.

\subsection{Time Reversal in continuous variable quantum mechanics 
(without spin)}

Take a single particle with no internal degree of freedom, described
by a wavefunction $\Psi(\nvect{r})$. Its dynamics is presribed by a
time independent Hamiltonian $\HH = (\hat{p} - e
\mathbf{A}(\hat{r}))^2 + V(\hat{r})$, where the functions $\mathbf{A}$
and $V$ are the vector and scalar potentials, respectively, and $e$ is
the charge of the particle.  The corresponding Schr\"odinger equation
for the wavefuncion $\Psi(r,t)$ reads
\begin{align} 
i \partial_t \Psi(\nvect{r},t) &= 
\left\{ (-i\partial_\nvect{r} -e\nvect{A}(\nvect{r}))^2 
+ V(\nvect{r}) \right\} \Psi(\nvect{r},t).  
\end{align} 
Any solution $\Psi$ of the above equation can be complex conjugated,
and gives a solution of the complex conjugate of the Schr\"odinger
equation, 
\begin{align} 
-i \partial_t \Psi(\nvect{r},t)^\ast &= 
\left\{ (-i\partial_\nvect{r} + e\nvect{A}(\nvect{r}))^2 
+ V(\nvect{r}) \right\} \Psi(\nvect{r},t)^\ast.  
\end{align} 

\subsubsection*{The operator of complex conjugation in real space basis}
We use $K$ to denote the operator that complex conjugates
everything to its right in real space basis,
\begin{align}
K f(\nvect{r}) &= f(\nvect{r})^\ast K;& K^2 &= 1,
\end{align}
for any complex valued function $f(\nvect{r})$ of position. 
The Schr\"odinger equation above can be rewritten using $K$ as
\begin{align}
K i\partial_t \Psi &= K i\partial_t K K \Psi = - i\partial_t \Psi^\ast = 
K \HH K K \Psi = \HH^\ast \Psi^\ast. 
\end{align}

Complex conjugation in real space basis conforms to intuitive
expectations of time reversal: it is local in space, takes
$\hat{x}\to\hat{x}$, and flips the momenta, $i\partial_x \to
-i\partial_x$.

\subsubsection*{Time reversal}

The above relation shows that for any closed quantum mechanical
system, there is a simple way to implement time reversal. This
requires to change \emph{both} the wavefunction $\Psi$ to $\Psi^\ast$
\emph{and} the Hamiltonian $\HH$ to $\HH^\ast$. The change of the
Hamiltonian involves $\mathbf{A} \to -\mathbf{A}$, i.e., flipping the
sign of the vector potential.

\subsubsection*{Time reversal symmetry}

In the special case where the Hamiltonian in real space basis is real,
$\HH^\ast = \HH$, we can implement time reversal by only
acting on the wavefunction. In that case, we say that the system has
time reversal symmetry. For the scalar Schr\"odinger equation above,
this happens if there is no vector potential, $\nvect{A}=0$.  To see
this more explicitly, consider time evolution for a time $t$, then
apply the antiunitary operator $K$, then continue time evolution
for time $t$, then apply $K$ once more:
\begin{align}
\hat{U} &= K e^{-i \HH t} K e^{-i \HH t} = e^{-K i \HH tK }
e^{-i \HH t} = e^{i \HH^\ast t } e^{-i \HH t}  
\end{align}
If $\HH^\ast = \HH$, then $\hat U = 1$, which means that $K$
acts like time reversal. 

\subsection{Lattice models with an internal degree of freedom}

In these notes we deal with models for solids which are lattice
Hamiltonians: the position (the external degree of freedom) is
discrete, and there can be an internal degree of freedom (sublattice,
orbital, spin, or other). 

\subsubsection*{Definition of the operator K}

For the operator of complex conjugation we need to fix not only the
external position basis, $\mathcal{E}_\text{external} = \{
\ket{\mathbf{\uci}} \}$, but also an internal basis,
$\mathcal{E}_\text{internal} = \{ \ket{\alpha} \}$. The property
defining $K$ then reads
\begin{multline}
\forall z \in \mathbb{C},\quad \forall 
\ket{\mathbf{\uci}},\ket{\mathbf{\uci}'}\in\mathcal{E}_\text{external},
\quad \forall \ket{\alpha},\ket{\alpha'}\in\mathcal{E}_\text{internal}:   \\
K z \ket{\mathbf{\uci},\alpha}\bra{\mathbf{\uci}',\alpha'} = 
 z^\ast \ket{\mathbf{\uci},\alpha}\bra{\mathbf{\uci}',\alpha'} K,
\end{multline}
where $z^\ast$ is the complex conjugate of $z$.  

We will use the shorthand $\ket{\Psi^\ast}$ and $\hat{A}^\ast$ to
represent $K\ket{\Psi}$ and $K\hat{A}K$, respectively. The 
defining equations are
\begin{align} 
\ket{\Psi} &= \sum_\uci \sum_\alpha \Psi_{\uci,\alpha}
\ket{\uci}\otimes \ket{\alpha};\\ 
\ket{\Psi^\ast} &= K \ket{\Psi} =
\sum_\uci \sum_\alpha \Psi_{\uci,\alpha}^\ast \ket{\uci}\otimes
\ket{\alpha};\\ 
\hat{A} &= 
\sum_{\uci'\uci} \sum_{\alpha'\alpha}
A_{\uci',\alpha',\uci,\alpha} \ket{\uci'}\bra{\uci}\otimes 
\ket{\alpha'} \bra{\alpha};\\ 
\hat{A}^\ast
&= K \hat{A} K =
\sum_{\uci'\uci} \sum_{\alpha'\alpha}
A_{\uci',\alpha',\uci,\alpha}^\ast \ket{\uci'}\bra{\uci}\otimes 
\ket{\alpha'} \bra{\alpha}.
\end{align}

\subsubsection*{Time-reversal affects external and internal degrees of freedom}

We look for a representation of time reversal symmetry $\TTT$ in terms
of a general antiunitary operator. Apart from complex conjugation,
which acts on both external and internal Hilbert space, we allow for
an additional unitary operation on the internal degrees of freedom
$\htau$, that is independent of position,
\begin{align}
\TTT &= \htau K.
\end{align}

We say that a Hamiltonian $\HH$ is time reversal invariant (or
time reversal symmetric) with respect to time reversal represented by
$\TTT$ if 
\begin{align}
\TTT \HH \TTT &= \HH.
\label{eq:bhz_trs_def}
\end{align}
In the same sense as for the chiral symmetry (cf.
Sect.~\ref{sec:ssh-chiral}), when we talk about a Hamiltonian, what we
really mean is a set of Hamiltonians $\HH(\underline{\xi})$, with
$\underline{\xi}$ representing parameters that are subject to
disorder. Thus, Eq.~\eqref{eq:bhz_trs_def} should hold for any of the
$\HH(\underline{\xi})$, with $\TTT$ independent of $\underline{\xi}$.

\subsection{Two types of time-reversal}

We can require that a time reversal operator $\TTT$, when squared,
should give at most a phase:
\begin{align}
\htau K \htau K = \htau \htau^\ast &= e^{i\phi} \II_\text{internal}.
\label{eq:bhz-tsquare}
\end{align}
If that was not the case, if the unitary operator $\htau \htau^\ast$
was nontrivial, then it would represent a unitary symmetry of a
time-reversal invariant Hamiltonian, since
\begin{align}
\htau \htau^\ast \HH (\htau \htau^\ast)^\dagger &= \htau K \htau K
\HH K \htau^\dagger K \htau^\dagger = \htau K \HH K \htau^\dagger
= \HH.
\label{eq:bhz-tsquare0}
\end{align}
As explained in Sect.~\eqref{sec:ssh-chiral}, when we want to
investigate topological phases, the usual first step is to get rid of
unitary symmetries one by one (except for the lattice translation
symmetry of the bulk Hamiltonian), by restricting our attention to a
single superselection sector of each symmetry. Thus, the only time
reversal symmetries that are left are those that fulfil
Eq.~\eqref{eq:bhz-tsquare}.

The phase factor $e^{i\phi} = \TTT^2$ turns out to have only two
possible values: $+1$ or $-1$.  Multiplying Eq.~\eqref{eq:bhz-tsquare}
from the left by $\htau^\dagger$, we get $\htau^\ast =
e^{i\phi}\htau^\dagger = e^{i\phi} (\htau^\ast)^T$, where the
superscript $T$ denotes transposition. Iterating this last relation
once more, we obtain $\htau^\ast = e^{2i\phi} \htau^\ast$, which means
$e^{i\phi} = \pm 1$, wherefore
\begin{align}
\TTT^2 &= \pm 1.
\end{align}

A Hamiltonian with no unitary symmetries can have only one type of
time-reversal symmetry: either $\TTT^2=+1$, or $\TTT^2=-1$, but not
both. 
Assume a Hamiltonian had two different time-reversal symmetries,
$\TTT$ and $\TTT_1$. Along the lines of Eq.~\eqref{eq:bhz-tsquare0},
the product of the two, the unitary operator $\TTT_1 \TTT$ would then
represent a unitary symmetry.  The only exception is if $\TTT_1 =
e^{i\chi} \TTT$, when they are not really different symmetries.
However, in this case, since $\TTT$ is antiunitary, these two
symmetries square to the same number, ${\TTT_1}^2= e^{i\chi} \TTT
e^{i\chi} \TTT = \TTT^2$.

An example for a time-reversal operator with $\TTT^2=+1$ is given by
the complex conjugation $K$. An example for a time-reversal operator
with $\TTT^2=-1$ is time reversal for a spin-$1/2$ particle. Since
time reversal should also flip the spin, it is achieved by
$\TTT=-i\sigma_y K$, with $K$ defined on the basis of the
eigenstates of $\sigma_z$. The fact that this works can be checked by
$\TTT \sigma_j \TTT^{-1} = -\sigma_j$ for $j=x,y,z$.

\subsubsection*{The operator $\htau$ is symmetric or antisymmetric}

Specifying the square of the time-reversal operation constrains the
operator $\htau$ to be symmetric or antisymmetric. Consider 
\begin{align}
\TTT^2 &= \htau K \htau K = \htau \htau^\ast = \pm 1;&
\htau^\ast &= \pm \htau^\dagger = (\pm \htau^T)^\ast,
\end{align}
where the subscript $T$ denotes transpose in the same basis where the
complex conjugate is defined. 
Therefore, 
\begin{align}
\TTT^2 &= +1 \quad \Longleftrightarrow \quad\htau = \htau^T \quad
\text{symmetric};\\
\TTT^2 &= -1 \quad \Longleftrightarrow \quad\htau = -\htau^T \quad
\text{antisymmetric}.
\end{align}

\subsection{Time reversal of type $\TTT^2=-1$ gives Kramers' degeneracy}

A defining property of an antiunitary operator $\TTT$ is that for any
pair of states $\ket{\Psi}$ and $\ket{\Phi}$, we have
\begin{align}
\label{eq:bhz-antinunitary_def2}
\sbraket{\TTT\Phi}{\TTT\Psi} &=
(\htau \ket{\Phi^\ast})^\dagger \htau \ket{\Psi^\ast}. 
= \ket{\Phi^\ast}^\dagger \htau^\dagger \htau \ket{\Psi^\ast} = 
\braket{\Phi^\ast}{\Psi^\ast} =
\braket{\Phi}{\Psi}^\ast.
\end{align}
Consider now this relation with $\ket\Phi = \TTT \ket\Psi$:
\begin{align}
\sbraket{\TTT \Psi}{\Psi}^\ast &=
\sbraket{\TTT^2 \Psi}{\TTT\Psi} = 
\sbraket{\pm \Psi}{\TTT \Psi} =
\pm \sbraket{\TTT \Psi}{\Psi}^\ast,
\end{align}
where the $\pm$ stands for the square of the time reversal operator
$\TTT$, which is $\pm 1$.  If $\TTT^2=+1$, the above line gives no
information, but if $\TTT^2=-1$, it leads immediately to $\braket{\TTT
  \Psi}{\Psi}=0$, which means that for every energy eigenstate, its
time-reversed partner, which is also an energy eigenstate with the
same energy, is orthogonal. This is known as Kramers degeneracy.

\subsection{Time-Reversal Symmetry of a Bulk Hamiltonian}

We now calculate the effect of time-reversal symmetry $\TTT = \htau K$
on the bulk momentum-space Hamiltonian $\HH(k)$.  This latter is
obtained, as in Sect.~\ref{sec:ssh-bulk}, by first
periodic boundary conditions, and defining a plane wave basis in the
corresponding external Hilbert space as
\begin{align}
\ket{\kk} &= \sum_{\kk} e^{i \vuci \kk} \ket{\vuci};&
\TTT \ket{\kk} &= \ket{-\kk} \TTT.
\end{align}
Next, $\Hbulk$ is the part of $\HH$ in the bulk, with periodic
boundary conditions, whose components in the plane wave basis define
the bulk momentum-space Hamiltonian,
\begin{align}
\HH (\kk) &= \bra{\kk} \HH_\text{bulk} \ket{\kk};&
\HH_\text{bulk} &= \sum_\kk \ket{\kk}\bra{\kk} \otimes \HH(\kk).
\end{align} 
The effect of time-reversal symmetry follows, 
\begin{align}
\TTT \HH_\text{bulk} \TTT^{-1} &
= \sum_\kk \ket{-\kk}\bra{-\kk} 
\otimes \htau \HH(\kk)^\ast \htau^\dagger 
= \sum_\kk \ket{\kk}\bra{\kk} 
\otimes \htau \HH(-\kk)^\ast \htau^\dagger. 
\end{align} 
We read off the action of $\TTT$ on the bulk
momentum-space Hamiltonian, and obtain the necessary requirement of
time-reversal symmetry as   
\begin{align}
\htau \HH(-\kk)^\ast \htau^\dagger &= \HH(\kk).
\label{eq:bhz-trs_brillouin}
\end{align}
Note that time-reversal symmetry of the bulk Hamiltonian is necessary,
but not sufficient, for time-reversal symmetry of the
system: perturbations at the edges can break time reversal. 

A direct consequence of time-reversal symmetry is that the dispersion
relation of a time-reversal symmetric Hamiltonian has to be symmetric
with respect to inversion in the Brillouin zone, $\kk \to -\kk$. Indeed,
take an eigenstate of 
$\HH(\kk)$, with
\begin{align}
\HH(\kk) \ket{u(\kk)} &= E(\kk) \ket{u(\kk)}.
\end{align}
Using time-reversal symmetry, Eq.~\eqref{eq:bhz-trs_brillouin}, we
obtain 
\begin{align}
\htau \HH(-\kk)^\ast \htau^\dagger \ket{u(\kk)} &= E(\kk)
\ket{u(\kk)}.
\end{align}
Multiplying from the left by $\htau^\dagger$ and complex conjugating,
we have 
\begin{align}
\HH(-\kk) \htau^T \ket{u(\kk)}^\ast &= E \htau^T \ket{u(\kk)}^\ast. 
\end{align}
This last line tells us that for every eigenstate $\ket{u(\kk)}$ of
$\HH(\kk)$, there is a time-reversed partner eigenstate of $\HH(-\kk)$
at the same energy, $\htau^T \ket{u(\kk)}^\ast$.  This implies
inversion symmetry of the energies, $E(\kk)=E(-\kk)$. Note, however,
that $E(\kk) = E(-\kk)$ is not enough to guarantee time-reversal
symmetry. 

It is especially interesting to look at points in the Brillouin zone
which map unto themselves under inversion: the Time-reversal invariant
momenta (TRIM). In $d$ dimensions there are $2^d$ such points, one of
which is at the center of the Brillouin zone (so-called $\Gamma$
point), and others at the edges. At such momenta, 
Eq.~\eqref{eq:bhz-trs_brillouin} implies that
\begin{align} 
\htau \HH(\kk_\TRIM)^\ast \htau^\dagger = \HH(\kk_\TRIM).
\end{align} 
If $\TTT^2=-1$, then because of Kramers degeneracy, at a TRIM, every
eigenvalue of the bulk momentum-space Hamiltonian is (at least) twice
degenerate.

\section{Doubling the Hilbert Space for Time-Reversal Symmetry} 
\label{sec:bhz-doubling}

There is a simple way to construct a system with Time-Reversal
Symmetry, $\HHtri$, starting from a lattice Hamiltonian $\HH$. 
We take another two copies of the system, and change the Hamiltonian
in one of them to $\HH^\ast = K \HH K$. We then couple them, much as
we did to layer Chern insulators on top of each other in
Sect.~\ref{subsec:qwz-constructing-layers}:
\begin{align}
\HHtri &= \ket{0}\bra{0} \otimes \HH + \ket{1}\bra{1} \otimes
\HH^\ast + \left( 
\ket{0}\bra{1}\otimes\II_\text{external}\otimes \hat{C} + h.c. \right),
\label{eq:bhz_smatrix-def_doubling}
\end{align}
where the hopping between the copies is accompanied by 
a position-independent operation $\hat{C}$ 
on the internal degree of freedom. 
In a matrix form, in real-space basis (and somewhat simplified
notation), this reads
\begin{align}
\Htri = 
\begin{bmatrix}
H & C \\
C^\dagger & H^\ast
\end{bmatrix}.
\label{eq:bhz-def_tri_twoblocks}
\end{align}

We will use $\hat{s}_{x,y,z}$ to denote the Pauli operators acting on the
``copy degree of freedom'', defined as 
\begin{align}
\hat{s}_{x/y/z} = \hat{\sigma}_{x/y/z} \otimes \II_\text{external}
\otimes \II_\text{internal}. 
\label{eq:bhz-ss_matrix_def}
\end{align}
Using these operators, the time-reversal invariant Hamiltonian reads 
\begin{multline}
\HHtri = \frac{1+\hs_z}{2}\otimes \HH + \frac{1-\hs_z}{2}
\otimes\HH^\ast \\ 
+\frac{\hs_x+i\hs_y}{2} \otimes \II_\text{external}
\otimes \hat{C} + \frac{\hs_x-i\hs_y}{2} \otimes \II_\text{external}
\otimes \hat{C}^\dagger.
\label{eq:bhz-def_ss_twoblocks}
\end{multline}


The choice of the coupling operator $\hat{C}$ is important, as it
decides which type of time-reversal symmetry $\HHtri$ will have.

\subsubsection*{Time reversal with $\TTT^2=-1$ 
requires antisymmetric coupling operator $\hat{C}$}

If we want a time-reversal symmetry that squares to $-1$, we can go
for
\begin{align}  
\TTT &= i\hs_y K;& \TTT^{-1} &= K (-i)\hs_y,
\end{align} 
with the factor of $i$ is included for
convenience, so that the matrix of $i\hs_y$ is real. 
The requirement of time-reversal symmetry can be obtained using
\begin{align}
(i\hs_y K ) \Htri (i\hs_y K)^{-1} &= 
\begin{bmatrix} 0 & 1\\ -1 & 0 \end{bmatrix}
\cdot
\begin{bmatrix} H^\ast & C^\ast \\  C^T & H \end{bmatrix}
\cdot
\begin{bmatrix} 0 & -1\\ 1 & 0 \end{bmatrix} \nonumber\\
 \quad &= 
\begin{bmatrix} C^T & H \\ -H^\ast & -C^\ast \end{bmatrix}
\cdot
\begin{bmatrix} 0 & -1\\ 1 & 0 \end{bmatrix}
=
\begin{bmatrix} H & -C^T \\ -C^\ast & H^\ast \end{bmatrix}.
\end{align} 

We have time-reversal symmetry represented by $\TTT=i\hs_y K$, if 
\begin{align}
i\hs_y K \HHtri (i\hs_y K)^{-1} &= \HHtri& \quad \Leftrightarrow 
\quad \quad \hat{C} &= -\hat{C}^T, 
\end{align}
where the subscript $T$ denotes transposition in the same fixed
internal basis that is used to define complex conjugation $K$.

\subsubsection*{Symmetric coupling operator $\hat{C}$ gives 
time reversal with $\TTT^2=+1$}

If the coupling operator is symmetric, $\hat{C} = \hat{C}^T$, then the
same derivation as above shows that we have time-reversal symmetry
represented by $\TTT=\hs K$,
\begin{align}
\hs_x K \HHtri K \hs_x &= \HHtri& \quad \Leftrightarrow 
\quad \quad \hat{C} &= \hat{C}^T.
\end{align}
This time-reversal operator squares to $+1$.

If all we want is a lattice Hamiltonian with a time-reversal symmetry
that squares to $+1$, we don't even need to double the Hilbert
space. We can just take
\begin{align}
\TTT &= K;& \HHtri &= \frac{\HH+\HH^\ast}{2} = K \HHtri K.
\end{align}
Colloquially, this construction is referred to as taking the real part of the
Hamiltonian. 

\subsection{A concrete example: the Bernevig-Hughes-Zhang model}

To have an example, we use the construction above to build a toy model
-- called BHZ model -- for a time-reversal invariant topological
insulator starting from the QWZ model of
Chapt.~\ref{chap:qiwuzhang}. We follow the construction through using
the bulk momentum-space Hamilonian, and obtain
\begin{align}
\HHbhz(\kk) &= \hs_0 \otimes 
[(\uu + \cos k_x +\cos k_y) \hsigma_z + \sin k_y \hsigma_y )] 
+ \hs_z  \otimes \sin k_x \hsigma_x + 
\hs_x \otimes \hat{C},
\label{eq:bhz_Hamiltonian_def2}
\end{align}
where $\hat{C}$ is a Hermitian coupling operator acting on the
internal degree of freedom. 
For $\hat{C}=0$, the Hamiltonian $\HHbhz$ reduces to the 4-band toy
model for HgTe, introduced by Bernevig, Hughes and Zhang
\cite{Bernevig-qshe}.
 
\subsubsection*{Two Time-Reversal Symmetries if there is no coupling}

If there is no coupling between the copies, $\hat{C}=0$, the BHZ model has
two different time-reversal symmetries, $\TTT=i\hs_y K$ and
$\TTT_2=\hs_x K$, due to its block diagonal structure reflecting a
unitary symmetry, $\hs_z \HHbhz \hs_z^\dagger = \HHbhz$.  In this
situation, the type of predictions we can make will depend on which of
these symmetries is robust against disorder. We will in the following
require the $\TTT^2=-1$ symmetry. If this symmetry is robust, then
everything we do will apply to the $\hat{C}=0$ case of $\HHbhz$. The extra
time reversal symmetry in that case is just a reminder that 
most features could be calculated in a more simple way,
by working in the superselection sectors of $\hs_z$ separately.

\section{Edge States in 2-dimensional
  time-reversal invariant insulators}

We now consider the situation of edge states in a two-dimensional
lattice Hamiltonian with time-reversal symmetry, much in the same way
as we did for Chern insulators in
Sect.~\ref{sec:qiwuzhang_edgestates}.

\begin{figure}[!ht]
\centering
\includegraphics[width=0.9\linewidth]
{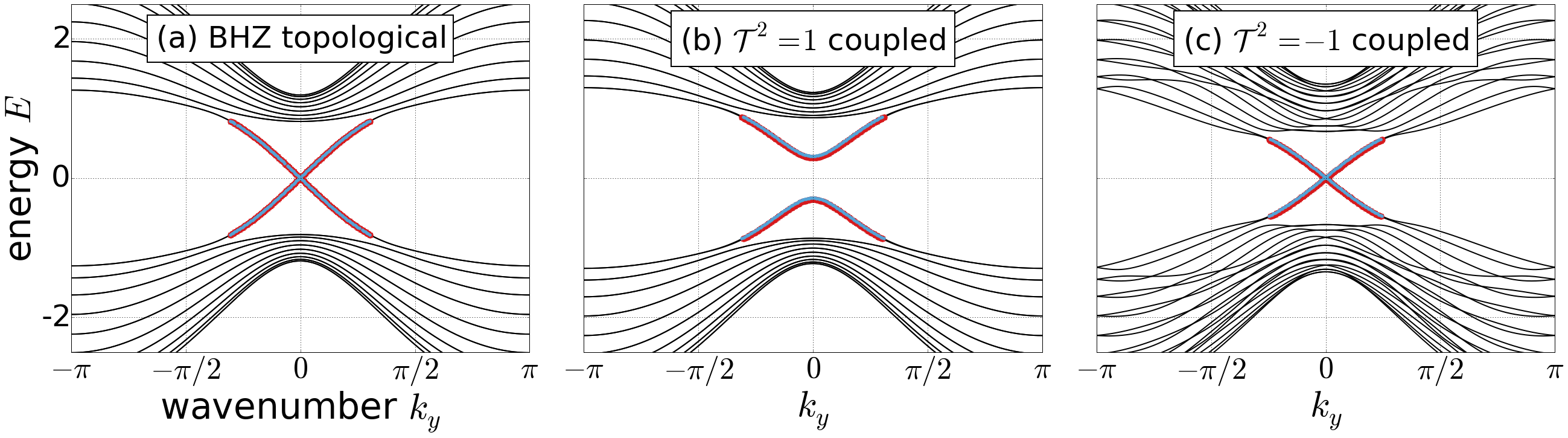}
\caption{\label{fig:bhz-edge_dispersions} Stripe dispersion relations
  of the BHZ model, with sublattice potential parameter
  $\uu=-1.2$. Right/left edge states (more than 60\% weight on the
  last/first two columns of unit cells) marked in dark red/light
  blue. (a): uncoupled layers, $\hat{C}=0$. (b): Symmetric coupling
  $\hat{C}=0.3 \sigma_x$ gaps the edge states out.  (c): Antisymmetric
  coupling $\hat{C}=0.3\sigma_y$ cannot open a gap in the edge
  spetrum.  }
\end{figure}

\subsection{An example: the BHZ model with different types of coupling}

We start with the concrete example of the BHZ model.  We
set the sublattice potential parameter $\uu=-1.2$, and plot the edge
dispersion relation, defined in the same way as for the Chern
insulators in Section \ref{sec:qiwuzhang_edgestates}.

As long as there is no coupling between the two copies, $\hat{C}=0$,
the system $\HHbhz$ is a direct sum of two Chern insulators, with
opposite Chern numbers. As Fig.~\ref{fig:bhz-edge_dispersions} (a)
shows, on each edge, there is a pair of edge state branches: a branch
on the layer with Hamiltonian $\hat{H}$, and a counterpropagating
branch on the layer with $\hat{H}^\ast$.  Although these two edge
state branches cross, this crossing will not turn into an
anticrossing: the states cannot scatter into each other since they are
on different layers.  The two edge state branches are linked by
time-reversal: they occupy the same position, but describe propagation
in opposite directions.  In fact, they are linked by both
time-reversal symmetries this system has, by $\hs_x K$ and $i\hs_y K$.

A coupling between the layers can can gap the edge states out, as
shown in Fig.~\ref{fig:bhz-edge_dispersions} (b).  We here used
$\hat{C}=0.3\sigma_x$, which respects the $\TTT^2=+1$ symmetry but breaks
the $\TTT^2=-1$ one. The crossings between counterpropagating edge
states on the same edge have turned into anticrossings,
as expected, since the coupling allows particles to hop between the
counterpropagating edge states (on the same edge, but in different
layers).


We see something different if we couple the layers while respecting
the $\TTT^2=-1$ time reversal symmetry, by, e.g., $\hat{C}=0.3\sigma_y$.
As Fig.~\ref{fig:bhz-edge_dispersions} (c) shows, the crossing at
$k_y=0$ between the edge state branches now does not turn into an
anticrossing.  As long as the coupling is not strong enough to close
the bulk gap, the edge states here appear to be \emph{protected}.

\subsection{Edge states in $\TTT^2=-1$}

\begin{figure}[!ht]
\centering
\includegraphics[width=0.5\linewidth]{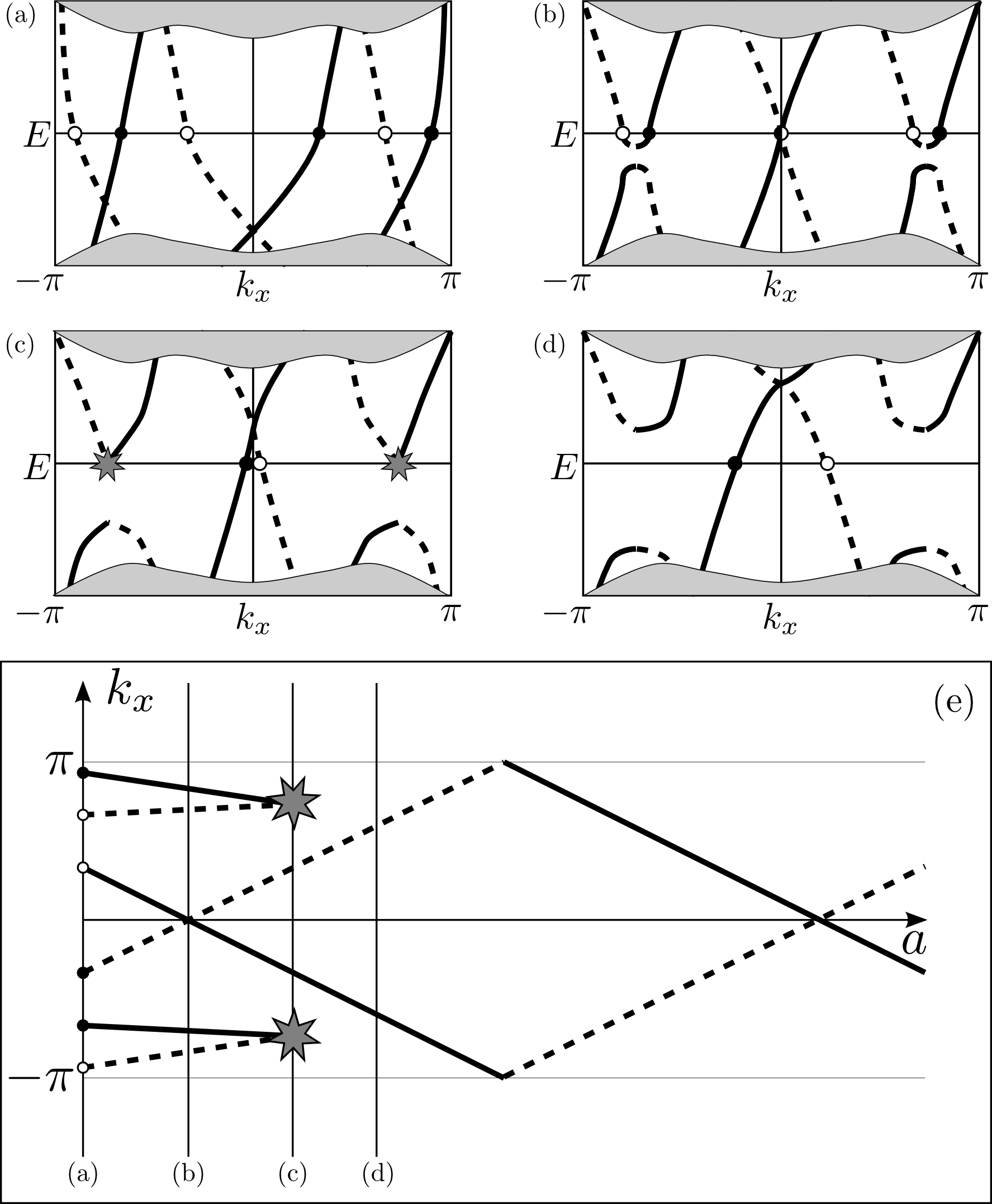}
\caption{\label{fig:a_Z2-surf-states} Edge states on a single edge of
  a 2-dimensional time-reversal invariant topological insulator with
  $\TTT^2=-1$, as the edge region undergoes a continuous deformation,
  parametrized by $a$, respecting the symmetry and the
  translational invariance along the edge. In (a)--(d), the edge state
  dispersions relations are shown, in the full edge Brillouin zone
  $k_x = -\pi , \ldots, \pi$, and in the energy window corresponding
  to the bulk gap. For clarity, right- (left-) propagating edge states
  are denoted by continuous (dashed) lines. Due to the deformation of
  the Hamiltonian, the edge state branches can move, bend, and couple,
  while the bulk remains unchanged. From (a) to (b), the crossing
  points between counterpropagating edge states become anticrossings,
  i.e., gaps open in these pairs of dispersion relation branches as a
  usual consequence of any parameter coupling them. Crossings at
  $k_x=0$ and $k_x=\pi$ cannot be gapped, as this would lead to a
  violation of the Kramers theorem. From (b) to (d), these gaps become
  so large that at energy $E=0$, the number of edge states drops from
  6 (3 Kramers pairs) to 2 (1 Kramers pair). The $k_x$ values of the
  edge states at 0 energy are plotted in (e), where this change in the
  number of edge states shows up as an ``annihilation'' of
  right-propagating and left-propagating edge states.}
\end{figure}

The states form
one-dimensional edge state bands in the one-dimensional Brillouin zone
$k_x=-\pi,\ldots,\pi$, shown schematically in
Fig.~\ref{fig:a_Z2-surf-states}. 
In general, an edge will host edge
states propagating in both directions. However, due to time-reversal
symmetry, the dispersion relations must be left-right symmetric when
plotted against the wavenumber $k_x$ along the edge direction. This
means that the number $N_+$ of right-moving edge states (these are
plotted with solid lines in Fig.~\ref{fig:a_Z2-surf-states}), and
$N_-$, the number of left-moving edge states (dashed lines) have to be
equal at any energy,
\begin{align}
N_+(E)&=N_-(E).
\end{align}

As with Chern insulators, we next consider the effect of adiabatic
deformations of the clean Hamiltonian on edge states. We consider
terms in the Hamiltonian that conserve translational invariance along
the edge, and respect Time Reversal Symmetry. The whole discussion of
Sect.~\ref{sect:qiwuzhang-edge_robustness} applies, and therefore
adiabatic deformations cannot change the signed sum of of the left-
and right-propagating edge states in the gap. However, time-reversal
symmetry restricts this sum to zero anyway.

Time reversal symmetry that squares to $\TTT^2=-1$, however, provides
a further restriction: adiabatic deformations can only change the
number of edge states by integer multiples of four (pairs of
pairs). To understand why, consider the adiabatic deformation
corresponding to Fig.~\ref{fig:a_Z2-surf-states} (a)-(d). Degeneracies
in the dispersion relation can be lifted by coupling the edge states,
as it happens in (b), and this can lead to certain edge states
disappearing at certain energies, as in (c). This can be visualized by
plotting the $k_x$ values at $E=0$ of the branches of the edge state
dispersion as functions of the deformation parameter (which is some
combination of the parameters of the Hamiltonian) $a$, as in
Fig.~\ref{fig:a_Z2-surf-states} (e). Due to the deformation, two
counterpropagating edge states can ``annihilate'', when the
corresponding modes form an avoided crossing. If this happens at a
generic momentum value $k$, as in (c), then, due to the time reversal
invariance, it also has to happen at $-k$, and so the number of edge
states decreases by 4, not by 2. The special momentum values of $k_x$
are the TRIM, which in this case are $k_x=0,\pm\pi$. If the edge state
momenta meet at a TRIM, as in (b) at $k_x=0$, their ``annihilation''
would change the number of edge states by 2 and not by 4. However,
this cannot happen, as it would create a situation that violates the
Kramers degeneracy: at the TRIM, energy eigenstates have to be doubly
degenerate. The deformations in Fig.~\ref{fig:a_Z2-surf-states} can be
also read from (d) to (a), and therefore apply to the introduction of
new edge states as well.

\subsection{$\mathbb{Z}_2$ invariant: parity of edge state pairs}

At any energy inside the bulk gap, the parity of the number of
edge-states Kramers pairs for a given dispersion relation is well
defined.  In Fig.~\ref{fig:a_Z2-surf-states}(a), there are 3
edge-state Kramers pairs for any energy in the bulk gap, i.e., the
parity is odd.  In Fig.~\ref{fig:a_Z2-surf-states}(c), there are 3 of
them for every energy except for energies in the mini-gap of the bands
on the left and right for which the number of edge-state Kramers pairs
is 1, and for the upper and lower boundaries of the mini-gap [the
  former depicted by the horizontal line in
  Fig.~\ref{fig:a_Z2-surf-states}(c)], where the number of Kramers
pairs is 2.  The parity is odd at almost every energy, except the two
isolated energy values at the mini-gap boundaries.

The general proposition is that the parity of the number of edge-state
Kramers pairs at a given edge for a given Hamiltonian at a given
energy is independent of the choice of energy, as long as this energy
is inside the bulk gap. Since in a time-reversal invariant system, all
edge states have counter-propagating partners, we can express this 
number as
\begin{align}
D &= \frac{N(E)}{2} \mathrm{mod}\,2 = \frac{N_+(E)+N_-(E)}{2}
\mathrm{mod}\,2,
\label{eq:D_mod_def}
\end{align}
where $N(E)=N_+(E)+N_-(E)$ is the total number of edge states at an
edge. A caveat is that there are a few
isolated energy values where this quantity is not well defined, e.g.,
the boundaries of mini-gaps in the above example, but these energies
form a set of zero measure. 

Since $D$ is a topological invariant, we can classify two-dimensional time-reversal
invariant lattice models according to it, i.e., the parity of the
number of edge-state Kramers pairs supported by a single edge of the
terminated lattice. Because it can take on two values, this `label'
$D$ is called the $\mathbb{Z}_2$ invariant, and is represented by a
bit taking on the value 1 (0) if the parity is odd (even).

As a final step, we should next consider disorder that breaks
translational invarance along the edge, in the same way as we did for
Chern insulators. Due to the presence of edge states propagating in
both directions along the edge, the treatment of disorder is a bit
trickier than it was for Chern states. 


\section{Absence of backscattering}

A remarkable property of Chern insulators is that
they support chiral edge states, i.e., edge states 
that have no counter-propagating counterparts. 
A simple fact implied by the chiral nature of these edge states is
that impurities are unable to backscatter a particle that occupies 
them. 
As we argue below, absence of backscattering 
is also characteristic of disordered two-dimensional time-reversal invariant topological insulators,
although the robustness is guaranteed only against time-reversal
symmetric scatterers.

In this Chapter, we introduce the scattering matrix, a simple concept that 
allows for a formal analysis of scattering at impurites,
and discuss the properties of edge state scattering 
in two-dimensional time-reversal invariant topological insulators.
The scattering matrix will also serve as a basic tool in the subsequent
chapter, where we 
give a simple theoretical description of electronic transport
of phase-coherent electrons,
and discuss observable consequences of the existence and
robustness of edge states.

\subsection{The scattering matrix}

Consider a phase-coherent two-dimensional conductor with a finite
width in the $y$ direction, and discrete
translational invariance along the $x$ axis.
Think of the system as having periodic boundary conditions
in the $x$ direction. 
As earlier, we describe the system in terms of a simple 
lattice model, where the unit cells form a square lattice of size
$N_x \times N_y$, and there might be an internal degree of freedom
associated to the unit cells. 

As the system has discrete translational invariance along $x$, 
we  can also think of it as a one-dimensional lattice, whose unit cell 
incorporates both the internal degree of freedom of the two-dimensional lattice
and the real-space structure along the $y$ axis. 
Using that picture, it is clear that 
the electronic energy eigenstates 
 propagating along the
$x$ axis at energy $E$ have a product structure,
as required by the one-dimensional Bloch's theorem:
\begin{equation}
\label{eq:a_propagatingwave}
\ket{\chl,\pm} = \ket{k_{\chl,\pm}} \otimes \ket{\Phi_{\chl,\pm}},
\end{equation}
On the right hand side, the first ket corresponds to 
a usual momentum eigenstate 
$\ket{k} = \frac{1}{\sqrt{N_x}} \sum_{\uci_x=1}^{N_x} e^{i k \uci_x} \ket{\uci_x}$ propagating along $x$,
whereas the second ket incorporates the shape of the 
transverse standing mode
as well as the internal degree of freedom.
The states appearing in Eq. \eqref{eq:a_propagatingwave}
are normalized to unity.
The integer $\chl = 1,2,\dots,N$ labels the propagating modes, 
also referred to as scattering channels. 
The $+$ and $-$ signs correspond to right-moving and 
left-moving states, respectively;
the direction of movement is assigned according to the
sign of the group velocity:
\bean
v_{\chl,\pm} = \left.\frac{d E(k)}{d k}\right|_{k=k_{\chl,\pm}},
\eean
where $E(k)$ is the dispersion relation of the one-dimensional band hosting
the state $\ket{\chl,\pm}$ .

Now we re-normalize the states above, such that 
different states carry the same particle current through 
an arbitrary vertical cross section of the system. 
We will use these current-normalized 
wave functions in the definition of the scattering matrix below, 
which guarantees that the latter is a unitary matrix. 
According to the one-dimensional relation we obtained between 
the current and the group velocity in Eq. \eqref{eq:current:currentvsgroupvelocity},
the current-normalized states can be defined using
the group velocity as 
\bean
\label{eq:absence:currentnormalized}
\ket{\chl,\pm}_c = \frac{1}{\sqrt{|v_{\chl,\pm}|}} \ket{\chl,\pm}.
\eean

\begin{figure}[!ht]
\begin{center}
\includegraphics[scale=0.5]{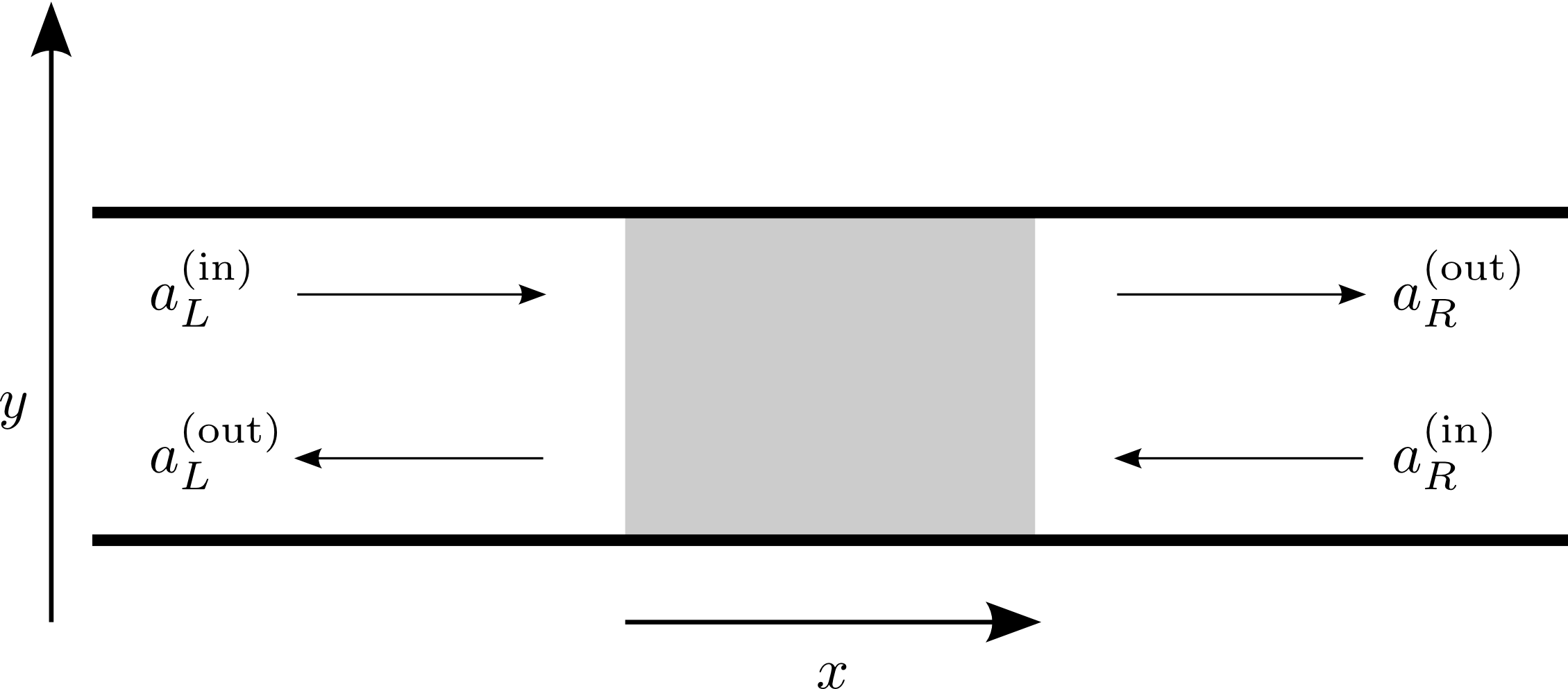}
\caption{\label{fig:a_smatrix}
Disordered region (gray) obstructing electrons in a 
two-dimensional phase-coherent conductor.
The scattering matrix $S$ relates the amplitudes 
$a^{\rm (in)}_L$ and $a^{\rm (in)}_R$ of incoming waves to
the amplitudes
$a^{\rm (out)}_L$ and $a^{\rm (out)}_R$ of outgoing waves.}
\end{center}
\end{figure}

Now consider the situation when the electrons are obstructed by
a disordered region in the conductor, as shown in Fig. \ref{fig:a_smatrix}.
A monoenergetic wave incident on the scattering region is
characterized by a vector of coefficients
\begin{equation}
\label{eq:absence:ain}
a^{\rm (in)} = \left(
	a^{\rm (in)}_{L,1}, 
	a^{\rm (in)}_{L,2},
	\dots,
	a^{\rm (in)}_{L,N},
	a^{\rm (in)}_{R,1},
	a^{\rm (in)}_{R,2},
	\dots,
	a^{\rm (in)}_{R,N}
\right).
\end{equation}
The first (second) set of $N$ coefficients correspond to propagating
waves \eqref{eq:absence:currentnormalized} in the left (right)
\emph{lead} $L$ ($R$), 
that is, the clean regions on the left (right) side of the disordered 
region.
The reflected and transmitted parts of the wave are described by the 
vector 
\begin{equation}
\label{eq:absence:aout}
a^{\rm (out)} = \left(
	a^{\rm (out)}_{L,1}, 
	a^{\rm (out)}_{L,2},
	\dots,
	a^{\rm (out)}_{L,N},
	a^{\rm (out)}_{R,1},
	a^{\rm (out)}_{R,2},
	\dots,
	a^{\rm (out)}_{R,N}
\right).
\end{equation}
The corresponding energy eigenstate reads
\bean
\label{eq:absence:scatteringstate}
\ket{\psi} = \sum_{\chl=1}^N 
a^\textrm{(in)}_{L,\chl} \ket{\chl,+,L}_c
+
a^\textrm{(out)}_{L,\chl} \ket{\chl,-,L}_c
+ 
a^\textrm{(in)}_{R,\chl} \ket{\chl,-,R}_c
+
a^\textrm{(out)}_{R,\chl} \ket{\chl,+,R}_c.
\eean
Here, the notation for the current-normalized states
introduced in Eq. \eqref{eq:absence:currentnormalized}
has been expanded by the lead index $L$/$R$. 

The scattering matrix $S$ relates the two vectors
introduced in Eqs. 
\eqref{eq:absence:ain} and \eqref{eq:absence:aout}:
\begin{equation}
\label{eq:smatrixdef}
a^{\rm (out)} = S a^{\rm (in)}.
\end{equation}
The size of the scattering matrix is $2N\times 2N$, and it has 
the following block structure:
\begin{equation}
S = \left(\begin{array}{cc}
r & t' \\
t & r'
\end{array}\right)
\end{equation}
where $r$ and $r'$ are $N\times N$ reflection matrices describing
reflection from left to left and from right to right, 
and $t$ and $t'$ are transmission matrices describing 
transmission from left to right and right to left.

Particle conservation, together with 
the current normalization Eq. \eqref{eq:absence:currentnormalized},
implies the unitarity of the scattering matrix $S$.
In turn, its unitary character implies that the Hermitian
matrices $tt^\dag$, $t't'^\dag$, $1-rr^\dag$, and
$1-r'r'^\dag$ all have the same set of real eigenvalues 
$T_1, T_2,\dots T_N$,
called transmission eigenvalues.

\subsection{A single Kramers pair of edge states}
\label{sec:a_onepair}

\begin{figure}[!ht]
\begin{center}
\includegraphics[scale=0.7]{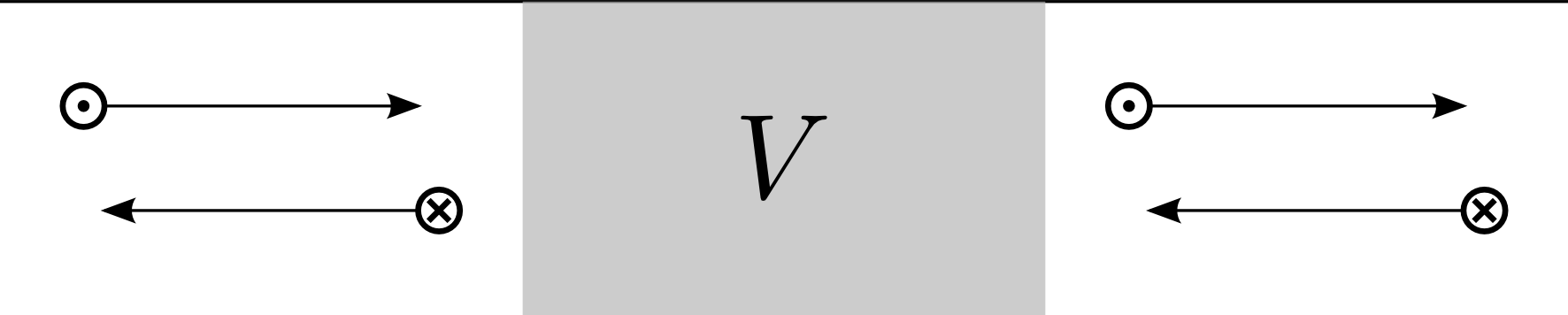}
\caption{\label{fig:a_scat}
Scattering of an edge state on a time-reversal symmetric defect $V$.
In a two-dimensional time-reversal invariant topological insulator with a single Kramers pair
of edge states,
the incoming electron is transmitted through such a defect
region with unit probability.}
\end{center}
\end{figure}

Now we use the scattering matrix $S$ to characterize defect-induced
scattering
of an electron occupying an edge state of a two-dimensional time-reversal invariant topological 
insulator.
Consider a half-plane of such a homogeneous lattice 
which supports
exactly one Kramers pair of edge states at a given energy $E$
in the bulk gap, as shown in Fig. \ref{fig:a_scat}.
Consider the scattering of the electron incident on the defect
from
the left side in Fig. \ref{fig:a_scat}.
The scatterer is characterized by the Hamiltonian $V$.
We will show that the impurity cannot backscatter the electron as long as 
$V$ is time-reversal symmetric.

We choose our propagating modes 
such that the incoming and outgoing states are
related by time-reversal symmetry, i.e., 
\begin{subequations}
\label{eq:absence:trs}
\bean
\ket{\chl,-,L}_c &=& \TTT \ket{\chl,+,L}_c \\
\ket{\chl,+,R}_c &=& \TTT  \ket{\chl,-,R}_c.
\eean
\end{subequations}
Also, recall that $\TTT^2=-1$.  
In the presence of the perturbation $V$, the edge states
of the disorder-free system 
are no longer energy eigenstates of the system.
A general scattering state $\ket{\psi}$ 
at energy $E$ is characterized by the vector 
$a^{\rm (in)}$ of incoming amplitudes.
According to  Eq. \eqref{eq:absence:scatteringstate}
and the  definition \eqref{eq:smatrixdef}
of the scattering matrix $S$,
the energy eigenstates outside the scattering region 
can be expressed as:
\bean
\label{eq:scatstate}
\ket{\psi} = 
& \sum_{\chl=1}^N &
\Big[
	a^{\rm (in)}_{L,\chl} \ket{\chl,+,L}_c+
	a^{\rm (in)}_{R,\chl} \ket{\chl,-,R}_c
\\ \nonumber
&&+
	\left(S a^{\rm (in)}\right)_{L,\chl} \ket{\chl,-,L}_c 
	+
	\left(S a^{\rm (in)}\right)_{R,\chl} \ket{\chl,+,R}_c
\Big].
\eean

Using Eq. \eqref{eq:absence:trs}, we find 
\bean
\label{eq:Tscatstate}
-\TTT \ket{\psi} = 
& \sum_{\chl=1}^N &
\Big[
	- a^{\rm (in)*}_{L,\chl} \ket{\chl,-,L}_c
	- a^{\rm (in)*}_{R,\chl} \ket{\chl,+,R}_c 
\\ \nonumber
&&+
	\left(S^* a^{\rm (in)*}\right)_{L,\chl} \ket{\chl,+,L}_c 
	+
	\left(S^* a^{\rm (in)*}\right)_{R,\chl} \ket{\chl,-,R}_c
\Big].
\eean
Due to time reversal symmetry, this state $-\TTT \ket{\psi}$ 
is also an energy eigenstate having the same energy
as $\ket{\psi}$.
Using the unitary character of the scattering matrix, 
the state $-\TTT \ket{\psi}$ can be rewritten as
\bean
\label{eq:Tscatstate2}
-\TTT \ket{\psi} = 
& \sum_{\chl=1}^N &
\Big[
	\left(S^* a^{\rm (in)*}\right)_{L,\chl} \ket{\chl,+,L}_c 
	+
	\left(S^* a^{\rm (in)*}\right)_{R,\chl} \ket{\chl,-,R}_c
\\ \nonumber
&&+
	\left(-S^T S^* a^{\rm (in)*}\right)_{L,\chl} \ket{\chl,-,L}_c
	+
	\left(-S^T S^* a^{\rm (in)*}\right)_{R,\chl} \ket{\chl,+,R}_c
\Big].
\eean
where $S^T$ denotes the transpose of $S$, 
is also an energy eigenstate having the same energy as $\ket{\psi}$. 
Comparing Eqs. \eqref{eq:scatstate}
and \eqref{eq:Tscatstate},
and knowing that the scattering matrix at a given energy
is uniquely defined, we conclude that
$S = -S^T$, that is
\begin{equation}
\left(\begin{array}{cc}
r & t' \\
t & r'
\end{array}\right)
= S  = -S^T =
\left(\begin{array}{cc}
- r & -t \\
- t' & - r'
\end{array}\right)
\end{equation}
 implying 
 \begin{equation}
 \label{eq:a_r0}
 r=r'=0,
 \end{equation} and hence
perfect transmission of each of the two incoming waves.

\begin{figure}[!ht]
\sidecaption
\includegraphics[width=0.5\linewidth]{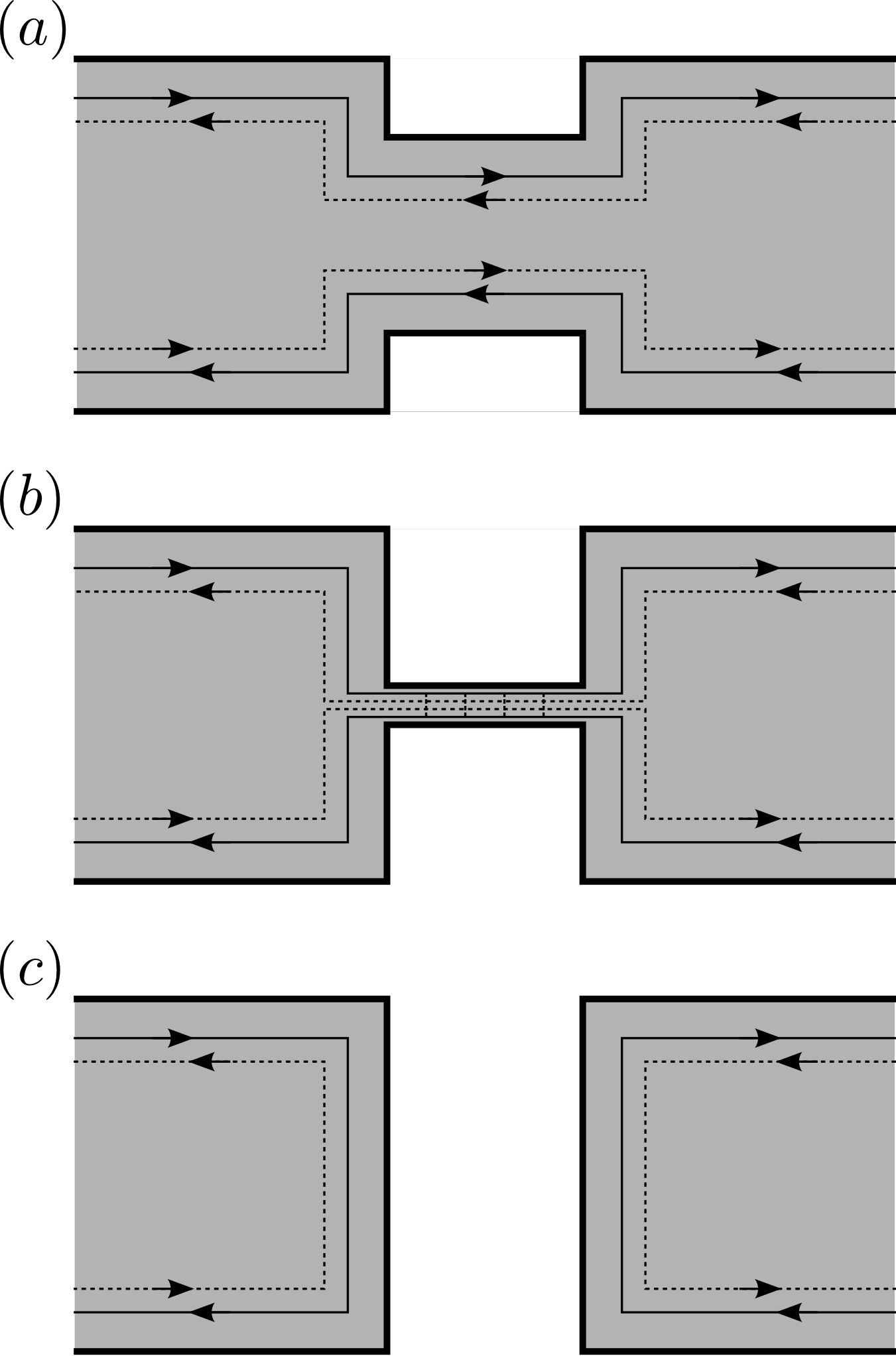}
\caption{\label{fig:a_backscattering}
Backscattering of edge states at a constriction. 
The states forming the edge-state Kramers pairs are depicted as
solid and dashed lines. 
(a) A time-reversal symmetric defect localized to the edges,
such as a small constriction shown here, is unable to backscatter
the incoming electron.
(b) Backscattering is possible between different edges, 
if the width of the constriction
is of the order of the decay length of the edge states. 
(c) A finite spatial gap between the left and right part of the wire
implies zero transmission.}
\end{figure}

If the  lattice has the geometry of a ribbon, and
the TRS scatterer extends to both edges, 
then the absence of backscattering is not guaranteed.
This is illustrated in Fig. \ref{fig:a_backscattering}, where we
compare three examples.
In (a), the defect is formed as 
a wide constriction on both edges, with a width much larger than 
 the 
characteristic length of the penetration of the edge states to the bulk
region of the ribbon.
Backscattering between  states at the same edge is forbidden 
due to TRS, and backscattering between states at different edges is
forbidden due to a large spatial separation of their
corresponding wave functions.
In (b), a similar but narrower constriction with a width comparable to
the penetration length of the edge states does allow for 
scattering between states on the lower and upper edges.
In this case, backscattering from a right-moving state on
one edge to a left-moving state at the other edge
is not forbidden. 
In (c), the constriction divides the ribbon to two
unconnected parts, resulting in zero transmission through
the constriction.

Naturally, backscattering is also allowed if the scatterer is not
time-reversal symmetric, or if the scattering process
is inelastic.
Backscattering is not forbidden for the `unprotected' edge states 
of topologically trivial ($D=0$) two-dimensional time-reversal invariant insulators.

\subsection{An odd number of Kramers pairs of edge states}
\label{sec:a_oddnumber}

The above statement \eqref{eq:a_r0} implying unit transmission, 
albeit in a somewhat weakened form, can be 
generalized for arbitrary two-dimensional time-reversal invariant topological 
insulator lattice models, 
including those where the number of edge-state Kramers pairs $N$ is
odd but not  one. 
The proposition is that in such a system,
given a time-reversal symmetric scatterer $V$
and an arbitrary energy $E$ in the bulk gap,
there exists at least one linear combination of the incoming states
of energy $E$ from each side of the defect that is perfectly transmitted
through the defect.

The proof follows that in the preceding section, with the difference 
that the quantities $r$ and $t$ describing reflection and transmission 
are $N \times N$ matrices, and that the antisymmetric 
nature of the $S$-matrix $S = -S^T$ implies
the antisymmetry of the reflection matrices $r = - r^T$.
According to Jacobi's theorem, every odd-dimensional 
antisymmetric matrix has a vanishing determinant, which 
is implied by 
\bean
\label{eq:jacobiproof}
\det ( r ) = \det (r ^T) = \det (-r ) = (-1)^N \det (r) = - \det (r),
\eean
where we used the antisymmetry of $r$ in the second
step and the oddness of $N$ in the last step. 
As a consequence of Eq. \eqref{eq:jacobiproof} we know  
$\det (r) = 0$, hence 
$\det(r^\dag r)= \det(r^\dag)\det( r) =0$.
Therefore, at least one eigenvalue of $r^\dag r$ is zero,
which implies that at least one transmission eigenvalue $T_\chl$ is
unity.



\subsection{Robustness against disorder}

\begin{figure}[!ht]
\sidecaption
\includegraphics[width=0.5\linewidth]{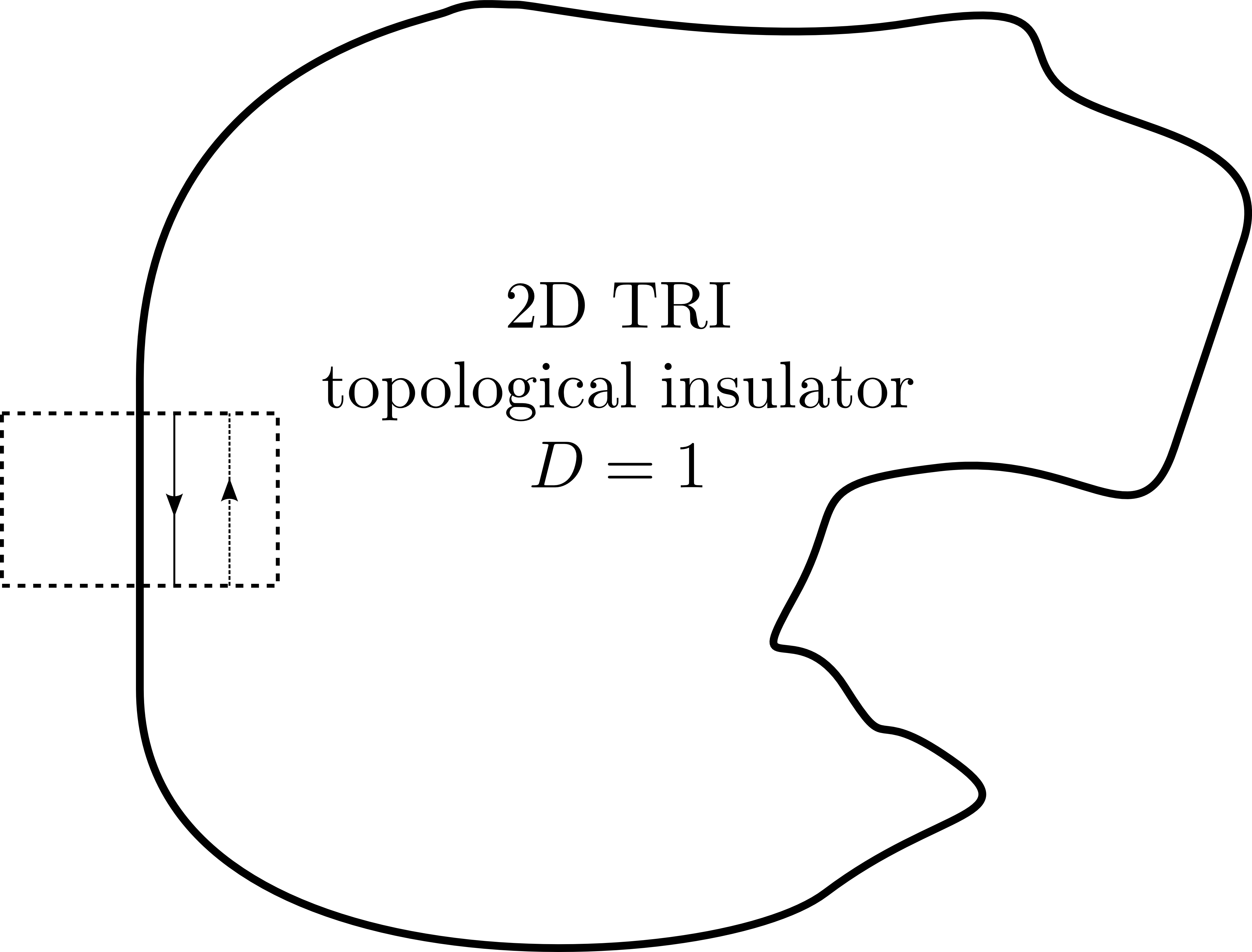}
\caption{\label{fig:a_potato}
A disordered two-dimensional time-reversal invariant topological insulator contacted with two electrodes.
Disorder is `switched off' and the edge is
`straightened out' within the dashed box, hence the edge modes
there resemble those of the disorder-free lattice.}
\end{figure}

The absence-of-backscattering 
result \eqref{eq:a_r0} 
implies a remarkable statement regarding 
the existence of (at least) one perfectly transmitting `edge state' 
in a finite-size disordered sample of a two-dimensional time-reversal invariant topological
insulator.

(See also the discussion about Fig.~\ref{fig:qiwuzhang_2contacts} in the
context of Chern Insulators).
Such a sample with an arbitrarily chosen geometry is shown in 
Fig.~\ref{fig:a_potato}.
Assume that the disorder is TRS and localized to the edge of the sample.
We claim that any chosen segment of the edge of this disordered 
sample supports, at any energy that is deep inside the bulk gap,
(at least) one counterpropagating 
Kramers pair of edge states that
are delocalized along the edge and able to
transmit electrons with unit probability.
This is a rather surprising feature in light of the fact that
in truly one-dimensional lattices, a small disorder is enough
to induce Anderson localization of the energy eigenstates, and hence
render the system an electronic insulator. 

To demonstrate the above statement, 
let us choose an edge segment of the disordered
sample for consideration, e.g., the edge segment running
outside the dashed box in Fig.~\ref{fig:a_potato}.
Now imagine that we `switch off' disorder in the 
complementer part of the edge of the sample,
and `straighten out' the geometry of that complementer part, the
latter being shown within the dashed box of Fig.~\ref{fig:a_potato}.
Furthermore, via an appropriate spatial adiabatic deformation 
of the Hamiltonian of the system in the vicinity of the complementer
part of the edge 
(i.e., within the dashed box in 
Fig.~\ref{fig:a_potato}),
we make sure that only a single edge-state Kramers pair
is present within this complementer part.
The existence of such an adiabatic deformation is guaranteed by the
topologically nontrivial character of the sample, see the discussion of 
Fig.~\ref{fig:a_Z2-surf-states}.
The  disordered edge segment, outside the dashed box in 
Fig.~\ref{fig:a_potato}, 
now functions as a scattering
region for the electrons in the straightened part of the edge.
From the result \eqref{eq:a_r0}
we know that such a TRS scatterer is unable to induce backscattering
between the edge modes of the straightened part of the edge, 
hence we must conclude that the disordered segment must indeed
support a perfectly transmitting edge state in each of the two
propagation directions. 

In the next chapter, we show that 
the electrical conductance of such a disordered sample is finite and 
`quantized', if it is measured through a source and a drain contact that
couple effectively to the edge states.

%% file: bhz_bulk.tex
\chapter{The $\mathbb{Z}_2$ invariant of two-dimensional topological insulators}
\label{chap:BHZbulk}

\abstract*{A time-reversal invariant topological insulator either has
  no topologically protected edge states, or one pair of such edge
  states. Thus, its bulk topological invariant is either 0 or 1: it is
  a $\mathbb{Z}_2$ number. Although obtaining a single yes/no answer
  might seem easier than the calculation of a Chern number, the
  $\mathbb{Z}_2$ invariant is notoriously difficult to calculate. In
  this Chapter we detail a way to calculate it that follows the same
  logic as before for the Chern number.}

In the previous Chapter, we have seen that two-dimensional insulators
can host topologically protected edge states even if time-reversal
symmetry is not broken, provided it squares to $-1$, i.e.,
$\TTT^2=-1$. Such systems fall into two
categories: no topologically protected edge states (trivial), or one
pair of such edge states (topological).  This property defines a
$\mathbb{Z}_2$ invariant for these insulators. In the spirit of the
bulk--boundary correspondence, we expect that the bulk momentum-space
Hamiltonian $\HH(\kk)$ should have a corresponding topological
invariant (generalized winding number). 

The bulk $\mathbb{Z}_2$ invariant is notoriously difficult to
calculate. The original definition of the
invariant\cite{Kane-Z2,z2_spinpump} uses a smooth gauge
in the whole Brillouin zone, that is hard to construct
\cite{soluyanov_smooth}, which makes the invariant difficult to
calculate.  An altogether different approach, which is robust and
calculatable, uses the scattering matrix instead of the
Hamiltonian\cite{scattering_top_2012}.

In this Chapter we review a definition of the bulk $\mathbb{Z}_2$
invariant\cite{rui_yu_equivalent} based on the dimensional reduction
to charge pumps. This is equivalent to the originally defined bulk
invariants\cite{rui_yu_equivalent}, but no smooth gauge is required to
calculate it. It can be outlined  as follows.
\begin{enumerate}
\item Start with a bulk Hamiltonian $\HH(k_x,k_y)$. Reinterpret $k_y$
  as time: $\HH(kx,ky)$ is a bulk one-dimensional Hamiltonian of an
  adiabatic pump. This is the dimensional reduction we used for Chern
  insulators in Chapt.~\ref{chap:qiwuzhang}.
\item Track the motion, with
  $k_y$ playing the role of time, of pumped particles in the bulk
  using Wannier states. We will call this the Wannier center flow.
\item Time-reversal symmetry restricts the
  Wannier center flow. As a result, in some cases the particle pump
  cannot be turned off adiabatically -- in those cases the insulator
  is topological. If it can be turned off, the insulator is trivial.
\end{enumerate}
In order to go through this argument, we will first gather the used
mathematical tools, i.e., generalize the Berry phase 
and the Wannier states. We will then define the Wannier center
flow, show that it can be calculated from the Wilson loop. Finally, we
will use examples to illustrate the $\mathbb{Z}_2$ invariant and argue
that it gives the number of topologically protected edge states. 




\section{Tools: Nonabelian Berry phase, multiband Wannier states}
\label{sec:bhz_bulk-tools}

To proceed to calculate the topological invariants, we need to
generalize the tools of geometric phases, introduced in
Chapt.~\ref{chap:berry_chern}, and of the Wannier states of
Chapt.~\ref{chap:polarization} to manifolds consisting of more bands.

\subsection{Preparation: Nonabelian Berry phase}
\label{subsec:bhz_bulk-nonabelian}

We defined the Berry phase in Chapt.~\ref{chap:berry_chern}, as the
relative phase around a loop $L$ of $N$ states $\ket{\Psi_j}$, with
$j=1,2,\ldots,N$.  Since the Berry phase is gauge independent, it is
really a property of the loop over $N$ one-dimensional projectors
$\ket{\Psi_j}\bra{\Psi_j}$.  In most physical applications -- in our
case as well -- the elements of the loop are specified as projectors
to a eigenstates of some Hamiltonian for $N$ different settings of
some parameters.

As a generalization of the Berry phase, we ask about the relative
phase around a loop on $N$ projectors, each of which is $\NF$
dimensional.  The physical motivation is that these are eigenspaces of
a Hamiltonian, i.e., projectors to subspaces spanned by degenerate
energy eigenstates\cite{wilczek_zee}. 



\subsubsection*{Discrete Wilson loop }

Consider a loop over $N \ge 3$ sets of states from the Hilbert space,
each consisting of $\NF\in\mathbb{N}$ orthonormal states, $\{
\ket{u_n(k)} | n=1,\ldots,\NF\}$, with $k=1,\ldots,N$.  We quantify
the overlap between set $k$ and set $l$ by the $\NF \times \NF$
\emph{overlap matrix} $M^{(kl)}$, with elements
\begin{align}
M^{(kl)}_{nm} &= \braket{u_n(k)}{u_m(l)},
\end{align}
with $n,m = 1\,\ldots,\NF$.

The \emph{discrete Wilson loop} is
the product of the overlap matrices along the loop, 
\begin{align}
\WW &= \MM^{(12)} \MM^{(23)} \ldots \MM^{(N-1,N)} \MM^{(N1)}. 
\label{eq:bhz_bulk-wilsondef}
\end{align}
We will be interested in the eigenvalues $\lambda_n$ of the Wilson
loop, with $n=1,\ldots,\NF$,
\begin{align}
\WW \,\wilsonv_n &= \lambda_n \wilsonv_n,
\label{eq:bhz_bulk-wilsoneigs-def}
\end{align}
where $\wilsonv_n$ is the $n$th eigenvector, with $n=1,\ldots,\NF$. 

Note that we could have started the Wilson
loop, Eq.~\eqref{eq:bhz_bulk-wilsondef}, at the $k$th group instead of
the first one, 
\begin{align}
\WW^{(k)} &= \MM^{(k,k+1)} \MM^{(k+1,k+2)} \ldots \MM^{(N,1)} \MM^{(1,2)} \ldots \MM^{(k-1,k)}. 
\label{eq:bhz_bulk-wilsondef_k}
\end{align}
Although the elements of the matrix $\WW^{(k)}$ depend on the starting
point $k$, the eigenvalues $\lambda_n$ do not. Multiplying 
Eq.~\eqref{eq:bhz_bulk-wilsoneigs-def} from the left by 
$\MM^{(k,k+1)}\ldots \MM^{(N,1)}$, we obtain 
\begin{align}
\WW^{(k)} \left( \MM^{(k,k+1)}\ldots \MM^{(N,1)} \right) \wilsonv_n 
&= \lambda_n \left( \MM^{(k,k+1)}\ldots \MM^{(N,1)} \right) \wilsonv_n.
\end{align}

\subsubsection*{$U(\NF)$ gauge invariance of the Wilson loop}

We will now show that the eigenvalues of the discrete Wilson loop over
groups of Hilbert space vectors only depend on the linear spaces
subtended by the vectors of each group. Each group can undergo an
independent unitary operation to redefine the vectors, this cannot
affect the eigenvalues of the Wilson loop. This is known as the
invariance under a $U(\NF)$ gauge transformation.

A simple route to prove the $U(\NF)$ gauge invariance is via the
operator $\hat{W}$ defined by the Wilson loop matrix $\WW$ in the
basis of group 1, 
\begin{align}
\hat{W} &= \sum_{n=1}^\NF \sum_{m=1}^\NF \ket{u_n(1)} W_{nm} \bra{u_m(1)}. 
\end{align}
This operator has the same eigenvalues as
the Wilson loop matrix itself. It can be expressed using the
projectors to the subspaces spanned by the groups of states,
\begin{align}
\PP_k &= \sum_n \ket{u_n(k)}\bra{u_n(k)}.
\end{align}
The Wilson loop operator reads,
\begin{align}
\hat{W} &= \PP_1 \PP_2 \PP_3 \ldots \PP_N \PP_1.
\label{eq:bhz_bulk-wilson_with_p}
\end{align}
We show this explicitly for $N=3$, 
\begin{multline}
\hat{W} = \sum_{n=1}^{\NF}\sum_{n_2=1}^\NF\sum_{n_3=1}^\NF
\sum_{m=1}^\NF \ket{u_n(1)} \\ \braket{u_n(1)}{u_{n_2}(2)}
\braket{u_{n_2}(2)}{u_{n_3}(3)} \braket{u_{n_N}(N)}{u_{m}(1)}
\bra{u_m(1)} = \PP_1 \PP_2 \PP_3 \PP_1.
\end{multline}
The generalization to arbitrary $N\ge 3$ is straightforward.  

Equation \eqref{eq:bhz_bulk-wilson_with_p} makes it explicit that the
Wilson loop operator, and hence, the eigenvalues of the Wilson loop,
are $U(\NF)$ gauge invariant.








\subsection{Wannier states for degenerate multiband one-dimensional insulators}
\label{subsec:bhz_bulk-wannier}

We now generalize the Wannier states of
Sect.~\ref{sec:polarization-wannier} to a one-dimensional insulator
with $\NF$ occupied bands. In case of nondegenerate bands, a simple
way to go would be to define a set of Wannier states for each band
separately. However, time reversal symmetry forces degeneracies in the
bands, at least at time reversal invariant momenta, and so this is not
possible. Moreover, even in the nondegenerate case it could be
advantegeous to mix states from different bands to create more tightly
localized Wannier states.

To be specific, and to obtain efficient numerical protocols, we take a
finite sample of a one-dimensional insulator of $N = 2M$ unit cells, with
periodic boundary conditions.  
An orthonormal set of negative energy bulk eigenstates reads
\begin{align}
\ket{\Psi_\eii(k)} &= \ket{k} \otimes \ket{u_\eii(k)} = 
\frac{1}{\sqrt{N}} \sum_{\uci=1}^N 
e^{i\uci k} \ket{\uci}\otimes \ket{u_\eii(k)}
\end{align}
with, as before, $k\in \{ \delta_k, 2\delta_k,\ldots, N \delta_k\}$,
and $\delta_k = 2\pi/N$.  The index $\eii$ labels the eigenstates,
with $\eii = 1,\ldots,\NF$ for occupied, negative energy states.  The
$\ket{u_\eii(k)}$ are the negative energy eigenstates of the bulk
momentum-space Hamiltonian $\HH(k)$.  We will not be interested in the
positive energy eigenstates. 

Although we took a specific set of energy eigenstates above, because
of degeneracies at the time-reversal invariant momenta, we really only
care about the projector $\PP$ to the negative energy subspace. This
is defined as 
\begin{align}
\PP &= \sum_k \sum_{\eii=1}^\NF \ket{\Psi_\eii(k)}\bra{\Psi_\eii(k)} = 
\sum_k \ket{k}\bra{k} \otimes \PP(k); \\
\PP(k) &= \sum_{\eii=1}^\NF \ket{u_\eii(k)}\bra{u_\eii(k)}. 
\end{align}

\subsubsection*{Defining properties of Wannier states}

We will need a total number $\NF N$ of Wannier states to span the
occupied subspace, $\ket{w_\eii(\jjj)}$, with $\jjj=1,\ldots,N$, and 
$\eii=1,\ldots,\NF$. These are 
defined by the usual properties:
\begin{subequations}
\label{eq:bhz_bulk-wannier_requirements}
\begin{align} 
\braket{w_{\eii'}(\jjj')}{w_\eii(\jjj)} &= 
\delta_{\jjj'\jjj} \delta_{\eii'\eii}&  &\text{Orthonormal set}\\
\sum_{\jjj=1}^N 
\sum_{\eii=1}^\NF
\ket{w_\eii(\jjj)}\bra{w_\eii(\jjj)} &= \PP &  
&\text{Span occupied subspace}
\\
\label{eq:bhz_bulk-wannier_translation_requirement} 
\forall \uci:\,
\braket{\uci+1}{w_\eii(\jjj+1)} &= \braket{\uci}{w_\eii(\jjj)}& &\text{Related by translation} \\
\vphantom{\sum_{\jjj=1}^N} 
\lim_{N\to\infty} \bra{w_\eii(N/2)}
(\hat{x}-N/2)^2 &\ket{w_\eii(N/2)} < \infty&   & \text{Localization} 
\label{eq:bhz_bulk-wannier_localization_requirement}
\end{align}
\end{subequations}
with the addition in
Eq.~\eqref{eq:bhz_bulk-wannier_translation_requirement} defined
modulo $N$.

The Ansatz of Sect.~\ref{sec:polarization-wannier} for the Wannier
states, Eq.~\eqref{eq:polarization_wn_fourier}, generalizes to the
multiband case as
\begin{align}
\ket{w_n(\jjj)} &= \frac{1}{\sqrt{N}} \sum_{k=\delta_k}^{N \delta_k} 
e^{-i\jjj k} \sum_{p=1}^\NF U_{np}(k) \ket{\Psi_p(k)}.
\end{align}
Thus, each Wannier state can contain contributions from all of the
occupied bands, the corresponding weights given by a $k$-dependent
unitary matrix $U(k)$.

\subsubsection*{The projected unitary
  position operator}

As we did in Sect.~\ref{sec:polarization-wannier}, we will specify the
set of Wannier states as the eigenstates of the unitary position
operator restricted the occupied bands,
\begin{align}
\hat{X}_P &= \PP e^{i \delta_k \hat{x}} \PP.
\end{align}
To obtain the Wannier states, we go through the same steps as in
Sect.~\ref{sec:polarization-wannier}, with an extra index $\eii$. We
outline the derivations and detail some of the steps below.  You can
then check whether the properties required of Wannier states,
Eq.~\eqref{eq:bhz_bulk-wannier_requirements}, are 
fulfilled, in the same way as in the single-band case.

We note that for finite $N$, the projected unitary position
$\hat{X}_P$ is not a normal operator, i.e., it does not commute with
its adjoint. As a result, its eigenstates form an orthonormal set only
in the thermodynamic limit of $N\to\infty$. 
Just as in the single-band case, this can be seen as a discretization
error, which disappears in the limit $N\to\infty$.


The first step is to rewrite the operator $\hat{X}_P$. For this, consider
\begin{equation}
\bra{\Psi_{\eii'}(k')} \hat{X} \ket{\Psi_\eii(k)} = 
\delta_{k+\delta_k,k'} \,\braket{u_{\eii'}(k+\delta_k)}{u_\eii(k)}
\end{equation}
where $\delta_{k+\delta_k,k'}=1$ if $k'=k+\delta_k$, and 0 otherwise.
Using this, the projected unitary position operator can be rewritten as
\begin{multline}
\hat{X}_P = 
\sum_{k' k} 
\sum_{\eii',\eii=1}^\NF 
\ket{\Psi_{\eii'}(k')} 
\bra{\Psi_{\eii'}(k')} \hat{X} 
\ket{\Psi_{\eii}(k)}\bra{\Psi_{\eii}(k)} \\ 
= \sum_{k} 
\sum_{\eii',\eii=1}^\NF 
\braket{u_{\eii'}(k+\delta_k)}{u_{\eii}(k)} \cdot 
\ket{\Psi_{\eii'}(k+\delta_k)}\bra{\Psi_\eii(k)}.
\label{eq:bhz_bulk-xp2}
\end{multline}

\subsubsection*{Spectrum of the projected unitary
  position operator and the Wilson loop}

As in the single-band case, the next step is to consider $\hat{X}_P$
raised to the $N$th power. This time, it will not be simply
proportional to the projector $\PP$, however. 
Bearing in mind the orthonormality of the energy eigenstates,
$\braket{\Psi_\eii(k)}{\Psi_{\eii'}(k')}=\delta_{k'k}
\delta_{\eii'\eii}$, we find
\begin{align}
\left(\hat{X}_P\right)^N &= \sum_{k} \sum_{mn} W_{mn}^{(k)}
\ket{\Psi_m(k)}\bra{\Psi_n(k)}. 
\end{align}
The Wilson loop matrices $\WW^{(k)}$, as per
Eq.~\eqref{eq:bhz_bulk-wilsondef_k}, are all unitary equivalent, and
have the same set of complex eigenvalues, 
\begin{align}
\lambda_\eii &= \abs{\lambda_\eii} e^{i\theta_\eii} \quad \text{ with}
\quad \eii=1,\ldots,\NF,\\
\abs{\lambda_\eii} &\le 1, \quad \theta_\eii \in [-\pi,\pi).
\end{align}
The spectrum of eigenvalues of $\hat{X}_P$ is
therefore composed of the $N$th roots of these eigenvalues, for $j =
1,\ldots,N,$ and $\eii = 1,\ldots,\NF$,
\begin{align}
\lambda_{\eii,\jjj} &= e^{i \theta_\eii/N  + i j \delta_k + \log
  (\abs{\lambda_\eii})/N}, \quad
\Longrightarrow \quad (\lambda_{\eii,\jjj})^N = \lambda_\eii. 
\end{align}

\subsubsection*{Wannier centers identified through the eigenvalues 
of the Wilson loop}

As in the single-band case, Sect.~\ref{sec:polarization-wannier}, we
identify the phases of the eigenvalues $\lambda_{\eii,\jjj}$ of the
projected position operator $\hat{X}_P$ with the centers of the
Wannier states.
There are $\NF$ sets of Wannier states, each set
containing states that are spaced by distances of $1$,   
\begin{align}
\expect{x}_{\eii,\jjj} &= \frac{N}{2\pi} \im \log \lambda_{\eii,\jjj}
= \expect{x}_\eii + \jjj;\\ 
\expect{x}_\eii &= \frac{\theta_\eii}{2\pi}.
\end{align}
The phases $\theta_\eii$ of the $\NF$ eigenvalues of the Wilson loop
$W$ are thus identified with the Wannier centers, more precisely, with
the amount by which the $\NF$ sets are displaced from the integer
positions.

\section{Time-reversal restrictions on Wannier centers}
\label{sec:bhz_bulk-wannierk}

We will now apply the prescription for Wannier states above to the
one-dimensional insulators obtained as slices of a two-dimensional
$\TTT^2=-1$ time-reversal invariant insulator at constant $k_y$.  We
will use the language of dimensional reduction, i.e., 
talk of the bulk Hamiltonian $\HH(k_x,k_y)$ as describing an adiabatic
particle pump with $k_y$ playing the role of time. We will use the
Wannier center flow, i.e., the quantities
$\expect{x}_\eii=\theta_\eii(k_y)/(2\pi)$ 
to track the motion of the particles in the bulk during a fictitious pump
cycle, $k_y = -\pi \to \pi$.

Time-reversal symmetry places constrains the Wannier center flow in
two ways: it enforces $k_y\leftrightarrow -k_y$ symmetry, i.e.,
$\theta_\eii(k_y) = \theta_{\eii'}(-k_y)$, and it ensures that for
$k_y=0$ and for $k_y=\pi$, the $\theta_\eii$ are doubly degenerate. 
In this Section we see how these constraints arise.


\subsection{Eigenstates at $\kk$ and $-\kk$ are related}

A consequence of time-reversal symmetry is that energy eigenstates at
$\kk$ can be transformed to eigenstates at $-\kk$.  One might think
that because of time-reversal symmetry, energy eigenstates come in
time-reversed pairs, i.e., that
$\htau \HH(-\kk)^\ast \htau^\dagger = \HH(\kk)$ would automatically ensure
that $\ket{u_n(-\kk)} = e^{i\phi(\kk)} \htau
\ket{u_n(\kk)^\ast}$. However, because of possible degeneracies, this
is not necessarily the case. The most we can say is that the state
$\ket{u_n(-\kk)}$ is some linear combination of time-reversed
eigenstates,
\begin{align}
\ket{u_n(-\kk)} &= 
\htau \sum_{m=1}^{\NF} \left(B_{nm}(\kk) \ket{u_m(\kk)}\right)^\ast = 
\sum_{m=1}^{\NF} B_{nm}(\kk)^\ast \htau \ket{u_m(\kk)^\ast}.
\label{eq:bhz_bulk-bmn1}
\end{align}
The coefficients $B_{nm}(\kk)$ define the unitary \emph{sewing
  matrix}. An explicit formula for its matrix elements is obtained by
multiplying the above equation from the left by $\bra{u_a(\kk)^\ast}
\htau^\dagger$, with some $a=1,\ldots,\NF$.  This has the effect on
the left- and right-hand side of Eq.~\eqref{eq:bhz_bulk-bmn1} of
\begin{align}
\bra{u_a(\kk)^\ast} \htau^\dagger \ket{u_n(-\kk)} &=
\left(\bra{u_n(-\kk)} \htau \ket{u_a(\kk)^\ast} \right)^\ast;\\
\bra{u_a(\kk)^\ast} \htau^\dagger 
\sum_{m=1}^{\NF} B_{nm}(\kk)^\ast \htau \ket{u_m(\kk)^\ast} &= 
B_{na}(\kk)^\ast,
\end{align}
where for the last equation we used the unitarity of $\htau$ and the
orthonormality of the set $\ket{u_m(\kk)}$. Comparing the two lines
above (and relabeling $a \to m$), we obtain
\begin{align}
B_{nm}(\kk) &= \bra{u_n(-\kk)} \htau \ket{u_m(\kk)^\ast}. 
\end{align}
Using this definition it is straightforward to show that the sewing
matrix is unitary, and that $B_{mn}(-\kk) = - B_{nm}(\kk)$.




\subsection{Wilson loops at $k_y$ and $-k_y$ have the same eigenvalues}

To see the relation between the Wilson loops at $k_y$ and $-k_y$, we
first relate the projectors to the occupied subspace at these momenta.
We use a shorthand, 
\begin{align}
\PP_{j}(k_y) &= 
\begin{cases}
    \PP(2\pi+j\delta_k,k_y) & \text{if $j \le 0$};\\
    \PP(j\delta_k,k_y), & \text{if $j > 0$}.
\end{cases}
\end{align}
Using Eq.~\eqref{eq:bhz_bulk-bmn1}, and the unitarity of the sewing
matrix $B$, we find
\begin{multline}
\PP_{-j}(-k_y) = \PP(-\kk) = \sum_{n=1}^\NF \ket{u_n(-\kk)}\bra{u_n(-\kk)}  \\
= \sum_{n=1}^\NF \sum_{m=1}^\NF\sum_{m'=1}^\NF
B_{nm}(\kk)^\ast \htau \ket{u_m(\kk)^\ast}
 B_{nm'}(\kk) \bra{u_{m'}(\kk)^\ast} \htau^\dagger \\
= 
\htau \PP_j(k_y)^\ast \tau^\dagger = 
\htau \PP_j(k_y)^T \tau^\dagger.
\label{eq:bhz_bulk-pk-k}
\end{multline}



The consequence of Eq.~\eqref{eq:bhz_bulk-pk-k} for the Wilson loop is 
\begin{align}
\hat{W}(-k_y) &= \htau \hat{W}(k_y)^T \htau^\dagger.
\label{eq:bhz_bulk-wilsonloop_tr}
\end{align}
We write down the proof explicitly for
$N=6$, 
\begin{multline}
\hat{W}(-k_y) = \PP_3(-k_y) \PP_2(-k_y) \PP_1(-k_y) \PP_0(-k_y) 
\PP_{-1}(-k_y) \PP_{-2}(-k_y) \PP_3(-k_y) \\ 
= \htau \PP_3(k_y)^T \PP_{-2}(k_y)^T \PP_{-1}(k_y)^T 
\PP_0(k_y)^T \PP_{1}(k_y)^T
\PP_{2}(k_y)^T \PP_3(k_y)^T \htau^\dagger\\
= \htau \hat{W}(k_y)^T \htau^\dagger,
\label{eq:bhz_bulk-wilsonloop_tr}
\end{multline}
the generalization to arbitrary even $N$ follows the same lines.
The set of eigenvalues of $\hat{W}$ is the same as that of its
tranpose $\hat{W}^T$, as this holds for any matrix. Moreover,
the unitary transformation of $\hat{W}^T$ to $\htau \hat{W}^T \htau^\dagger$ 
does not change the eigenvalues either. 
To summarize, we find that the eigenvalues of the Wilson loop at
$-k_y$ are the same as of the Wilson loop at $k_y$,
\begin{align}
\theta_n(k_y) = \theta_n(-k_y).
\label{eq:bhz_bulk-lambdapkmk}
\end{align}

From Eq.~\eqref{eq:bhz_bulk-lambdapkmk}, we have that the Wannier
center flow is symmetric around $k_y=0$, and hence, also symmetric
around $k_y=\pi$. This means that it is enough to examine the Wannier
centers from $k_y=0$ to $k_y=\pi$.

\subsection{Wilson loops at $k_y=0$ and $k_y=\pi$ are doubly degenerate}

We now concentrate on the two special values of the $y$ wavenumber,
$k_y=0$ and $k_Y=\pi$, which are mapped unto themselves by time
reversal.  The one-dimensional Hamiltonians $\Hbulk(0)$ and
$\Hbulk(\pi)$ are time-reversal invariant, and thus, their every
eigenstate must come with its time-reversed partner, obtained by the
local time-reversal operation $\TTT$. 
Due to Kramers
theorem, an energy eigenstate and its time-reversed partner are
orthogonal, $\bra{\Psi}\TTT\ket{\Psi}=0$.

The Wilson loop $\hat{W}$ at $k_y=0$ and $k_y=\pi$ is doubly degenerate.  To
show this, take an eigenstate of the Wilson loop,
$\hat{W}\ket{\Psi} = \lambda \ket{\Psi}$.  Using
Eq.~\eqref{eq:bhz_bulk-wilsonloop_tr}, we find
\begin{align}
\lambda \ket{\Psi} = \hat{W}\ket{\Psi} &= 
 \tau \tau^\dagger \hat{W} \tau \tau^\dagger \ket{\Psi}= 
\htau \hat{W}^T \htau^\dagger \ket{\Psi};\\
\label{eq:bhz_bulk-wilsonloop-2}
\hat{W}^\dagger \htau \ket{\Psi^\ast} &=
 \lambda^\ast \htau \ket{\Psi^\ast}.
\end{align}
We obtained line \eqref{eq:bhz_bulk-wilsonloop-2} by multiplication
from the left by $\htau$ and complex conjugation, and using the
antisymmetry of $\htau$.  In the final line, we have obtained that the
Wilson loop $\hat{W}$ has a left eigenvector with eigenvalue
$\lambda^\ast$. Since this is orthogonal to $\ket{\Psi}$, however, the
right eigenvalue $\lambda$ must be at least twice degenerate. 

\section{Two types of Wannier center flow}
\label{sec:bhz_bulk-adiabatic}

We now examine the Wannier center flow, i.e., the functions
$\theta_n(k_y)$, in time-reversal invariant two-dimensional insulators
with $\TTT^2=-1$. Due to the restrictions of 
$k_y \leftrightarrow -k_y$ symmetry and degeneracy at $k_y=0,\pi$,
we will find two classes of Wannier center flow. In the trivial class,
the center flow can be adiabatically (i.e., continuously,
while respecting the restrictions) deformed to the trivial case, with
$\theta_n=0$ for every $n$ and every $k_y$. The topological class is
the set of cases where this is not possible.

\begin{figure}[!ht]
\centering
\includegraphics[width=0.9\linewidth]
{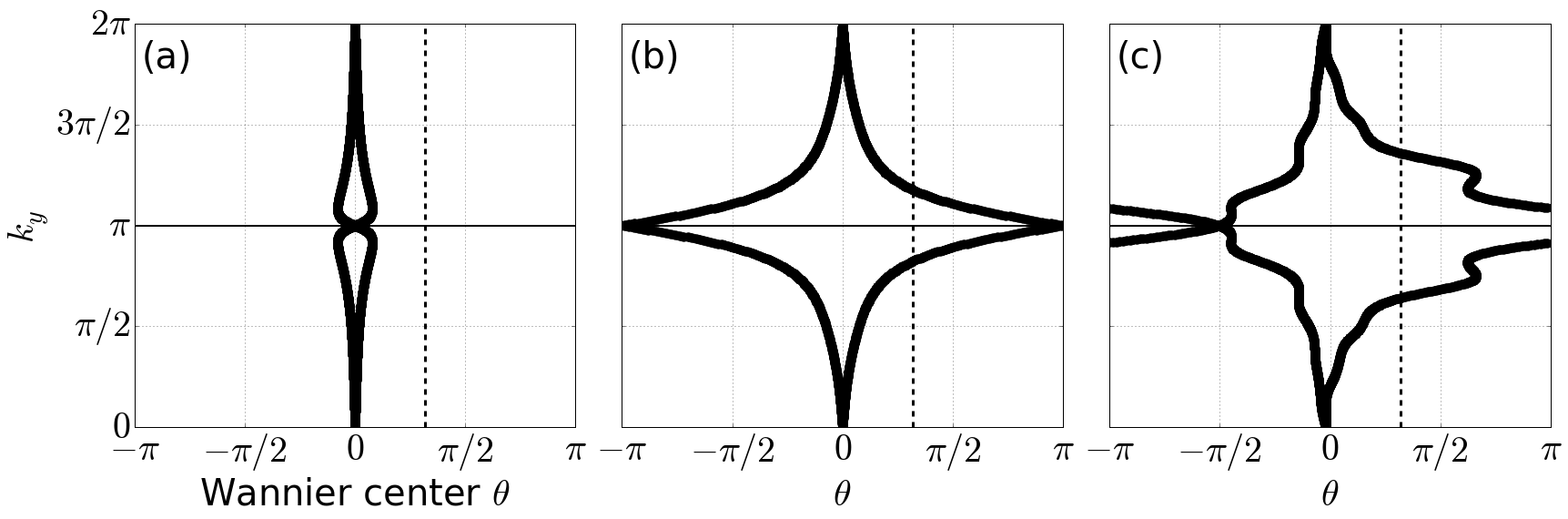}
\caption{\label{fig:bhz_bulk-wilson-loops} 
Wilson loop examples.  }
\end{figure}

To have a concrete example at hand, we examine the Wannier center flow
for the BHZ model of the previous Chapter,
Eq.~\eqref{eq:bhz_Hamiltonian_def2}, in a trivial (a) and in a
topological (b) case, and a third, more general topological (c) model.
All three cases are covered by a modified BHZ Hamiltonian,
\begin{multline}
\HH(\kk) = \hs_0 \otimes 
[(\uu + \cos k_x +\cos k_y) \hsigma_z + \sin k_y \hsigma_y )] 
+ \hs_z  \otimes \sin k_x \hsigma_x + 
\hs_x \otimes \hat{C} + \\
g \hat{s}_z \otimes \hsigma_y (\cos k_x + \cos 7k_y - 2).
\label{eq:bhz_bulk-Hamiltonian_crazy}
\end{multline}
For case (a), we set $g=0$, take the sublattice potential parameter
$\uu = 2.1$, and coupling operator $\hat{C} = 0.02
\hat{\sigma}_y$. This is adiabatically connected to the trivial limit
of the BHZ model at $\uu = +\infty$.  In case (b), we set $g=0$, take
sublattice potential parameter $\uu = 1$, and coupling operator
$\hat{C} = 0.3 \hat{\sigma}_y$, deep in the topological regime. In
case (c) we add the extra term to the BHZ model to have a more generic
case, with $g=0.1$, and use $\uu = 1$, coupling $\hat{C} = 0.1
\hsigma_y$. The Wannier center flows for the three cases are shown in
Fig.~\ref{fig:bhz_bulk-wilson-loops}.

We consider what
adiabatic deformations of the Hamiltonian can do to the center
flow. Focusing to $k_y=0\to\pi$, the center flow consists of branches
$\theta_n(k_y)$, that are continuous functions of $k_y$, beginning at
$\theta_n(0)$ and ending at $\theta_n(\pi)$. Due to an adiabatic
deformation,
\begin{itemize}
\item A branch can bend while $\theta_n(0)$ and $\theta_n(\pi)$ are
  fixed;
\item The endpoint at $k_y=0$ (or $k_y=\pi$) of a branch can shift: in
  that case, the endpoint of the other branch, the Kramers partner at
  $k_y=0$ (or $k_y=\pi$) is shifted with it;
\item Branches $\theta_n$ and $\theta_m$ can recombine: a crossing between them at some $k_y$ can turn into an avoided crossing. 
\end{itemize} 

Consider the example of Fig.~\ref{fig:bhz_bulk-wilson-loops}. Bending
of the branches and shifting of the endpoints can bring case (a) to a
trivial case, where all branches are vertical, $\theta_n(k_y)=0$ for
every $n$ and $k_y$. Case (b) can be deformed to
case (c). Notice, however, that neither cases (b) nor (c) can be
deformed to the trivial case. 

\subsection{Bulk topological invariant}

We define the bulk topological invariant $\Nbulk$, by choosing
some fixed $\tilde{\theta} \in [-\pi,\pi)$, and asking for the parity
  of the number of times the Wannier center flow crosses
  this $\tilde\theta$. In formulas,
\begin{align}
\tilde\theta &\in [-\pi,\pi);\\ N_n(\tilde\theta) &= \text{Number of
    solutions $k_y$ of } \,\,\theta_n(k_y)=\tilde\theta;\\ \Nbulk &=
  \left( \sum_{n=1}^\NF N_n(\tilde\theta) \right) \,\,\,\text{mod}\,\,
  2 \quad\quad (\text{independent of }\tilde\theta) .
\end{align}
The number $\Nbulk$ is invariant under adiabatic deformations
of the bulk Hamiltonian, as can be shown by considering the possible
changes. Bending of a branch $\theta_n$ can create or destroy
solutions of $\theta_n(k_y)=\tilde\theta$, but only pairwise. Shifting
of the endpoint can create or destroy single solutions of
$\theta_n(k_y)=\tilde\theta$, but in that case, a single solution of
$\theta_m(k_y)=\tilde\theta$, is also created/destroyed, where
$\theta_m$ is the Kramers partner of $\theta_n$ at the
endpoint. Finally, recombination of branches cannot change the number
of crossings.  The number $N(\tilde\theta)$ is also invariant under a
shift of $\tilde\theta$, as already announced. A shifting of
$\tilde\theta$ is equivalent to a shifting of the Wannier center flow,
whose effects we already considered above.

\subsubsection*{The bulk topological invariant is the $\mathbb{Z}_2$ 
invariant of the previous Chapter}


The full proof that the bulk invariant $\Nbulk$ is the same as the
parity $D$ of the number of edge state pairs, Eq.\eqref{eq:D_mod_def},
is quite involved\cite{rui_yu_equivalent,grusdt_measuring14}. We
content ourselves with just pointing out here that both
$\Nbulk$ and $D$ represent obstructions to deform the Hamiltonian
adiabatically to the so-called atomic limit, when the unit cells are
completely disconnected from each other. Clearly, switching of a
charge pump requires that there are no edge states present, and
therefore, $\Nbulk=0$ requires $D=0$. To show that the converse is
true is more complicated, and we do not discuss it here.  





\section{The $\mathbb{Z}_2$ invariant for
  systems with inversion symmetry}

For two-dimensional time-reversal invariant insulators with inversion
(i.e., parity) symmetry, the $\mathbb{Z}_2$ topological invariant
becomes very straightforward. We state the result below, and leave the
proof as an exercise for the reader.

\subsubsection*{Definition of inversion symmetry}
As introduced in
Sect.~\ref{sec:polarization-inversion}, the operation of inversion,
$\Inv$, acts on the bulk momentum-space Hamiltonian using
an operator $\Invi$, by
\begin{align}
\Inv \HH(\kk) \Inv^{-1} &= \Invi \HH(-\kk) \Invi^\dagger.  
\end{align}
We now require the operator $\Invi$ not only to be independent of the
wavenumber $\kk$, to be unitary, Hermitian, but also to commute with
time reversal, i.e.,
\begin{align}
\Inv^\dagger \Inv &= 1;&
\Inv^2 &= 1;&
\TTT \Inv &= \Inv \TTT. 
\end{align}

\subsubsection*{At a time-reversal invariant momentum, the Kramers
  pairs have the same inversion eigenvalue}

Consider the time-reversal invariant momenta (TRIM), $\Gamma_j$. In
the BHZ model, these are
$(k_x,k_y)=(0,0),(0,\pi),(\pi,0),(\pi,\pi)$. In general there are
$2^d$ such momenta in a $d$-dimensional lattice model. Each eigenstate
$\ket{u(\Gamma_j)}$ of the bulk momentum-space Hamiltonian at these
momenta has an orthogonal Kramers pair $\TTT \ket{u(\Gamma_j)}$,
\begin{align} 
\HH(\Gamma_j) \ket{u(\Gamma_j)} &= E \ket{u(\Gamma_j)}&\quad 
\Longrightarrow \quad \quad
\HH(\Gamma_j) \TTT \ket{u(\Gamma_j)} &= E \TTT \ket{u(\Gamma_j)}.
\end{align} 
If $\HH$ is inversion symmetric, $\ket{u}$ can be chosen to be an
eigenstate of $\Invi$ as well, since   
\begin{align} 
\Invi \HH(\Gamma_j) \Invi  &= \HH(-\Gamma_j) = \HH(\Gamma_j).
\end{align} 
Therefore, 
\begin{align}
  \Invi \ket{u(\Gamma_j)}  &= \pm \ket{u(\Gamma_j)}.
\end{align} 
The Kramers pair of $\ket{u(\Gamma_j)}$ has to have the same inversion
eigenvalue as $\ket{u(\Gamma_j)}$, 
\begin{align}
  \Invi \TTT\ket{u(\Gamma_j)} &= \TTT \Invi \ket{u(\Gamma_j)} = \pm
  \TTT\ket{u(\Gamma_j)}.
\end{align}

In a system with both time-reversal and inversion symmetry, we get
$2^d$ topological invariants of the bulk Hamiltonian, one for each
time-reversal invariant momentum $\Gamma_j$.  These are the products
of the parity eigenvalues $\xi_m(\Gamma_j)$ of the occupied Kramers
pairs at $\Gamma_j$. 
However, inversion symmetry is usually broken at the edges, 
and so these invariants do not give rise to robust edge states. 

The product of the inversion eigenvalues of
all occupied Kramers pairs at all the time-reversal invariant momenta
$\Gamma_j$ is the same as the $\mathbb{Z}_2$
invariant, 
\begin{align}
(-1)^\Nbulk &= \prod_j \prod_m \xi_m(\Gamma_j).
\label{eq:D_kramers}
\end{align}
We leave the proof of this useful result as an exercise for the
reader.

\subsection{Example: the BHZ model}

A concrete example for inversion symmetry is given by the BHZ model of
Sec.~\ref{sec:bhz-doubling}, with no coupling $\hat{C} = 0$. It
can be checked directly that this has inversion symmetry, with 
\begin{align}
\text{BHZ: }\quad \Invi &=\hs_0 \otimes \hsigma_z.
\end{align}

To calculate the $\mathbb{Z}_2$ invariant of the BHZ model, we take
the four time-reversal invariant momenta, $\kk_1,\kk_2,\kk_3,\kk_4$,
are the combinations of $k_x,k_y$ with $k_x=0,\pi$ and $k_y=0,\pi$.
The Hamiltonian $H_{BHZ}(k_x,k_y)$ at these momenta is proportional to
the inversion operator, 
\begin{align}
H_{BHZ}(\kk_1 = 0,0) &= (\uu + 2) \Invi;&
H_{BHZ}(\kk_4 = \pi,\pi) &= (\uu - 2) \Invi;\\
H_{BHZ}(\kk_2 = 0,\pi) &= 
\uu \Invi;&
H_{BHZ}(\kk_3 = \pi,0) &= 
\uu \Invi.
\end{align}
In these cases the Hamiltonian and the inversion operator obviously
have the same eigenstates.  At each TRIM, two of these states form one
occupied Kramers pair and the two others one empty Kramers pair. If
$\uu>2$, at all four TRIM, the occupied Kramers pair is the one with
inversion eigenvalue (parity) of $-1$, and so Eq.~\eqref{eq:D_kramers}
gives $\Nbulk=0$. Likewise, if $\uu<-2$, the eigenvalues are all $+1$,
and we again obtain $\Nbulk=0$. For $0<\uu<2$, we have $P$ eigenvalues
$-1, -1,+1,-1$ at the four TRIM $\kk_1,\kk_2,\kk_3,\kk_4$,
respectively, whereas if $-2<\uu<0$, we have $-1,+1,+1,+1$. In both
cases, Eq.~\eqref{eq:D_kramers} gives $\Nbulk=1$. This indeed is the
correct result, that we obtained via the Chern number earlier.

\section*{Problems}
\addcontentsline{toc}{section}{Problems}


\begin{prob}
\label{prob:bhz_bulk-proof}
\textbf{Inversion symmetry and interlayer coupling in the BHZ
  model}\\ 
Consider the BHZ model with layer coupling $\hat{C} = C
\hs_x \otimes \hsigma_y$. This breaks the inversion symmetry $\Invi =
\hs_0 \otimes \hsigma_z$. Nevertheless, the Wannier centers of the
Kramers pairs at $k_y=0$ and $k_y=\pi$ are stuck to $\theta=0$ or
$\theta=\pi$, and are only shifted by the extra term $\propto \hs_z
\hsigma_y$ added to the BHZ model in
Eq.~\eqref{eq:bhz_bulk-Hamiltonian_crazy}. Can you explain why? 
(hint: extra inversion symmetry)
\end{prob}

\begin{prob}
\label{prob:bhz_bulk-proof}
\textbf{Proof of the formula for the $\mathbb{Z}_2$ invariant of an
  inversion-symmetric topological insulator}\\ 
Show, using the results
of Sect.~\ref{sec:polarization-inversion}, that the $\mathbb{Z}_2$
invariant of a two-dimensional time-reversal invariant and inversion
symmetric insulator can be expressed using Eq.~\eqref{eq:D_kramers}. 
\end{prob}


%% file: Landauer-v2.tex

\chapter{Electrical conduction of edge states}
\label{chap:a_landauer}


\abstract*{Electrical conduction in clean, impurity-free
nanostructures at  low temperatures 
qualitatively deviate from the behavior of Ohmic conductors.
We demonstrate such deviations using a simple 
zero-temperature model of a clean and phase-coherent metallic wire, 
leading us toward the Landauer-B\"uttiker description
of phase-coherent electrical conduction.
We also discuss how scattering at static impurities affects
electrical conduction in general.
As the main subject of the chapter, we show how
the presence of edge states in two-dimensional topological insulators 
can have an easily measurable physical consequence: 
a nonvanishing, quantized conductance, even 
in the presence of disorder.}

It is well known that 
the electrical conduction of ordinary metallic samples at room temperature 
shows the following two characteristics.
First, there is a linear relation between the electric current $I$ 
that flows through the sample and the voltage $V$ that drops between the 
two ends of the sample: $I/V = G \equiv R^{-1}$,
where $G$ ($R$) is the conductance (resistance) of the sample.
Second, the conductances $G_i$ of different
samples made of the same metal
but with different geometries 
show the regularity $G_i L_i / A_i = \sigma$ for $\forall i$,
where $L_i$ is the length of the sample and $A_i$ is the 
area of its cross section.  
The material-specific quantity $\sigma$ is called the conductivity.
Conductors obeying both of these relations are 
referred to as Ohmic.

Microscopic theories describing the above behavior 
(e.g., Drude model, Boltzmann equation)
rely on models involving impurities, lattice vibrations,
and electron scattering within
the material. 
Electrical conduction in clean (impurity-free)
nanostructures at  low temperature 
might therefore qualitatively deviate from the Ohmic case.
Here, we demonstrate such deviations on a simple zero-temperature 
model 
of a two-dimensional, 
perfectly clean, constant-cross-section
metallic wire, depicted in Fig. \ref{fig:a_cleanwire}a.
Then 
we describe how scattering at
static impurities affects the conduction in general.
Finally, as a central result in the field of topological insulators, 
we point out that electrical conduction via the edge states
of two-dimensional topological insulators shows a 
strong robustness against such impurities.

\section{Electrical conduction in a clean quantum wire}
\label{sec:a_quantumwire}

\begin{figure}[!ht]
\begin{center}
\includegraphics[width=0.9\linewidth]{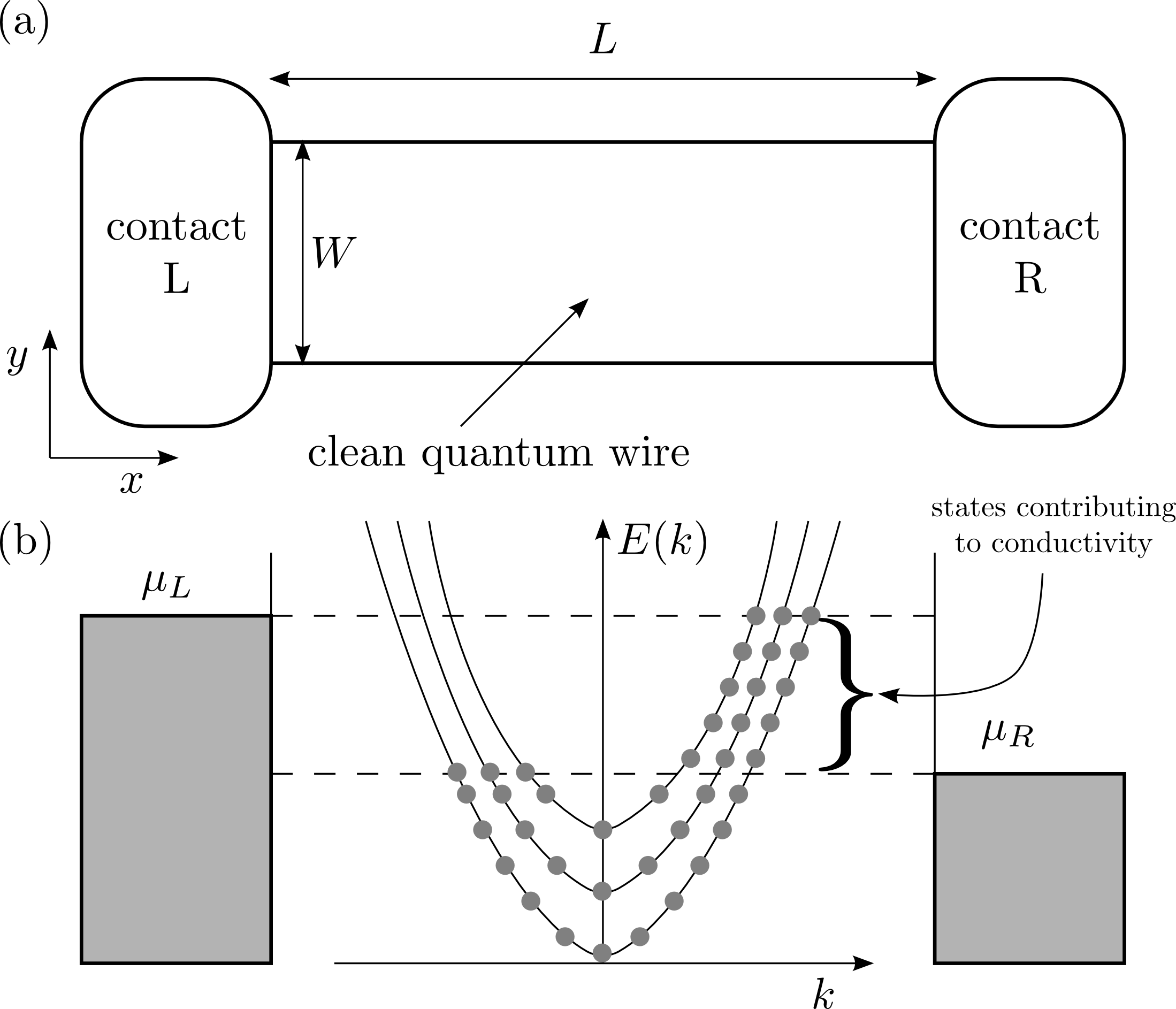}
\caption{\label{fig:a_cleanwire}
(a) Schematic representation of a clean quantum wire contacted to two
electron reservoirs (contacts).
(b) Occupations of electronic states in the contacts and the 
quantum wire in the nonequilibrium situation 
when a finite voltage $V$ is applied between the left and
right reservoirs.}
\end{center}
\end{figure}

As shown in Fig. \ref{fig:a_cleanwire}a, take a wire that lies along
the $x$ axis and has a finite width in the $y$ direction.  The latter
might be defined by an electric confinement potential or the
termination of the crystal lattice.  Each electronic energy
eigenfunction $\ket{\chl,k}$ in such a wire is a product of a standing
wave along $y$, labeled by a positive integer $\chl$, and a plane wave
propagating along $x$, labeled by a real wave number $k$ (see
Eq. \eqref{eq:a_propagatingwave}).  A typical set of dispersion
relations $E_{\chl k}$ (`subbands') for three different $\chl $
indices is shown in Fig. \ref{fig:a_cleanwire}b.

We also make assumptions on the two metallic contacts that serve as
source and drain of electrons.  We assume that the electrons in each
contact are in thermal equilibrium, but the Fermi energies in the
contacts differ by $\mu_L - \mu_R = |e|V>0$.  (Note that in this
chapter, proper physical units are used, hence constants such as the
elementary charge $|e|$, reduced Planck's constant $\hbar$, lattice
constant $a$ are reinstated.)  We consider the \emph{linear
  conductance}, that is, the case of an infinitesimal voltage $|e| V
\to 0$.  We further assume that both contacts absorb every incident
electron with unit probability, and that the energy distribution of
the electrons they emit is the thermal distribution with the
respective Fermi energy.

These assumptions guarantee that the right-moving (left-moving)
electronic states in the wire are occupied according to the thermal
distribution of the left (right) contact, as illustrated in
Fig. \ref{fig:a_cleanwire}b.  Now, we work with electron states
normalized to the area of the channel.  It is a simple fact that with
this normalization convention, a single occupied state in the $\chl$th
channel, with wave number $k$ carries an electric current of
$\frac{-|e| v_{\chl k}}{N_x a}$, where $N_x a$ is the length of the
wire, and $v_{\chl k} = \frac{1}{\hbar} \frac{dE_{\chl k}}{dk}$ is the
group velocity of the considered state.  Therefore, the current
flowing through the wire is
\begin{equation}
I = -|e| \frac 1 {N_x a} \sum_{\chl k} v_{\chl k} \left[
	f(E_{\chl k} - \mu_L) - f(E_{\chl k} - \mu_R)
\right],
\end{equation}
where $f(\epsilon) = \left(\exp\frac{\epsilon}{k_{\rm B} T} + 1\right)^{-1}$
is the Fermi-Dirac distribution. 
Converting the $k$ sum to an integral via
$\frac 1 {N_x a} \sum_k \dots \mapsto \int_{-\pi/a}^{\pi/a} 
\frac{dk}{2\pi} \dots$
yields
\begin{equation}
I =
- |e| \sum_n \int_{-\pi/a}^{\pi/a} \frac{dk}{2\pi} \frac 1 \hbar \frac{dE_{\chl k}}{dk}
\left[
	f(E_{\chl k} - \mu_L) - f(E_{\chl k} - \mu_R)
\right].
\end{equation}
The Fermi-Dirac distribution has a sharp edge at zero temperature, 
implying
\begin{equation}
\label{eq:a_currentclean}
I = 
- \frac{|e|}{h} M \int_{\mu_R}^{\mu_L} d E
\\
=
- \frac{|e|}{h} (\mu_L - \mu_R) M
\\
=
M \frac{e^2}{h} V
\end{equation}
Note that the first equality in
\eqref{eq:a_currentclean} 
holds only if the number of subbands intersected by $\mu_L$ and $\mu_R$ 
are the same,
which is  indeed the case if the voltage $V$ is small enough.
The number of these subbands, also called
`open channels', is denoted by the integer $M$.
From \eqref{eq:a_currentclean} it follows that the
conductance of the wire is 
an integer multiple of $e^2/h$ 
(commonly referred to as `quantized conductance'):
\begin{equation}
\label{eq:a_conductanceclean}
G = \frac{e^2}{h}M.
\end{equation}
The numerical value of $e^2/h$ is
approximately 40 $\mu$S (microsiemens), which corresponds to 
a resistance of approximately 26 k$\Omega$.
Note the the conductance quantum is defined as $G_0 = 2e^2/h$,
i.e., as the conductance of a single open channel with twofold spin 
degeneracy.

It is instructive to compare the conduction in 
 our clean quantum wire to the ordinary
 Ohmic conduction summarized above. 
According to \eqref{eq:a_currentclean},
the proportionality between voltage and current holds for
a clean quantum wire as well as for an ordinary metal.
However, the dependence of the conductance on the length of 
the sample differs qualitatively in the two cases:
in an ordinary metal, a twofold increase in the length of the wire
halves the conductance,
whereas the conductance of a clean quantum wire is insensitive to
length variations. 

Whether the conductance of the clean quantum wire is sensitive
to variations of the wire width depends on the nature of 
the transversal modes. 
Conventional quantum wires that are created by a transverse confinement
potential have a subband dispersion similar to that in 
Fig. \ref{fig:a_cleanwire}b.
There, the energy separation between the subbands decreases
as the width of the wire is increased, therefore
the number of subbands available for conduction increases.
This leads to an increased conductance for an increased width,
similarly to the case of ordinary metals.
If, however, we consider a topological insulator, 
where the current is carried by  states localized to the edges of the
wire, the conductance of the wire will be insensitive 
to the width of the wire.

\section{Phase-coherent electrical conduction in the presence
of scatterers}
\label{sec:a_scatterers}

\begin{figure}[!ht]
\begin{center}
\includegraphics[width=0.9\linewidth]{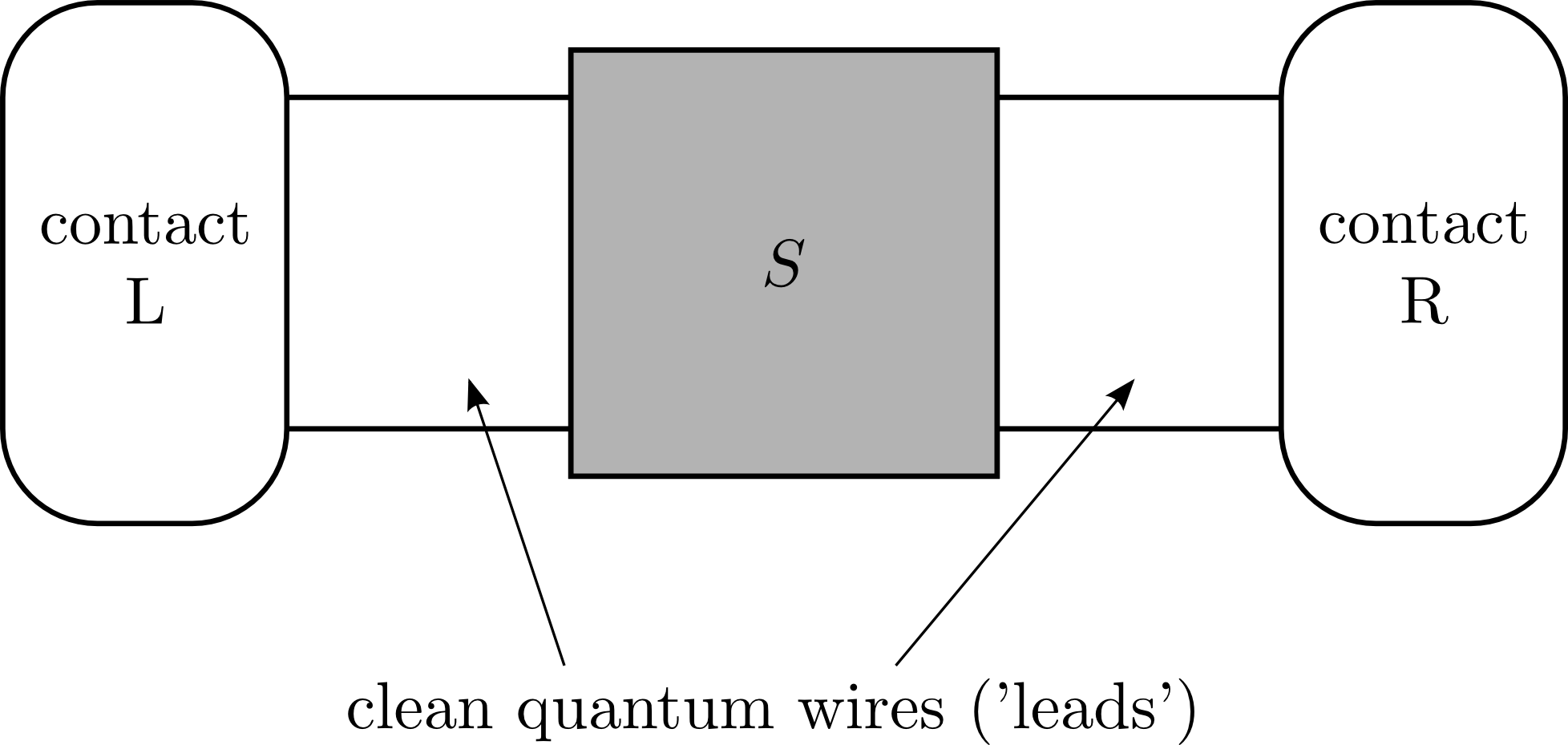}
\caption{\label{fig:a_landauer}
Simple model of a phase-coherent conductor in the presence of 
scatterers. The ideal contacts L and R are connected 
via ideal leads
to the disordered
region represented by the scattering matrix $S$.}
\end{center}
\end{figure}

Having calculated the conductance 
\eqref{eq:a_conductanceclean}
of a clean quantum wire, 
we now describe how this conductance is changed by the presence
of impurities. 
We analyze the model shown in Fig. \ref{fig:a_landauer},
where the disordered region, described by a scattering matrix $S$,
is connected to the two contacts by two identical clean quantum wires,
also called `leads' in this context. 

First, we consider the case when each lead supports a single
open channel.
The current in the lead connecting contact $L$ and
the scattering region consists of a contribution from 
right-moving states arriving from contact $L$ and partially backscattered
with probability $R = |r|^2$, and 
from left-moving states arriving from contact $R$ and partially
transmitted with probability $T'=|t'|^2$:
\begin{equation}
I = -|e| \frac 1 {N_x a} \sum_k v_k \left[
	\left(1- R(E_k)\right) f_L (E_k) - T'(E_k) f_R(E_k)
\right]
\end{equation}
Converting the $k$ sum to an integral, assuming that the
transmission and reflection probabilities are independent of energy in 
the small energy window between $\mu_R$ and $\mu_L$, 
and using $1-R = T = T'$, we arrive at
\begin{equation}
I=
- \frac{|e|}{h} T \int_{\mu_R}^{\mu_L} dE 
\left[ f_L(E) - f_R(E) \right]
=
\frac{e^2}{h} T V,
\end{equation}
which implies that the conductance can be expressed through
the transmission
coefficient $T$:
\begin{equation}
\label{eq:a_singlechannel}
G=\frac{e^2}{h} T.
\end{equation}

The result \eqref{eq:a_singlechannel} can be straightforwardly
generalized to the case when the leads support more than
one open channel. 
The generalized result for the conductance, 
also known as the Landauer formula, 
reads:
\begin{equation}
\label{eq:a_landauer}
G=\frac{e^2}{h} \sum_{n=1}^M T_n,
\end{equation}
where $T_n$ are the transmission eigenvalues of the scattering matrix
i.e., the real eigenvalues of the Hermitian matrix $tt^\dag$,
as defined in the preceding chapter.

\section{Electrical conduction in two-dimensional topological insulators}
\label{sec:a_conduction2DTI}

After presenting the Landauer formula as a generic tool to describe  
electrical conduction of a phase-coherent metal, 
we will use it know to characterize the conductances 
of various two-dimensional topological insulator samples.

\subsection{Chern Insulators}


In Sect.~\ref{sec:qiwuzhang_edgestates}, 
we have seen that an impurity-free straight strip 
of a topologically nontrivial Chern Insulator supports edge states.
The relation between the Chern number $Q$ of the
Chern Insulator and the 
numbers of edge states at a single edge at a given energy $E$,
propagating `clockwise' ($N_+(E)$) and
`anticlockwise' ($N_-(E)$), is $Q = N_+(E) - N_-(E)$.
In addition, in Sect.~\ref{sect:qiwuzhang-edge_robustness}
it was shown that any
segment of the edge of a disordered
Chern Insulator with Chern number $Q$ and an arbitrary
geometry supports $|Q|$ chiral edge modes.
Here we show that existence of these edge modes
leads to experimentally detectable
effects in the electrical transport through Chern Insulator
samples.

\begin{figure}
\sidecaption
\includegraphics[width=0.6\linewidth]
{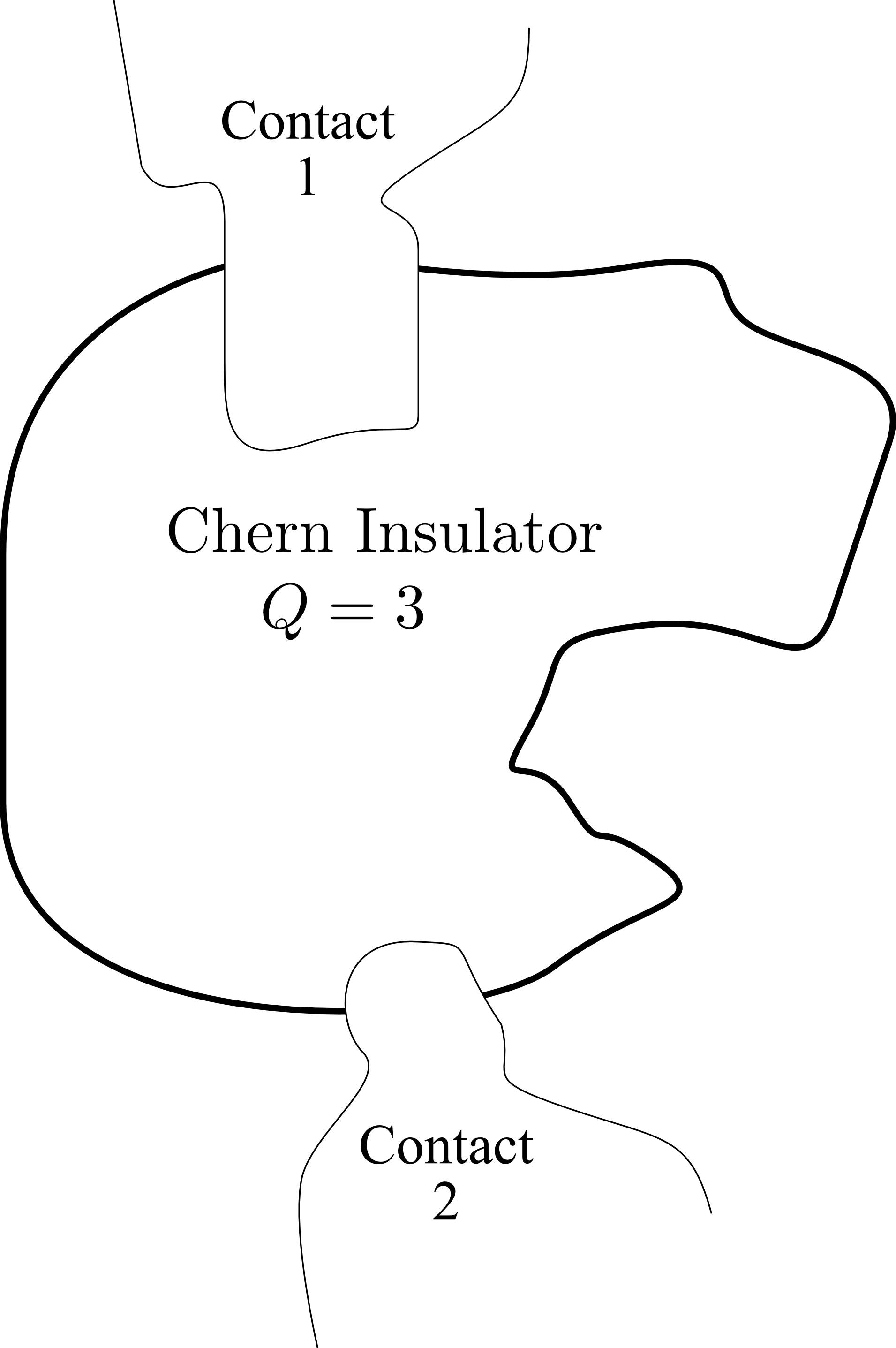}
\caption{A disordered sample of Chern insulator, with contacts 1 and
  2, that can be used to pass current through the sample in order to
  detect edge states. 
 \label{fig:a_chern_2contacts}}
\end{figure}

We consider a transport setup where the Chern Insulator is 
contacted with two metallic electrodes, as shown in 
Fig. \ref{fig:a_chern_2contacts}.
In this discussion, we rely on the 
usual assumptions behind the Landauer formula:
phase-coherence of the 
electrons, good contact between contacts and sample, 
and large spatial separation of the two electrodes
ensuring the absence of tunneling contributions to the
conductance.

In the following list, we summarize how the phase-coherent 
electrical conductance
of a Chern Insulator varies with the sample geometry, 
absence or presence of disorder, and the value of the 
electronic Fermi energy. 
\begin{enumerate}
\item 
	Disorder-free sample with a strip geometry 
	(see Fig.~\ref{fig:a_cleanwire}a)
	\begin{enumerate}
	\item \emph{Fermi energy lies in a band.} 
		In this case, the sample is a clean quantum wire 
		(see Sect.~\ref{sec:a_quantumwire})
		with an integer number of open channels.
		The corresponding 
		transversal wave functions might or might not be localized
		to the sample edges, and therefore 
		the number of channels might be different from
		any combination of 
		$Q$, $N_+$ or $N_-$.
		According to Eq. \ref{eq:a_conductanceclean},
		the conductance of such a clean quantum wire is
		quantized and insensitive to the length of the sample. 
		Furthermore, the conductance grows in a step-like fashion
		if the width of the sample is increased. 
	\item \emph{Fermi energy lies in the gap.}
		The sample is a clean quantum wire with open channels
		that are all localized to the sample edges. 
		The number of those channels is $N_+(E) + N_-(E)$,
		where $E$ is the Fermi energy:
		in each of the two possible direction of current flow, 
		there are $N_+$ channels on one edge and $N_-$ on
		the other edge that contribute to conduction.
		Conductance is finite and quantized, a behavior 
		rather unexpected from an insulator. 
		The conduction is not Ohmic, as the conductance is
		insensitive to both the length and the width of the sample.
	\end{enumerate}
\item
	Disordered sample with an irregular shape 
	(see Fig.~\ref{fig:a_chern_2contacts}):
	\begin{enumerate}
	\item \emph{Fermi energy lies in a band.} 
		Because of the presence of disorder, 
		the electrical conduction of such a sample might be Ohmic.
		There are no protected edge states at the Fermi energy.
	\item \emph{Fermi energy lies well within the gap.}
		According to Sect.~\ref{sect:qiwuzhang-edge_robustness},
		any edge segment of such a sample
		supports $Q$ reflectionless chiral edge modes
		at the Fermi energy.
		Therefore, conductance is typically quantized,
		$G= |Q| e^2/h$,
		although, disorder permitting,
		it might in principle be larger than this value. 
		The quantized conductance is insensitive to changes in 
		the geometry or the disorder configuration. 
		This transport property, unexpected for an insulator, 
		let alone for one with disorder, 
		is a hallmark of Chern Insulators.
	\end{enumerate}
\end{enumerate}
In the case of two-dimensional samples there is often an experimental 
		possibility of tuning the electronic Fermi energy \emph{in situ} by
		controlling the voltage applied between the sample and a 
		nearby metallic plate (\emph{gate electrode}), 
		as discussed in Sect.~\ref{sec:hgte}.
		This allows, in principle, to observe the changes in the 
		electrical conduction of the sample as the
		Fermi energy is tuned across the gap.

\subsection{Two-dimensional time-reversal invariant topological insulators with $\TTT^2=-1$}

In the following list, we summarize the predictions of the Landauer
formalism for the conductance of 
two-dimensional time-reversal invariant topological insulators with $\TTT^2=-1$
(`$D=1$ insulators' for short).
\begin{enumerate}
\item Disorder-free sample with a strip geometry:
	\begin{enumerate}
	\item \emph{Fermi energy lies in a band.}
	A simple consequence of the Landauer formula is that 
	phase-coherent conductance of an impurity-free $D=1$
	topological insulator of the strip geometry shown in
	Fig. \ref{fig:a_cleanwire}a is quantized.
	The conductance grows if the width of the strip is increased,
	but insensitive to change in the length. 
	\item \emph{Fermi energy lies in the gap.}
	Only edge channels are open in this case. 
	These also provide conductance quantization. 
	As the number of edge-state Kramers pair per edge is odd, 
	the conductance might be $2e^2/h$, $6e^2/h$, $10e^2/h$, etc.
	Conductance is insensitive to width or length changes of the sample.
	\end{enumerate}
\item Disordered sample with an irregular shape and TRS disorder:
	\begin{enumerate}
	\item \emph{Fermi energy lies in a band.}
	The electrical conduction might be Ohmic. 
	\item \emph{Fermi energy lies in the gap.}  We have shown in
          Chapt.~\ref{chap:BHZ} that a $D=1$ insulator supports one
          protected edge-state Kramers pair per edge, which allows for
          reflectionless electronic transmission if only TRS defects
          are present.  The Landauer formula \eqref{eq:a_landauer}
          implies, for typical cases, $G = 2e^2/h$ for such a sample,
          as one edge state per edge contributes to conduction.  The
          conductance might also be larger, provided that the number
          of edge-state Kramers pairs is larger than 1 and disorder is
          ineffective in reducing the transmission of the
          topologically unprotected pairs.
	\end{enumerate}
\end{enumerate}

We note that in real materials with $\mathbb{Z}_2$ invariant $D=1$,
various mechanisms might lead to backscattering
and, in turn, to $G < 2e^2/h$.
Examples include TRS-breaking impurities, 
TRS impurities that bridge the spatial distance  
between the edges (see Chapt.~\ref{chap:BHZ}),
hybridization of edge states from opposite edges
in narrow samples, and inelastic scattering on phonons or
spinful impurities.

\section{An experiment with HgTe quantum wells}
\label{sec:hgte}

Electrical transport measurements \cite{Konig-qshe}
on appropriately designed layers of the  
semiconductor material mercury-telluride (HgTe) show signatures of 
edge-state conduction in the absence of magnetic field. 
These measurements are in line with the theoretical prediction that 
a HgTe layer with a carefully chosen thickness can realize 
a topologically nontrivial ($D=1$) two-dimensional time-reversal invariant insulator with $\TTT^2=-1$. 
In this section, we outline the main findings of this
experiment, as well as  
its relation to the BHZ model introduced and discussed in chapter
\ref{chap:BHZ}. 

\begin{figure}[!ht]
\begin{center}
\includegraphics[width=0.9\linewidth]{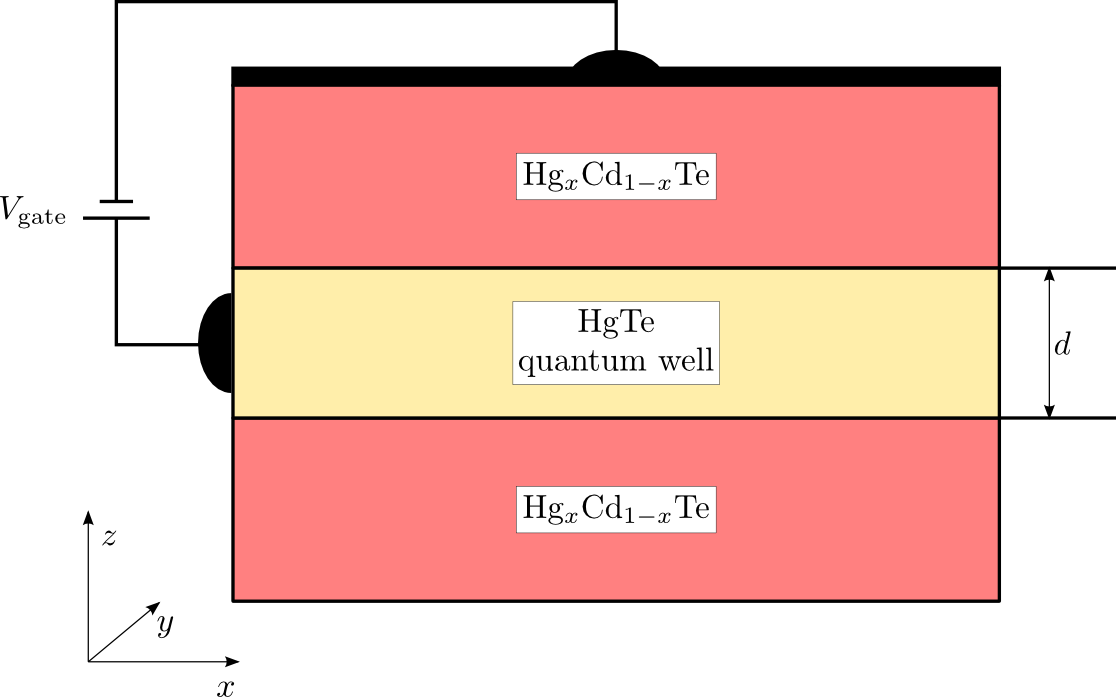}
\caption{
	\label{fig:a_HgTe-qwell}
	Schematic representation of a HgTe quantum well of width $d$, 
	sandwiched between two Hg$_x$Cd$_{1-x}$Te layers.
	Electrons are confined to the HgTe layer, and their Fermi energy can
	be tuned \emph{in situ} by adjusting the voltage $V_{\rm gate}$
	of the metallic electrode on the top of the sample (black).
	For a more accurate description of the experimental arrangement,
	see \cite{Konig-qshe-review}.}
\end{center}
\end{figure}

\begin{figure}[!ht]
\begin{center}
\includegraphics[width=0.9\linewidth]{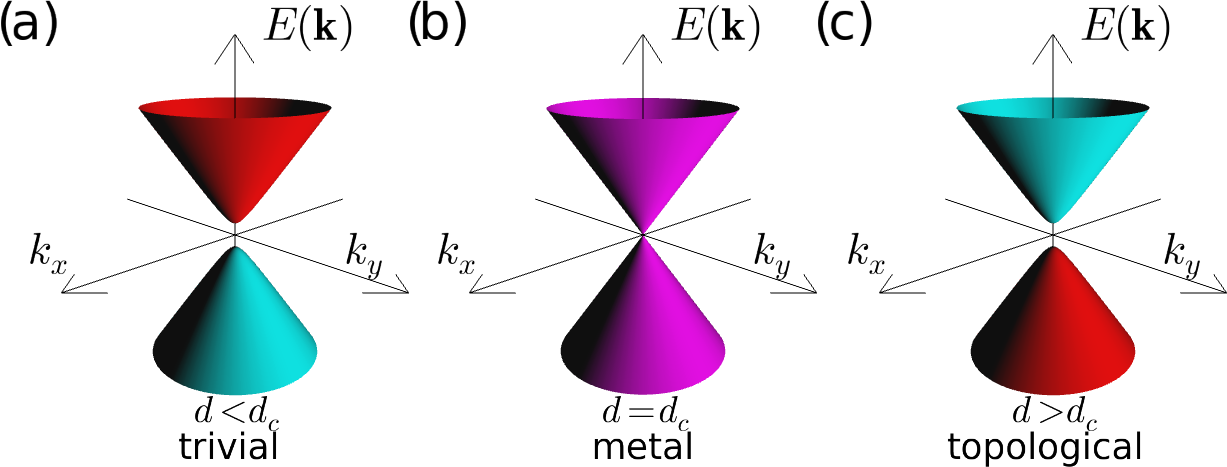}
\caption{\label{fig:a_BHZ-fig2}
Evolution of the two-dimensional band structure of a HgTe
quantum well as a function of its thickness $d$.
(a) For a thin quantum well below the critical thickness $d<d_c$, 
the band structure has a gap and the system
is a trivial insulator.
(b) At a critical thickness $d=d_c$, the band gap closes and
the system is metallic. 
(c) For a thick quantum well with $d>d_c$, 
the band gap reopens and the 
system becomes a two-dimensional topological insulator. }
\end{center}
\end{figure}

The experiments are performed on sandwich-like structures
formed by a few-nanometer thick HgTe layer (\emph{quantum well})
embedded 
between two similar layers of the alloy Hg$_x$Cd$_{1-x}$Te,
as shown in Fig.~\ref{fig:a_HgTe-qwell}.
(In the experiment reported in \cite{Konig-qshe}, 
the alloy composition $x=0.3$ was used.)
In this structure, the electronic states with energies close the 
Fermi energy are confined to the HgTe layer that is parallel
to the $x$-$y$ plane in Fig.~\ref{fig:a_HgTe-qwell}.
The energy corresponding to the confinement direction
$z$ is quantized.
The carriers are free to move along the HgTe layer, i.e., 
parallel to the $x$-$y$ plane, therefore two-dimensional subbands are 
formed in the HgTe quantum well.
Detailed band-structure calculations
of Ref.~\cite{Bernevig-qshe} show that as the 
the thickness $d$ of the HgTe layer is decreased, the
lowermost conduction subband and the uppermost 
valence subband touch at a critical thickness $d=d_c$,
and the gap is reopened for even thinner HgTe layers. 
(For the alloy composition $x=0.3$ used in the experiment, the 
critical thickness is $d_c \approx 6.35$ nm.)
This behavior is illustrated schematically
in Fig.~\ref{fig:a_BHZ-fig2}, 
which illustrates the electronic dispersions
of the uppermost valence subband and the
the lowermost conduction subband,
at the center of the BZ, 
for three different thicknesses of the quantum well.

Band-structure calculations have also revealed 
a  connection between the subbands 
depicted in Fig.~\ref{fig:a_BHZ-fig2}
and the BHZ model introduced and discussed 
in Chapt.~\ref{chap:BHZ}.
The $4\times 4$ effective Hamiltonian describing the
two spinful two-dimensional subbands around their extremum point at the centre of the
HgTe Brillouin zone resembles the low-energy continuum Hamiltonian
derived from the BHZ lattice model in the vicinity of the
$u \approx -2$ value.
Changing the thickness $d$ of the HgTe layer corresponds to
a change in the parameter $u$ of the BHZ model,
and the critical thickness $d = d_c$ corresponds to $u = -2$ and,
consequently, a zero mass in the corresponding two-dimensional Dirac equation.

As a consequence of the strong analogy of the band structure of the
HgTe quantum well and that of the BHZ model,
it is expected that either for $d<d_c$ or for $d>d_c$ the material is
a $D=1$ insulator with a single Kramers pair of edge states.
Arguments presented in \cite{Bernevig-qshe} suggest that the
thick quantum wells with $d>d_c$ are topologically nontrivial.

\begin{figure}[!ht]
\begin{center}
\includegraphics[width=0.9\linewidth]{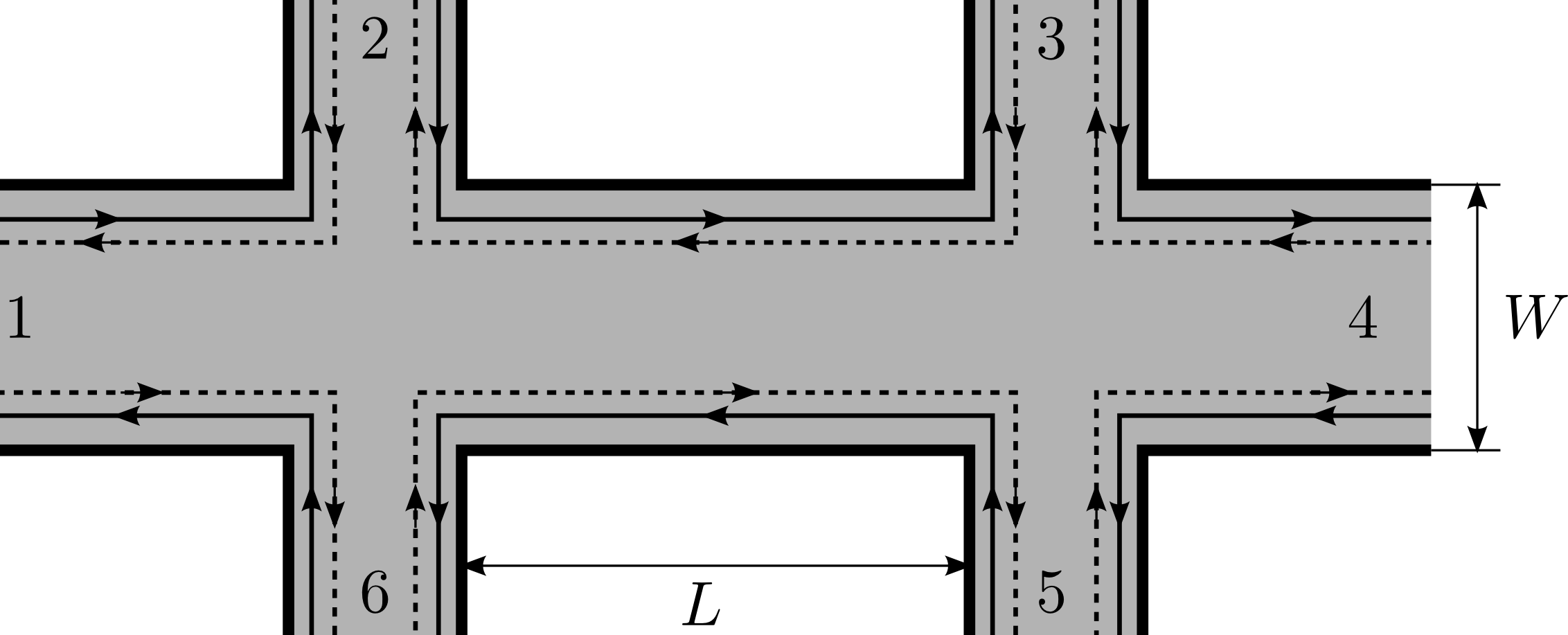}
\caption{\label{fig:a_TRI_insulator_Hallbar}
HgTe quantum well patterned in the Hall-bar geometry (gray area).
Numbered terminals lead to metallic contacts.
Solid and dashed lines depict counterpropagating edge states.}
\end{center}
\end{figure}

Electrical transport measurements were carried out in 
HgTe quantum wells patterned in the Hall bar geometry
shown in Fig.~\ref{fig:a_TRI_insulator_Hallbar}.
The quantity that has been used in this experiment
to reveal edge-state transport is the 
four-terminal resistance $R_{14,23} = V_{23}/I_{14}$, where
$V_{23}$ is the voltage between contacts 2 and 3, and
$I_{14}$ is the current flowing between contacts 1 and 4.
This quantity $R_{14,23}$ was measured for various devices with
different thicknesses $d$, below and above the critical thickness $d_c$,
of the HgTe layer, and for different values of the Fermi energy.
The latter can be tuned \emph{in situ} by controlling the voltage between the
HgTe layer and a metallic `gate' electrode on the top of the layered 
semiconductor structure, as shown in Fig.~\ref{fig:a_HgTe-qwell}.

To appreciate the experimental result, let us first derive 
the four-terminal resistance $R_{14,23}$ for such a
device. 
To this end, we express $V_{23}$ with $I_{14}$.
Ohm's law implies $V_{23} = I_{23}/G_{23}$, 
where $I_{23}$ is the current flowing through the edge segment between
contacts 2 and 3, whereas $G_{23}$ is the conductance of that
edge segment.
Furthermore, as the current $I_{14}$ flowing through 
terminals 1 and 4 is equally divided between the upper and lower edges,
the relation $I_{23} = I_{14}/2$ holds, implying the result
$R_{14,23} = 1/(2G_{23})$. 

If the Fermi energy lies in the bands neighboring the gap, then 
irrespective of the topological invariant of the system, 
the HgTe quantum well 
behaves as a good conductor with $G_{23} \gg e^2/h$, 
implying $R_{14,23} \ll h/e^2$.
If the Fermi energy is tuned to the gap in the topologically
nontrivial case $d>d_c$, then 
$G_{23} = e^2/h$ and therefore $R_{14,23} = h/(2e^2)$.
This holds, of course, only at a temperature low enough and a
sample size small enough such that phase coherence is 
guaranteed. 
The presence of static time-reversal invariant defects is included.  
If the system is topologically trivial ($d<d_c$), then
there is no edge transport, and the quantum well is a good insulator with
 $R_{14,23} \gg h/e^2$.

The findings of the experiments are consistent with the above
expectations. 
Furthermore, the four-terminal resistance of topologically nontrivial
HgTe layers with different widths were measured, with 
the resistance found to be an approximately constant function 
of the width $W$ of the Hall bar.
This is a further indication that the current in these samples
is carried by edge states. 

To wrap up this chapter, we note that 
InAs/GaSb bilayer quantum wells are an alternative semiconductor
material system where two-dimensional topological insulators can be realized
\cite{ChaoXingLiu-typeII,LingjieDu}. 
Graphene is believed to be a two-dimensional topological insulator as well
\cite{Kane-Z2},
even though its energy gap between the valence and 
conduction band, induced by spin-orbit interaction and
estimated to be of the order of $\mu$eV, 
seems to be too small to allow for the detection of 
edge-state transport even at the lowest available temperatures.
The concept of a time-reversal invariant topological insulator can be extended to 
3D crystals as well, where the role of the edge states is played
by states localized to the two-dimensional surface of the 3D material. 
The description of such systems is out of the scope of the present course;
the interested reader might consult, e.g., 
Refs.~\cite{Hasan-topinsreview,Bernevig-book}.

%
